\documentclass[	DIV=calc,%
							paper=a4,%
chapterprefix,
							fontsize=12pt,%
twoside, openright
]{scrbook}

\usepackage{lipsum}								
\usepackage[english]{babel}							
\usepackage[protrusion=true,expansion=true]{microtype}	
\usepackage{amsmath,amsfonts,amsthm}				
\usepackage[pdftex]{graphicx}						
\usepackage[svgnames]{xcolor}					
\usepackage[hang, small,labelfont=bf,up,textfont=it,up]{caption}	
\usepackage{epstopdf}							
\usepackage{subfigure}							
\usepackage{booktabs}							
\usepackage{fix-cm}	
\usepackage[square,numbers,comma,sort&compress]{natbib}
\usepackage{slashbox}
\usepackage{amssymb}
\usepackage{rotating}
\usepackage{multirow}
\usepackage{paralist}
\usepackage{hyperref}
\usepackage{sectsty}
\usepackage{geometry}
\usepackage{setspace}
\usepackage{titlesec}
\usepackage{fancyhdr}
\newcommand{\PublicationEntry}[3]{
		\noindent \textbf{#1} \\					
		\noindent #2, 					
		\noindent \textit{#3}. 	
		\normalsize \\}

\fancypagestyle{plain}{ 
     \fancyhead{} 
      
     } 
 
\fancypagestyle{chapter}{ 
     \fancyhead{}

     \fancyfoot[LE, RO]{\footnotesize  \thepage}} 

\fancypagestyle{carlos}{ 
     \fancyhead{}

     \fancyfoot[LE, RO]{\footnotesize  \thepage}} 

\usepackage[Glenn]{fncychap} 
 
\ChNameVar{\color{Olive} \Large} 
\ChNumVar{\Huge} 
\ChTitleVar{\color{Olive} \Huge} 
\ChRuleWidth{0.5pt} 
\ChNameUpperCase

\usepackage{lettrine}
\newcommand{\initial}[1]{%
     \lettrine[lines=3,lhang=0.3,nindent=0em]{
     				\color{Olive}
     				{\textsf{#1}}}{}}

\newcommand{\ben}{\begin{eqnarray}}
\newcommand{\een}{\end{eqnarray}}

\newcommand{\be}{\begin{equation}}
\newcommand{\ee}{\end{equation}}
\newcommand{\bea}{\begin{eqnarray}} 
\newcommand{\eea}{\end{eqnarray}}

\usepackage{dsfont}
\begin{document}

\begin{titlepage}
\thispagestyle{empty} 
\begin{figure}[h]
{\centering
{\includegraphics[width=0.45\textwidth]{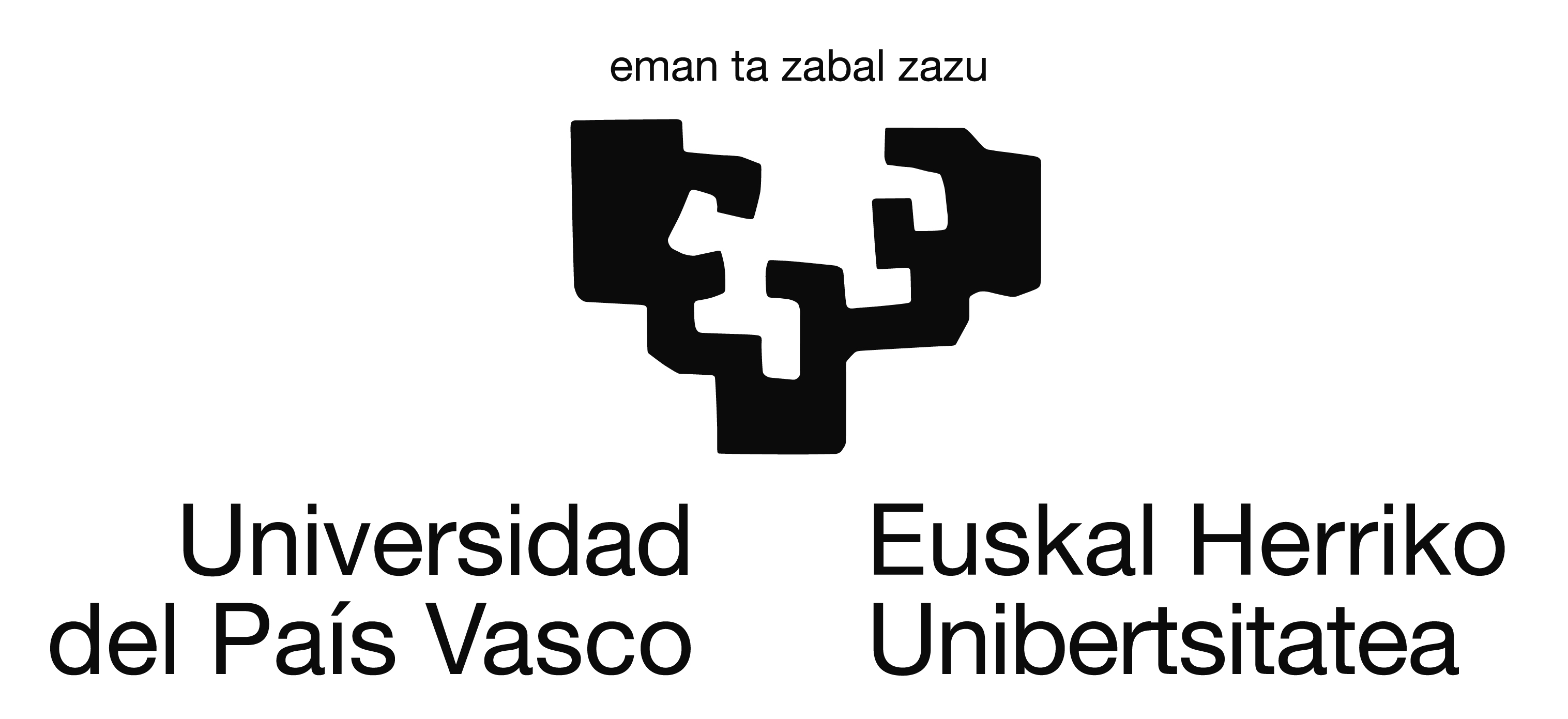}}\par}
\end{figure}
\begin{center}
{\LARGE \bf On Supermembrane in D=4, multiple M0--brane in D=11 and Supersymmetric Higher Spin Theories}\\
\vspace{2.5cm}
{\LARGE \bf{Carlos Meliveo Garc\'ia}}\\
\vspace{10.00cm}
Ph.D. Thesis\\
Department of Theoretical Physics and History of Science\\
University of the Basque Country (UPV/EHU)\\
Leioa, December 2013\\
\end{center} 
\end{titlepage}

\begin{titlepage}
\thispagestyle{empty} 
 \pagebreak 
\end{titlepage}

\begin{titlepage}
\thispagestyle{empty} 
\begin{figure}[h]
{\centering
{\includegraphics[width=0.45\textwidth]{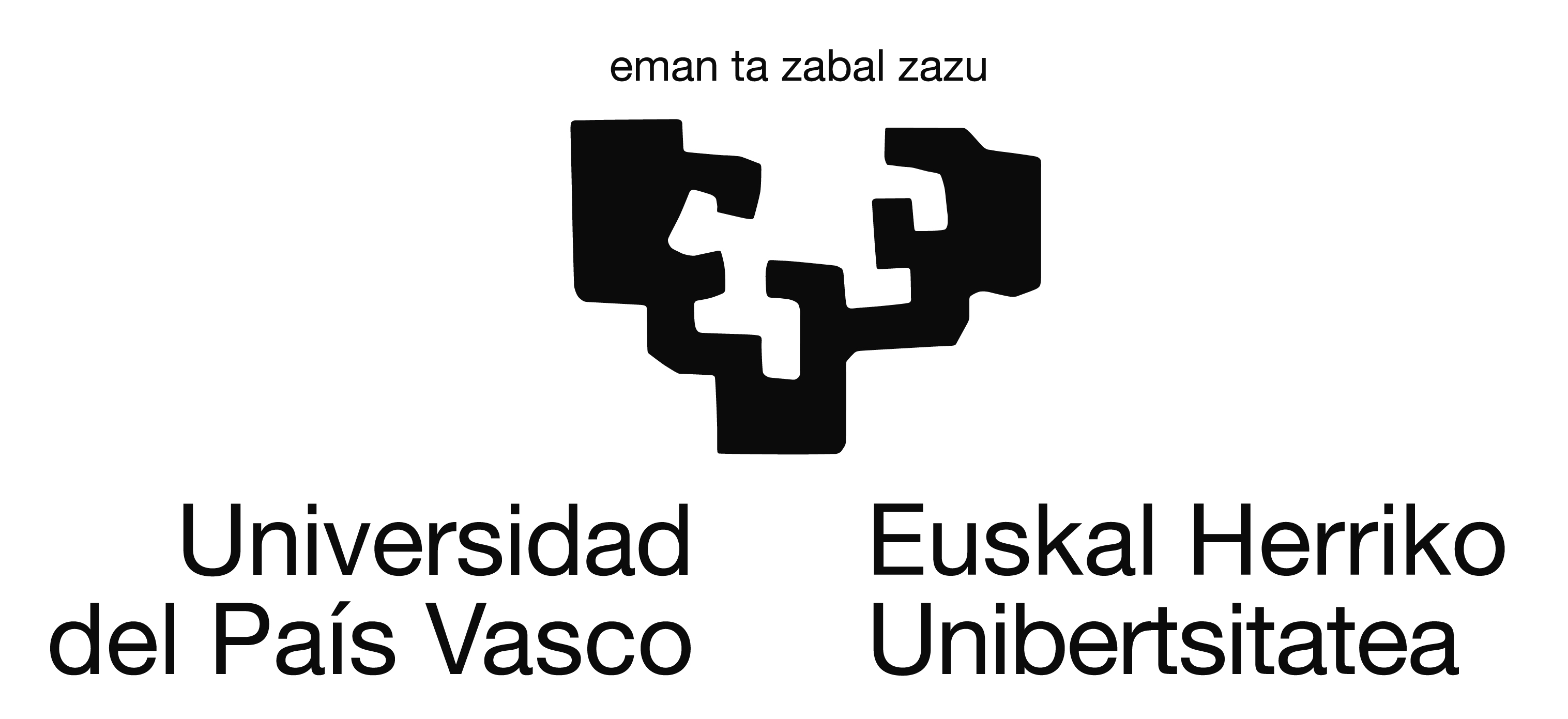}}\par}
\end{figure}

\begin{center}
Department of Theoretical Physics and History of Science\\ 
\vspace{1.5cm} 
\vspace{1.75cm}
{\LARGE \bf On Supermembrane in D=4, multiple M0--brane in D=11 and Supersymmetric Higher Spin Theories}\\
\vspace{1.5cm} 
\vspace{1.75cm}
\end{center}


\vspace{1.75cm}
\begin{center}
\large{Supervised by Professor Igor A. Bandos\\
from the University of the Basque Country and Ikerbasque Foundation\\}
\end{center}

\vspace{3.75cm}
\begin{flushright}
\large{ Submitted by Carlos Meliveo Garc\'ia\\
for the degree of Doctor of Physics}
\end{flushright}

\end{titlepage}

\begin{titlepage}
\thispagestyle{empty} 
 \pagebreak 
\end{titlepage}

\pagestyle{empty}
\begin{flushright}
\vspace*{1cm} \large\emph{{A mi familia.}}
\end{flushright}


%
%


\begin{titlepage}
\thispagestyle{empty} 
 \pagebreak 
\end{titlepage}

\pagestyle{fancy}
\pagenumbering{roman}
\setcounter{page}{1} 
\addcontentsline{toc}{section}{Acknowledgements}
\chapter*{Acknowledgements}
\thispagestyle{chapter} 
\initial{T}his thesis is based on the research carried out at the Department of Theoretical Physics and History of Science at the 
University of the Basque Country in the period 2009-2013.

During these years, I have had the chance to meet many people who have helped me in many different
ways towards the completion of this thesis. For that reason, I would like to express my deepest gratitude to each of them.

First of all, I  want to thank the person who has really made this thesis possible, my supervisor \textbf{Igor A. Bandos}.  
Thanks Igor for the patience and support during these years.  
 
I thank all members of the Department for their support and pleasant working atmosphere: \textbf{Juan M. Aguirregabiria}, 
\textbf{Montserrat Barrio}, \textbf{Jos\'e Juan Blanco}, \textbf{Mariam Bouhmadi}, \textbf{David Brizuela}, \textbf{Tom Broadhurst}, \textbf{Alberto Chamorro}, 
\textbf{Alexander Feinstein}, \textbf{Luis Herrera}, \textbf{Jes\'us Ib\'anez}, \textbf{Ruth Lazcoz}, \textbf{Michele Modugno}, \textbf{Mart\'in Rivas}, \textbf{Gunar Schnell}, 
\textbf{Vincenzo Salzano}, \textbf{Jos\'e M. Mart\'in Senovilla}, \textbf{G\'eza T\'oth}, \textbf{Jon Urrestilla}, 
\textbf{Manuel \'A. Valle}, \textbf{Lianao Wu} and \textbf{Zolt\'an Zimbor\'as}. 
Specially, I would like to thank 
\textbf{Ra\"{u}l Vera} because without him my computer never worked properly and 
I thank \textbf{I\~nigo L. Egusquiza} for encouraging me during the hard times.

On the other hand, I would like to express my gratitude to the Ph.D. students and Postdocs who have accompanied me 
along this thesis. 
I would like to mention \textbf{Kepa Sousa}, \textbf{Irene Sendra}, \textbf{Diego S\'aez}, \textbf{Charlotte Van Hulse}, 
\textbf{I\~nigo Urizar}, \textbf{Giussepe Vitagliano}, \textbf{Borja Reina}, \textbf{Ariadna Montiel}, \textbf{Lluc Garc\'ia}, \textbf{Asier L\'opez}, 
\textbf{Joanes Lizarraga},  \textbf{Iagoba Apell\'aniz} and \textbf{Pablo Jimeno}, 
thank you very much, you have became really good friends.  

I would also like to thank the support from my whole family; \textbf{Ama}, \textbf{Aita}, \textbf{Bego}, \textbf{Sergio}, \textbf{Nadia} 
including my nephews \textbf{Ane}, \textbf{Jon} and \textbf{Markel} 
and my old friends; \textbf{\'Alvaro}, \textbf{Bolu}, \textbf{Cubas}, 
\textbf{Depra}, \textbf{Javi}, \textbf{Kraig}, \textbf{Mako} and \textbf{Ote}.

Finally I want to thank \textbf{Amaia} for her patience, support and making this life easier.

\addcontentsline{toc}{section}{Resumen}
\chapter*{Introducci\'on y Resumen}
\thispagestyle{chapter} 

\initial{E}l modelo est\'andar de la f\'isica de part\'iculas elementales ha demostrado tener la habilidad de describir una 
multitud de fen\'omenos que tienen lugar en la naturaleza, siendo ampliamente respaldado por 
experimentos realizados en aceleradores de part\'iculas que confirman sus predicciones. 
El ejemplo m\'as reciente es el anuncio del descubrimiento del bos\'on de Higgs por parte de 
las colaboraciones ATLAS \cite{Atlas2012} y CMS \cite{CMS2012}.

\bigskip

El modelo est\'andar esta basado en la simetr\'ia ${\it gauge}$ $SU(3) \times SU(2) \times U(1)$ \cite{Pati1973,Georgi1974} donde 
$SU(2) \times U(1)$ es la simetr\'ia de la teor\'ia de campos ${\it gauge}$ electrod\'ebiles $W^{\pm}_{\mu}, Z_{\mu}, A_{\mu}$ que provee la 
descripci\'on unificada de la interacci\'on electromagn\'etica y la d\'ebil.
Fue propuesta por Sheldon L. Glashow \cite{Glashow1961}, Steven Weinberg \cite{Weinberg1967} y 
Abdus Salam \cite{Salam1968} \footnote{Galardonados con el Premio Nobel en 1979.}. 

\bigskip

La simetr\'ia de ${\it gauge}$ $SU(2)\times U(1)$ est\'a espont\'aneamente rota hasta un subgrupo $U(1)$ ''diagonal'' de 
$SU(2)\times U(1)$. Como resultado, los bosones $W^{\pm}_{\mu}, Z_{\mu}$, que corresponden a los generadores de las simetr\'ias rotas, 
obtienen una masa gracias al mecanismo de Englert--Brout--Guralnik--Hagen--Kibble--Higgs \cite{Higgs1964,Englert1964,Guralnik1964,Kibble1967}. 
En cambio, el campo electromagn\'etico $A_{\mu}$ se corresponde al generador de la simetr\'ia preservada y, 
entonces, describe una part\'icula sin masa, el fot\'on.

\bigskip

La unificaci\'on de las interacciones electromagn\'eticas y d\'ebiles no es la primera en la historia de la f\'isica. 
Recordemos que en el siglo XIX James C. Maxwell demostr\'o que fen\'omenos 
el\'ectricos y magn\'eticos, que anteriormente eran considerados como no relacionados, pod\'ian ser descritos por 
un conjunto \'unico de ecuaciones que a d\'ia de hoy llevan su nombre. Estas ecuaciones describen la teor\'ia del electromagnetismo y
realizan la unificaci\'on de los fen\'omenos el\'ectricos y magn\'eticos \cite{MaxwellEq}.

\bigskip

La idea de unificaci\'on puede ser interpretada como un camino hacia un entendimiento m\'as profundo de la naturaleza, incluso como 
un principio filos\'ofico de unidad de la naturaleza:\\
la naturaleza es \'unica y el entendimiento m\'as profundo puede --y tiene que-- revelar la explicaci\'on com\'un de fen\'omenos 
aparentemente diferentes.

\bigskip

Adem\'as de teor\'ia de interacciones electrod\'ebiles, el modelo est\'andar contiene la teor\'ia de campos ${\it gauge}$ $A^{r}_{\mu}, r=1,...,8$ 
del grupo $SU(3)$, que se conoce como Cromodin\'amica Cu\'antica (QCD por sus siglas en ingl\'es).
Los cuantos de este campo, que describe las interacciones fuertes, se llaman ''gluones''.

\bigskip

En el modelo est\'andar la QCD est\'a aparte de la interecci\'on electrod\'ebil, incluso la constante de acoplo 
de los gluones es uno de los par\'ametros independientes.
Eso se refleja en la aparici\'on del grupo $SU(3)$ en la simetr\'ia ${\it gauge}$ del modelo est\'andar en producto directo con 
$SU(2)\times U(1)$ y, en la falta de un mecanismo de mezcla con este \'ultimo.
(Por otro lado, en la teor\'ia electrod\'ebil la mezcla de $SU(2)$ con $U(1)$ aparece como el resultado de una rotura 
espont\'anea de simetr\'ia con preservaci\'on del subgrupo $U(1)$ diagonal de $SU(2)\times U(1)$.)

\bigskip

Ya en los a\~nos 70 se empez\'o la b\'usqueda de la teor\'ia que unificara m\'as estas interacciones tomando como base teor\'ias de
campos ${\it gauge}$ de grupos semisimples que contengan los grupos $SU(3)$, $SU(2)$ y $ U(1)$ como subgrupos.
Entre estas teor\'ias, que se conocen como Teor\'ias de Gran Unificaci\'on (GUT por sus siglas en ingl\'es), las m\'as investigadas 
son las que tienen grupos de simetr\'ia ${\it gauge}$ $SU(5)$, $SO(10)$ y $E_6$.

\bigskip

Todas las GUT tienen en com\'un ciertas propiedades, entre ellas:
\begin{itemize}
 \item La existencia de una \'unica constante de acoplo, en vez de las tres del
       modelo est\'andar: la separaci\'on de efectos de interacciones a bajas energ\'ias provee de los mecanismos din\'amicos.
 \item En todas las teor\'ias GUT el prot\'on es inestable \cite{Langacker1981}: se pueden producir decaimientos del tipo 
       $p\rightarrow \pi^{0}e^{+}$ en los cuales se viola la conservaci\'on del n\'umero de bariones. 
       Esto permite estudiar la viabilidad de estas teor\'ias contrastando sus predicciones del tiempo de vida medio del prot\'on
       con el l\'imite encontrado en el experimento super-Kamiokande $\tau_{p\rightarrow \pi^{0}e^{+}} >  10^{33}$ a\~nos.
\end{itemize}

\bigskip

La realizaci\'on posible del programa de unificaci\'on en el marco de las teor\'ias GUT es aparentemente incompleta: 
no pretenden unificar las interacciones fuertes y electrod\'ebiles con la gravedad, que se puede describir en t\'erminos del campo 
de la m\'etrica del espaciotiempo $g_{\mu\nu}$.
La base natural para buscar la unificaci\'on de todas las interacciones fundamentales $W^{\pm}_{\mu}, Z_{\mu}, A_{\mu},A^{r}_{\mu},g_{\mu\nu}$ 
est\'a relacionada con la supersimetr\'ia \cite{Golfand1971,Volkov73,Volkov73+1,Wess1974} y la teor\'ia de cuerdas (hoy en d\'ia tambi\'en conocida 
como teor\'ia M).
Adem\'as estas podr\'ian dar cuenta de otros problemas.
Por ejemplo, algunos candidatos a part\'icula de materia oscura son part\'iculas que surgen de manera natural en 
teor\'ias supersim\'etricas. Mencionamos el neutralino, la part\'icula hipot\'etica, el\'ectricamente neutra, 
que s\'olo interacciona a trav\'es de la  interacci\'on gravitatoria y la d\'ebil 
y que en algunos modelos aparece como mezcla (combinaci\'on lineal) de supercompa\~neras (''${\it superpartners}$'') de part\'iculas del modelo est\'andar, 
como el ''Higgsino''(del Higgs), los ''Winos'' (de $W^{\pm}_{\mu}$) y ''Zinos''(de $Z_{\mu}$). 

\bigskip

La gravedad cl\'asica se describe mediante la Teor\'ia de la Relatividad General. 
El grupo ${\it gauge}$ se puede identificar como difeomorfismos, es decir, transformaciones generales de coordenadas o en el 
formalismo de tetradas (''${\it moving}$ ${\it frame}$'') con el grupo de Lorentz $SO(1,3)$.
El campo ${\it gauge}$ (de esp\'in 2) es la m\'etrica $g_{\mu\nu}$, o  las tetradas (''${\it vielbeins}$'') $e^{a}_{\mu}$, que componen la 
m\'etrica $g_{\mu\nu}=e^{a}_{\mu} \eta_{ab} e^{b}_{\nu}$.
La teor\'ia tambi\'en contiene la conexi\'on de Christoffel $\Gamma_{\mu\nu}^{\rho}(g)$ o/y un campo ${\it gauge}$ para 
la simetr\'ia de Lorentz, la conexi\'on de spin $\omega_{\mu}^{ab}(e)$, que son campos compuestos construidos de la m\'etrica o de 
las tetradas, 
respectivamente.
Resumiendo, la Relatividad General es una teor\'ia ${\it gauge}$ para las simetr\'ias del espaciotiempo \cite{Kibble1961}.
 
\bigskip

Como se ha mencionado anteriormente, el modelo est\'andar tambi\'en es una teor\'ia de campos ${\it gauge}$, pero con 
grupo $SU(3) \times SU(2) \times U(1)$ de simetrias internas.
Es decir, es una teor\'ia de campos ${\it gauge}$ del tipo de Yang-Mills \cite{Yang1954}.

El teorema de Coleman y Mandula \cite{Coleman:1967ad} establece que el \'algebra de Lie m\'as general de las simetr\'ias de la 
matriz S de una teor\'ia cu\'antica de campos
contiene el operador energ\'ia--momento $P_m$, los generadores de rotaciones de Lorentz $M_{mn}$ y un n\'umero finito 
de operadores $B_l$ que son escalares del grupo de Lorentz. 
Adem\'as estos \'ultimos deben pertenecer al \'algebra de Lie de un grupo compacto de Lie.\\
Se puede tratar este como un teorema ${\it no-go}$ que prohibe unificar las simetr\'ias del espaciotiempo con las simetr\'ias internas.

\bigskip

La supersimetr\'ia evita las restriciones del teorema de Coleman y Mandula \cite{Haag:1974qh} generalizando la noci\'on de 
\'algebra de Lie. 
Sus generadores $Q_{\alpha}$, $\bar{Q}_{\dot{\alpha}}$ son fermi\'onicos y satisfacen unas relaciones de anticonmutaci\'on, y no 
conmutaci\'on como es el caso de \'algebras de Lie habituales.
Estas nuevas ''\'algebras'' que, involucran tanto conmutadores como anticonmutadores, se pueden considerar como construidas con 
generadores bos\'onicos y fermi\'onicos, se llaman super\'algebras o \'algebras de Lie graduadas.

\bigskip

Todas las representaciones lineales de la supersimetr\'ia contienen el mismo n\'umero de estados bos\'onicos y fermi\'onicos. Esta 
es la raz\'on de que la b\'usqueda de la supersimetr\'ia en la naturaleza se relacione con la b\'usqueda de las supercompa\~neras  
de las 
part\'iculas elementales conocidas. En caso de supersimetr\'ia preservada tienen las mismas caracter\'isticas que sus an\'alogas 
''habituales'' pero la estad\'istica opuesta: las supercompa\~neras de los bosones son fermiones y viceversa.
Los datos experimentales sugieren claramente que en la naturaleza la supersimetr\'ia tiene que ser una simetr\'ia rota, presumiblemente 
de manera espont\'anea. La diferencia de masa entre supercompa\~neras tiene entonces que estar relacionada con la escala de energ\'ia 
a la que la supersimetr\'ia se rompe.

\bigskip

Las teor\'ias supersim\'etricas tienen unas propiedades muy atractivas.
A nivel cu\'antico algunos modelos supersim\'etricos exhiben un mejor comportamiento ultravioleta. Por ejemplo, ya en el 
modelo propuesto por Wess y Zumino en 1974 \cite{Wess1974} debido a las cancelaciones entre las contribuciones 
fermi\'onicas y bos\'onicas no aparecen divergencias a un lazo (''${\it loop}$'').\\ 
Pero sin duda el mejor ejemplo es la teor\'ia extendida de super--Yang--Mills (SYM) con supersimetr\'ia 
extendida ${\cal N}=4$, que nos provee con el \'unico ejemplo conocido de teor\'ia de campos 4--dimensionales que es finita 
a todos los \'ordenes de la teor\'ia de perturbaciones \cite{Mandelstam1983}.\\

\bigskip

Como se hab\'ia notado ya en uno de los primeros art\'iculos de supersimetr\'ia \cite{Volkov73,Volkov73+1}, invariancia bajo supersimetr\'ia 
local (''${\it gauge}$'') implica invariancia bajo difeomorfismos, y entonces, es una teor\'ia de gravedad.
En efecto, el \'algebra de supersimetr\'ia sencilla\\ 
\begin{eqnarray}
\{Q_{\alpha},Q_{\beta}\}=0,\qquad \{\bar{Q}_{\dot{\alpha}},\bar{Q}_{\dot{\beta}}\}=0,\nonumber\\
\{Q_{\alpha},\bar{Q}_{\dot{\alpha}}\}\sim\sigma^{a}_{\alpha\dot{\alpha}}P_{a},\nonumber 
\end{eqnarray}
implica que dos transformaciones de supersimetr\'ia producen una traslaci\'on. 
Entonces, el \'algebra de supersimetr\'ia local se cierra en la traslaci\'on local, es decir en transformaciones de coordenadas 
locales (invertibles) del tipo m\'as general, que se llaman difeomorfismos.

\bigskip

En los a\~nos setenta se cre\'ia que la gravedad supersim\'etrica (supergravedad) \cite{Volkov1973,Volkov1974,new-mSG75-78,Freedman1976,Deser1976,VanNieuwenhuizen1981}
pod\'ia resolver la tensi\'on aparente que exist\'ia entre la teor\'ia de la Relatividad General de Einstein y la 
mec\'anica cu\'antica.
La esperanza original fue que la supersimetr\'ia extendida m\'axima, ${\cal N}=8$, pod\'ia prohibir los contrat\'erminos y hacer que la 
supergravedad ${\cal N}=8$ \cite{Cremmer1979} no tuviera divergencias \cite{Deser1977}. Investigaciones m\'as detalladas mostraron la 
existencia de posibles contrat\'erminos en lazos m\'as altos \cite{Deser1978,Kallosh1981,Howe1981,Howe1984,Howe2003}. 
Curiosamente, una t\'imida esperanza en la cancelaci\'on de todas las divergencias en supergravedad ${\cal N}=8$ ha reaparecido 
recientemente \cite{Bossard2011a,Bern2011,Bern2007,Bern2008,Bern2009,Green2010,Bossard2011,Elvang2010}.

\bigskip

A\'un as\'i, la mayor\'ia de cient\'ificos creen$\backslash$esperan m\'as que la supergravedad cu\'antica necesita una completitud 
ultravioleta, es decir que unos grados de libertad adicionales tienen que manifiestarse en procesos de alta energ\'ia para corregir propiedades 
de las amplitudes.
La teor\'ia de cuerdas supersim\'etrica se considera actualmente como dicha completitud ultravioleta, siendo su l\'imite de bajas energ\'ias 
la supergravedad.
Lo que hoy en d\'ia conocemos como teor\'ia de supercuerdas o teor\'ia M es, de un lado, el resultado de incluir supersimetr\'ia en la teor\'ia 
de cuerdas bos\'onicas, incorporando as\'i fermiones y mejorando su espectro de estados cu\'anticos al eliminar el estado 
taqui\'onico caracter\'istico de la cuerda bos\'onica.          

\bigskip

De otro lado, la teor\'ia M apareci\'o como unificaci\'on de los diferentes modelos de supercuerda. En los a\~nos ochenta del 
siglo pasado, tras la primera revoluci\'on de supercuerdas, se hab\'ian desarrollado cinco modelos consistentes de cuerdas supersim\'etricas:

\begin{itemize}
\item Supercuerda tipo I
\item Supercuerda tipo IIA
\item Supercuerda tipo IIB
\item Supercuerda heter\'otica $SO(32)$
\item Supercuerda heter\'otica $E_{8}\otimes E_{8}$.
\end{itemize}

Todas estas teor\'ias no contienen ning\'un estado taqui\'onico (a diferencia del modelo de cuerdas bos\'onicas) y adem\'as son 
consistentes en 10 dimensiones, es decir, tienen dimensi\'on cr\'itica $D=10$ en vez de las 26 del modelo bos\'onico.
El hecho de que fueran cinco las teor\'ias consistentes se pod\'ia considerar como un problema si lo que se pretende es 
encontrar una teor\'ia \'unica.

\bigskip

Investigaciones m\'as extensas resultaron en el descubrimiento de las dualidades que conectan los diferentes modelos de cuerdas, 
y las supergravedades que aparecen en sus l\'imites de bajas energ\'ias, entre ellos y con supergravedad once-dimensional \cite{M-theory}.

As\'i por ejemplo, las teor\'ias de tipo IIA y IIB, tras realizar una compactificaci\'on de una de las dimensiones del espaciotiempo 
en un c\'irculo, son equivalentes.
Las transformaciones que las identifican se conocen como dualidad T. Curiosamente, si el radio de la dimensi\'on compacta de la 
teor\'ia IIA es $R$, el radio del c\'irculo en la teor\'ia equivalente del tipo IIB es $\alpha'\over R$, donde $\alpha'$ es la 
pendiente de las trayectorias de Regge $J=\hbar\alpha'M^2+\alpha$ en las que se situan los estados cu\'anticos de la cuerda con 
esp\'in $J$ y masa $M$.

La otra dualidad, conocida como dualidad S, es \footnote{Veas\'e por ejemplo \cite{Zwiebach2004}} una generalizaci\'on de la dualidad que existe en 
la electrodin\'amica cl\'asica entre el campo el\'ectrico y el magn\'etico. A diferencia de la dualidad T, que se ve en teor\'ia de 
perturbaciones de la cuerda, la dualidad S es no perturbativa. Se puede observar 
su manisfestaci\'on como una simetr\'ia $SL(2,\mathds{R})$ de la supergravedad del tipo IIB.
En la teor\'ia cu\'antica de la cuerda IIB la simetr\'ia $SL(2,\mathds{R})$ est\'a rota hasta su subgrupo $SL(2,\mathds{Z})$ debido a la 
existencia de otros objetos supersim\'etricos extendidos, super--p--branas de diferentes tipos, y por la cuantizaci\'on de sus tensiones 
$T_p$ \cite{Nepomechie1985} como consecuencia de una generalizaci\'on de la ''condici\'on de cuantizaci\'on de Dirac'' 
\cite{Blagojevic1988}. Este \'ultimo, en su forma cl\'asica, resulta en la cuantizaci\'on de la carga el\'ectrica en el caso de la 
presencia de al menos un monopolo magn\'etico\footnote{Veas\'e \cite{Blagojevic1988,Duff94}}.
La unidad de todas las dualidades de la teor\'ia de cuerdas fue apreciada en \cite{M-theory} y resulta en la noci\'on de dualidad $U$.

\bigskip

Estos y otros descubrimientos \cite{M-theory+3,Witten:1995im,Townsend1995,Polchinski1995} sugirieron la conjetura de la teor\'ia M, 
una teor\'ia hipot\'etica que reune los cinco modelos de cuerdas y la supergravedad once--dimensional y tiene las dualidades que 
relacionan los modelos de cuerdas como sus simetr\'ias.
Las cinco teor\'ias de supercuerdas aparecen como distintos l\'imites perturbativos de esta teor\'ia subyacente (''${\it underlying}$'')
que ''vive'' en 11 dimensiones \cite{Horava1996,Schwarz1996}; y en otro l\'imite de bajas energ\'ias se reduce a la supergravedad en 11 dimensiones formulada por 
E. Cremmer, B. Julia y J. Scherk en 1978 \cite{CJS}.

\bigskip

Tenemos que mencionar que son once las dimensiones ''elegidas'' por la teor\'ia M ya que once dimensiones es el n\'umero m\'aximo que puede 
tener una teor\'ia supersim\'etrica sin que en su reducci\'on a 4 dimensiones aparezcan part\'iculas con esp\'in mayor que 2.

En este aspecto cabe subrayar que la cuerda s\'i que contiene campos de esp\'in altos en su espectro de estados cu\'anticos y que, 
en relaci\'on con este hecho, existen especulaciones sobre que las teor\'ias de cuerdas pueden ser una fase rota de un 
modelo m\'as sim\'etrico de campos conformes de espines altos sin masa \cite{Misha88-89+1}.

 \bigskip
En el estudio del r\'egimen no perturbativo de la teor\'ia de cuerdas$\backslash$teor\'ia M la noci\'on de super--p--brana, 
objeto supersim\'etrico extendido con dimensi\'on de volumen--mundo $d=p+1$, es fundamental.
Son estados BPS (Bogomol'nyi-Prasad-Sommerfield), es decir son estables porque saturan un cota BPS (''${\it BPS}$ ${\it bound}$'') que, a su vez, 
significa que su energ\'ia (o densidad de energ\'ia) tiene un valor minimal entre 
los estados con valores fijos de algunas cargas. 
La cota BPS, es decir, la restricci\'on por abajo del valor de la energ\'ia por el valor de una carga topol\'ogica, suele aparecer 
en la teor\'ia de cuerdas como resultado de preservar (una parte de) la supersimetr\'ia.
En teor\'ia de cuerdas$\backslash$teor\'ia M los estados de p--branas BPS pueden ser descritos por soluciones supersim\'etricas 
de las ecuaciones de supergravedad en 10 y 11 dimensiones.
La estabilidad mencionada anteriormente, hace que sean similares a soluciones solit\'onicas de las famosas ecuaciones no lineales de 
KdV (Korteweg--de Vries) y Sine--Gordon \cite{Faddeev1987}.

La relaci\'on de la estabilidad de los estados BPS con la topolog\'ia y con la preservaci\'on de supersimetr\'ia sugieren que las 
super--p--branas descritas por soluciones supersim\'etricas de las ecuaciones de supergravedad cl\'asica \cite{Duff94,Stelle98} son 
objetos de la teor\'ia completa, es decir de la hipot\'etica teor\'ia M cu\'antica.

Las soluciones supersim\'etricas de la teor\'ia de supergravedad en $D$ dimensiones \cite{Duff94,Stelle98} describen los estados 
fundamentales de las super--p--branas. 
La din\'amica de las excitaciones sobre estos estados se describe mediante acciones efectivas de super--p--branas 
\cite{Hughes:1986dn,Hughes:1986fa,BST87,Townsend1996,Cederwall1997,Aganagic1997,Eric+Paul:1996,Cederwall1997a,Aganagic1997a,Bandos1997,Aganagic1997b}, 
similares a la acci\'on de Green y Schwarz para la supercuerda \cite{G+S84}.
Los l\'imites bos\'onicos de estas acciones se pueden considerar como fuentes de la teor\'ia de la Relatividad General y de los l\'imites 
bos\'onicos de supergravedad en $D$ dimensiones.

La descripci\'on din\'amica completa de los sistemas de super--p--branas en interacci\'on con 
supergravedad din\'amica es un problema m\'as complicado y no resuelto de una forma completa\footnote{V\'ease 
\cite{BdAI,BdAIL03,B+I03,IB+JdA:05} para un progreso parcial en esta direcci\'on.}.

En esta tesis estudiamos, entre otros, sistemas en interacci\'on de supergravedad din\'amica y super--p--brana en superespacio 
simple 4--dimensional. El estudio de este tipo de sistemas es importante ya que pueden ayudarnos a comprender sistemas m\'as 
complicados en espaciotiempo de 10 y 11 dimensiones de la teor\'ia de cuerdas$\backslash$teor\'ia M.

\bigskip
\section*{Supergravedad en interacci\'on con super--p--brana}

La acci\'on para la M2--brana en 11 dimensiones se contruy\'o en \cite{BST87}.
M\'as a\'un, en \cite{BST87} se mostr\'o que la consistencia del acoplo de la supermembrana en un fondo (''${\it background}$'') de 
supergravedad (es decir, la existencia de simetr\'ia--$\kappa$ en espacio curvo) impone al fondo un conjunto 
de ligaduras del superespacio que resultan en las ecuaciones del movimiento para los campos f\'isicos de supergravedad 
en 11 dimensiones, del mismo modo en que las condiciones de consistencia para acoplar una supercuerda a supergravedad 
en supercampos en 10 dimensiones produce las ecuaciones del movimiento para supergravedad $D=10$ \cite{Grisaru:1985fv}.
Es decir, la consistencia del modelo en superespacio curvo de 11 dimensiones requiere que la curvatura y la torsi\'on de 
este obedezcan las ligaduras de supergravedad que a su vez, resultan en las ecuaciones del movimiento de supergravedad.
En este sentido se puede decir que la din\'amica de la supergravedad est\'a gobernada por la M2--brana.

\bigskip

Poco despu\'es de los art\'iculos pioneros \cite{BST87} se estudi\'o \cite{4Dsuperm} el hom\'ologo no trivial m\'as simple 
de la M2--brana, la supermembrana $D=4$ ${\cal N}=1$. Su autoconsistencia en superespacio curvo tambi\'en necesita de un conjunto de 
ligaduras del superespacio. 
Sin embargo, al contrario que en el caso de 11 dimensiones estas ligaduras $D=4$ ${\cal N}=1$ son {\it off-shell} 
en el sentido de que como consecuencia suya no se producen ecuaciones del movimiento.
Esto implica que es posible construir la descripci\'on Lagrangiana manifiestamente covariante supersim\'etrica en supercampos 
del sistema en interacci\'on de supergravedad y supermembrana $D=4$ ${\cal N}=1$.
Curisosamente este sistema no se hab\'ia contruido, por lo que ser\'a parte del estudio de esta tesis\footnote{V\'ease cap\'itulo \ref{chaptersugra+membrane}.}.

\bigskip

Como se ha mencionado anteriormente, los estados fundamentales de las super--p--branas en $D$ dimensiones est\'an descritos por
las soluciones supersim\'etricas de la teor\'ia de supergravedad en $D$ dimensiones \cite{Duff94,Stelle98}. 
A pesar de que estas soluciones son puramente bos\'onicas preservan un medio de la supersimetr\'ia local caracter\'istica de la 
teor\'ia de supergravedad.
Esto significa que hay una rotura espont\'anea parcial de supersimetr\'ia debido a la presencia de una super--p--brana \cite{Hughes:1986dn,Hughes:1986fa}.
La relaci\'on entre las supersimetr\'ias preservadas (${1/2}$) por la soluci\'on de la p--brana de supergravedad con la 
simetr\'ia--$\kappa$ de la acci\'on en volumen--mundo para la correspondiente super--p--brana se puso de manifiesto en \cite{Bergshoeff:1997kr}.


\bigskip

En \cite{BdAI} se mostr\'o que el l\'imite puramente bos\'onico de la acci\'on de la supermembrana, donde el fermi\'on de Goldstone 
de la supermembrana (el hom\'ologo en el volumen-mundo del Goldstonion de Volkov y Akulov \cite{Volkov73,Volkov73+1}) es 
puesto a cero, 
$\hat{\theta}^{\check\beta}(\xi)=0$, sigue preservando un medio de la supersimetr\'ia local de supergravedad.
Esta parte preservada (${1/2}$) de la supersimetr\'ia del espaciotiempo est\'a en correspondencia uno a uno con la 
simetr\'ia--$\kappa$ de la acci\'on completa de la supermembrana \cite{BST87}
y es la simetr\'ia ${\it gauge}$ del sistema en interacci\'on descrito por la suma de la acci\'on 
bos\'onica para la membrana y la acci\'on para supergravedad en 11 dimensiones sin campos auxiliares.

\bigskip

El origen de esta propiedad poco evidente (que no est\'a restringido \'unicamente a las acciones de 
supergravedad--supermembrana en interacci\'on en 11 dimensiones, si no que es v\'alido para una larga colecci\'on de sistemas 
din\'amicos de super--p--branas y supergravedad) se aclar\'o en \cite{BdAIL03,B+I03} donde se mostr\'o que la suma del l\'imite 
puramente bos\'onico de la acci\'on de la super--p--brana y la acci\'on en componentes espaciotemporales para supergravedad 
se puede obtener fijando un ${\it gauge}$ en la acci\'on completa en supercampos para el sistema en interacci\'on de 
supergravedad y super--p--brana. Esta \'ultima acci\'on es la suma de la acci\'on completa de la super--p--brana y la acci\'on en 
supercampos para supergravedad (cuando esta existe y es conocida).

\bigskip

Es por esto que la descripci\'on del sistema en interacci\'on de supergravedad--super--p--brana mediante la suma del l\'imite 
puramente bos\'onico de la acci\'on de la p--brana y la acci\'on en componentes espaciotemporales de supergravedad sin campos 
auxiliares \cite{BdAI,BdAIL03,B+I03,IB+JdA:05} se llama descripci\'on ''completa pero con ${\it gauge}$ fijo'', donde ''completa'' 
hace referencia al hecho de que reproduce todas las ecuaciones del sistema en interacci\'on, aunque en su versi\'on con un ${\it gauge}$ fijo. 
Incluso reproduce el l\'imite $\hat{\theta}^{\check\beta}(\xi)=0$ de la ecuaci\'on para el fermi\'on de Goldstone de la 
super--p--brana \cite{BdAIL03,B+I03,IB+JdA:05}.
La descripci\'on de un sistema din\'amico de este tipo para supermembrana $D=4$ ${\cal N}=1$ en interacci\'on con supergravedad y 
multipletes de materia se desarroll\'o en t\'erminos de campos ''componentes'' en \cite{Tomas+}.

\bigskip

As\'i pues el sistema en interacci\'on de supergravedad din\'amica y super--p--brana puede ser estudiado en el marco de la 
descripci\'on covariante pero con ${\it gauge}$ fijo incluso cuando no se conoce la formulaci\'on en supercampos para 
supergravedad.

\bigskip

Sin embargo, el estudio en dimensiones m\'as bajas, en particular 
en $D=4$, de las ecuaciones en supercampos para los sistemas en interacci\'on de 
supergravedad y superbranas, cuando es posible, tiene tambi\'en gran inter\'es ya que podr\'ia proveer nuevos datos sobre las 
propiedades de sistemas m\'as complicados de la teor\'ia M. En particular, como veremos en el cap\'itulo \ref{chaptersugra+membrane}, 
ayuda a comprender el papel de los campos auxiliares de supergravedad en 
algunos sistemas en interacci\'on de supergravedad y super--p--branas.
Adem\'as estos sistemas son de inter\'es en si mismos ya que podr\'ian servir de base para construir modelos fenomenol\'ogicos de 
supergravedad cuadridimensional.
En el cap\'itulo \ref{chaptersugra+membrane} de esta tesis presentamos un estudio completo de la descripci\'on Lagrangiana en 
supercampos del sistema en interacci\'on de supergravedad minimal $D=4$ ${\cal N}=1$ y supermembrana.
Este estudio puede ser considerado como el desarrollo de la l\'inea de investigaci\'on de \cite{BdAIL03,B+I03} y \cite{Bandos2011}.

\bigskip

La descripci\'on Lagrangiana en supercampos para el sistema din\'amico de supergravedad $D=4$ ${\cal N}=1$ y superpart\'icula se 
desarroll\'o en \cite{BdAIL03}.
Las ecuaciones en supercampos para el sistema din\'amico de supergravedad, supercuerda y supermultiplete tensorial se 
obtuvieron en \cite{B+I03}.

\bigskip

Ambos modelos se han usado para estudiar el origen as\'i como las propiedades de la descripci\'on Lagrangiana completa 
pero con un ${\it gauge}$ fijo del sistema de supergravedad y super--p--brana.
Esta descripci\'on propuesta y desarrollada en \cite{BdAI,BdAIL03,B+I03,IB+JdA:05} puede ser usada tambi\'en en sistemas de 
supergravedad m\'as brana(s) en interacci\'on en dimensiones m\'as altas.

\bigskip

Debido a que el sistema de ecuaciones en supercampos para el sistema de supergravedad, supercuerda y supermultiplete tensorial 
obtenido en \cite{B+I03} resulta ser demasiado complicado para ser pr\'actico,
probablemente se podr\'ia obtener un sistema de ecuaciones menos complicado si se usara la formulaci\'on en supercampos de la 
llamada supergravedad minimal ''nueva'' \cite{new-mSG75-78,new-mSG81} en vez de la supergravedad minimal ''vieja'' 
\cite{minSG,WZ77} usada en \cite{B+I03}.
Esta suposici\'on est\'a relacionada con el hecho de que la formulaci\'on minimal ''nueva'' incluye un tensor antisim\'etrico 
auxiliar que tiene un acoplo natural al modelo de cuerda por lo que no es necesario introducir a mano, como en \cite{B+I03}, 
un multiplete tensorial adem\'as del de supergravedad.

\bigskip

Por otro lado, se pueden usar los resultado de \cite{B+I03} para extraer las ecuaciones en supercampos para la supercuerda 
en interacci\'on con el multiplete tensorial en superespacio plano $D=4$ ${\cal N}=1$.
La existencia de esa interacci\'on no trivial est\'a relacionada con el hecho de que, de acuerdo con \cite{Gates:1980ay}, 
se puede usar el multiplete tensorial para construir una 3--forma supersim\'etrica y cerrada en superespacio plano 
$D=4$ ${\cal N}=1$.
Es natural empezar estudiando las ecuaciones en supercampos para el sistema formado por la supercuerda y el multiplete 
tensorial antes de pasar al estudio de supergravedad en interacci\'on con supercuerda pero el sistema de ecuaciones en supercampos 
en interacci\'on resulta ser incluso m\'as sencillo si lo escribimos para un objeto supersim\'etrico extendido en interacci\'on con 
un supermultiplete escalar.

\bigskip

No es posible hacerlo con la supercuerda pero si con la supermembrana ya que no es posible construir una 3--forma 
${\it field}$ ${\it strength}$ a partir del multiplete escalar, pero una 4--forma s\'i.
As\'i que como primer paso hacia el estudio de la supermembrana en interacci\'on con supergravedad, estudiaremos en 
el cap\'itulo \ref{chapterscalar} la descripci\'on Lagrangiana y obtendremos las ecuaciones del movimiento en supercampos 
para el sistema en interacci\'on de supermembrana $D=4$ ${\cal N}=1$ y multiplete escalar.

\bigskip

El estudio de la descripci\'on Lagrangiana en supercampos del sistema de supermembrana $D=4$ ${\cal N}=1$ y supergravedad comenz\'o 
en \cite{IB+CM:2011}, donde se desarroll\'o un formalismo del tipo Wess--Zumino para la {\it supergravedad minimal especial} de 
Grisaru--Siegel--Gates--Ovrut--Waldram \cite{Grisaru:1981xm,Gates:1980az,Ovrut:1997ur} y se encontraron las expresiones para 
las corrientes (en supercampos) asociadas a la supermembrana que aparecen en el lado derecho de las ecuaciones en supercampos 
de supergravedad.

\bigskip

Como veremos en el cap\'itulo \ref{chaptersugra+membrane} las ecuaciones en supercampos de supergravedad con contribuciones de la 
membrana son muy complicadas pero se simplifican dr\'asticamente en un ${\it gauge}$ especial que llamamos ${\it gauge}$ 
``WZ$_{\hat{\theta}=0}$''. Se llega a este ${\it gauge}$ fijando el ${\it gauge}$ usual de Wess--Zumino (WZ) para supergravedad y 
despu\'es usando un medio de las supersimetr\'ias locales del volumen--mundo de la supermembrana para fijar el ${\it gauge}$ en el 
cual el fermi\'on de Goldstone de la supermembrana se hace cero, $\hat{\theta}^{\check\beta}(\xi)=0$.

\bigskip

En este ${\it gauge}$ resolvemos las ecuaciones para los campos auxiliares y mostramos que hay tres tipos de contribuciones 
provinientes de la supermembrana que aparecen en las ecuaciones de Einstein del sistema en interacci\'on.
Adem\'as de los t\'erminos singulares relacionados con el volumen--mundo de la supermembrana $W^3$, la supermembrana produce 
dos t\'erminos regulares que pueden ser considerados como contribuciones a la constante cosmol\'ogica en las dos regiones del 
espaciotiempo separadas por el volumen--mundo de la membrana.

\bigskip

La primera de estas contribuciones, conocida por el estudio \cite{Ovrut:1997ur} (y anteriormente en \cite{Aurilia:1978qs,OS80,Aurilia:1980xj,Duff1980}, 
ver \cite{IB+CM:2011} para m\'as 
referencias y discusi\'on sobre ella), es la constante cosmol\'ogica generada din\'amicamente.
La segunda contribuci\'on no singular de la supermembrana, cambia el valor de la constante cosmol\'ogica en una de las dos 
regiones de espaciotiempo haciendo que el valor de las constante cosmol\'ogica en las dos ramas del espaciotiempo $M^{4}_{+}$ y 
$M^{4}_{-}$, separadas por el volumen mundo de la supermembrana sea, en general, distinto.
Este efecto, que puede ser llamado ${\it shift}$ o renormalizaci\'on de la constante cosmol\'ogica debida a las contribuciones 
de la supermembrana, fue discutido en \cite{Aurilia:1978qs} y \cite{Brown:1988kg} en una perspectiva puramente bos\'onica.

\bigskip

As\'i pu\'es gen\'ericamente el estado fundamental de nuestro sistema en interacci\'on describe a la 
supermembrana separando dos espacios de Anti-deSitter con diferentes valores de la constante cosmol\'ogica.
En el contexto puramente bos\'onico las soluciones de ese tipo fueron estudiadas (adem\'as de \cite{Aurilia:1978qs} y \cite{Brown:1988kg}) 
en \cite{Gogberashvili:1998iu,JMMS+:2001,shellUni}.
La soluci\'on particular en la que ambas constantes cosmol\'ogicas tienen el mismo valor se puede encontrar en \cite{Ovrut:1997ur}.
En esta tesis tambi\'en se presenta una discusi\'on de posibles soluciones supersim\'etricas del 
sistema en interacci\'on con una supermembrana separando dos espacios asint\'oticamente Anti--deSitter con diferentes valores 
de la constante cosmol\'ogica.

\bigskip
\section*{Teor\'ias de campos de alto esp\'in}

Es conocido que las teor\'ias conformes de campos libres de alto esp\'in en $D=4$ 
pueden ser formuladas como una teor\'ia de campos en un espacio tensorial de diez dimensiones, 
$\Sigma^{(10|0)}$, parametrizado por 10 coordenadas bos\'onicas $(x^m,y^{mn})$ \cite{Frons:85,BLS99,V01s,V01c,Dima,O+Misha03,BPST04,BBdAST05},
\begin{equation*}
\label{x1} X^{\alpha\beta}=X^{\beta\alpha}= {1\over 4}x^m\gamma_m^{\alpha\beta} + {1\over 8}
y^{mn}\gamma_{mn}^{\alpha\beta}\;  \qquad \alpha , \beta =1,2,3,4\; , \quad m,n=0,1,2,3 \; .
\quad
\end{equation*}
y en un superespacio tensorial ${\cal N}=1$, $\Sigma^{(10|4)}$ con coordenadas $(X^{\alpha\beta},\theta^\alpha)$ \cite{BLS99,V01s,V01c,Dima,O+Misha03,BPST04,BBdAST05}.

\bigskip

Este espacio tensorial bos\'onico se propuso como base natural para contruir teor\'ias de campos conformes 
de alto esp\'in en $D=4$ en \cite{Frons:85}.

\bigskip

En \cite{Holten1982,Curtright1988},\cite{BdAPV05}, \cite{Frons:85,BLS99,V01s,V01c,Dima,O+Misha03,BPST04} y \cite{BBdAST05,BdAIPV:03,D'Au-Fre:82} se han presentado espacios tensoriales 
m\'as generales de dimensi\'on ${n(n+1)\over 2}$ en el sentido de 
matrices $n\times n$ sim\'etricas $X^{\alpha\beta}$ $(\alpha,\beta=1,\dots,n)$ que, para $n$ par ($n=4,8,16$ y $32$), 
determinan la extensi\'on del espaciotiempo est\'andar de dimension $D$=4,6,10 ($D={n \over 2}+2)$ y $D=11$.

A\~nadiendo $n$ coordenadas fermi\'onicas $\theta^\alpha$ se pueden obtener los superespacios tensoriales ''simples'' (${\cal N}=1$) 
$\Sigma^{({n(n+1)\over 2}|n)}$,
\begin{equation*}
\label{x+th} 
\Sigma^{({n(n+1)\over 2}|n)}\; : \qquad  Z^{\cal M}:= (X^{\alpha\beta}\, , \,
\theta^\alpha) \; , \qquad \begin{cases}\alpha,\beta =1,...,n , \cr
X^{\alpha\beta}=X^{\beta\alpha}\;  . \end{cases}
\end{equation*}
que, en su versi\'on plana, tambi\'en tienen estructura de supergrupo. 

\bigskip

Tomar $n$ par no es una restricci\'on si uno piensa en el origen espinorial que subyace en los \'indices $\alpha$;
es m\'as esto motiva la resticci\'on a $n=2^k=2,4,8,16,...$ suponiendo que $\theta^\alpha$ son espinores.
Aunque las coordenadas fermi\'onicas en $\Sigma^{({n(n+1)\over 2}|n)}$ se suponen reales normalmente, en el caso 
$n=4$ $D=4$ es conveniente considerar $\theta^\alpha$ como un espinor de Majorana en la realizaci\'on de Weyl de las matrices de 
Dirac por lo que $\theta^\alpha=\left(\theta^A, \bar{\theta}_{\dot{A}}\right)$.

\bigskip

Para $n=2$ las coordenadas esp\'in-tensoriales $X^{\alpha\beta}$ est\'an expresadas en t\'erminos de las coordenadas del 
3--vector espaciotemporal, $X^{\alpha\beta}\propto x^a\tilde{\gamma}^{\alpha\beta}_a$ as\'i que $\Sigma^{(3|2)}$ es simplemente 
el superespacio usual $D=3$ ${\cal N}=1$.
El caso $n=32$ da como resultado la extensi\'on del superespacio de 11 dimensiones $\Sigma^{(528|32)}$, importante en el 
contexto de la hip\'otesis de los preones BPS \cite{BPS01,BdAIPV:03} y tambi\'en en el an\'alisis de la estructura ${\it gauge}$ 
oculta de la supergravedad en $D=11$ \cite{D'Au-Fre:82,BdAPV05}.

\bigskip

En discusiones de las teor\'ias de espines altos, incluso en el cap\'itulo \ref{chapterspin} de esta tesis, se consideran los casos 
$n=4,8,16$ que se usan para describir teor\'ias 
conformes de campos de alto esp\'in sin masa en $D=4,6,10$.
Casi todas nuestras ecuaciones en el cap\'itulo \ref{chapterspin} ser\'an v\'alidas para esas dimensiones, aunque haremos especial \'enfasis en el caso 
$n=4$ que corresponde a $D=4$.

\bigskip

El primer sistema mec\'anico en el superespacio tensorial  ${\cal N}=1$ $D=4$ $\Sigma^{(10|4)}$ y sus generalizaciones de 
dimensiones m\'as altas $\Sigma^{({n(n+1)\over 2}|n)}$ con $n>4$ se propusieron en \cite{BL98},
donde se observ\'o que el estado fundamental de ese modelo de superpart\'icula describe un estado BPS que preserva todas las 
supersimetr\'ias excepto una.
El posible papel como ''contituyentes'' de estos estados en la teor\'ia de cuerdas/M se introdujo y se discuti\'o en general en 
\cite{BPS01}, donde se les llam\'o ''preones BPS'' ( v\'ease tambi\'en \cite{BdAIPV:03}).
As\'i que desde ese punto de vista, la superpart\'icula en \cite{BL98} podr\'ia ser llamada ''pre\'onica''.
Su cuantizaci\'on se desarroll\'o en \cite{BLS99}, donde se mostr\'o que el espectro de la superpart\'icula pre\'onica $n=4$ 
cuantizada se puede describir con una torre de campos conformes de alto esp\'in sin masa de todas las helicidades posibles y se presentaron 
evidencias de que los modelos con $n=8$ y $n=16$ describen teor\'ias conformes de alto esp\'in en espaciotiempos de 6 y 10 dimensiones. 
  
\bigskip

En \cite{V01s,V01c} se present\'o y estudi\'o una forma elegante de las ecuaciones de alto esp\'in bos\'onicas y fermi\'onicas en el 
espacio tensorial $\Sigma^{(10|0)}$. 
Mientras que en \cite{BBdAST05} se dio la forma expl\'icita de las ecuaciones conformes de alto esp\'in en espacios tensoriales con
$D=6,10$.
Las ecuaciones en supercampos para los supermultipletes de campos conformes sin masa de dimensi\'on $D=6,10$ en los superespacios 
tensoriales ${\cal N}=1$ $\Sigma^{(36|8)}$ y $\Sigma^{(136|16)}$,
 \begin{equation*}
 \label{x+th=10D}
D=10\,\quad\Sigma^{(136|16)}\; : \quad  Z^{\cal M } := (X^{\alpha\beta}\, , \, \theta^\alpha)
\; , \quad \begin{cases} \alpha,\beta =1,...,16 , \cr X^{\alpha\beta}= {1\over 16}
x^m\tilde{\sigma}{}_m^{\alpha\beta}+{1\over 2\cdot 5!}
y^{m_1...m_5}\tilde{\sigma}{}_{m_1...m_5}^{\alpha\beta} \; , \cr m=0,1,\ldots 9\end{cases}
\end{equation*}
fueron propuestas en \cite{BPST04}. 

\bigskip

En particular, los supermultipletes ${\cal N}=1$ de campos conformes de alto esp\'in en $D=4,6,10$ est\'an descritos por supercampos 
escalares en los correspondiente superespacios $n$=4,8,16 $\Sigma^{({n(n+1)\over 2}|n)}$,
\begin{equation*}
\label{scalar} \Phi (X^{\alpha\beta} , \theta^\gamma)=b(X)+f_\alpha(X)\,\theta^\alpha+
\sum_{i=2}^{n}\phi_{\alpha_1\cdots\alpha_i}(X)\,\theta^{\alpha_1}\cdots\theta^{\alpha_i}\;,
\end{equation*}
obedeciendo la ecuaci\'on \cite{BPST04}
\begin{eqnarray*}
\label{hsSEq=1N} D_{[\alpha}D_{\beta ]} \Phi (X, \theta ) = 0 \; .
\end{eqnarray*}
Aqu\'i,
\begin{eqnarray*}
\label{DfN=1} D_\alpha= {\partial \over \partial \theta^\alpha} + i \theta^{\beta}
\partial_{\beta\alpha}\; , \qquad
D_{\alpha\beta}=\partial_{\alpha\beta}:={\partial \over \partial X^{\alpha\beta}}\; , \qquad
 \end{eqnarray*}
son derivadas covariantes en el superespacio tensorial r\'igido $\Sigma^{({n(n+1)\over 2}|n)}$ que satisfacen
\begin{eqnarray*}
\label{(DD)=2id} \{D_\alpha ,D_\beta\}=2i\partial_{\alpha\beta}
\end{eqnarray*}
donde se puede ver que exhiben la estructura de extensi\'on central de las super\'algebras de los superespacios tensoriales 
\cite{BdAPV:30-32} (v\'ease \cite{CAIPB:00}).

\bigskip

Las ecuaciones de alto esp\'in en superespacio tensorial con supersimetr\'ia $\cal{N}$-extendida,
han sido estudiadas en \cite{Bandos2011a,Bandos2012}\footnote{Veas\'e \cite{KuzenkoHS,GatesHS}  
para una descripci\'on de las ecuaciones supersim\'etricas de alto esp\'in libres en superespacio usual y \cite{S+S:98,ESS:02} 
para la versi\'on en supercampos de las ecuaciones de campos de alto esp\'in de Vasiliev en interacci\'on \cite{Misha88-89,Misha92,Misha03s}.} 
que constituyen la base del cap\'itulo \ref{chapterspin} de esta tesis.

\bigskip

En el cap\'itulo \ref{chapterspin} presentamos las ecuaciones supersim\'etricas conformes de alto esp\'in 
libres con ${\cal N}=2$, ${\cal N}=4$ y ${\cal N}=8$ en superespacios tensoriales ${\cal N}$-extendidos $\Sigma^{({n(n+1)\over 2}|{\cal N}\,n)}$.
Entre otros, vamos a mostrar que las ecuaciones conformes de alto esp\'in de los supermultipletes ${\cal N}=2$ de $D=4,6,10$ est\'an 
descritas por supercampos escalares quirales $\Phi (X^{\alpha\beta} , \Theta^\gamma, \bar{\Theta}{}^\gamma)$ en el superespacio 
tensorial ${\cal N}=2$ $\Sigma^{({n(n+1)\over 2}|2n)}$, que obedecen el siguiente conjunto de ecuaciones lineales en supercampos,  
\begin{eqnarray*}
\bar{{\cal D}}_\alpha \Phi =0\; , \qquad {\cal D}_{[\alpha}{\cal D}_{\beta
]} \Phi  = 0 \; ,
\end{eqnarray*}
donde
\begin{eqnarray*}
\label{Df-N=2} {\cal D}_\alpha &=& {\partial \over \partial \Theta^\alpha} + i
\bar{\Theta}^{\beta}
\partial_{\beta\alpha}= {1\over 2} (D_{\alpha 1}
+ i D_{\alpha 2})\, , \qquad \nonumber \\ \bar{{\cal D}}_\alpha &=& {\partial \over \partial
\bar{\Theta}^\alpha} + i {\Theta}^{\beta} \partial_{\beta\alpha}= - ({\cal D}_\alpha)^*\, ,
\qquad  {\cal D}_{\alpha\beta}=\partial_{\alpha\beta}:={\partial \over \partial
X^{\alpha\beta}}\; , \qquad
 \end{eqnarray*}
son las derivadas covariantes en el superespacio r\'igido $\mathcal{N}=2$--extendido $\Sigma^{({{n(n+1)}\over 2}|2n)}$, 
que obedecen el super\'algebra
\begin{eqnarray*}
\label{(DD)=2id} \{{\cal D}_\alpha , {\cal D}_\beta \}= 0\; , \qquad \{{\cal D}_\alpha
, \bar{{\cal D}}_\beta \}=2i\partial_{\alpha\beta} \; ,  \qquad \{\bar{{\cal D}}_\alpha ,
\bar{{\cal D}}_\beta \}=0 \; .
 \end{eqnarray*}
Presentamos tambi\'en las ecuaciones en supercampos para superespacios tensoriales extendidos con ${\cal N}$ par y mayor que $2$ 
que generalizan las ecuaciones de ${\cal N}=2$.

\bigskip

Tambi\'en presentaremos la generalizaci\'on supersim\'etrica extendida ${\cal N}>1$ del modelo de superpart\'icula pre\'onica 
de \cite{BLS99} y mostraremos como se pueden obtener las ecuaciones de los supercampos de alto esp\'in cuantizando el modelo de 
superpart\'icula en $\Sigma^{({n(n+1)\over 2}|{\cal N}\,n)}$ para ${\cal N}\geq 2$ con ${\cal N}$ par.

\bigskip

Aunque nuestras ecuaciones son v\'alidas para cualesquiera ${\cal N}$ y $n$ par, elaboraremos en detalle los casos ${\cal N}=2,4,8$, 
que, para $n=4$, corresponden a los supermultipletes sin masas de campos de alto esp\'in en $D=4$, los cuales est\'an en 
correspondencia clara con las teor\'ias est\'andar de campos de ''bajo esp\'in''. Estas son el hipermultiplete para 
${\cal N}=2$, el supermultiplete supersim\'etrico de Yang--Mills para ${\cal N}=4$ y el multiplete de supergravedad maximal 
para ${\cal N}=8$, el cual en su versi\'on linearizada se puede describir mediante supercampos escalares en superespacios 
est\'andar extendidos en $D=4$ $\Sigma^{(4|4{\cal N})}$ con ${\cal N}=2,4,8$.

\bigskip

Una de las razones de nuestro inter\'es en sistemas $\mathcal{N}$-extendidos supersim\'etricos de teor\'ias de alto esp\'in ven\'ia de la 
observaci\'on de que la supersimetr\'ia $\mathcal{N}$-extendida con $\mathcal{N}=4$ unifica los campos ${\it gauge}$ escalares y vectoriales.
Por otro lado, todas las ecuaciones de alto esp\'in han sido formuladas en t\'erminos de campos escalares y espinoriales en espacio 
tensorial as\'i que el estudio de supersimetr\'ias $\mathcal{N}$-extendidas podr\'ia ser interesante para buscar posibles generalizaciones 
de las ecuaciones de Maxwell y Einstein en superespacios tensoriales.

\bigskip

De hecho, en cierto punto de nuestro estudio de las ecuaciones en supercampos en superespacio tensorial ${\cal N}=4$ 
$\Sigma^{(10|16)}$ aparecen los an\'alogos de las ecuaciones de Maxwell en espacio tensorial .
Sin embargo, un an\'alisis m\'as detallado muestra que estos campos (esp\'in)--tensoriales se pueden expresar como derivadas de otros 
campos escalares en superespacio tensorial as\'i que, por ejemplo, el sector bos\'onico del multiplete de alto esp\'in conforme 
${\cal N}=4$ est\'a expresado por dos campos escalares complejos en el espacio tensorial, $\phi$ y $\tilde{\phi}$.

\bigskip
 
Del mismo modo, cuando estudiamos las ecuaciones en supercampos en superespacio tensorial 
${\cal N}=8$, $\Sigma^{(10|32)}$, a pesar de que en cierta etapa aparecen las generalizaciones en espacio tensorial de las 
ecuaciones de (super)gravedad conforme, se demuestra que se reducen a las ecuaciones para campos escalares (y espinoriales) en espacio 
tensorial presentadas por primera vez por Vasiliev \cite{V01s}.
En cierto sentido se puede decir que el resultado de aumentar ${\cal N}$, es que aparecen m\'as campos escalares y espinoriales.

\bigskip

Sin embargo, para ${\cal N}=4$ aparece un nuevo fen\'omeno. Como el nuevo campo escalar aparece en la teor\'ia \'unicamente 
a trav\'es de campos de tipo de Maxwell, es decir bajo derivadas bos\'onicas $\partial_{\alpha\beta}$, la teor\'ia se hace 
invariante bajo desplazamientos constantes de ese campo bos\'onico que hace que el segundo campo escalar 
sea similar a los axiones (para los cuales dicha simetr\'ia se llama simetr\'ia de Peccei--Queen 
\cite{Pe-Qui77}\footnote{Dado que en la teor\'ia de cuerdas de tipo IIB y en supergravedad IIB el axi\'on aparece como un 
miembro de la familia de los campos ${\it gauge}$ RR, su simetr\'ia de Peccei--Queen puede ser considerada como el hom\'ologo de 
la simetr\'ia ${\it gauge}$ caracter\'istica de los potenciales ${\it gauge}$ RR altos.}).
En el multiplete ${\cal N}=8$ reformulado en t\'erminos de campos, la simetr\'ia de Peccei--Queen se hace m\'as 
complicada para los campos escalares e incluso est\'a presente en los campos espinoriales que entran en el modelo bajo la acci\'on de 
una derivada simulando la estructura de los campos de Rarita-Schwinger.

\bigskip
\section*{Sistema de m\'ultiples M0--branas}

En \cite{Witten:1995im} se motiv\'o que una descripci\'on aproximada de un sistema de $p$--branas de Dirichlet (D$p$--branas) 
quasi coincidentes viene dada por una teor\'ia de Yang--Mills maximal supersim\'etrica (SYM) con grupo de ${\it gauge}$ 
$U(N)$, la cual puede ser obtenida mediante reducci\'on dimensional de la teor\'ia de SYM en D=10 con $U(N)$ a $d=p+1$.
Esta incluye $D-p-1$ matrices Herm\'iticas con campos escalares cuyos elementos diagonales describen las diferentes posiciones de 
las D$p$-branas mientras que los elementos no diagonales dan cuenta de las cuerdas que unen las diferentes D$p$-branas.

\bigskip

Como es sabido que una sola D$p$-brana est\'a descrita por la suma de la acci\'on supersim\'etrica de Dirac--Born--Infeld, 
proporcionando una generalizaci\'on no lineal de la acci\'on de Yang--Mills con U(1), y un t\'ermino de Wess-Zumino (v\'ease \cite{Bandos1997a,Eric+Paul:1996,Cederwall1997,Cederwall1997a,Aganagic1997,Aganagic1997a,Aganagic1997c}),  era natural buscar una generalizaci\'on no lineal de la acci\'on de SYM no--abeliana que proporcione 
una descripci\'on no lineal m\'as completa del sistema de D$p$-branas quasi coincidentes.
Para el l\'imite bos\'onico de m\'ultiples D$p$-branas quasi coincidentes (sistema mD$p$) la descripci\'on m\'as popular est\'a 
dada por la acci\'on de ''branas diel\'ectricas'' de Myers \cite{Myers:1999ps}.
Esta se obtuvo a trav\'es de una serie de transformaciones de dualidad T a partir de la 10D acci\'on no--abeliana 
de Born--Infeld con traza sim\'etrica, propuesta por Tseytlin \cite{Tseytlin:DBInA} para el l\'imite puramente bos\'onico del sistema 
de m\'ultiples D$9$--branas (sistema mD$9$) que llenan todo el espaciotiempo. 
Ambas acciones \cite{Myers:1999ps} y \cite{Tseytlin:DBInA} han resistido todos los intentos de construir sus generalizaciones 
supersim\'etricas durante muchos a\~nos. Adem\'as, la acci\'on de Myers no tiene simetr\'ia de Lorentz.

\bigskip

La descripci\'on supersim\'etrica y covariante Lorentz del sistema de mD$p$ se obtuvo en \cite{Howe+Linstrom+Linus}
en el marco del llamado ''enfoque de fermiones de frontera''.
Sin embargo, esta descripci\'on viene dada en lo que se llama ''cuantizaci\'on en nivel menos uno'' que significa que 
para llegar a una descripci\'on del sistema de mD$p$ similar al del de D$p$--branas (como por ejemplo el de \cite{Eric+Paul:1996}), 
se tiene que realizar una cuantizaci\'on del sistema din\'amico. 
Esta tarea no es trivial y no se ha resuelto de manera completa a\'un, lo que ha motivado un gran n\'umero de intentos de 
obtener una descripci\'on aproximada pero covariante bajo el grupo de Lorentz y supersim\'etrica del sistema de mD$p$ que vaya m\'as all\'a de la 
aproximaci\'on de SYM (v\'ease por ejemplo \cite{IB09:D0}). S\'olo para el caso del sistema de mD$0$ existe un candidato 
no lineal, supersim\'etrico e invariante Lorentz en $D=10$ para la acci\'on de mD$0$ \cite{Dima+JHEP03,Dima+JHEP03+1}.

\bigskip

Debido a que las D$p$--branas con $p=0,2,4$ pueden ser obtenidas mendiante una reducci\'on dimensional de las branas 
M$0$, M$2$ y M$5$ en $D=11$, es l\'ogico suponer que se pueda obtener el sistema de mD$p$ de su respectivo sistema de 
mM$p$. 

\bigskip

Sin embargo, para el caso del sistema de mM$5$ incluso la cuesti\'on de cual es el an\'alogo de la descripci\'on aproximada 
de muy bajas energ\'ias de SYM a\'un no se conoce a ciencia cierta (v\'ease por ejemplo \cite{SSWW,Bandos2013,Bandos2013a} para estudios relacionados y referencias).
Para el caso del sistema de mM$2$ branas dicho problema ha permanecido sin soluci\'on muchos a\~nos pero recientemente el $d=3$ ${\cal N}=8$ modelo 
 de Bagger, Lambert y Gustavsson (BLG) supersim\'etrico \cite{BLG} basado en 3--\'algebras (v\'ease \cite{deAzcarraga:2010mr} y referencias all\'i) en vez de 
\'algebras de Lie y un modelo de Aharony, Bergman, Jafferis y Maldacena (ABJM) m\'as convencional \cite{ABJM} (con simetr\'ia de ${\it gauge}$ $SU(N)\times SU(N)$ y s\'olo ${\cal N}=6$ 
supersimetr\'ias manifiestas) han sido propuestos para dicho papel.

\bigskip

Si se trata del sistema de m\'ultiples M$0$ branas, se construy\'o un candidato puramente bos\'onico en \cite{YLozano+0207,YLozano+0207+1} 
generalizando la acci\'on de Myers de la D$0$--brana diel\'ectrica de 11 dimensiones. 
Por otro lado, se obtuvieron las ecuaciones del movimiento aproximadas pero supersim\'etricas e invariantes bajo el grupo de Lorentz 
para el sistema de mM$0$ en \cite{mM0:PLB} en el marco del enfoque de {\it superembedding} (v\'ease \cite{bpstv,hs2} as\'i como 
\cite{Dima99,IB09:M-D} y referencias all\'i).

\bigskip

La generalizaci\'on de esas ecuaciones para el sistema de mM$0$ en superespacio curvo de supergravedad 11 dimensional que 
describe la generalizaci\'on de la ''teor\'ia de M(atrices)'' \cite{Banks:1996vh} (v\'ease \cite{Wit1988}) para el caso de su 
interacci\'on con un fondo de supergravedad arbitrario, se present\'o y estudi\'o en \cite{mM0:PRL}.
En \cite{mM0-pp:PRD} se mostr\'o que en el caso de que el fondo fuera de ondas pp ({\it pp--wave}), esas ecuaciones 
reproducen (en cierta aproximaci\'on) el llamado modelo BMN de matrices propuesto para ese fondo por 
Berenstein, Maldacena y Nastase en \cite{BMN}.

\bigskip

Este resultado confirma que las ecuaciones de \cite{mM0:PRL} describen la teor\'ia de Matrices interaccionando con un fondo de
supergravedad. Sin embargo, debido al origen de superespacio de dichas ecuaciones, sus apliaciones incluso en un fondo de supergravedad 
puramente bos\'onico son extremadamente complicadas.
Para este fin se necesita por lo primero encontrar la soluci\'on completa en supercampos de las ligaduras de supergravedad 11D \cite{BrinkHowe80} 
que representan la soluci\'on bos\'onica supersim\'etrica de las ecuaciones de supergravedad en el espaciotiempo.
Esto hizo que fuera deseable encontrar una acci\'on que reprodujera las ecuaciones del modelo de Matrices de \cite{mM0:PRL} o 
sus generalizaciones.

\bigskip

Para el caso del sistema de mM0 en superespacio plano esta acci\'on se propuso en \cite{mM0:action},
donde se mostr\'o que posee supersimetr\'ia local ${\cal N}=16$ 1d.
En el cap\'itulo \ref{chapter11D} se derivar\'an y estudiar\'an las ecuaciones del movimiento del sistema de mM0 descrito por
dicha acci\'on. Se estudiar\'an las soluciones supersim\'etricas de dichas ecuaciones mostrando que su sector relacionado 
con el centro de energ\'ia es similar a la soluci\'on de las ecuaciones para una \'unica M$0$ brana y se presentar\'an tambi\'en dos 
ejemplos de soluciones no supersim\'etricas con diferentes propiedades del movimiento del centro de energ\'ia. 

\section*{Contenido de la tesis}

\bigskip 
El primer cap\'itulo contiene una introducci\'on al  superespacio as\'i como a los supercampos y superformas, 
ya que a lo largo de la tesis se har\'a uso de dichos conceptos. Tras esta breve exposici\'on se explicar\'a la geometr\'ia de los 
superespacios plano y curvo, para pasar a describir la supergravedad.
Se da una descripci\'on de la supergravedad minimal, su acci\'on en superespacio, as\'i de como obtener sus ecuaciones del movimiento 
a partir de esta acci\'on en supercampos.
Aunque a priori no es trivial encontrar las variaciones de los ${\it supervielbeins}$, ya que estos est\'an restringidos por las ligaduras 
de supergravedad, es posible encontrar variaciones admisibles que preservan dichas ligaduras \cite{BdAIL03,WZ77}.
Terminamos el cap\'itulo \ref{chapterintroduccion} presentando estas variaciones admisibles para la supergravedad minimal, que usaremos 
en el cap\'itulo \ref{chaptersugra+membrane}.

\bigskip 

En el cap\'itulo \ref{chapterscalar} presentamos la acci\'on en supercampos para el sistema din\'amico de supermembrana $D=4$ 
${\cal N}=1$ en interacci\'on con multiplete escalar y la usamos para obtener las ecuaciones del movimiento en supercampos de este 
sistema.

Estas incluyen las ecuaciones de la supermembrana que coinciden formalmente con las ecuaciones de esta en un fondo 
de campo escalar ({\it off-shell}) y las ecuaciones para el supercampo quiral especial con fuente producida por la 
supermembrana.

En el caso en el cual la parte de la acci\'on correspondiente al supermultiplete escalar contiene \'unicamente el t\'ermino cin\'etico
m\'as simple, extraemos las ecuaciones en componentes a partir de las ecuaciones en supercampos y las resolvemos a orden principal
en la tensi\'on de la supermembrana.

Tambi\'en se discute la inclusi\'on de un superpotencial no trivial  y su relaci\'on con las soluciones supersim\'etricas del tipo 
${\it domain}$ ${\it wall}$ conocidas.

\bigskip

El cap\'itulo \ref{chaptersugra+membrane} est\'a dedicado a obtener el conjunto completo de ecuaciones del movimiento para 
el sistema en interacci\'on de supermembrana y supergravedad din\'amica $D=4$ ${\cal N}=1$ variando la acci\'on completa en supercampos. 
Una vez obtenidas, escribimos estas ecuaciones en supercampos en el ${\it gauge}$ especial ``WZ$_{\hat{\theta}=0}$'' donde el campo de 
Goldstone de la supermembrana se ha puesto a cero ($\hat{\theta}=0$) y adem\'as est\'a fijado el ${\it gauge}$ de Wess--Zumino en los 
supercampos de supergravedad.

En este ${\it gauge}$ resolvemos las ecuaciones para los campos auxiliares y discutimos el efecto de la generaci\'on din\'amica de la constante 
cosmol\'ogica en la ecuaci\'on de Einstein del sistema en interacci\'on y su renormalizaci\'on debida a contribuciones regulares de 
la supermembrana. Estos dos efectos (descritos por primera vez en los a\~nos 70 y 80 en un contexto bos\'onico y en la literatura de 
supergravedad) resultan en que la soluci\'on describe en el caso gen\'erico, dos espaciotiempos con distintas constantes cosmol\'ogicas separados por el 
volumen--mundo de la supermembrana.

\bigskip

Continuamos en el cap\'itulo \ref{chapterspin} proponiendo las ecuaciones en supercampos en superespacios tensoriales 
${\cal N}$-extendidos para describir las generalizaciones supersim\'etricas con ${\cal N}=2,4,8$ de las teor\'ias conformes libres de 
alto esp\'in. Describimos tambi\'en como se puede obtener estas ecuaciones cuantizando un modelo de superpart\'icula en superespacios tensoriales ${\cal N}$-extendidos.

Mostramos que los supermultipletes de alto esp\'in ${\cal N}$-extendidos contienen \'unicamente campos escalares y espinoriales en el espacio 
tensorial, as\'i que a diferencia del enfoque est\'andar de supercampos en superespacios habituales, no aparecen generalizaciones no triviales de las 
ecuaciones de Maxwell y Einstein cuando ${\cal N}>2$. Para ${\cal N}=4,8$ las componentes de tipo esp\'in-tensor m\'as altas del supercampo tensorial extendido se expresan a trav\'es de 
campos escalares y espinoriales adicionales que obedecen las mismas ecuaciones libres de alto esp\'in, pero estos son del
tipo axi\'on en el sentido de que poseen simetr\'ias de tipo Peccei--Quinn.

\bigskip

En el \'ultimo cap\'itulo \ref{chapter11D} estudiamos las propiedades de la acci\'on covariante supersim\'etrica y con simetr\'ia--$\kappa$  
de $N$ M0--branas quasi coincidentes (sistema mM0) en superespacio plano de once dimensiones y obtenemos las ecuaciones 
supersim\'etricas para este sistema din\'amico.

A pesar de que una \'unica M0--brana se corresponde con la superpart\'icula sin masa en 11 dimensiones, el movimiento del centro 
de energ\'ia del sistema de mM0 est\'a caracterizado por una masa $M$ no negativa. Esta masa est\'a construida a partir de los campos 
matriciales que describen el movimiento relativo de los constituyentes del sistema de mM0.

Mostramos que cualquier soluci\'on bos\'onica del sistema de mM0 puede ser supersim\'etrica si y s\'olo si esta masa efectiva se anula,
$M^2=0$, y que todas las soluciones bos\'onicas supersim\'etricas preservan un medio de las supersimetr\'ias de once dimensiones. Presentamos 
tambi\'en unas soluciones no supersim\'etricas con $M^2\not= 0$ y discutimos unas propiedades peculiares del sistema de mM0.

\addcontentsline{toc}{section}{List of Publications}
\chapter*{List of Publications}
\thispagestyle{chapter} 


\PublicationEntry{Superfield equations for the interacting system of D=4 N=1 supermembrane and scalar multiplet}
{Igor A. Bandos, Carlos Meliveo}{Nucl. Phys. B849, 1-27 (2011)}

\PublicationEntry{Extended supersymmetry in massless conformal higher spin theory}{
Igor A. Bandos, Jos\'e A. de Azc\'arraga, Carlos Meliveo}{Nucl.Phys. B853, 760-776 (2011)}

\PublicationEntry{Three form potential in (special) minimal supergravity superspace and supermembrane supercurrent}{
Igor A. Bandos, Carlos Meliveo}{J. Phys. Conf. Ser. 343 012012 (2012)}

\PublicationEntry{Supermembrane interaction with dynamical D=4 N=1 supergravity.
 Superfield Lagrangian description and spacetime equations of motion}{
Igor A. Bandos, Carlos Meliveo}{JHEP 1208 (2012) 140}

\PublicationEntry{Conformal higher spin theory in extended tensorial superspace}
{Igor A. Bandos, Jos\'e A. de Azc\'arraga, Carlos Meliveo}{Fortsch.Phys. 60, 861-867 (2012)}

\PublicationEntry{On supermembrane supercurrent and special minimal supergravity}
{Igor A. Bandos, Carlos Meliveo}{Fortsch.Phys. 60, 868-874 (2012)}

\PublicationEntry{Covariant action and equations of motion for the eleven dimensional multiple M0-brane system}{
Igor A. Bandos, Carlos Meliveo}{Phys.Rev. D 87 126011 (2013)}


%
%


\addcontentsline{toc}{section}{Contents} \tableofcontents 
%

\begin{titlepage}
\thispagestyle{empty} 
 \pagebreak 
\end{titlepage}

\pagestyle{fancy}
\pagenumbering{arabic}
\setcounter{page}{1} 

\chapter{Introduction}
\label{chaptersuperspace}
\thispagestyle{chapter} 
\label{chapterintroduccion}
\initial{I}n field theory supersymmetry transforms bosonic fields into fermionic fields and vice versa. Hence, the irreducible representations of supersymmetry, 
supermultiplets, involve some number of bosonic and fermionic fields. A compact and elegant way to describe supermultiplets is provided 
by superfield approach.

Superspace is an extension of the ordinary spacetime including, besides the usual spacetime coordinates $x^\mu$, extra anticommutative (or fermionic) 
coordinates. In the case of flat ${\cal N}$--extended $D=4$ superspace these fermionic coordinates are collected in ${\cal N}$
two--component Weyl spinors $\theta^{\alpha i}=(\bar{\theta}^{\dot{\alpha}}_{i})^*$.

Chapters \ref{chapterscalar}, \ref{chaptersugra+membrane} are devoted to the study of the $D=4$ ${\cal N}=1$ supermembrane, 
which is a membrane moving in simple (${\cal N}=1$) $D=4$ superspace, so let us begin reviewing some properties of this 
superspace.

\section{Flat $D=4$ ${\cal N}=1$ superspace}

We denote the coordinates of flat $D=4$ ${\cal N}=1$ superspace $\Sigma^{(4|4)}$ by
\begin{eqnarray}\label{coordinates}
z^{M}= (x^\mu, \theta^{\alpha}, \bar{\theta}^{\dot{\alpha}})
\end{eqnarray}
where $\mu=0,1,2,3$ is a 4--vector index, ${\alpha}=1,2$ and $\dot{\alpha}=1,2$ are spinorial indices; the ''supervector'' index 
M represents both types of indices, $M=(\mu ,\alpha,\dot{\alpha})$.

The spinorial coordinates anticommute, $\theta^{\alpha}\bar{\theta}^{\dot{\alpha}}=-\bar{\theta}^{\dot{\alpha}}\theta^{\alpha}$, 
$\theta^{\alpha}\theta^{\beta}=-\theta^{\beta}\theta^{\alpha}$, 
$\bar{\theta}^{\dot{\alpha}}\bar{\theta}^{\dot{\beta}}=-\bar{\theta}^{\dot{\beta}}\bar{\theta}^{\dot{\alpha}}$,
while the bosonic coordinates commute among themselves, $x^\mu x^\nu=x^\nu x^\mu$, and also with fermionic coordinates
 $x^\mu\theta^{\alpha}=\theta^{\alpha}x^\mu$. These properties can be collected in
\begin{eqnarray}\label{product}
z^{M}z^{N}=(-)^{\epsilon(N)\epsilon(M)}z^{N}z^{M} 
\end{eqnarray}
where $\epsilon(M)\equiv\epsilon(z^M)$ is the so--called Grassmann parity, $\epsilon(\mu)=0$, $\epsilon(\alpha)=1=\epsilon(\dot{\alpha})$.

\section{Superfields on $D=4$ ${\cal N}=1$ superspace}\label{superfieldsintro}
Superfields are functions defined on superspace which means that superfields depend on both bosonic and fermionic coordinates.
As far as fermionic coordinates anticommute among themselves, they are nilpotent, $\theta'\theta'=-\theta'\theta'=0$.
This property implies that the series expansion of superfield in fermionic coordinates contains a finite number of terms.
The coefficients in that series, called superfield components, are the ordinary spacetime fields
\begin{eqnarray}
F(z)=F(x,\theta,\overline{\theta})=f(x)+\theta \phi(x)+\overline{\theta}\overline{\chi}(x)+\theta\theta m(x)+
\overline{\theta}\overline{\theta}n(x) \nonumber \\
+\theta\sigma^{m}\overline{\theta}v_{m}(x)+\theta\theta\overline{\theta}\overline{\lambda}(x)+
\overline{\theta}\overline{\theta}\theta\psi(x)+\theta\theta\overline{\theta}\overline{\theta}d(x).
\end{eqnarray}
The statistics (Grassmann parity) of the component fields appearing as a coefficient for even and odd powers of $\theta$, $\bar{\theta}$ are different 
(bosonic versus fermionic or vice versa) and depends on the statistics (Grassmann parity) of the superfield.

The rigid $D=4$ ${\cal N}=1$ supersymmetry is realized in superspace as constant translations of the fermionic coordinate, 
$\delta \theta=\epsilon$, $\delta\bar{\theta}=\bar{\epsilon}$, supplemented with the following transformations of bosonic coordinates, 
$\delta x^{\mu}=-i\epsilon\sigma^{\mu}\overline{\theta}+i\theta\sigma^{\mu}\overline{\epsilon}$ (see \cite{Volkov73,Volkov73+1}), to resume,
\begin{eqnarray}
&&\delta x^{\mu}=-i\epsilon\sigma^{\mu}\overline{\theta}+i\theta\sigma^{\mu}\overline{\epsilon}\nonumber\\
&&\delta \theta^{\alpha}=\epsilon^{\alpha}\nonumber\\
&&\delta\bar{\theta}^{\dot{\alpha}}=\bar{\epsilon}^{\dot{\alpha}}.
\end{eqnarray}
%

In general a superfield is a highly reducible representation of the supersymmetry. To extract an irreducible representation one usually 
needs to impose on superfield some equations in terms of fermionic covariant derivatives (see bellow) called constraints. 
One distinguishes on--shell and off--shell constraints. The on--shell constraints restrict the field content of superfield to physical 
fields and impose on these equations of motion.
The off--shell constraints do not impose on physical fields equations of motion and also leave nonvanishing not only physical component 
fields of the superfield, but also so--called auxiliary fields, the presence of which provides off-shell closure of the (super)algebra 
of supersymmetry transformations.

\section{Differential Superforms}\label{superforms}  
The convenience of differential forms is that they are manisfestly invariant under coordinate transformations. Let us introduce differentials of superspace 
coordinates, $dz^{M}=(dx^{\mu}, d\theta^{\alpha},d\bar{\theta}^{\dot{\alpha}})$ and the exterior product of these
\begin{eqnarray}
&&dz^{M}\wedge dz^{N}=-(-)^{(\epsilon(N)+1)(\epsilon(M)+1)}dz^{N}\wedge dz^{M}=(-)^{\epsilon(N)\epsilon(M)+1}dz^{N}\wedge dz^{M} \nonumber \\
&&dz^{M} z^{N}=-(-)^{\epsilon(N)\epsilon(M)}z^{N} dz^{M}, 
\end{eqnarray}
with $\epsilon(N)$ defined in (\ref{product}).
This implies 
\begin{eqnarray}
&&dx^{\mu}\wedge dx^{\nu}=-dx^{\nu}\wedge dx^{\mu},\quad d\theta^{\alpha}\wedge dx^{\mu}=-dx^{\mu}\wedge d\theta^{\alpha}, \nonumber\\ 
&&d\theta^{\alpha}\wedge d\theta^{\beta}=+d\theta^{\beta}\wedge d\theta^{\alpha},\quad etc.
\end{eqnarray}
The differential p--form in superspace has then the following structure
\begin{eqnarray}\label{p--form}
\Omega_p={1\over p!}dz^{M_{1}}\wedge ....\wedge dz^{M_{p}}\Omega_{M_{p}...M_{1}}(z)  
\end{eqnarray}
where $\Omega_{M_{p}...M_{1}}(z)$ is a superfield carrying supervector indices.
Being contracted with exterior product of $dz^{M}$, $\Omega_{M_{p}...M_{1}}(z)$ has to be graded--antisymmetric, 
$\Omega_{\{M_{p}...M_{1}]}(z)$, which implies
\begin{eqnarray}
&&\Omega_{...\mu...\nu...}(z)=-\Omega_{...\nu...\mu...}(z),\nonumber\\ 
&&\Omega_{...\mu...\beta...}(z)=-\Omega_{...\beta...\mu...}(z),\nonumber\\
&&\Omega_{...\alpha...\beta...}(z)=+\Omega_{...\beta...\alpha...}(z).
\end{eqnarray}
If the form $\Omega_p$ is bosonic, the superfields $\Omega_{M_{p}...M_{1}}(z)$ with odd number of spinorial indices will have a fermionic behavior while those with even number 
will have a bosonic one.

With these definitions the exterior product of differential superforms obey
\begin{eqnarray}\label{formulassuperforms}
&&(c_{1}\Omega_{p}+c_{2}{\Omega_{p}}')\wedge \Omega_{q}=c_{1}\Omega_{p}\wedge\Omega_{q}+c_{2}{\Omega_{p}}'\wedge\Omega_{q} \nonumber \\
&&\Omega_{p}\wedge\Omega_{q}=(-)^{pq+\epsilon(\Omega_{p})\epsilon(\Omega_{q})}\Omega_{q}\wedge\Omega_{p} \nonumber \\
&&\Omega_{p}\wedge(\Omega_{q}\wedge\Omega_{r})=(\Omega_{p}\wedge\Omega_{q})\wedge\Omega_{r}.
\end{eqnarray}

Once defined the superforms we introduce the exterior derivative, an operator which maps p-forms into (p+1)-forms 
\begin{eqnarray}
&&d\Omega_{p}={1\over p!}dz^{M_{1}}\wedge ....\wedge dz^{M_{p}}\wedge dz^{N}\frac{\partial}{\partial z^{N}}\Omega_{M_{p}...M_{1}}(z)= \nonumber \\ 
&&={1\over (p+1)!}dz^{M_{1}}\wedge ....\wedge dz^{M_{p+1}}(\partial_{M_{p+1}}\Omega_{M_{p}...M_{1}}(z)+(-)^{\epsilon(N)\epsilon(M)+1}\text{cyclic permutations}).\nonumber\\
\end{eqnarray}
Some useful properties involving exterior derivative one
\begin{eqnarray}\label{usefulderivative}
&&dd=0, \nonumber \\
&&d(c_{1}\Omega_{p}+c_{2}\Sigma_{p})=c_{1}d\Omega_{p}+c_{2}d\Sigma_{p}, \nonumber \\
&&d(\Omega_{p}\wedge\Sigma_{q})=\Omega_{p}\wedge d\Sigma_{q} +(-)^{q}d\Omega_{p}\wedge\Sigma_{q}.
\end{eqnarray}
\section{Vielbein}
Supervielbeins are 1-forms which define a supersymmetric generalization of local reference frame. We denote the 
bosonic and fermionic supervielbein one forms of  $\Sigma^{(4|4)}$ by
\begin{eqnarray}\label{4Ea}  E^a&=& dZ^M E_M^a(Z)\; , \quad
E^{{\alpha}}= dZ^M E_M^{{\alpha}}(Z)\; , \quad
 \bar{E}{}^{\dot\alpha}= dZ^M \bar{E}_M{}^{\dot\alpha}(Z)\; ,  \quad \\ \nonumber
 && a=0,1,2,3\,, \qquad \alpha =1,2\, , \qquad \dot{\alpha}=1,2\; .
\end{eqnarray}
Sometimes it is convenient to collect them in
\begin{eqnarray}\label{4EAM}
E^{A} = ( E^a, E^{\underline{\alpha}})  =( E^a, E^{{\alpha}}, \bar{E}{}^{{\dot\alpha}})=dZ^ME_M^A(Z)\; , \qquad
\end{eqnarray}
where ${\underline{\alpha}}=1,2,3,4$ can be understood as Majorana spinor index.

In superspace the torsion 2--forms are defined as the covariant exterior derivatives of the bosonic and fermionic 
supervielbein forms
\begin{eqnarray}\label{Ta:=}
T^a &:=& {\cal D}E^a=dE^a - E^b\wedge w_b{}^a= {1\over 2} E^B\wedge E^C T_{CB}{}^a
 \qquad \\
 \label{Talf:=}
T^\alpha &:=& {\cal D}E^\alpha=dE^\alpha - E^\beta\wedge w_\beta {}^\alpha= {1\over 2} E^B\wedge E^C T_{CB}{}^\alpha \; , \quad w_\beta {}^\alpha := {1\over 4} w^{ab} \sigma_{ab\beta} {}^\alpha\; , \quad \\
\label{Tdalf:=}
T^{\dot\alpha} &:=& {\cal D}E^{\dot\alpha}=dE^{\dot\alpha} - E^{\dot\beta}\wedge w_{\dot\beta} {}^{\dot\alpha}= {1\over 2} E^B\wedge E^C T_{CB}{}^{\dot\alpha} \; , \quad  w_{\dot\beta} {}^{\dot\alpha} := {1\over 4} w^{ab} \tilde{\sigma}_{ab} {}^{\dot\alpha}{}_{\dot\beta} \; , \quad
 \qquad
\end{eqnarray}
where $w^{ab}=-w^{ba}=dZ^{M}w_M^{ab}(Z)$ is the spin connection 1-form 
, $\sigma^{ab}{}_{\beta} {}^\alpha=\sigma^{[a}\tilde{\sigma}^{b]}$ 
and $\tilde{\sigma}{}^{ab}{}^{\dot\alpha}{}_{\dot\beta}=\tilde{\sigma}^{[a}{\sigma}^{b]} $ are antisymmetrized 
products of the relativistic Pauli matrices (see Appendix A) and $d$, $\wedge$ are exterior derivative and exterior
product of differential forms previously defined in section \ref{superforms}.
This later is antisymmetric for bosonic one forms, $E^a\wedge E^b=- E^b\wedge E^a$, 
symmetric for two fermionic one forms, $E^{ \alpha} \wedge E^{ \beta} =E^{ \beta} \wedge E^{ \alpha}$,  
and again antisymmetric for the product of bosonic and fermionic one  forms,  
$E^a\wedge E^{\beta}=- E^{ \beta}\wedge E^a$ (see \ref{formulassuperforms}).

The torsion 2-forms obey the Bianchi identities
\begin{eqnarray}\label{BI=DT}
{\cal D}T^a + E^b \wedge R_b{}^a=0 \; , \qquad {\cal D}T^\alpha + E^\beta \wedge R_\beta{}^\alpha=0 \; , \qquad
 {\cal D}T^{\dot{\alpha}} + E^{\dot{\beta}} \wedge R_{\dot{\beta}} {}^{\dot{\alpha}} =0 \; , \qquad
 \qquad
\end{eqnarray}
where
\begin{eqnarray}\label{Rab}
R^{ab}= (dw-w\wedge w)^{ab}={1\over 2} E^B\wedge E^C R_{CB}{}^{ab}\; \qquad
\qquad
\end{eqnarray}
is the curvature 2-form. Its Bianchi identities read $DR^{ab}=0$.
\subsection{Flat Superspace} 
 In flat $D=4$ ${\cal N}=1$ $\Sigma^{(4|4)}$ superspace, supervielbeins obey the constraints
 \begin{eqnarray}
 \label{Ta=4D} & T^{a}:=dE^{ {a}} =
 -2i{E}\wedge \sigma^{ {a}} {\bar E}  \; , \qquad  T^\alpha:= dE^{ \alpha} = 0  \; ,  \qquad  
T^{\dot{\alpha}}:=d\bar{E}{}^{ \dot{\alpha}} = 0 \;,
 \end{eqnarray}
which can be solved by
 \begin{eqnarray}\label{Eadx-i}
 {E}^{ {a}}= dx^a - i d\theta^{\alpha} \sigma^a_{\alpha\dot{\alpha}} \bar{\theta}{}^{\dot{\alpha}} + 
i\theta^{\alpha} \sigma^a_{\alpha\dot{\alpha}} d\bar{\theta}{}^{\dot{\alpha}}\; ,   \qquad E^{\alpha}= 
 d\theta^{\alpha}\; , \qquad \bar{E}{}^{\dot{\alpha}}=d\bar{\theta}{}^{\dot{\alpha}}\;\qquad
 \end{eqnarray}
expressing the supervielbein in terms of superspace coordinates (\ref{coordinates}).
Decomposing the exterior derivative on the supervielbein basis
\begin{eqnarray}\label{d=ED}
d= E^\alpha D_\alpha + \bar{E}{}^{\dot{\alpha}}\bar{D}_{\dot{\alpha}} + E^a D_a\; , \qquad
\end{eqnarray}
we obtain the expressions for supersymmetric covariant derivatives,
\begin{eqnarray} \label{Dalpha=}
 D_a=\partial_a \; , \qquad
 D_\alpha= \partial_\alpha + i (\sigma^a\bar{\theta})_\alpha \partial_a \; , \qquad
 \bar{D}_{\dot{\alpha}} = \bar{\partial}_{\dot{\alpha}} + i (\theta\sigma^a)_{\dot{\alpha}} \partial_a =- (D_\alpha)^* \; . \qquad
\end{eqnarray}
These obey the superalgebra with only one nontrivial (anti-)commutation relation,
\begin{eqnarray}\label{(DfbDf)=}
{}\{ D_\alpha ,  \bar{D}_{\dot{\alpha}} \}= 2i \sigma^a_{\alpha\dot{\alpha}} \partial_a \; , \qquad
\end{eqnarray}
while the other (anti-)commutators vanish
\begin{eqnarray}\label{(DfbDf)=}
&&\{D_{\alpha},D_{\beta}\}=0\; \qquad \{\overline{D}_{\dot{\alpha}},\overline{D}_{\dot{\beta}}\}=0 \nonumber\\
&&[\partial_{\mu},D_{\alpha}]=0=[\partial_{\mu},\overline{D}_{\dot{\alpha}}]\; \qquad [\partial_{\mu},\partial_{\nu}]=0\; . \qquad
\end{eqnarray}

\section{Supergravity}
Supergravity, a supersymmetric version of Einstein's gravity, has local supersymmetry as gauge symmetry.
As mentioned earlier in this chapter, supersymmetry can be realized on--shell or off--shell which means
the supersymmetry algebra closes with/without using the equations of motion. This also applies to local supersymmetry.
There are several different off--shell formulations of supergravity which differ by choices of auxiliary field sector. 
In this thesis we will use the so--called minimal supergravity.
In chapter \ref{chaptersugra+membrane} we will use minimal $N=1$ $D=4$ off--shell supergravity in its superspace formulation which 
we are going to describe now.

\subsection{Minimal Supergravity}
The list of superspace constraints of minimal supergravity \cite{WZ77,OS78,BW+1,Siegel79} is \footnote{A minimal complete set of 
superspace constraints for the minimal supergravity 
multiplet \cite{minSG} can be found, {\it e.g.}, in 
 \cite{1001,BW,BZ87}; see \cite{Duff1980,AlgC} for a discussion of the algebraic origin of the supergravity 
constraints.} 
\begin{eqnarray}\label{contrmin}
&&T_{\alpha \dot{\beta}}{}^a= -2i\sigma^a_{\alpha \dot{\beta}},\nonumber\\
&&T_{\alpha\beta}{}^A=0=T_{\dot{\alpha}\dot{\beta}}{}^A,\nonumber\\
&&T_{\alpha\dot{\beta}}{}^{\dot{\gamma}}=0,\nonumber\\
&&T_{\alpha b}{}^c=0,\nonumber\\
&&R_{\alpha\dot{\beta}}{}^{ab}=0.
\end{eqnarray}
With these constraints, the identities (\ref{BI=DT}) express the torsion 
and curvature forms through the set of main superfields
\begin{eqnarray}\label{4WGa}
& G_a:= 2i (T_{a\beta}{}^\beta -T_{a\dot{\beta}}{}^{\dot{\beta}})\; , 
\\ \label{4WR} 
& \bar{R}:= -{1\over 3} R_{{\alpha}{\beta}}{}^{{\alpha}{\beta}} = (R)^* \; , 
\qquad \\ 
\label{4WchW}
& W^{\alpha\beta\gamma} := 
4i \tilde{\sigma}^{c\dot{\gamma}\gamma} R_{\dot{\gamma}c}{}^{\alpha\beta}
= W^{(\alpha\beta\gamma)}
= (\bar{W}^{\dot{\alpha}\dot{\beta}\dot{\gamma}})^* \; . 
\end{eqnarray}
The final expressions for the superspace torsion 2--forms are,
\begin{eqnarray}\label{4WTa=:} 
T^a & 
=- 2i\sigma^a_{\alpha\dot{\alpha}} E^\alpha \wedge \bar{E}^{\dot{\alpha}} 
-{1\over 8} E^b \wedge E^c \varepsilon^a{}_{bcd} 
G^d \; ,  
\\ 
\label{4WTal=:}
T^{\alpha} & 
= {i\over 8} E^c \wedge E^{\beta} (\sigma_c\tilde{\sigma}_d)_{\beta}
{}^{\alpha} G^d  - \hspace{2.3cm} 
\nonumber \\ &  -{i\over 8} E^c \wedge \bar{E}^{\dot{\beta}} 
\epsilon^{\alpha\beta}\sigma_{c\beta\dot{\beta}}R 
+
 {1\over 2} E^c \wedge E^b \; T_{bc}{}^{\alpha}\; , \\ 
\label{4WTdA=:}
T^{\dot{\alpha}} 
& = {i\over 8} E^c \wedge E^{\beta} \epsilon^{\dot{\alpha}\dot{\beta}}
\sigma_{c\beta\dot{\beta}}
\bar{R} - \hspace{2.3cm}  
\;   
\nonumber \\ &   
-{i\over 8} E^c \wedge \bar{E}^{\dot{\beta}} 
(\tilde{\sigma}_d\sigma_c)^{\dot{\alpha}}{}_{\dot{\beta}}
\, G^d +
 {1\over 2} E^c \wedge E^b \; T_{bc}{}^{\dot{\alpha}}\; .
\end{eqnarray}
The Bianchi identities also imply the following equations for the main superfields,
\begin{eqnarray}
\label{chR:}
& {\cal D}_\alpha \bar{R}=0\; , \qquad \bar{{\cal D}}_{\dot{\alpha}} {R}=0\; ,
 \\ 
\label{chW:} & \bar{{\cal D}}_{\dot{\alpha}} W^{\alpha\beta\gamma}= 0\; , 
\qquad 
{{\cal D}}_{{\alpha}} \bar{W}^{\dot{\alpha}\dot{\beta}\dot{\gamma}}= 0\;,
\\ 
\label{DG=DR:} &
\bar{{\cal D}}^{\dot{\alpha}}G_{{\alpha}\dot{\alpha}}= {\cal D}_{\alpha} R 
\; , \qquad 
{{\cal D}}^{{\alpha}}G_{{\alpha}\dot{\alpha}}= 
\bar{{\cal D}}_{\dot{\alpha}} \bar{R} \; , 
\\ \label{DW=DG:} &
{{\cal D}}_{{\gamma}}W^{{\alpha}{\beta}{\gamma}}= 
\bar{{\cal D}}_{\dot{\gamma}} {{\cal D}}^{({\alpha}}G^{{\beta})\dot{\gamma}}
\; , \qquad \nonumber \\ & 
\bar{{\cal D}}_{\dot{\gamma}} \bar{W}^{\dot{\alpha}\dot{\beta}\dot{\gamma}}
= {{\cal D}}_{{\gamma}} \bar{{\cal D}}^{(\dot{\alpha}|} 
G^{{\gamma}|\dot{\beta})} \; . \qquad
\end{eqnarray}
The superspace Riemann curvature 2-form can be decomposed on the anti-self-dual and self-dual parts enclosed in symmetric spin--tensor 
2--forms $R^{\alpha\beta}$ and $R^{\dot{\alpha}\dot{\beta}}=(R^{\alpha\beta})^*$. As a consequence of the constraints (\ref{contrmin})
\begin{eqnarray}\label{4WR=:}
R^{\alpha\beta} & \equiv  dw^{\alpha\beta} - w^{\alpha\gamma} 
\wedge w_\gamma{}^\beta \equiv {1 \over 4} 
R^{ab} (\sigma_a\tilde{\sigma}_b)^{\alpha\beta}= \nonumber
\\ 
& 
= -{1\over 2} E^\alpha \wedge E^\beta \bar{R} 
-{i\over 8} E^c \wedge E^{(\alpha}\,  
\tilde{\sigma}_c{}^{\dot{\gamma}\beta)} \bar{{\cal D}}_{\dot{\gamma}}\bar{R}+ 
\nonumber \\ 
& 
+{i\over 8} E^c \wedge E^{\gamma} 
(\sigma_c\tilde{\sigma}_d)_{\gamma}{}^{(\beta} {\cal D}^{\alpha)} G^d -  
\nonumber \\ & 
-{i\over 8}  E^c \wedge \bar{E}^{\dot{\beta}} \sigma_{c\gamma\dot{\beta}} 
W^{\alpha\beta\gamma} + {1\over 2} E^d \wedge E^c R_{cd}{}^{\alpha\beta} \; .
\end{eqnarray}
In our conventions the spinor covariant derivatives are defined by the following decomposition of 
the covariant differential 
${D}$ 
\begin{eqnarray}\label{4WD}
&&{D}:=  E^A {\cal D}_A = E^a {\cal D}_a +  E^{\underline{\alpha}} 
{\cal D}_{\underline{\alpha}}= 
\nonumber \\
&& =E^a {\cal D}_a +  E^{\alpha} {\cal D}_{\alpha}+ 
\bar{E}^{\dot{\alpha}} \bar{{\cal D}}_{\dot{\alpha}} \; . 
\end{eqnarray}
(hence, 
$({D}_{\alpha})^{*}=-\bar{D}_{\dot{\alpha}}$).
The algebra of covariant derivatives ${D}_A$, Eq. (\ref{4WD}), 
is encoded in the Ricci identities
\begin{eqnarray}
\label{RI}
& {D}{D}V_{A}= 
R_{A}^{\; B} \, V_{B} \; \leftrightarrow 
\begin{cases} {D}{D}V_{a}= 
R_{a}^{\; b} \, V_{b} \; , \cr 
{D}{D}V_{\alpha}= 
R_{\alpha}^{\; \beta} \, V_{\beta} \; , \cr 
{D}{D}V_{\dot{\alpha}}= 
R_{\dot{\alpha}}^{\; \dot{\beta}} \, V_{\dot{\beta}} \; , \cr \end{cases}
\end{eqnarray} 
where $V_A = (V_{a}, V_{\alpha},  V_{\dot{\alpha}})$ is an arbitrary 
supervector with tangent superspace Lorentz indices. 
If we decompose them on the basic 2-forms $E^A \wedge E^B$, one finds 
(see \cite{1001,BW}) 
\begin{eqnarray}\label{algD} & 
 [ {\cal D}_{A}\; , {\cal D}_{B}\}V_{C} = - 
T_{{A}{B}}{}^{{D}} {{\cal D}}_{{D}}V_{{C}}+ 
R_{{A}{B}\;{C}}{}^{{D}}\, V_{{D}} \; . 
\end{eqnarray} 
When the constraints 
(\ref{4WTa=:}), (\ref{4WTal=:}),  (\ref{4WTdA=:}), 
(\ref{chR:}), (\ref{chW:}), (\ref{DG=DR:}), (\ref{4WR=:}) are taken into account, 
Eqs. (\ref{RI}) (or   (\ref{algD}))   
 imply 
\begin{eqnarray}
\label{(DD)}
& \{ {\cal D}_\alpha , {\cal D}_\beta \} \; V_\gamma = - 
\bar{R} \epsilon_{\gamma (\alpha } V_{\beta)}\; , 
\\ 
\label{(DD)1}
& \{ {\cal D}_\alpha , {\cal D}_\beta \} \; V^\gamma = - 
\bar{R} V_{(\alpha}\delta_{\beta)}{}^{\gamma}\; , 
\\ 
\label{(DbD)} 
& \{ {\cal D}_\alpha , \bar{\cal D}_{\dot{\beta}}\}  = 
2i \sigma^a_{\alpha\dot{\beta}}{\cal D}_a \equiv 
2i {\cal D}_{\alpha\dot{\beta}}
\; ,  \qquad etc. 
\end{eqnarray}
The Bianchi identities also allow to find the expression for superfield generalization of the gravitino field strength, $T_{bc}{}^{\alpha}(Z)$,
\begin{eqnarray}\label{Tabg} T_{{\alpha}\dot{\alpha}\; \beta \dot{\beta }\; {\gamma}} &
\equiv   \sigma^a_{{\alpha}\dot{\alpha}} \sigma^b_{\beta \dot{\beta }} \epsilon_{{\gamma}{\delta}}
T_{ab}{}^{{\delta}}=  -{1\over 8}  \epsilon_{{\alpha}{\beta}} {\bar{{\cal D}}}_{(\dot{\alpha}|} G_{\gamma
|\dot{\beta})} - {1\over 8} \epsilon_{\dot{\alpha}\dot{\beta}}[W_{\alpha \beta\gamma} -
2\epsilon_{\gamma (\alpha}{\cal D}_{\beta)} R]
 \;. \end{eqnarray}
As a result, the superfield generalization of the left hand side of the supergravity Rarita--Schwinger 
equation reads
\begin{eqnarray}\label{SGRS=off}
\epsilon^{abcd}T_{bc}{}^{\alpha}\sigma_{d\alpha\dot{\alpha}} ={i\over 8}
\tilde{\sigma}^{a\dot{\beta}\beta} \bar{{\cal D}}_{(\dot{\beta}|} G_{\beta|\dot{\alpha})} + {3i\over 8}
{\sigma}^a_{\beta \dot{\alpha}} {\cal D}^{\beta}R \;.
\end{eqnarray}
The superfield generalization of the Ricci tensor is
\begin{eqnarray} \label{RRicci}
&&R_{bc}{}^{ac} = {1\over 32} ({{\cal D}}^{{\beta}} \bar{{\cal D}}^{(\dot{\alpha}|}
G^{{\alpha}|\dot{\beta})} - \bar{{\cal D}}^{\dot{\beta}} {{\cal D}}^{({\beta}}G^{{\alpha})\dot{\alpha}})
\sigma^a_{\alpha\dot{\alpha}}\sigma_{b\beta\dot{\beta}} -\\\nonumber
&&- {3\over 64} (\bar{{\cal D}}\bar{{\cal D}}\bar{R} + {{\cal D}}{{\cal D}}{R}- 4 R\bar{R})\delta_b^a\; .
\end{eqnarray}
This suggests that superfield supergravity equation should have the form
 \begin{eqnarray}\label{SGeqmG:}
 && G_a =0 \; , \qquad
\\ \label{SGeqmR:}
 && R=0 \; ,  \qquad \bar{R}=0\; . \qquad
\end{eqnarray}
\subsection{Superfield supergravity action and admissible variation of constrained supervielbein}

The superfield action of the minimal off-shell formulation of $D=4$, ${\cal N}=1$  supergravity \cite{WZ78} is given by 
the superdeterminant (or Berezinian) of the matrix of supervielbein coefficients,  $E_M^A(Z)$ in (\ref{4EAM}), 
which obey the set of supergravity constraints (\ref{4WTa=:}), (\ref{4WTal=:}), (\ref{4WTdA=:}),
\begin{eqnarray}\label{SGact:} S_{SG} = \int d^8Z \;
E :=  \int d^4 x \tilde{d}^4\theta \; sdet(E_M^A) \; . \qquad
\end{eqnarray}


One can obtain the supergravity superfield  equations (\ref{SGeqmG:}), (\ref{SGeqmR:}) by varying the superspace action 
(\ref{SGact:}).  This is not straightforward because the supervielbein superfields are restricted by the constraints (\ref{4WTa=:}), 
(\ref{4WTal=:}),(\ref{4WTdA=:}). There are two basic ways to solve this problem. The first consists in solving the  superspace supergravity 
constraints in terms of unconstrained superfields-- {\it pre-potentials} \cite{OS78,Siegel:1978nn,Siegel79,OS80}  -- the set 
of which in the case of minimal supergravity can be  restricted to the axial vector superfield 
${\cal H}^\mu (Z)= {\cal H}^\mu (x,\theta,\bar{\theta})$ \cite{OS78}  and the chiral compensator 
superfield $\Phi(x+i{\cal H}(Z), \theta)$ \cite{Siegel:1978nn}. 

Another way, called the Wess--Zumino approach to superfield supergravity \cite{WZ77,GWZ78,WZ78,WZ79,BW}, does not imply to 
solve the constraints but rather to solve the equations which appear as a result of requiring that the variations of the 
supervielbein and spin connection
\begin{eqnarray}\label{varEMA} \delta
E_M^{\, A}(Z) = E_M^{\, B} {\cal K}_{B}^{\, A} (\delta )\; , \quad \delta w_M^{ab}(Z) = E_M^{\, C}
u_{{C}}^{ab} (\delta )\; , \end{eqnarray}
preserve the superspace supergravity constraints  \cite{WZ78}, Eqs. (\ref{4WTa=:}), (\ref{4WTal=:}), (\ref{4WTdA=:}).

Actually this procedure of finding ${\it admissible}$ variations of the supervielbein and superspace spin connection is a 
linearized but covariantized version of the constraint solution used in pre-potential approach. The independent parameters 
of admissible variations clearly reflect the pre-potential structure of the off-shell supergravity. In the case of minimal 
supergravity their set contains \cite{WZ78} (see also \cite{BdAIL03}) $\delta H^a$, corresponding to the variation of the axial 
vector pre-potential of \cite{OS78}, and complex scalar variations $\delta {\cal U}$ and $\delta \bar{\cal U}$  entering  (\ref{varEMA}) only under the 
action of chiral projectors (as $({\cal D}{\cal D}-\bar{R})\delta {\cal U}$ and $(\bar{\cal D}\bar{\cal D}-{R})\delta \bar{{\cal U}}$) 
and thus corresponding to the variations of complex pre-potential of the chiral compensator of the minimal supergravity.
The admissible variations of supervielbein read \cite{WZ78,BdAIL03}
\begin{eqnarray}\label{varEa1} 
&&\delta E^{a} =E^a (\Lambda (\delta ) + \bar{\Lambda} (\delta ))
 - {1\over 4} E^b \tilde{\sigma}_b^{ \dot{\alpha} {\alpha} }
[{\cal D}_{{\alpha}}, \bar{{\cal D}}_{\dot{\alpha}}] \delta H^a +
 i E^{\alpha} {\cal D}_{{\alpha}}\delta H^a   - i \bar{E}^{\dot{\alpha}}\bar{{\cal
D}}_{\dot{\alpha}} \delta H^a \; ,\nonumber \\
\\
\label{varEal1}
&& \delta E^{\alpha} = E^a \Xi_a^{\alpha}(\delta ) +
E^{\alpha} \Lambda (\delta ) + {1\over 8} \bar{E}^{\dot{\alpha}} R \sigma_a{}_{\dot{\alpha}}{}^{\alpha}
\delta H^a \; ,  \end{eqnarray}
where
\begin{eqnarray}
\label{2Lb+*Lb} & 2\Lambda (\delta ) + \bar{\Lambda} (\delta )  = {1\over 4} \tilde{\sigma}_a^{
\dot{\alpha} {\alpha} } {\cal D}_{{\alpha}} \bar{{\cal D}}_{\dot{\alpha}}\delta H^a + {1\over 8} G_a
\delta H^a +   3 ( {\cal D}{\cal D}- \bar{R})\delta {\cal U} \;
 \end{eqnarray}
and the explicit expression for $\Xi_a^{\alpha}(\delta )$ in (\ref{varEal1}) can be found in \cite{BdAIL03}. Neither that 
nor the explicit expressions for the admissible variations of spin connection in (\ref{varEMA}) will be needed for our 
discussion along this thesis.

Indeed, the variation of superdeterminant of the supervielbein of the minimal supergravity superspace can be calculated 
using (\ref{varEa1}), (\ref{varEal1}) only and reads
 (see \cite{WZ78})
\begin{eqnarray}\label{varsdE}
&&\delta E = E [ - {1\over 12} \tilde{\sigma}_a^{\dot{\alpha}\alpha} [ {\cal D}_{\alpha}, \bar{\cal
D}_{\dot{\alpha}}] \delta H^a + {1\over 6} G_a \; \delta H^a +\nonumber\\
&&+  2(\bar{\cal D} \bar{\cal D} - R) \delta \bar{{\cal U}} +  2({\cal D} {\cal D} - \bar{R}) \delta {\cal U}] \; . 
\end{eqnarray}
Taking into account the identity \cite{WZ78}
\begin{eqnarray} \label{idd80} \int d^8Z\,   E \; ({\cal D}_{A}\xi^{A} + \xi^B T_{BA}{}^A) (-1)^{A}\; \equiv 0 \; ,
\end{eqnarray}  one  finds the variation  of the minimal supergravity action  (\ref{SGact:})
 \begin{eqnarray}\label{vSGsf} & \delta S_{SG} =  \int
d^8Z E\;  [{1\over 6} G_a \; \delta H^a -  2 R\; \delta \bar{{\cal U}} -2 \bar{R}\; \delta {\cal U}] \; . \quad
\end{eqnarray}
This clearly produces the superfield supergravity equations of the form (\ref{SGeqmG:}), (\ref{SGeqmR:}) which result in the 
Rarita--Schwinger equation and Einstein equation without cosmological constant
\begin{eqnarray}
\label{SGRS=0}
&& \epsilon^{abcd}T_{bc}{}^{\alpha}\sigma_{d\alpha\dot{\alpha}} = 0 \; , \qquad
\\ \label{RRicci=0}
&& R_{bc}{}^{ac} = 0 \; .
\end{eqnarray}

\chapter{$D=4$ ${\cal N}=1$ Supermembrane interaction with dynamical Scalar Superfield}\label{chapterscalar}
\thispagestyle{chapter} 

\initial{I}n this chapter we will present the action for the dynamical $D=4$ ${\cal N}=1$ supermembrane in interaction with 
a dynamical scalar multiplet. We will show that this action is invariant under fermionic transformations called $\kappa$--symmetry  
reflecting that its ground state preserves a part of target space supersymmetry. 
The possibilty for constructing such an action is related to the existence of a nontrivial Chevalley-Eilenberg 3--cocycle ({\it i.e}  a 
supersymmetric invariant closed 4--form \cite{JdA+PKT89,Jose+PaulPRL:89}) constructed from the scalar supermultiplet which enters 
the Wess-Zumino part of the action allowing us to couple the $D=4$ ${\cal N}=1$ supermembrane to scalar supermultiplet.
To this end we will review the well known description of the scalar supermultiplet by chiral superfield in superspace.
We present the supermembrane action in the {\it off--shell} scalar supermultiplet background and find the equations of motion.
Then we discuss the special scalar supermultiplet contructed from real (instead complex) prepotential, which allows a dual description 
by 3--form potential $C_{3}{}^{\prime}$ in flat $D=4$ ${\cal N}=1$ superspace. Just this special scalar supermultiplet can be coupled 
to supermembrane (beyond the background field approximation) as far as the above $C_{3}{}^{\prime}$ pulled--back to supermembrane 
worldvolume $W^3$ serves to construct the Wess--Zumino term of the supermembrane action.

We describe the dynamical system of the special scalar supermultiplet interacting with supermembrane and obtain the equations of 
motion with supermembrane source.
In the simplest case in which the scalar multiplet part of the action contains only the simplest kinetic term we also 
extract the equations of motion for the physical fields. 
We solve these dynamical field equations for the physical fields at leading order in supermembrane tension.
We conclude this chapter by disscusing the inclusion of nontrivial superpotential and the relation with known domain wall solutions.

\bigskip

\section{Free supermembrane in flat $D=4$ ${\cal N}=1$ superspace}\label{freesupermembrane}

The possibility to construct the $D=4$, ${\cal N}=1$ supermembrane action is related to that in $D=4$, ${\cal N}=1$ superspace 
there exists the following supersymmetric invariant closed 4-form\footnote{Recall that in our notation the exterior derivative 
acts from the right, so that for any $p$-form $\Omega_p$ and $q$-form $\Omega_q$, $\; d(\Omega_p\wedge \Omega_q)= \Omega_p\wedge d\Omega_q + (-)^q d\Omega_p\wedge \Omega_q$. 
See \ref{usefulderivative}}
\begin{eqnarray}\label{h4:=C-E}
h_4=dc_3 :=- {i\over 4} E^b\wedge E^a\wedge E^\alpha\wedge E^\beta \sigma_{ab}{}_{\alpha\beta} + {i\over 4} E^b\wedge E^a \wedge \bar{E}^{\dot{\alpha}}\wedge \bar{E}^{\dot{\beta}} \tilde{\sigma}_{ab}{}_{\dot{\alpha}\dot{\beta}}\; .  \qquad
\end{eqnarray}
This describes a 3-cocycle which is nontrivial in Chevalley-Eilenberg (CE) cohomology \cite{JdA+PKT89,Jose+PaulPRL:89}, which 
implies that $h_4$ is a supersymmetric invariant closed four form, $dh_4=0$ and, despite it can be expressed as an exterior 
derivative of a 3-form, $h_4=dc_3$ (and, hence, is trivial cocycle of de Rahm cohomology), the corresponding 3-form $c_3$ is not 
invariant under supersymmetry.

The action for a free supermembrane in $D=4$, ${\cal N}=1$  superspace reads \cite{4Dsuperm}
\begin{eqnarray}\label{SM2=4D}
 S_{p=2} &=&  {1\over 2}\int d^3 \xi \sqrt{ g} -   \int\limits_{W^3} \hat{c}_3 = \; \qquad \nonumber \\ &=&   -{1\over 6}\int\limits_{W^3}  *\hat{E}_a\wedge \hat{E}^a  -   \int\limits_{W^3} \hat{c}_3\;   ,  \qquad
\qquad
\end{eqnarray}
where, in the first line $g= det(g_{mn})$ is the determinant of the induced metric,
\begin{eqnarray}\label{g=EE}
g_{mn}= \hat{E}_m^{a}\eta_{ab} \hat{E}_n^{b}\; , \qquad
\hat{E}_m^{a}:= \partial_m \hat{x}^a - i \partial_m\hat{\theta}{}^{\alpha} \sigma^a_{\alpha\dot{\alpha}} \hat{\bar{\theta}}{}^{\dot{\alpha}} + i\hat{\theta}{}^{\alpha} \sigma^a_{\alpha\dot{\alpha}} \partial_m\hat{\bar{\theta}}{}^{\dot{\alpha}}\; , \qquad
\end{eqnarray}
$W^3$ is the supermembrane worldvolume the embedding of which into the target superspace $\Sigma^{(4|4)}$ is defined 
parametrically by the coordinate functions
$\hat{z}{}^{{ {M}}}(\xi)= (\hat{x}{}^{a}(\xi)\, ,
\hat{\theta}^{{\alpha}}(\xi), \hat{\bar{\theta}}{}^{\dot{{\alpha}}}(\xi))$; $\xi^m= (\xi^0, \xi^1, \xi^2)$ are local 
coordinates on $W^3$,
\begin{eqnarray}\label{W3inS44}
W^3\; \subset  \Sigma^{(4|4)}\; : \qquad z^M= \hat{z}{}^{{ {M}}}(\xi)= (\hat{x}{}^{a}(\xi)\, ,
\hat{\theta}^{{\alpha}}(\xi), \hat{\bar{\theta}}{}^{\dot{{\alpha}}}(\xi))\; . \qquad
\end{eqnarray}
Finally,
\begin{eqnarray}\label{hc3=4D}
\hat{c}_3:= {1\over 3!} \hat{E}^{a_3} \wedge \hat{E}^{a_2} \wedge   \hat{E}^{a_1}
c_{a_1a_2a_3}(\hat{Z}) =
{1\over 3!}
d\xi^{m_3} \wedge
d\xi^{m_2} \wedge
d\xi^{m_1} \hat{c}_{m_1m_2m_3}= \qquad \nonumber \\ = -{1\over 6} d^3\xi \epsilon^{m_1m_2m_3}\hat{c}_{m_1m_2m_3} \;  \qquad
\end{eqnarray}
is the pull--back of the 3-form defined in Eq. (\ref{h4:=C-E}) to $W^3$, so that  the second, Wess--Zumino part of the action 
can be written in the form of (see \cite{4Dsuperm})  
$\int\limits_{W^3} \hat{c}_3= -{1\over 6}\int d^3\xi \epsilon^{m_1m_2m_3}\hat{c}_{m_1m_2m_3}$.

Here we consider only the case of closed supermembrane so that the worldvolume $W^3$ has no boundary, 
$\partial W^3= 0\!\!\! /$, and $\int_{W^3}d(...)=0$.
Then we do not need in the explicit form of $\hat{c}_{m_1m_2m_3}$ in (\ref{hc3=4D}) as far as variation of its integral 
in (\ref{SM2=4D}) can be calculated (using the Lie derivative formula, $\delta c_3= i_\delta dc_3+ di_\delta c_3$) through 
its exterior derivative, the pull--back $\hat{h}_4:= h_4(\hat{Z})$ of the CE cocycle (\ref{h4:=C-E}), $h_4=dc_3$.

In the second line of Eq. (\ref{SM2=4D}) we have written the first, Nambu-Goto term of the action as an integral of a 
differential three form. This is constructed from the pull--back of the bosonic vielbein form
\begin{eqnarray}\label{hEa=dxiE}
\hat{E}{}^a= d\xi^m \hat{E}_m^{a}\; , \qquad
\hat{E}_m^{a}:= \partial_m \hat{x}^a - i \partial_m\hat{\theta}{}^{\alpha} \sigma^a_{\alpha\dot{\alpha}} \hat{\bar{\theta}}{}^{\dot{\alpha}} + i\hat{\theta}{}^{\alpha} \sigma^a_{\alpha\dot{\alpha}} \partial_m\hat{\bar{\theta}}{}^{\dot{\alpha}}\; , \qquad
\end{eqnarray}
using the worldvolume Hodge star operation,
\begin{eqnarray}\label{*Ea:=}
*\hat{E}^a:= {1\over 2}d\xi^m\wedge d\xi^n\sqrt{g}\epsilon_{mnk}g^{kl}\hat{E}_l^a \; . \qquad
\end{eqnarray}

The action (\ref{SM2=4D}) is invariant under the local fermionic $\kappa$--symmetry transformations. These have the form of
\begin{eqnarray}\label{kappaX=}
\delta_\kappa x^\mu =
i \kappa^{\alpha} \sigma^\mu_{\alpha\dot{\alpha}} \bar{\theta}{}^{\dot{\alpha}} - i\theta^{\alpha} \sigma^\mu_{\alpha\dot{\alpha}}  \bar{\kappa}{}^{\dot{\alpha}} \; ,  \qquad \delta_\kappa \theta^{\alpha}= \kappa^{\alpha}\; , \qquad \delta_\kappa \bar{\theta}{}^{\dot{\alpha}} = \bar{\kappa}{}^{\dot{\alpha}} \; , \qquad
\end{eqnarray}
where the spinorial fermionic parameter $\kappa^\alpha =\kappa^\alpha (\xi) = (\bar{\kappa}{}^{\dot{\alpha}})^*$ has actually only 
two independent components because it obeys the equations
\begin{eqnarray}\label{bk=kbg}
 \bar{\kappa}_{\dot{\alpha}}= \kappa^\beta {\bar{\gamma}}_{\beta\dot{\alpha}}\qquad \Leftrightarrow \qquad
 \kappa^\alpha= \bar{\kappa}_{\dot{\alpha}}
 \tilde{\bar{\gamma}}{}^{\dot{\beta}\alpha} \;  \quad
\end{eqnarray}
with
\begin{eqnarray}\label{bg=}
 {\bar{\gamma}}_{\beta\dot{\alpha}}= \epsilon_{\beta\alpha} \epsilon_{\dot{\alpha}\dot{\beta}}
 \tilde{\bar{\gamma}}{}^{\dot{\beta}\alpha}  = {i\over 3! \sqrt{g}}\sigma^a_{\beta\dot{\alpha}}\epsilon_{abcd}
 \epsilon^{mnk} \hat{E}_m^b\hat{E}_n^c\hat{E}_k^d\;  . \qquad
\end{eqnarray}
By construction, the matrix ${\bar{\gamma}}$ obeys
\begin{eqnarray}\label{gbg=I}
 {\bar{\gamma}}_{\beta\dot{\beta}}
 \tilde{\bar{\gamma}}{}^{\dot{\beta}\alpha}  = \delta_{\beta}
{}^{\alpha}\;  \qquad
\end{eqnarray}
which makes two equations in (\ref{bk=kbg}) equivalent.

To prove the $\kappa$--symmetry one has to use the identities
\begin{eqnarray}\label{EEEs=}
{1\over 2} \hat{E}{}^c \wedge \hat{E}{}^b \wedge \hat{E}{}^\alpha \sigma_{bc\alpha\beta}=
*\hat{E}_a\wedge \hat{E}{}^\alpha (\sigma^a\tilde{\bar{\gamma}})_{\alpha\beta}
\;  \qquad
\end{eqnarray}
which allows to present the variation of the kinetic, Nambu-Goto type, and the Wess--Zumino terms in similar form.

It is convenient to write the $\kappa$--symmetry transformations in the form of
\begin{eqnarray}\label{ikE0=}
i_\kappa \hat{E}{}^a := \delta_\kappa \hat{Z}^M E_M{}^a(\hat{Z}) =0 \; , \qquad \begin{cases} i_\kappa \hat{E}{}^\alpha := \delta_\kappa \hat{Z}^M E_M{}^\alpha (\hat{Z})
 = \kappa^\alpha= \bar{\kappa}_{\dot{\alpha}}
 \tilde{\bar{\gamma}}{}^{\dot{\beta}\alpha} \; , \cr i_\kappa \hat{\bar{E}}{}^{\dot\alpha}:= \delta_\kappa \hat{Z}^M E_M{}^{\dot{\alpha}}(\hat{Z})= \bar{\kappa}_{\dot{\alpha}}= \kappa^\beta {\bar{\gamma}}_{\beta\dot{\alpha}} \; .\qquad\end{cases}
\end{eqnarray}

\bigskip

\section{Scalar supermultiplet as described by chiral superfield}
\label{scalar1}
\label{scalarM}
\label{scalarSec}

In this section we review the well known description of scalar supermultiplet by chiral superfield in superspace 
\cite{Ogievetsky:1975nu,BW,1001}.

As noted in section \ref{superfieldsintro} superfields are highly reducible representations of the supersymmetry algebra but it is 
possible to extract irreducible representations from them by imposing suitable covariant constraints.
The simplest irreducible representation  of the  $D=4$, ${\cal N}=1$  supersymmetry,
the scalar supermultiplet, is described by the chiral superfield, this is to say by complex superfield obeying the so-called 
chirality equation
\begin{eqnarray}\label{bDP=0}
\bar{D}_{\dot{\alpha}}\Phi =0 \; . \qquad
\end{eqnarray}
The complex conjugate  ({\it c.c.}), $\overline{\Phi}= (\Phi)^*$, obeys
\begin{eqnarray}\label{DbP=0}
D_{{\alpha}}\overline{\Phi} =0 \;  \qquad
\end{eqnarray}
and is called anti--chiral superfield.  The free equations of motion for the physical fields of a massless scalar supermultiplet 
($\phi(x)= \Phi\vert_{\theta=0}$ and  $i\psi_\alpha (x)= D_\alpha\Phi\vert_{\theta=0}$) are collected in the superfield equation
\begin{eqnarray}\label{DDP=0}
DD{\Phi}:= D^{{\alpha}}D_{{\alpha}}{\Phi}=0  \; . \qquad
\end{eqnarray}
This equation and its c.c. can be derived from the action
\begin{eqnarray}\label{S=PhibPhi}
S_{kin}= \int d^8z\, \Phi \bar{\Phi}  \; , \qquad
\end{eqnarray}
where the superspace integration measure $d^8z=d^4xd^2\theta d^2\bar{\theta}$ is normalized as\footnote{Notice that, although 
the {\it r.h.s.} of this equation  is not manifestly hermitian, its imaginary part is integral of complete derivative
(as far as $\bar{D}\bar{D}\, DD=  DD\bar{D}\bar{D}\, -4i \tilde{\sigma}{}^{a\, \dot{\alpha}\alpha } \partial_a[D_\alpha , \bar{D}_{\dot{\alpha}}]$)
and, as such, can be ignored in our discussion. }
\begin{eqnarray}\label{S=PhibPhi}
d^8z=d^4x \bar{D}\bar{D}\; DD\, := d^4x \bar{D}_{\dot{\alpha}}\bar{D}{}^{\dot{\alpha}}\; D^\alpha D_\alpha \, \; .
\end{eqnarray}
Indeed, the variation of this functional reads $\delta S_{kin}= \int d^8z\, (\Phi \delta \bar{\Phi} + \delta\Phi \,  \bar{\Phi} )$. 
As far as the variation of chiral superfield should be chiral, $\bar{D}_{\dot{\alpha}}\delta\Phi=0$, and 
${D}_{{\alpha}}\delta\bar{\Phi}=0$, we can equivalently write the action variation as
$\delta S_{kin}= \int d^4x\, \bar{D}_{\dot{\alpha}} \bar{D}{}^{\dot{\alpha}} (( DD\Phi)  \delta \bar{\Phi}) + c.c.$, which 
results in the equations of motion (\ref{DDP=0}).

The most general selfinteraction of the scalar supermultiplet is described by the superfield action
\begin{eqnarray}\label{S=PhibPhi}
S_{s-int}{[\Phi ; \bar{\Phi}]}&=& \int d^8z\, {\cal K}(\Phi , \bar{\Phi}) +  \int d^6\zeta_L \, W(\Phi )  +
\int d^6\zeta_R \, \bar{W}(\bar{\Phi}) = \nonumber \\
&=& \int d^4x \, \bar{D}\bar{D}\, DD\; {\cal K}(\Phi , \bar{\Phi}) +  \int d^4x DD \, W(\Phi )  + \int d^4x \, \bar{D}\bar{D}\,  \, \bar{W}(\bar{\Phi})
 \;  , \qquad
\end{eqnarray}
where   ${\cal K}(\Phi , \bar{\Phi})$ is an  arbitrary function of chiral superfield and its complex conjugate called 
{\it K\"ahler potential} and $W(\Phi)$ ($=(\bar{W}(\bar{\Phi})^*$) is an arbitrary holomorphic function of the complex 
scalar superfield $\Phi$ called {\it superpotential}. This latter is chiral, 
$\bar{D}_{\dot\alpha}W(\Phi)=W^\prime(\Phi)\bar{D}_{\dot\alpha}\Phi =0$, and hence is integrated with chiral measure 
defined by $d^6\zeta_L =d^4x DD$ (and $d^6\zeta_R =d^4x \bar{D}\bar{D}$).
To have the standard kinetic term for the scalar field of the supermultiplet, the K\"ahler potential is usually chosen to obey
 \begin{eqnarray}\label{K''=}
{\cal K}^{\prime\prime}_{\varphi \,  \bar{\varphi}}(\varphi , \bar{\varphi}):=  {\partial \over \partial \varphi}  {\partial \over \partial \bar{\varphi}}{\cal K}(\varphi , \bar{\varphi})\not=0\; .  \qquad
\end{eqnarray}

The superfield equations of motion following from the action $S_{s-int}[\Phi ; \bar{\Phi}]$ (\ref{S=PhibPhi}) are
\begin{eqnarray}\label{EqM=sc0b}
\bar{{\cal E}}&:= & DD {\cal K}^\prime_{\bar{\Phi}} +  \bar{W}^\prime_{\bar{\Phi} } = \qquad \nonumber \\  &=& DD\Phi \; {\cal K}^{\prime\prime}_{\Phi \bar{\Phi} }(\Phi , \bar{\Phi}) +
D^\alpha\Phi \, D_\alpha\Phi \; {\cal K}^{\prime\prime\prime}_{\Phi  \Phi \bar{\Phi}}(\Phi , \bar{\Phi}) + \bar{W}^\prime_{\bar{\Phi} } (\bar{\Phi})  =0 \; ,
 \qquad \\ \label{EqM=sc0}
{{\cal E}}&:= & \bar{D}\bar{D} {\cal K}^{\prime}_{{\Phi}} +   W^\prime_{\Phi }  = \qquad \nonumber \\  &=&
\bar{D}\bar{D}\bar{\Phi} \; {\cal K}^{\prime\prime}_{\Phi  \bar{\Phi}}(\Phi , \bar{\Phi}) +
  \bar{D}_{\dot\alpha}\bar{\Phi} \, \bar{D}^{\dot\alpha}\bar{\Phi} \; {\cal K}^{\prime\prime\prime}_{\Phi \bar{\Phi}\bar{\Phi}}(\Phi , \bar{\Phi}) + W^\prime_{\Phi }(\Phi ) =0 \; ,
 \qquad
\end{eqnarray}
where prime  denotes the derivative with respect to argument,  ${\cal K}^\prime_{\bar{\Phi}} := {\partial {\cal K}({\Phi},\bar{\Phi})\over \partial {\bar{\Phi}}}$, $\;{\cal K}^{\prime\prime}_{\bar{\Phi}\bar{\Phi}} := {\partial^2 \over \partial {\bar{\Phi}}^2}{\cal K}({\Phi},\bar{\Phi})$, 
{\it etc.} These equations can be obtained by solving the chirality conditions (\ref{bDP=0}) and (\ref{DbP=0}) in terms 
of prepotential, generic complex superfield $P$ ($=(\bar{P})^*$),
\begin{eqnarray}\label{Phi=bDbDP}
\Phi=\bar{D}\bar{D}P\; , \qquad \bar{\Phi}=DD\bar{P}\; , \qquad
\end{eqnarray}
and vary with respect to this prepotential and its complex conjugate,
\begin{eqnarray}\label{EqM=S/P}
{{\cal E}}= {\delta S_{s-int}[\Phi ; \bar{\Phi}]\over \delta {P}}\; , \qquad \bar{{\cal E}}= {\delta S_{s-int}[\Phi ; \bar{\Phi}]\over\delta \bar{P}}\; .  \qquad
\end{eqnarray}

\subsection{Four form field strength constructed from  the scalar  supermultiplet}
\label{fourform}
Having a chiral superfield $K$,
\begin{eqnarray}\label{bDK=0}
\bar{D}_{\dot{\alpha}}K =0 \; , \qquad {D}_{{\alpha}}\bar{K} =0 \; , \qquad
\end{eqnarray}
one can construct the following supersymmetric invariant closed four
form (CE cocycle) in flat $D=4$, ${\cal N}=1$ superspace
\cite{Gates:1980ay}
\begin{eqnarray}\label{F4:=dC3=}
F_4=dC_3 :={1\over 4} E^b\wedge E^a \wedge E^\alpha\wedge E^\beta \sigma_{ab}{}_{\alpha\beta} \; \bar{K} + {1\over 4} E^b\wedge E^a \wedge \bar{E}^{\dot{\alpha}}\wedge \bar{E}^{\dot{\beta}} \tilde{\sigma}_{ab}{}_{\dot{\alpha}\dot{\beta}}\; K + \qquad \nonumber \\
+ {1\over 4!} E^c\wedge E^b \wedge E^a \wedge E^\alpha\epsilon_{abcd} \sigma^d_{\alpha\dot{\beta}} \;\bar{D}{}^{\dot{\beta}} \bar{K} + {1\over 4!} E^c\wedge E^b \wedge E^a \wedge \bar{E}{}^{\dot{\beta}}\epsilon_{abcd} \sigma^d_{\alpha\dot{\beta}} \;{D}{}^{\alpha} {K} - \qquad \nonumber \\
 + {1\over 4!}  E^d \wedge E^c\wedge E^b \wedge E^a {i\over 4}\epsilon_{abcd}\left(\bar{D}\bar{D}\bar{K}- DDK\right) .  \qquad
\end{eqnarray}
Notice that we intentionally have not used the notation $\Phi$ for
chiral superfield to stress that, {\it e.g.} having a free chiral
superfield of Eqs. (\ref{bDP=0}) satisfying equations of motion
(\ref{DDP=0}), one can construct the three form using some holomorphic
functions $K=K(\Phi)$, $\bar{K}=\bar{K}(\overline{\Phi})$ which obey
$DDK= K^{\prime\prime}(\Phi) \, D^\alpha \Phi  \, D_\alpha \Phi  $
instead of (\ref{DDP=0}).

Interestingly enough, $F_4$ in (\ref{F4:=dC3=}) can be considered as real part of the complex closed
form  ${\cal F}^{L}_4$, \begin{eqnarray}\label{F4=cF4L+cF4R}
& {F}_4= \Re e ({\cal F}^{L}_4):= {1\over 2}
\left({\cal F}^{L}_4 + {\cal F}^{R}_4\right)\; ,   \qquad
\\ \label{cF4L=ch}
& {\cal F}^L_4 = {1\over  4 }E^b\wedge E^a \wedge  E^\alpha \wedge E^\beta \;
\sigma_{ab \; \alpha\beta}\, \bar{K}\; +  {1\over  4!} E^c\wedge E^b\wedge E^a \wedge  E^\alpha \;
\epsilon_{abcd} \sigma^d_{\alpha\dot{\beta}}\, \bar{D}{}^{\dot{\beta}}\bar{K}\;  + \nonumber \\ & +
{1\over  4!} E^d\wedge E^c\wedge E^b\wedge E^a \;
{i\over 16} \epsilon_{abcd} \, \bar{D}\bar{D}\bar{K}\; , \; \qquad {D}_\alpha\bar{K}=0\; , \; \qquad
\\ \label{cF4R=ch}
& {\cal F}^R_4 = {1\over  4 }E^b\wedge E^a \wedge  {\bar{E}}{}^{\dot{\alpha}} \wedge {\bar{E}}{}^{\dot{\beta}} \;
\tilde{\sigma}_{ab \; \dot{\alpha}\dot{\beta}}\, {K}\; +
{1\over  4!} E^c\wedge E^b\wedge E^a \wedge  {\bar{E}}{}^{\dot{\alpha}} \;
\epsilon_{abcd} \sigma^d_{{\beta}\dot{\alpha}}\, {D}{}^{{\beta}}{K}\;  + \nonumber \\ & +
{1\over  4!} E^d\wedge E^c\wedge E^b\wedge E^a \;
{i\over 16}\epsilon_{abcd} \, {D}{D}{K}\; , \; \qquad \bar{D}_{\dot{\alpha}}{K}=0\; . \qquad
\end{eqnarray}
The fact that these forms are closed as a consequences of (\ref{bDK=0}),
\begin{eqnarray}\label{dcF4L=0}
 d{\cal F}^L_4 = 0 \qquad &\Leftrightarrow & \qquad {D}_A\bar{K}=0\; , \; \qquad
\\ \label{dcF4R=0}
d{\cal F}^R_4 = 0 \qquad &\Leftrightarrow & \qquad  \bar{D}_{\dot{A}}{K}=0\; , \qquad
\end{eqnarray}
suggests the existence of the complex 3-form potentials  $C^L_3$ and ${C}^R_3=(C^L_3)^*$ such that
${\cal F}^L_4 =dC^L_3$ and ${\cal F}^R_4 =dC^R_3$.

To study a  supermembrane in the background of scalar multiplet, which will be the subject of the next section,  we 
do not need an explicit expression for $C_3$. However, we do need it to obtain the equations for the scalar multiplet 
fields with a source from supermembrane, so that we will come back to discussing the problem of constructing potentials 
in section \ref{IntEqs}.

\bigskip

\section{Supermembrane action in the scalar multiplet background}\label{background}

The action of supermembrane in the background of a scalar multiplet can be written in the form
\begin{eqnarray}\label{SM2+=4D}
S_{p=2} &=&  {1\over 2}\int d^3 \xi \sqrt{K\bar{K}}\sqrt{ g} -   \int\limits_{W^3} \hat{C}_3\;   ,  \qquad
\nonumber \\ &=&  -{1\over 6}\int\limits_{W^3}  *\hat{E}_a\wedge \hat{E}^a  \sqrt{K\bar{K}} -   \int\limits_{W^3} \hat{C}_3\;   ,  \qquad
\end{eqnarray}
where $\hat{C}_3$ is the pull--back of the $C_3$ potential defined by Eq. (\ref{F4:=dC3=}) involving the chiral superfields 
$K$ and $\bar{K}$ (\ref{bDK=0}). For simplify  we omit the hat symbol from the pull--backs of superfields here and below in 
the places where this cannot produce a confusion.

The action (\ref{SM2+=4D}) is invariant under the $\kappa$--symmetry
transformations
\begin{eqnarray}\label{ikE=}
i_\kappa \hat{E}{}^a =0 \; , \qquad i_\kappa \hat{E}{}^\alpha =
 \kappa^\alpha \; , \qquad i_\kappa \hat{\bar{E}}{}^{\dot\alpha}= \bar{\kappa}_{\dot{\alpha}} \; ,  \; \qquad
\end{eqnarray}
with $i_\kappa d\hat{Z}^M := \delta_\kappa \hat{Z}^M$, similar to ones in (\ref{ikE0=}) and (\ref{kappaX=}) but with 
the spinorial parameter obeying the reducibility conditions
\begin{eqnarray}\label{bk=Kkbg}
 \bar{\kappa}_{\dot{\alpha}}= - i  \kappa^\beta
 \bar{\gamma}_{\beta\dot{\alpha}}\; \sqrt{{\bar{K}/ K}}\; \qquad \Leftrightarrow \qquad
 \kappa^\alpha= i \bar{\kappa}_{\dot{\alpha}}
 \tilde{\bar{\gamma}}{}^{\dot{\beta}\alpha} \;  \sqrt{{K/{\bar{K}}}}\;  \qquad
\end{eqnarray}
defined by a projector which  differs from the one in (\ref{bk=kbg}) by a (super)field dependent phase factor
$ i \sqrt{{K^{}/{\bar{K}}}^{}}$.

Notice that, if we write the counterpart of the action (\ref{SM2+=4D}) with an arbitrary function ${\cal S}(K,\bar{K})$ instead 
of $ \sqrt{K\bar{K}}$ and perform the fermionic variation  (\ref{ikE=}) of such an action,  we find that the local 
fermionic  $\kappa$--symmetry parameter should obey the equations
$\kappa^\alpha \partial {\cal S}/\partial K = i/2 \bar{\kappa}_{\dot{\beta}}\tilde{\gamma}{}^{\dot{\beta}\alpha}$ and 
$\bar{\kappa}_{\dot{\beta}} \partial {\cal S}/\partial \bar{K} = \kappa^\alpha {\gamma}_{\alpha\dot{\beta}}$. 
This system of equations has a nontrivial solution when 
$\partial {\cal S}/\partial \bar{K} = { 1\over  4 \partial{\cal S}/\partial {K}}$. 
This latter equation is solved by ${\cal S}(K,\bar{K})=\sqrt{K\bar{K}}$ so that the 
action (\ref{SM2+=4D}) for scalar multiplet in supergravity background can be constructed 
from the requirement of the $\kappa$--symmetry.

\subsection{Equations of motion for supermembrane in a background of an off--shell scalar supermultiplet}

The supermembrane equations of motion can be obtained by varying the action (\ref{SM2+=4D}) with respect to coordinate 
functions $\hat{Z}{}^M(\xi)$, so that we can write them in the form of ${\delta S_{p=2}\over \delta \hat{Z}{}^M(\xi)}=0$. 
The convenient form of the bosonic and fermionic equations can be extracted by multiplying this on the inverse 
supervielbein, $E_\alpha^M(\hat{Z}){\delta S_{p=2}\over \delta \hat{Z}{}^M(\xi)}=0$ and 
$E_a^M(\hat{Z}){\delta S_{p=2}\over \delta \hat{Z}{}^M(\xi)}=0$. These combinations appear as 
the coefficients for $i_\delta \hat{E}^\alpha:=  \delta \hat{Z}{}^M(\xi)E_M{}^\alpha (\hat{Z})$, $i_\delta \hat{\bar{E}}{}^{\dot\alpha}:=  \delta \hat{Z}{}^M(\xi)\bar{E}_M{}^{\dot\alpha} (\hat{Z})$ 
and $i_\delta \hat{E}^a:=  \delta \hat{Z}{}^M(\xi)E_M{}^a (\hat{Z})$ in the integrand of the action variation. This implies 
the possibility to write the formal expression for supermembrane equations of motion in the form
\begin{eqnarray}\label{smEqm=f-gen}
  {{\delta S_{p=2}}\over {i_\delta \hat{E}{}^\alpha} }:= E_\alpha^M(\hat{Z}){\delta S_{p=2}\over \delta \hat{Z}{}^M(\xi)}=0\; &,&  \quad  {{\delta S_{p=2}}\over {i_\delta \hat{\bar{E}}{}^{\dot{\alpha}}}}:= E_{\dot{\alpha}}^M(\hat{Z}){\delta S_{p=2}\over \delta \hat{Z}{}^M(\xi)}=0\; ,  \qquad \\ \label{smEqm=b-gen} {{\delta S_{p=2}}\over {i_\delta \hat{E}{}^a }}&:=& E_a^M(\hat{Z}){\delta S_{p=2}\over \delta \hat{Z}{}^M(\xi)}=0\; . \qquad
\end{eqnarray}
The straightforward calculation gives the following explicit form of these equations of motion
\begin{eqnarray}\label{smEqm=f}
&& *\hat{E}_a \wedge \left( i \hat{\bar{E}}{}^{\dot{\alpha}} \sigma^a_{\alpha\dot{\alpha}} \sqrt{K\bar{K}} - \hat{E}_\beta (\bar{\gamma}\tilde{\sigma}{}^a)_\alpha{}^\beta \bar{K} \right) + {1\over 12} *\hat{E}_a \wedge \hat{E}^a \left(\sqrt{\bar{K}/K} D_\alpha K- i\bar{\gamma}_{\alpha\dot{\alpha}} \bar{D}{}^{\dot{\alpha}}\bar{K} \right)=0\; ,  \nonumber \\ && \\ \label{smEqm=bf}
&& *\hat{E}_a \wedge \left( i \hat{{E}}{}^{{\alpha}} \sigma^a_{\alpha\dot{\alpha}} \sqrt{K\bar{K}} - \hat{\bar{E}}{}_{\dot\beta} (\tilde{\sigma}{}^a\bar{\gamma}){}^{\dot\beta}{}_{\dot\alpha} {K} \right) + {1\over 12} *\hat{E}_a \wedge \hat{E}^a \left(\sqrt{K/\bar{K}} \bar{D}_{\dot\alpha} \bar{K}- iD^\alpha K \bar{\gamma}_{\alpha\dot{\alpha}} \right)=0\; , \nonumber \\ && \end{eqnarray}
\begin{eqnarray} \label{smEqm=b} \; D(*\hat{E}_a) &- & {1\over 6} *\hat{E}_b \wedge \hat{E}{}^b \left({D_a\ln\bar{K}} +  {D_a\ln K} \right)+ {1\over 2}  *\hat{E}_a \wedge  ( d\ln \hat{\bar{K}} +  d\ln \hat{K}) - \qquad \nonumber \\ && - {i\over 12\sqrt{K\bar{K}}}\hat{E}{}^d\wedge\hat{E}{}^c\wedge  \hat{E}{}^b \,\epsilon_{abcd} \,  (\bar{D}\bar{D}\bar{K}-DDK) - \qquad \nonumber \\ && - {1\over 4\sqrt{K\bar{K}}}\hat{E}{}^c\wedge  \hat{E}{}^b \wedge \epsilon_{abcd} \sigma^d_{\alpha\dot{\alpha}}\,  (\hat{E}{}^\alpha \bar{D}{}^{\dot{\alpha}}\bar{K}+ \hat{\bar{E}}{}^{\dot{\alpha}}{D}{}^\alpha K) -  \qquad \nonumber \\  && - \hat{E}{}^b \wedge \hat{E}{}^\alpha \wedge \hat{E}{}^\beta \sigma_{ab}{}_{\alpha\beta}\; \sqrt{{\bar{K}}/{K}}  - \hat{E}{}^b \wedge \hat{\bar{E}}{}^{\dot\alpha} \wedge \hat{\bar{E}}{}^{\dot\beta} \tilde{\sigma}_{ab}{}_{\dot{\alpha}\dot{\beta}}\; \sqrt{K/{\bar{K}}}  = 0\; . \qquad
\end{eqnarray}

Notice that the above equations of motion are not independent. According to the second Noether theorem, the gauge symmetries of a 
dynamical system result in the so-called Noether identities relating the left-hand sides of equations of motion of this system. 
The supermembrane possesses a number of gauge symmetries, including the local fermionic $\kappa$--symmetry (\ref{ikE=}), 
(\ref{bk=Kkbg}). This is reflected by the fact that contracting our fermionic equation (\ref{smEqm=f}) with
 $i\sqrt{K/\bar{K}} \tilde{\bar{\gamma}}{}^{\dot{\beta}\alpha}$ we arrive at Eq. (\ref{smEqm=bf}).
Denoting the left hand sides of equations (\ref{smEqm=f}) and (\ref{smEqm=bf}) by $\Psi_\alpha$ and $\bar{\Psi}_{\dot\alpha}$, 
respectively, we can write the above described Noether identity for the  $\kappa$--symmetry in the form of
\begin{eqnarray} \label{NI=kappa}
\tilde{\bar{\gamma}}{}^{\dot{\beta}\alpha}\Psi_\alpha \equiv -i\sqrt{\bar{K}/K}\epsilon^{\dot{\beta\dot{\alpha}}} \bar{\Psi}_{\dot\alpha}\; . \qquad
\end{eqnarray}

\bigskip

\section{Superfield equations for the dynamical system of special scalar supermultiplet interacting with supermembrane}
\label{IntEqs}

 \subsection{Special scalar multiplet and its dual three form potential}
\label{threeform}

In our discussion below we will be considering not generic but   {\it special scalar multiplet} described by the chiral superfield 
constructed from the  {\it real prepotential} $V=(V)^*$,
\begin{eqnarray}\label{Phi=DDV}
\Phi=\bar{D}\bar{D}V\; , \qquad \bar{\Phi}=DDV\;  . \qquad
\end{eqnarray}
On the level of auxiliary fields the distinction of this special case is that one of the real auxiliary scalars 
of the generic scalar multiplet is replaced in it by a divergence of a real vector, $\partial_\mu k^\mu$ or, equivalently, 
by the field strength of a three form potential $k_{\nu\rho\sigma}= k^\mu \epsilon_{\mu\nu\rho\sigma}$ (in this latter form 
it was described in \cite{Gates:1980ay} and, as one of 'variant superfield representations', in \cite{Gates:1980az}).

Indeed, the complex prepotential $P$ of the generic chiral multiplet,  $\Phi=\bar{D}\bar{D}P$, $ \bar{\Phi}=DD\bar{P}$, is defined 
up to the gauge transformations, $\bar{P}\mapsto \bar{P} + D_\alpha \Xi^\alpha$. These imply that the imaginary part of the generic 
prepotential is transformed by
$\Im m\, P := (P-\bar{P})/2i\mapsto \Im  m\, P + (D_\alpha \Xi^\alpha - D_{\dot\alpha} \bar{\Xi}^{\dot\alpha})/2i$.
Hence not-pure gauge parts of the superfield parameter $\Im m\, P$ are the ones which do not have their exact counterparts in 
the composed superfield $(D_\alpha \Xi^\alpha - D_{\dot\alpha} \bar{\Xi}^{\dot\alpha})/2i$.
One can check that the superfield parameter $D_\alpha \Xi^\alpha - D_{\dot\alpha} \bar{\Xi}^{\dot\alpha}$ has all the components 
but one having contributions of different independent functions without derivatives. The only exception is the highest component 
in its decomposition  which reads $- 4i \theta\theta\, \bar{\theta}\bar{\theta} \partial_a(k^a+\bar{k}{}^a)$ and includes the 
divergence of the real part of the complex vector  $k_a=\tilde{\sigma}_a^{\dot{\alpha}\alpha}(\bar{D}_{\dot{\alpha}}\Xi_\alpha)\vert_{\theta=0}= (\bar{k}_a)^*$
versus an arbitrary function in a generic real scalar superfield, like $P$. Then one can guess that the equations of motion for 
the special scalar supermultiplet will differ from the set of equations for a generic scalar supermultiplet by that one of the 
algebraic auxiliary field equations of the latter ($({\cal E}-{\cal E})\vert_0=0$) will be replaced by its derivative 
($\partial_a({\cal E}-{\cal E})\vert_0=0$). In other words, the general solution of the (auxiliary) field equations of 
the special scalar supermultiplet involves one additional (with respect to the generic case) arbitrary real constant. Furthermore, 
this indicates that the above mentioned auxiliary field equation of the special scalar supermultiplet 
($\partial_a({\cal E}-{\cal E})\vert_0=0$) is dependent, {\it i.e.} can be obtained as a consequence of other equations; 
this implies that  the only effect of the use of the complex prepotential in the generic case is vanishing of a real constant 
which is indefinite in the case of special scalar multiplet 
(where $\partial_a({\cal E}-{\cal E})\vert_0=0\;\Rightarrow \; ({\cal E}-{\cal E})\vert_0=-2ic$). 
We will see that this is indeed the case.

 \subsubsection{Equations of motion of special scalar multiplet}

The variation of the general action (\ref{S=PhibPhi}) for the special chiral superfields 
(\ref{Phi=DDV}) with respect to real prepotential $V$ apparently produces only the real part of the complex 
equation (\ref{EqM=sc0}),
\begin{eqnarray}\label{EqM=sc0+b}
{\delta S_{s-int}[\bar{D}\bar{D}V ; DDV]\over  \delta {V}}=0 \; \Rightarrow \quad  {{\cal E}}+\bar{{\cal E}}&:= & \bar{D}\bar{D} {\cal K}^{\prime}_{{\Phi}} + DD {\cal K}^\prime_{\bar{\Phi}}  +  W^\prime_{\Phi } +  \bar{W}^\prime_{\bar{\Phi} }  = 0
\; .
 \qquad
\end{eqnarray}

However, as far as the left hand sides of the equations of motion for generic scalar multiplet, Eqs. 
(\ref{EqM=sc0}) and (\ref{EqM=sc0b}), are, respectively, anti-chiral and chiral, 
${D}_{\alpha} \bar{{\cal E}}= 0$ and $\bar{D}_{\dot\alpha} {\cal E}=0$, 
Eq. (\ref{EqM=sc0+b}) implies that the imaginary part of the complex equation 
(\ref{EqM=sc0}) is equal to a constant,
\begin{eqnarray}\label{EqM-bEqM=c}
{{\cal E}}+\bar{{\cal E}}=0 \; \qquad \Rightarrow \qquad  D_\alpha ({{\cal E}}-\bar{{\cal E}})=0\qquad \Rightarrow \qquad  \partial_a ({{\cal E}}-\bar{{\cal E}})=0 \; . \qquad
\end{eqnarray}
Hence the only effect of the use of the special chiral superfields 
(\ref{Phi=DDV}) instead of the generic scalar superfield (\ref{Phi=bDbDP}) 
is that the equation ${{\cal E}}=0$ is replaced by
${{\cal E}}=-ic$ where $c$ is an arbitrary real constant.

\subsubsection{On spontaneous supersymmetry breaking}

The presence of this arbitrary constant in the right hand side of the superfield equations of motion, 
${{\cal E}}=-ic$,  actually suggests a possible spontaneous supersymmetry breaking in the theory of special chiral multiplet.
To clarify this, let us discuss the simple case of a free massless special scalar multiplet, in which the action reads
$\int d^8z \Phi\bar{\Phi}= \int d^8z DDV \bar{D}\bar{D}V$ so that ${\bar{\cal E}}=DD\Phi $ and the equations of 
motion (\ref{EqM=sc0+b}) simplify to
\begin{eqnarray}\label{EqM0=sc} DD\Phi  + \bar{D}\bar{D}\bar{\Phi}=0\; . \qquad
\end{eqnarray}
As it has been discussed above (Eq. (\ref{EqM-bEqM=c})) these equations lead  to 
$D_\alpha (DD\Phi  - \bar{D}\bar{D}\bar{\Phi}) =0$ and $\partial_a (DD\Phi  - \bar{D}\bar{D}\bar{\Phi}) =0$. 
Algebraically all this set of equations is solved by  $DD\Phi   =-ic$  with the above mentioned arbitrary real constant $c$,
\begin{eqnarray}\label{EqM=-ic} DD\Phi  + \bar{D}\bar{D}\bar{\Phi}=0 \qquad \Rightarrow \qquad
DD\Phi   =-ic\; , \qquad c=const\, . \qquad
\end{eqnarray}
In particular, this constant enters the solution of auxiliary field equations which now reads
\begin{eqnarray}\label{auxEq=-ic}
DD\Phi \vert_0  =-ic\; , \qquad c=const\, . \qquad
\end{eqnarray}
As the on-shell supersymmetry transformations of the fermionic fields $\psi_\alpha= -iD_\alpha\Phi\vert_0$ are obtained 
from the off-shell ones,
$\delta \psi_\alpha =  {i\over 2}\varepsilon_\beta\, DD\Phi\vert_0 + 2(\sigma^a\bar{\varepsilon})_\alpha \partial_a \phi$, by 
inserting the above solution  of the auxiliary field equations, they read 
\begin{eqnarray}\label{susy=ic}
\delta \psi_\alpha = {c\over 2} \, \varepsilon_\beta\, + 2(\sigma^a\bar{\varepsilon})_\alpha \partial_a \phi\, . \qquad
\end{eqnarray}
Hence, for nonvanishing value of $c$, the on-shell supersymmetry transformations of $\delta \psi_\alpha$ contains the additive 
contribution of supersymmetry parameter $\varepsilon_\beta\,$ characteristic of the transformation rules of the 
Volkov-Akulov Goldstone fermion \cite{Volkov73, Volkov73+1} the presence of which may be considered as an indication of the spontaneous 
supersymmetry breaking.

However, studying more carefully the case of free special scalar multiplet, one finds that such a spontaneous symmetry 
breaking actually does not occur if nontrivial boundary conditions are not introduced. 
Indeed, the constant in the superfield equations (\ref{EqM=-ic}) can be reproduced from the generic scalar supermultiplet 
action which includes the superpotential linear in chiral superfield, $W(\Phi)=-ic\Phi$. As it was observed already in 
\cite{Zumino+Iliopoulos:74},  such a term can be removed from the action by a field redefinition. However, the boundary 
term contribution may change the situation; this role can be also played by supermembrane contribution. Further discussion 
on spontaneous supersymmetry breaking in the interacting system of scalar multiplet and supermembrane goes beyond the scope of this 
thesis. We turn to the three form potential presentation of the special chiral supermultiplet.

\subsubsection{Dual three form potential}

The four form field strength constructed with the use of special scalar multiplet (\ref{Phi=DDV}) is obtained from 
(\ref{F4:=dC3=}) by substituting
\begin{eqnarray}\label{K=DDV}
\bar{K}=\bar{\Phi}=DDV\;  , \qquad K= {\Phi}= \bar{D}\bar{D}V\; . \qquad
\end{eqnarray}
It reads
\begin{eqnarray}\label{F4=F4(V)}
F_4 = dC^\prime_3 &= {1\over  4 }E^b\wedge E^a \wedge  E^\alpha \wedge E^\beta \;
\sigma_{ab \, \alpha\beta} DDV\, + {1\over  4}E^b\wedge E^a \wedge  \bar{E}{}^{\dot{\alpha}} \wedge \bar{E}{}^{\dot{\beta}} \;
\tilde{\sigma}_{ab \, \dot{\alpha}\dot{\beta}} \bar{D}\bar{D}V\, + \nonumber \\ & +
{1\over  4!} E^c\wedge E^b\wedge E^a \wedge \left( E^\alpha \;
\epsilon_{abcd} \sigma^d_{{\alpha}\dot{\beta}}\, \bar{D}{}^{\dot{\beta}}DDV\; +   \bar{E}{}^{\dot{\alpha}} \;
\epsilon_{abcd} \sigma^d_{{\beta}\dot{\alpha}}\, {D}{}^{{\beta}}\bar{D}\bar{D}V\;\right) \;  + \nonumber \\ & +
{1\over  4!} E^d\wedge E^c\wedge E^b\wedge E^a \;
{i\over 4} \epsilon_{abcd} \, (\bar{D}\bar{D}DDV - DD\bar{D}\bar{D}V)\; . \; \qquad
\end{eqnarray}
The corresponding 3-form potential $C^\prime_3$ can be written in terms of the real prepotential as follows
\cite{Gates:1980ay}
\begin{eqnarray}\label{C3'=}
 C^\prime_3 &=& 2i E^c \wedge  E^\alpha \wedge \bar{E}{}^{\dot{\beta}} \; {\sigma}_{c \; {\alpha}\dot{\beta}}\; V+  {1\over 2}E^c\wedge E^b \wedge  E{}^{{\alpha}}  \sigma_{bc \; {\alpha}}{}^{\beta}\, D_\beta V\; - \nonumber \\ &&
 -  {1\over  2}E^b\wedge E^a \wedge  \bar{E}{}^{\dot{\beta}} \;
\tilde{\sigma}_{ab}{}^{\dot{\alpha}}{}_{\dot{\beta}}\, \bar{D}_{\dot{\alpha}}V\; -
 {1\over 4!} E^c\wedge E^b\wedge E^a
 \epsilon_{abcd}  \tilde{\sigma}^{d \; \dot{\beta}\alpha} \,  [{D}_\alpha \, , \, \bar{D}_{\dot{\beta}}] V \; . \; \qquad
\end{eqnarray}
Of course, this expression can be changed on an equivalent one using gauge transformations
$\delta C_3=d\alpha_2$. These do not change the field strength (\ref{F4=F4(V)}) and are responsible for the possibility to do 
not have the lower dimensional form contributions ($\propto E^\alpha\wedge E^\beta \wedge E^\gamma$ {\it etc.}) in the 
above $C_3^\prime$.

The existence  of this simple three form $C_3^\prime$ giving a dual description of the special chiral 
supermultiplet (\ref{Phi=DDV}) is the main reason to restrict our discussion below by this special case.

\subsection{Superfield equations of motion for interacting system}

Let us consider the most general interaction of the special scalar supermultiplet with supermembrane
as described by the action  (\ref{SM2+=4D}) with $K= \Phi= \bar{D}_{\dot{\alpha}}\bar{D}{}^{\dot{\alpha}}V$ and
$\bar{K}=  \bar{\Phi }= {D}^{{\alpha}}{D}_{{\alpha}}V $ as in (\ref{K=DDV}), {\it i.e.} by
\begin{eqnarray}\label{Sm+SM2=4D}
S&=& \int d^8z\, {\cal K}(\Phi , \bar{\Phi}) +  \int d^6\zeta_L \, W(\Phi )  + c.c.
+  {1\over 2}\int d^3 \xi \, \sqrt{ g}\, \sqrt{\hat{\Phi}\hat{\bar{\Phi}}} -   \int\limits_{W^3} \hat{C}^\prime_3= \qquad
 \\ &=& \int d^4x\, \bar{D}\bar{D}DD{\cal K}(\Phi , \bar{\Phi}) +  \int d^4x (DDW(\Phi )  + c.c.)   -{1\over 6}\int\limits_{W^3}  *\hat{E}_a\wedge \hat{E}^a  \sqrt{\hat{\Phi}\hat{\bar{\Phi}}} -   \int\limits_{W^3} \hat{C}_3^\prime\; \nonumber
\end{eqnarray}
with special chiral superfield (\ref{Phi=DDV}),
\begin{eqnarray}\label{Phi=DDV2}
 \Phi= \bar{D}_{\dot{\alpha}}\bar{D}{}^{\dot{\alpha}}V \; ,  \qquad   \bar{\Phi }= {D}^{{\alpha}}{D}_{{\alpha}}V \; .  \qquad
\end{eqnarray}

The variation of the interacting action (\ref{Sm+SM2=4D}) with respect to supermembrane variables
gives formally the same equations of motion as for the supermembrane in the background, (\ref{smEqm=f})--(\ref{smEqm=b}), 
but with $K=\Phi=DDV$, \footnote{To simplify the expressions, we omitted the hat symbol from the pull--backs of 
superfields and their derivatives in equations (\ref{smEqm=bfs}) and (\ref{smEqm=bs}), but, in contrast, left all the pull--back 
symbols in equation (\ref{smEqm=fs}) so that one can appreciate simplification comparing this with its complex conjugate 
Eq. (\ref{smEqm=bfs}).}
 \begin{eqnarray}\label{smEqm=fs}
&& *\hat{E}_a \wedge \left( i \hat{\bar{E}}{}^{\dot{\alpha}} \sigma^a_{\alpha\dot{\alpha}} \sqrt{\hat{\Phi}\hat{\bar{\Phi}}} - \hat{E}_\beta (\bar{\gamma}\tilde{\sigma}{}^a)_\alpha{}^\beta \hat{\bar{\Phi}} \right) + {1\over 12} *\hat{E}_a \wedge \hat{E}^a \left(\sqrt{\hat{\bar{\Phi}}/\hat{\Phi}} \, \widehat{D_\alpha \Phi }- i\bar{\gamma}_{\alpha\dot{\alpha}} \widehat{\bar{D}{}^{\dot{\alpha}}\bar{\Phi}} \right)=0\; ,  \nonumber \\ && \\ \label{smEqm=bfs}
&& *\hat{E}_a \wedge \left( i \hat{{E}}{}^{{\alpha}} \sigma^a_{\alpha\dot{\alpha}} \sqrt{\Phi\bar{\Phi}} - \hat{\bar{E}}{}_{\dot\beta} (\tilde{\sigma}{}^a\bar{\gamma}){}^{\dot\beta}{}_{\dot\alpha} {\Phi} \right) + {1\over 12} *\hat{E}_a \wedge \hat{E}^a \left(\sqrt{\Phi/\bar{\Phi}} \bar{D}_{\dot\alpha} \bar{\Phi}- iD^\alpha \Phi \bar{\gamma}_{\alpha\dot{\alpha}} \right)=0\; , \nonumber \\ && \end{eqnarray}
\begin{eqnarray} \label{smEqm=bs} \; D(*\hat{E}_a) &- & {1\over 6} *\hat{E}_b \wedge \hat{E}{}^b \left({D_a\ln\bar{\Phi}} +  {D_a\ln \Phi} \right)+ {1\over 2}*\hat{E}_a \wedge  ( d\ln {\bar{\Phi}} +  d\ln {\Phi}) - \qquad \nonumber \\ && - {i\over 12\sqrt{\Phi\bar{\Phi}}}\hat{E}{}^d\wedge\hat{E}{}^c\wedge  \hat{E}{}^b \,\epsilon_{abcd} \,  (\bar{D}\bar{D}\bar{\Phi}-DD\Phi) - \qquad \nonumber \\ && - {1\over 4\sqrt{\Phi\bar{\Phi}}}\hat{E}{}^c\wedge  \hat{E}{}^b \wedge \epsilon_{abcd} \sigma^d_{\alpha\dot{\alpha}}\,  (\hat{E}{}^\alpha \bar{D}{}^{\dot{\alpha}}\bar{\Phi}+ \hat{\bar{E}}{}^{\dot{\alpha}}{D}{}^\alpha \Phi) -  \qquad \nonumber \\  && - \hat{E}{}^b \wedge \hat{E}{}^\alpha \wedge \hat{E}{}^\beta \sigma_{ab}{}_{\alpha\beta}\; \sqrt{{\bar{\Phi}}/{\Phi}}  - \hat{E}{}^b \wedge \hat{\bar{E}}{}^{\dot\alpha} \wedge \hat{\bar{E}}{}^{\dot\beta} \tilde{\sigma}_{ab}{}_{\dot{\alpha}\dot{\beta}}\; \sqrt{\Phi/{\bar{\Phi}}}  = 0\; . \qquad
\end{eqnarray}
However, the target superspace superfields the pull--backs of which enter these equations have to be the solutions of 
interacting equations with the source terms from the supermembrane. These superfield interacting equations read
\begin{eqnarray}\label{Eqm=J}
 {\cal E}+\bar{\cal E} = J\; ,  \qquad
\end{eqnarray}
where
\begin{eqnarray}\label{Eqm=J}
 {\cal E}=
 \bar{D}\bar{D}\bar{\Phi} \; {\cal K}^{\prime\prime}_{\Phi  \bar{\Phi}}(\Phi , \bar{\Phi}) +
  \bar{D}_{\dot\alpha}\bar{\Phi} \, \bar{D}^{\dot\alpha}\bar{\Phi} \; {\cal K}^{\prime\prime\prime}_{\Phi \bar{\Phi}\bar{\Phi}}(\Phi , \bar{\Phi}) + W^\prime_{\Phi }(\Phi )\;  \nonumber \\
\end{eqnarray}
(see  (\ref{EqM=sc0}) and (\ref{EqM=sc0+b}))  and
\begin{eqnarray}\label{J=dS/dV}
J(z) = - {\delta S_{p=2}\over \delta V(z)}
\;   \qquad
\end{eqnarray}
is the current superfield from the supermembrane. The problem of obtaining the complete set of interacting equations for 
the dynamical system of supermembrane and special chiral supermultiplet is now reduced to the problem of calculating this 
supermembrane current.

\subsection{Supermembrane current}\label{Supermembranecurrent}
The supermembrane current is split naturally on the contributions from the Nambu--Goto and the Wess--Zumino terms of 
the action (\ref{SM2+=4D}) with (\ref{Phi=DDV})
\begin{eqnarray}\label{J=JNG+JWZ}
J(z) = J^{NG}(z)+ J^{WZ}(z) = - {\delta S_{p=2}\over \delta V(z)}
\;   \qquad
\end{eqnarray}
The Nambu-Goto part of the current
\begin{eqnarray}\label{JNG:=}
 & J^{NG}(z) := - {\delta \over \delta V(z)}  {1\over 2}\int d^3 \xi \, \sqrt{ g} \sqrt{\widehat{\bar{D}\bar{D}V}\, \widehat{DDV}}
\end{eqnarray}
($\widehat{DDV}:= D^\alpha D_\alpha V(z)\vert_{z^M=\hat{z}{}^M(\xi)}$)
is calculated by first using the properties of the superspace delta function
\begin{eqnarray}\label{d8:=}
 \delta^8(z):={1\over 16}\,\delta^4(x) \, \theta\theta \, \bar{\theta}\bar{\theta}\; , \qquad
 \int d^8z\, \delta^8(z-z') f(z) = f(z') \;
\end{eqnarray}
to present (\ref{JNG:=}) in the form
\begin{eqnarray}\label{JNG=intd8z}
 & J^{NG}(z) = - {\delta \over \delta V(Z)}  {1\over 2} \int d^8z'\, \sqrt{{\bar{D}\bar{D}V(z')}\, {DDV}(z')} \, \int d^3 \xi \, \sqrt{ g} \delta^8(z'-\hat{z})\; . \qquad
\end{eqnarray}
Then the calculation reduces to using the definition of variation
${\delta V(z')\over \delta V(z)}=\delta^8(z'-{z})$ and performing the superspace integration. In such a way one arrives at
\begin{eqnarray}\label{JNG=}
 J^{NG}(Z) &=& - {1\over 4} \, \int d^3 \xi \, \sqrt{g} \, \sqrt{{\hat{\Phi}\over \hat{\bar{\Phi}}}}\; {D}{D} \delta^8(z-\hat{z})  - {1\over 4} \, \int d^3 \xi \, \sqrt{g}\, \sqrt{{\hat{\bar{\Phi}}\over\hat{ \Phi}}}\; \bar{D}\bar{D}\delta^8(z-\hat{z}) \, , \qquad
\end{eqnarray}
where
$\Phi=\bar{D}\bar{D}V$, $\bar{\Phi}=DDV$ (see Eqs. (\ref{Phi=bDbDP})) and
$\hat{\Phi}:= {\Phi}(\hat{z}(\xi))$ {\it etc.}
Similarly one can present   the Wess--Zumino current in the form of
\begin{eqnarray}\label{JWZ=}
J^{WZ}(Z) &=& \left(2i \int_{W^3} \hat{E}^c\wedge \hat{E}{}^\alpha \wedge \hat{E}{}^{\dot\alpha} \sigma_{c\alpha\dot\alpha}  + \right. \nonumber \\ && \left. + {1\over 2}  \int_{W^3} \hat{E}^c\wedge \hat{E}^b\wedge \hat{E}{}^\alpha \sigma_{bc\alpha}{}^{\beta}D_\beta  -  {1\over 2}  \int_{W^3} \hat{E}^c\wedge \hat{E}^b \wedge \hat{E}{}^{\dot\alpha} \tilde{\sigma}_{bc}{}^{\dot{\beta}}{}_{\dot\alpha} \bar{D}_{\dot{\beta}} - \right. \nonumber \\ && \left. - {1\over 4!}  \int_{W^3} \hat{E}^c\wedge \hat{E}^b \wedge \hat{E}^a\epsilon_{abcd} \tilde{\sigma}^{d\dot{\alpha}\alpha} [D_\alpha, \bar{D}_{\dot{\alpha}}] \right)  \delta^8(z-\hat{z})\; . \qquad
\end{eqnarray}

\section{Simplest equations of motion for spacetime fields interacting with dynamical supermembrane}\label{simplest}

Having the superfield equations with supermembrane current contributions, the next stage is to extract the equations of 
motion for the physical fields of the supermultiplet. We will do this for the  simplest case when the 
special scalar multiplet part of the interacting action is given by
the kinetic term (\ref{S=PhibPhi}) only, this is to say for the interacting system described by the action
\begin{eqnarray}\label{S=S0+SM2}
S=S_{kin}+ S_{p=2} = \int d^8z\, \Phi \bar{\Phi}  + {1\over 2}\int d^3 \xi \sqrt{\hat{\Phi}\hat{\bar{\Phi}}}\sqrt{ g} -   \int\limits_{W^3} \hat{C}_3{}^\prime\;     \qquad
\end{eqnarray}
where $\Phi=\bar{D}\bar{D}V$,  $\bar{\Phi}=DDV$ (\ref{Phi=DDV2}) and $ \hat{C}^\prime_3$ is the pull--back to $W^{3}$ of 
the 3-form ${C}^\prime_3$ defined in (\ref{C3'=}). The interacting equations of motion for the bulk superfields, 
Eqs.  (\ref{Eqm=J}), in this case simplifies to
\begin{eqnarray}\label{Eqm0=J}
  DD\Phi+\bar{D}\bar{D}\bar{\Phi} = J(z)\;   \qquad
\end{eqnarray}
where the current $J(z)$ is given by (\ref{J=JNG+JWZ}), (\ref{JNG=}) and (\ref{JWZ=}).

\subsection{General structure of the simplest special scalar multiplet equations with a superfield source}

Superfield equation (\ref{Eqm0=J}) encodes the dynamical equations for the physical fields of the scalar multiplet, 
$\phi(x)=\Phi\vert_0$ and $\psi_\alpha (x)=- i(D_\alpha\Phi )\vert_0$, as well as algebraic equations for auxiliary 
fields  $DD\Phi\vert_0$ and $\bar{D}\bar{D}\bar{\Phi}\vert_0 $. These latter include the leading component of the real 
superfield equation (\ref{Eqm0=J})
\begin{eqnarray}\label{Eqm0=Jv0}
 DD\Phi \vert_0 +\bar{D}\bar{D}\bar{\Phi} \vert_0 = J(z)\vert_0\;   \qquad
\end{eqnarray}
as well as the first order equation
\begin{eqnarray}\label{d(aux-aux)=}
 \partial_a( DD\Phi \vert_0 -\bar{D}\bar{D}\bar{\Phi} \vert_0 ) = - {i\over 4} \tilde{\sigma}_a^{\dot{\alpha}\alpha}[D_\alpha \, , \, \bar{D}_{\dot\alpha}] J(z)\vert_0\;  . \qquad
\end{eqnarray}
The set of dynamical field equations include the Dirac (actually Weyl) equation with the source from supermembrane,
\begin{eqnarray}\label{Dirac=J}
{\sigma}^a_{\alpha\dot{\alpha}}\partial_a\psi^\alpha:= - i \partial_{\alpha\dot{\alpha}}D_\alpha\Phi \vert_0 = {1\over 4} \bar{D}_{\dot{\alpha}}J(z)\vert_0 \; , \qquad
\end{eqnarray}
and the Klein-Gordon equation, also with the source,
\begin{eqnarray}\label{K-G=J}
\Box \phi(x):= \Box \Phi\vert_0 = -{1\over 16} \bar{D}\bar{D}J(z)\vert_0 \;  . \qquad
\end{eqnarray}
Now, to specify supermembrane contributions to the scalar multiplet field equations we have to calculate the derivatives of 
the supermembrane current.

\subsection{Dynamical scalar multiplet equations with supermembrane source contributions}

\subsubsection{Auxiliary field equations}

The leading components $J\vert_0$  of the current $J$ in (\ref{Eqm0=Jv0}) is the sum of
\begin{eqnarray}\label{JNG0=}
 J^{NG}\vert_{0} &=&  {1\over 16} \,\sqrt{{\phi\over \bar{\phi}}}\; \int d^3 \xi \, \sqrt{g} \, \hat{\bar{\theta}}\hat{\bar{\theta}} \delta^4(x-\hat{x}) + {1\over 16} \sqrt{{\bar{\phi}\over \phi}}\; \int d^3 \xi \, \sqrt{g}\, \hat{{\theta}}\hat{{\theta}}\delta^4(x-\hat{x})  \;  \qquad
\end{eqnarray}
and
\begin{eqnarray}\label{JWZ0=}
J^{WZ}(Z)\vert_{0}  &=& {1\over 48}  \int_{W^3} \hat{E}^c\wedge \hat{E}^b \wedge \hat{E}^a\epsilon_{abcd}
 \hat{{\theta}}{\sigma}^{d} \hat{\bar{\theta}}\, \delta^4(x-\hat{x})+ {\cal O}(f^4)
 \; , \qquad
\end{eqnarray}
where $ {\cal O}(f^4)$ denotes the terms of the fourth order in fermions (in this case, these are worldvolume 
fermionic fields $\hat{\theta}$, $\hat{\bar{\theta}}$ and their worldvolume derivatives,  
$\partial_m\hat{\theta}:=\partial \hat{\theta}/\partial\xi^m$ and {\it c.c.}); the explicit form of 
these one can find in the Appendix B (Eq.   (\ref{JWZ0=App})).

Substituting the above expressions  into Eq. (\ref{Eqm0=Jv0}), one finds that the real part of the auxiliary fields 
of the chiral multiplet has quite a complex form in terms of supermembrane variables
\begin{eqnarray}\label{Eqm0=Jv0+O}
 DD\Phi \vert_0 +\bar{D}\bar{D}\bar{\Phi} \vert_0 = {1\over 16} \,\sqrt{{\phi\over \bar{\phi}}}\; \int d^3 \xi \, \sqrt{g} \, \hat{\bar{\theta}}\hat{\bar{\theta}} \delta^4(x-\hat{x})  + {1\over 16} \sqrt{{\bar{\phi}\over \phi}}\; \int d^3 \xi \, \sqrt{g}\, \hat{{\theta}}\hat{{\theta}}\delta^4(x-\hat{x}) + \nonumber \\
 + {1\over 48}  \int_{W^3} \hat{E}^c\wedge \hat{E}^b \wedge \hat{E}^a\epsilon_{abcd}
 \hat{{\theta}}{\sigma}^{d} \hat{\bar{\theta}}\, \delta^4(x-\hat{x}) + {\cal O}(f^4)\; ,  \qquad
\end{eqnarray}
where $ {\cal O}(f^4)$ are the same as in Eq. (\ref{JWZ0=}) (and thus can be read off Eq. (\ref{JWZ0=App})).

The second auxiliary field equation, Eq. (\ref{d(aux-aux)=}), reads  \begin{eqnarray}\label{d(aux-aux)=b+}
 \partial_a( DD\Phi \vert_0 -\bar{D}\bar{D}\bar{\Phi} \vert_0 ) = - {i\over 8\cdot 4!} \int\limits_{W^3}
\hat{E}^d\wedge \hat{E}^c\wedge \hat{E}^b \epsilon_{abcd} \delta^4(x-\hat{x}) + {\cal O}(f^2) \;  , \qquad
\end{eqnarray}
where the terms of higher order in fermions, ${\cal O}(f^2)$ can be found in Eqs. (\ref{DbDJWZ0=f2}) and (\ref{DbDJWZ0=f4}) of 
Appendix B (multiplying the expressions presented there by $- {i\over 4} \tilde{\sigma}_a^{\dot{\alpha}\alpha}$). On the first 
look it might seem that Eq. (\ref{d(aux-aux)=b+}) imposes additional restrictions on the supermembrane motion. Such possible 
restrictions might come from the selfconsistency condition of Eq. (\ref{d(aux-aux)=b+}); at zero order in fermions that 
reads \footnote{
$ \int\limits_{W^3}
\hat{E}^d\wedge \hat{E}^c\wedge \hat{E}^b \epsilon_{abcd} \delta^4(x-\hat{x})= \int\limits_{W^3} d\hat{x}^d\wedge d\hat{x}^c\wedge d\hat{x}^b \epsilon_{abcd} \delta^4(x-\hat{x})+ fermionic \; contributions $.  }
\begin{eqnarray}\label{dd(aux-aux)=0}
 \partial_{[a} \epsilon_{b]c_1c_2c_3} \int\limits_{W^3} d\hat{x}^{c_3}\wedge d\hat{x}^{c_2}\wedge d\hat{x}^{c_1}  \delta^4(x-\hat{x})=0
  \;  . \qquad
\end{eqnarray}
However, one can check that  this equation is satisfied identically. Indeed, using the identity 
$ \epsilon_{c_1c_2c_3[a} \partial_{b]}\equiv   -{3\over 2}\epsilon_{ab[c_1c_2}\partial_{c_3]} $ one can write the 
{\it l.h.s.} of Eq. (\ref{dd(aux-aux)=0}) in the form of 
$-{3\over 2}  \epsilon_{abc_1c_2}\int\limits_{W^3} d\hat{x}^{c_2}\wedge d\hat{x}^{c_1}\wedge d \delta^4(x-\hat{x})=
-{3\over 2}  \epsilon_{abc_1c_2}\int\limits_{W^3} d\left(d\hat{x}^{c_2}\wedge d\hat{x}^{c_1}\,  \delta^4(x-\hat{x}) \right)$ 
which vanishes as an integral of total derivative in the case of closed supermembrane which we are studying here 
($\partial W^3 =0\!\!\!/\quad \Rightarrow\quad \int_{W^3}d(...)=0$) .

As we have discussed in section \ref{threeform}, the on-shell transformations are obtained from the off-shell ones,
$\delta \psi_\alpha = 2(\sigma^a\bar{\varepsilon})_\alpha \partial_a \phi + {i\over 2}\varepsilon_\beta\, DD\Phi\vert_0$ for 
the case of fermions, by substituting the solution of the equations for the auxiliary fields. This implies that the on-shell 
supersymmetry transformation of fermions will be quite complicated due to the complicated structure of the auxiliary field 
equations (\ref{Eqm0=Jv0+O}) and  (\ref{d(aux-aux)=}).
As the on-shell fermionic supersymmetry transformations can be used to extract BPS conditions for the supersymmetric solutions, 
their further study in our simple system might lead to useful suggestions for the investigation of the backreaction of D=10,11 
super-$p$-branes on the  BPS solutions of supergravity equations.

\subsubsection{Dynamical field equations}

Fortunately, the dynamical equations for the physical fields of the special scalar multiplet following from the simplest 
interacting action Eq. (\ref{S=S0+SM2}) do not obtain contributions from the auxiliary fields of the scalar multiplet,
which on the mass shell are expressed by quite complicated Eqs. (\ref{Eqm0=Jv0+O}) and (\ref{d(aux-aux)=}).

Specifying the current contributions to (\ref{Dirac=J}) and (\ref{K-G=J}), we find the massless Dirac equation with the 
supermembrane contributions,
\begin{eqnarray}\label{Dirac=J+O}
{\sigma}^a_{\alpha\dot{\alpha}}\partial_a\psi^\alpha &=&
 {1\over  32}\,\sqrt{{\phi\over \bar{\phi}}}\; \int d^3 \xi \, \sqrt{g} \, \hat{\bar{\theta}}_{\dot{\alpha}}\delta^4 (x-\hat{x})- \qquad \nonumber \\
&& - {1\over 8\cdot 4!}  \int_{W^3} \hat{E}^c\wedge \hat{E}^b \wedge \hat{E}^a\epsilon_{abcd}
 (\hat{{\theta}}{\sigma}^{d})_{\dot{\alpha}}\, \delta^4(x-\hat{x}) +{\cal O}(f^3)
   \; , \qquad
\end{eqnarray}
and the Klein-Gordon equation, also with the source from supermembrane,
\begin{eqnarray}\label{K-G=J+O}
&& \Box \phi(x)=   {1\over 64} \,\sqrt{{{\phi}\over {\bar{\phi}}}}\; \int d^3 \xi \, \sqrt{g} \delta^4 (x-\hat{x}) - \qquad \nonumber \\
&& -{1\over 64}\int\limits_{W^3} \hat{E}^c\wedge \hat{E}^b\wedge d\hat{\theta} \sigma_{bc}\hat{\theta} \, \delta^4(x-\hat{x}) -
 {i\over 64}  \int\limits_{W^3} \hat{E}^c\wedge \hat{E}^b \wedge \hat{E}^a\epsilon_{abcd} (\hat{\theta})^2\partial^d\delta^4(x-\hat{x})+ \qquad \nonumber \\ &&
+ {i\over 64} \, \int d^3 \xi \, \sqrt{g}  (\hat{{\theta}}\sigma^a\hat{\bar{\theta}}) \sqrt{{\hat{\phi}\over {\hat{\bar{\phi}}}}}\; \partial_a\delta^4 (x-\hat{x}) + {\cal O}(f^4)
\;  . \qquad
\end{eqnarray}
The explicit form of the terms of higher order in fermions, ${\cal O}(f^4)$ in  (\ref{K-G=J+O}) and
${\cal O}(f^3)$ in  (\ref{Dirac=J+O}), can be extracted from the Eqs.  (\ref{bDbDJNG0=App-2}) and (\ref{bDbDJWZ0=}) in Appendix B.

\subsection{Simplest solution of the dynamical equations at leading order in supermembrane tension}\label{simplestsolution}

The above equations can be formally solved by
\begin{eqnarray}\label{psi=J+O}
\psi^\alpha &=& \psi_0^\alpha
 +{1\over  32}\, \int d^3 \xi \, \sqrt{g} \, \sqrt{{\hat{\phi}\over \hat{\bar{\phi}}}}\; (\hat{\bar{\theta}}\tilde{\sigma}^a)^{\alpha}\partial_a G_0 (x-\hat{x})+\qquad \nonumber \\
&& + {1\over 8\cdot 4!}  \int_{W^3} \hat{E}^c\wedge \hat{E}^b \wedge \hat{E}^a\epsilon_{abcd}
 (\hat{{\theta}}{\sigma}^{d}\tilde{\sigma}^a)^{\alpha}\partial_a\, G_0 (x-\hat{x})  +{\cal O}(f^3)
   \; , \qquad
\\ \label{phi=J+O}
\phi(x)&=& \phi_0(x) + {1\over 64} \, \int d^3 \xi \, \sqrt{g}  \sqrt{{\hat{\phi}\over \hat{\bar{\phi}}}}\; \left(G_0 (x-\hat{x}) + i   (\hat{{\theta}}\sigma^a\hat{\bar{\theta}}) \partial_aG_0 (x-\hat{x})\right)   - \qquad \nonumber \\
&& -{1\over 64}\int\limits_{W^3} \hat{E}^c\wedge \hat{E}^b\wedge d\hat{\theta} \sigma_{bc}\hat{\theta} G_0(x-\hat{x}) -
 {i\over 64}  \int\limits_{W^3} \hat{E}^c\wedge \hat{E}^b \wedge \hat{E}^a\epsilon_{abcd} (\hat{\theta})^2\partial^d G_0(x-\hat{x})+ \nonumber \\  && + {\cal O}(f^4)
\;  , \qquad
\end{eqnarray}
where $\psi_0^\alpha$ and $\phi_0(x)$ are solutions of the free equations and $G_0 (x-\hat{x})$ is the Green function of the 
free $D=4$ Klein-Gordon operator $\Box :=\partial_a\partial^a$,
\begin{eqnarray}\label{BoxG0=d}
\Box \phi_0(x)=0\; , \qquad {\sigma}^a_{\alpha\dot{\alpha}}\partial_a\psi_0^\alpha =0\,  \qquad \Box G_0 (x-\hat{x})= \delta^4(x-\hat{x})\;  .  \qquad
\end{eqnarray}
Eqs. (\ref{psi=J+O}) and (\ref{phi=J+O}) give only formal solutions as far as the pull--back of the phase of
the complex scalar superfield enters their {\it r.h.s.} through $ \sqrt{{\hat{\phi}\over \hat{\bar{\phi}}}}$ multipliers in 
the integrants.

Assuming the solution of the homogeneous equation to be real, $\phi_0(x)=(\phi_0(x))^*$ one can solve 
Eqs. (\ref{Dirac=J+O}) and (\ref{K-G=J+O}) in the first order in the supermembrane tension $T$ 
(this is set to unity in our equations above and below, but can be easily restored by 
$\int d^3 \xi \, \sqrt{g}  \mapsto T\int d^3 \xi \, \sqrt{g}  $ and $\int\limits_{W^3}\mapsto T\int\limits_{W^3}$). 
This reads (setting back $T=1$)
\begin{eqnarray}\label{psi=J+O+}
\psi^\alpha &=& \psi_0^\alpha
 +{1\over  32}\, \int d^3 \xi \, \sqrt{g} \, (\hat{\bar{\theta}}\tilde{\sigma}^a)^{\alpha}\partial_a G_0 (x-\hat{x})+ \qquad \nonumber \\
&& + {1\over 8\cdot 4!}  \int_{W^3} \hat{E}^c\wedge \hat{E}^b \wedge \hat{E}^a\epsilon_{abcd}
 (\hat{{\theta}}{\sigma}^{d}\tilde{\sigma}^a)^{\alpha}\partial_a\, G_0 (x-\hat{x})  +{\cal O}(f^3)
   \; , \qquad
\\ \label{phi=J+O+}
\phi(x)&=& \phi_0(x) + {1\over 64} \, \int d^3 \xi \, \sqrt{g} \left(G_0 (x-\hat{x}) + i   (\hat{{\theta}}\sigma^a\hat{\bar{\theta}}) \partial_aG_0 (x-\hat{x})\right)   + \qquad \nonumber \\
&& -{1\over 64}\int\limits_{W^3} \hat{E}^c\wedge \hat{E}^b\wedge d\hat{\theta} \sigma_{bc}\hat{\theta} G_0(x-\hat{x}) -
 {i\over 64}  \int\limits_{W^3} \hat{E}^c\wedge \hat{E}^b \wedge \hat{E}^a\epsilon_{abcd} (\hat{\theta})^2\partial^d G_0(x-\hat{x})+ \nonumber \\  && + {\cal O}(f^4)
\;  , \qquad  \phi_0(x)= (\phi_0(x))^* \;  . \qquad
\end{eqnarray}
The contribution of higher order in string tension would include the product of distributions
(of the type $G_0(x-\hat{x}(\xi_1))\, \delta^4(x-\hat{x}(\xi_2))$) and their accounting requires a careful study 
of a classical counterpart of the renormalization procedure, similar to the one developed for the radiation 
reaction problem \cite{Lyakhovich+Sharapov+2002} and for general relativity \cite{Infeld+Plebanski}. The generalization of 
such a technique for the case of $p$--brane has been developed in very recent \cite{Lechner:2010dr}.

\subsection{Superfield equations with nontrivial superpotential}\label{domainwall}

As a first stage in searching for solution of the superfield equations with a nontrivial superpotential, let us consider the 
relation with known domain wall solutions of the Wess--Zumino model \cite{Abraham+Townsend:1990,Cvetic+:1991}. To this end let us 
consider our dynamical system with nontrivial superpotential and the  simplest kinetic term. This is described by the interacting 
action
\begin{eqnarray}\label{S=SWZ+SM2}
S = \int d^8z\, \Phi \bar{\Phi}  + \int d^6\zeta_L W(\Phi) + \int d^6\zeta_R \bar{W}(\bar{\Phi})  + {1\over 2}\int d^3 \xi \sqrt{\hat{\Phi}\hat{\bar{\Phi}}}\sqrt{ g} -   \int\limits_{W^3} \hat{C}_3{}^\prime\;  .   \qquad
\end{eqnarray}
with $\Phi$ and $\bar{\Phi}$ and $\hat{C}_3{}^\prime$ expressed in terms of real pre-potential $V(z)$ by Eqs. (\ref{Phi=DDV}) 
and (\ref{C3'=}).

The superfield equations of motion (\ref{Eqm0=J}) acquire now the superpotential contributions,
\begin{eqnarray}\label{Eqm0+W=J}
  DD\Phi+\bar{D}\bar{D}\bar{\Phi} + W^\prime_\Phi (\Phi) + \bar{W}^\prime_{\bar{\Phi}}(\bar{\Phi}) = J(z)\;   \qquad
\end{eqnarray}
The auxiliary field equations read
\begin{eqnarray}\label{EqmW=Jv0}
&& DD\Phi \vert_0 +\bar{D}\bar{D}\bar{\Phi}\vert_0  + W^\prime_\phi (\phi) + \bar{W}^\prime_{\bar{\phi}}(\bar{\phi}) = J(z)\vert_0\; ,  \qquad \\
 \label{d(aux-aux)=W}
 && \partial_a( DD\Phi \vert_0 + \bar{W}^\prime_{\bar{\phi}}(\bar{\phi}) -\bar{D}\bar{D}\bar{\Phi} \vert_0  - W^\prime_\phi (\phi) ) = - {i\over 4} \tilde{\sigma}_a^{\dot{\alpha}\alpha}[D_\alpha \, , \, \bar{D}_{\dot\alpha}] J(z)\vert_0\;  , \qquad
\end{eqnarray}
and the dynamical field equations are
\begin{eqnarray}\label{Dirac-W=J}
&& {\sigma}^a_{\alpha\dot{\alpha}}\partial_a\psi^\alpha + {i\over 4} \bar{\psi}{}_{\dot{\alpha}} \, \bar{W}^{\prime\prime}_{\bar{\phi}\bar{\phi}}(\bar{\phi}) = {1\over 4} \bar{D}_{\dot{\alpha}}J(z)\vert_0 \; , \qquad
\\
\label{K-G-W=J}
&& \Box \phi(x) - {1\over 16} \bar{D}\bar{D}\bar{\Phi}\vert_0 \bar{W}^{\prime\prime}_{\bar{\phi}\bar{\phi}}(\bar{\phi}) +
{1\over 16} \bar{\psi}{}_{\dot{\alpha}}\bar{\psi}{}^{\dot{\alpha}} \bar{W}^{\prime\prime\prime}_{\bar{\phi}\bar{\phi}\bar{\phi}}(\bar{\phi})  = -{1\over 16} \bar{D}\bar{D}J(z)\vert_0 \;  \qquad
\end{eqnarray}
with the same supermembrane current (\ref{J=JNG+JWZ}), (\ref{JNG=}), (\ref{JWZ=}).

{\it In the absence of supermembrane current, $J=0$}, the auxiliary field equations are solved by
$DD\Phi \vert_0 =- \bar{W}^\prime_{\bar{\phi}}(\bar{\phi}) +ic$,
$\bar{D}\bar{D}\bar{\Phi}\vert_0 =- W^\prime_\phi (\phi) -ic$ and the constant $c$ can be removed by redefining superpotential 
$W^\prime_\phi (\phi)\mapsto W^\prime_\phi (\phi)+ ic $. Thus, without lost of generality, one can simplify notation and 
substitute $- \bar{W}^\prime_{\bar{\phi}}(\bar{\phi})$ for $DD\Phi \vert_0$ in the dynamical equations (\ref{K-G-W=J}).
Domain wall ansatz of \cite{Abraham+Townsend:1990,Cvetic+:1991} implies that all the fields are static and depend on only one 
spatial coordinate which we chose to be $x^2=y$.  Then Eqs. (\ref{Dirac-W=J}) and (\ref{K-G-W=J}) with $J=0$ becomes
\begin{eqnarray}\label{Dirac-Wy=J}
&& {\sigma}^2_{\alpha\dot{\alpha}} \partial_y \psi^\alpha (y)+ {i\over 4} \bar{\psi}{}_{\dot{\alpha}}(y) \, \bar{W}^{\prime\prime}_{\bar{\phi}\bar{\phi}}(\bar{\phi}(y)) = 0 \; , \qquad
\\
\label{K-G-Wy=J}
&& \partial_y^2 \phi(y) - {1\over 16} {W}^{\prime}_{{\phi}}({\phi})  \bar{W}^{\prime\prime}_{\bar{\phi}\bar{\phi}}(\bar{\phi})-
{1\over 16} \bar{\psi}{}_{\dot{\alpha}}\bar{\psi}{}^{\dot{\alpha}} \bar{W}^{\prime\prime\prime}_{\bar{\phi}\bar{\phi}\bar{\phi}}(\bar{\phi})  = 0 \;  \qquad
\end{eqnarray}
Notice that Eq. (\ref{Dirac-Wy=J}) split into the pair of equations for $\psi_1 ,\bar{\psi}_1$ and $\psi_2 ,\bar{\psi}_2$,
\begin{eqnarray}\label{Dirac-Wy1=J}
\partial_y \psi_1 (y)+ {1\over 4} \bar{\psi}{}_{\dot{1}}(y) \, \bar{W}^{\prime\prime}_{\bar{\phi}\bar{\phi}}(\bar{\phi}(y)) = 0 \; , \qquad \partial_y \psi_2 (y)+ {1\over 4} \bar{\psi}{}_{\dot{2}}(y) \, \bar{W}^{\prime\prime}_{\bar{\phi}\bar{\phi}}(\bar{\phi}(y)) = 0 \;
\; , \qquad
\end{eqnarray}
such that the solution of the second can be constructed from the solution of the first as $\psi_2= \psi_1$, $
\bar{\psi}{}_{\dot{2}}=\bar{\psi}{}_{\dot{1}}$.  For such a solution of the fermionic equation $\bar{\psi}{}_{\dot{\alpha}}\bar{\psi}{}^{\dot{\alpha}}=0$ 
and the bosonic equation simplifies to
$\partial_y^2 \phi(y) - {1\over 16} {W}^{\prime}_{{\phi}}({\phi})  \bar{W}^{\prime\prime}_{\bar{\phi}\bar{\phi}}(\bar{\phi})=0$. 
This, in its turn, is solved by any solution of the following first order BPS equations \cite{Abraham+Townsend:1990,Cvetic+:1991}
\begin{eqnarray}
\label{dphi=}
&& \partial_y\phi(y) - {e^{i\alpha}\over 4} \bar{W}^{\prime}_{\bar{\phi}}(\bar{\phi}(y))  = 0 \; , \qquad
\partial_y\bar{\phi}(y)- {e^{-i\alpha}\over 4} {W}^{\prime}_{{\phi}}({\phi}(y))  = 0 \; . \qquad
\end{eqnarray}
Abraham and Townsend \cite{Abraham+Townsend:1990} studied the intersecting domain wall solutions of
(\ref{dphi=}) with $W=\Phi^4-4\Phi$. The generalization of the above equations for the case of supergravity was studied in 
\cite{Cvetic+:1991}. For $W= a^2\Phi - {\Phi^3\over 3}$  Eq. (\ref{dphi=}) has kink solution 
$\phi= a\, \tanh (ya)$ \cite{Cvetic+:1991}. Notice that in the case of special chiral multiplet such a potential 
would be deformed by the contribution of an arbitrary imaginary constant, $a^2 \mapsto a^2+ic$.

When the BPS equations (\ref{dphi=})  are satisfied, the solution of fermionic equations can be written in the form
\cite{Abraham+Townsend:1990}
\begin{eqnarray}
\label{psi1=}
\psi_1 =2 \chi \, e^{-i\alpha /2} \partial_y\phi(y) =\psi_2 \; ,  \qquad  \bar{\psi}_{\dot{1}} = 2 \chi e^{i\alpha /2} \partial_y\phi(y) =\bar{\psi}_{\dot{2}}\;  \qquad
 \;  \qquad
\end{eqnarray}
with a real Grassmann (fermionic) constant  $\chi$,
\begin{eqnarray}
\label{chi=chi*}
 \chi =\chi^* \; ,  \qquad \chi \chi=0 \; .
\end{eqnarray}

An interesting problem is to study the influence of the supermembrane source on the above discussed 
nonsingular domain wall solutions. First observation is that, to maintain the general structure of the solution (\ref{psi1=}) of 
the fermionic equations, the source contribution in the {\it r.h.s.} of (\ref{Dirac-W=J}),
${\sigma}^2_{\alpha\dot{\alpha}}\partial_y\psi^\alpha (y)+ {i\over 4} \bar{\psi}{}_{\dot{\alpha}} (y)\, \bar{W}^{\prime\prime}_{\bar{\phi}\bar{\phi}}(\bar{\phi})(y) = {1\over 4} \bar{D}_{\dot{\alpha}}J(z)\vert_0$, 
should be proportional to the same real Grassmann constant  $\bar{D}_{\dot{\alpha}}J(z)\vert_0 \propto \chi$. 
This in its turn suggests the following ansatz for the fermionic coordinates functions
\begin{eqnarray}
\label{th=}
\hat{\theta}{}^{\alpha} (\xi) = u^{\alpha}(\xi) \chi \; , \qquad \hat{\bar{\theta}}{}^{\dot\alpha} (\xi) = \bar{u}^{\dot\alpha}(\xi) \chi
 \;   \qquad
\end{eqnarray}
with some bosonic functions $u^{\alpha}(\xi)= (\bar{u}^{\dot\alpha}(\xi))^*$. Such an ansatz results in 
that $\hat{\theta}{}^{\alpha} \hat{\theta}{}^{\beta} =0= \hat{\theta}{}^{\alpha} \hat{\bar{\theta}}{}^{\dot\alpha} $ and, hence, 
in that the pull--back of bosonic vielbein simplifies to $\hat{E}^a=d\hat{x}^a$. Furthermore, assuming that the normal to the 
supermembrane worldvolume cannot be orthogonal to the $y=x^2$ axis, we can chose the `static gauge' 
where $\hat{x}^0(\xi) =\xi^0=\tau$, $\hat{x}^1(\xi)=\xi^1$, $\hat{x}{}^3=\xi^2$ so that the only nontrivial bosonic coordinate 
function (supermembrane Goldstone field) is identified with $\hat{x}{}^2=\hat{y}(\xi )$. In this gauge
\begin{eqnarray}
\label{hEa=static}
\hat{E}^0=d\xi^0\; , \qquad \hat{E}^1=d\xi^1
 \; ,  \qquad  \hat{E}^2=d\hat{y}(\xi)= d\hat{y}(\xi^0,\xi^1,\xi^2)\; ,  \qquad  \hat{E}^3=d\xi^2 \; . \qquad
\end{eqnarray}
Furthermore, with such an ansatz $J\vert_0=0$ and all the (quite complicated) components of current superfield 
(see appendix \ref{appendixB}) simplify drastically reducing to their leading terms. Then the problem of finding (particular) solutions of 
the system of interacting equations looks manageable.

\chapter{Supermembrane interaction with dynamical D=4 N=1 supergravity.
Superfield Lagrangian description and spacetime equations of motion}\label{chaptersugra+membrane}
\thispagestyle{chapter} 

 \initial{I}n this chapter we study the interacting system of supermembrane and $D=4$ ${\cal N}=1$ dynamical 
supergravity.
We obtain the complete set of equations of motion for this system by varying its complete superfield action.
These include the supermembrane equations, which are formally the same as in the case of supermembrane in supergravity background, and 
the superfield supergravity equations with supermembrane contributions.
The existence of a three form potential $C_3$ allowing for a Wess--Zumino term in the supermembrane part of the action imposes a 
restriction on the prepotential structure of minimal supergravity, making its chiral compensator being constructed from a real rather than 
complex prepotential.
We will develop the Wess--Zumino type approach for this {\it special minimal supergravity} and present its basic variations which 
are characterized, besides $\delta H^a$, by one real variation ($\delta V$) instead a 
complex ones ($\delta U$, $\delta\bar{U}$).
The most important consecuence of this modification in the prepotential structure is that the right hand side of the 
Einstein equation acquires a term proportional to an integration constant, a dynamically generated cosmological constant. 
We also present the supergravity superfield equations with the supermembrane contributions obtained by varying the action with respect to the 
superfields of {\it special minimal supergravity} and write these resulting superfield equations in the special
''WZ$_{\hat{\theta}=0}$'' gauge, which is the standard Wess--Zumino gauge completed by the condition that the supermembrane 
Goldstone field is set to zero ($\hat{\theta}=0$).
We solve the auxiliary field equations and show that these result in the effect of dynamical generation of cosmological 
constant and its 'renormalization'(due to supermembrane contributions) in such a way that the cosmological constant values in the branches of spacetime 
separated by the supermembrane worldvolume are generically different.

\section{Superfield supergravity and supermembrane in curved D=4, ${\cal N}$=1 superspace}
\label{sugra2}
\subsection{Superfield supergravity action, superspace constraints and equations of motion}\label{sugra2-2}

The superfield action of the minimal off-shell formulation of $D=4$, ${\cal N}=1$  supergravity \cite{WZ78}
\begin{eqnarray}\label{SGact} S_{SG} = \int d^8Z \;
E :=  \int d^4 x \tilde{d}^4\theta \; sdet(E_M^A) \; , \qquad
\end{eqnarray}
is given by the superdeterminant (or Berezinian) of the matrix of supervielbein coefficients, 
$E_M^A(Z)$ in (\ref{4EAM}), which obey the set of supergravity constraints.
These can be collected together with their consequences in the following expressions for the superspace 
torsion 2-forms (\ref{Ta:=}), (\ref{Talf:=}) (see \cite{BdAIL03} and refs. therein)
\begin{eqnarray}\label{4WTa=}T^a &=&- 2i\sigma^a_{\alpha\dot{\alpha}} E^\alpha \wedge \bar{E}^{\dot{\alpha}} - {1\over 8} E^b \wedge E^c
\varepsilon^a{}_{bcd} G^d \; ,\;   \\ \label{4WTal=} T^{\alpha} &:=& (T^{\dot{\alpha}})^*= {i\over 8} E^c \wedge E^{\beta}
(\sigma_c\tilde{\sigma}_d)_{\beta} {}^{\alpha} G^d   -{i\over 8} E^c
\wedge \bar{E}^{\dot{\beta}} \epsilon^{\alpha\beta}\sigma_{c\beta\dot{\beta}}R +
 {1\over 2} E^c \wedge E^b \; T_{bc}{}^{\alpha}\; .\;\nonumber\\
\end{eqnarray}
The {\it  main superfields}, real vector $G_a=(G_a)^*$ and complex scalar $R=(\bar{R})^*$, entering 
(\ref{4WTa=}) and (\ref{4WTal=}),  obey
\begin{eqnarray} \label{chR} & {\cal D}_\alpha \bar{R}=0\;
, \qquad \bar{{\cal D}}_{\dot{\alpha}} {R}=0\; ,
\\
\label{DG=DR} & \bar{{\cal
D}}^{\dot{\alpha}}G_{{\alpha}\dot{\alpha}}= - {\cal D}_{\alpha} R \; , \qquad {{\cal
D}}^{{\alpha}}G_{{\alpha}\dot{\alpha}}= - \bar{{\cal D}}_{\dot{\alpha}} \bar{R} \; .
 \qquad \end{eqnarray}
These relations can be obtained by studying the Bianchi identities (\ref{BI=DT}), which also allow to find 
the expression for  superfield generalization of the gravitino field strength, $T_{bc}{}^{\alpha}(Z)$,
\begin{eqnarray}\label{Tabg} T_{{\alpha}\dot{\alpha}\; \beta \dot{\beta }\; {\gamma}} &
\equiv   \sigma^a_{{\alpha}\dot{\alpha}} \sigma^b_{\beta \dot{\beta }} \epsilon_{{\gamma}{\delta}}
T_{ab}{}^{{\delta}}=  -{1\over 8}  \epsilon_{{\alpha}{\beta}} {\bar{{\cal D}}}_{(\dot{\alpha}|} G_{\gamma
|\dot{\beta})} - {1\over 8} \epsilon_{\dot{\alpha}\dot{\beta}}[W_{\alpha \beta\gamma} -
2\epsilon_{\gamma (\alpha}{\cal D}_{\beta)} R]
 \; , \end{eqnarray}
involving one more main superfield, $W_{\alpha \beta\gamma}=W_{(\alpha \beta\gamma )}=: (\bar{W}_{\dot{\alpha}\dot{\beta}\dot{\gamma}})^*$. 
This  obeys
 \begin{eqnarray}
 \label{chW} & \bar{{\cal D}}_{\dot{\alpha}} W^{\alpha\beta\gamma}= 0\; , \qquad
{{\cal D}}_{{\alpha}}\bar{W}^{\dot{\alpha}\dot{\beta}\dot{\gamma}}= 0\;,
 \\ \label{DW=DG} & {{\cal D}}_{{\gamma}}W^{{\alpha}{\beta}{\gamma}}= \bar{{\cal D}}_{\dot{\gamma}} {{\cal
D}}^{({\alpha}}G^{{\beta})\dot{\gamma}} \; . \qquad
\end{eqnarray}

Studying the Bianchi identities with the constraints   (\ref{4WTa=}), (\ref{4WTal=})  one also finds that 
the superfield generalization of  the left hand side of the supergravity Rarita--Schwinger equation reads
\begin{eqnarray}\label{SGRS=off}
\epsilon^{abcd}T_{bc}{}^{\alpha}\sigma_{d\alpha\dot{\alpha}} ={i\over 8}
\tilde{\sigma}^{a\dot{\beta}\beta} \bar{{\cal D}}_{(\dot{\beta}|} G_{\beta|\dot{\alpha})} + {3i\over 8}
{\sigma}^a_{\beta \dot{\alpha}} {\cal D}^{\beta}R \; ,
\end{eqnarray}
and the superfield generalization of the Ricci tensor is
\begin{eqnarray} \label{RRicci}
R_{bc}{}^{ac}& = {1\over 32} ({{\cal D}}^{{\beta}} \bar{{\cal D}}^{(\dot{\alpha}|}
G^{{\alpha}|\dot{\beta})} - \bar{{\cal D}}^{\dot{\beta}} {{\cal D}}^{({\beta}}G^{{\alpha})\dot{\alpha}})
\sigma^a_{\alpha\dot{\alpha}}\sigma_{b\beta\dot{\beta}} - {3\over 64}
(\bar{{\cal D}}\bar{{\cal D}}\bar{R} + {{\cal D}}{{\cal D}}{R}- 4 R\bar{R})\delta_b^a\; .
\end{eqnarray}
This suggests that superfield supergravity equation should have the form\footnote{See \cite{BdAIL03,IB+CM:2011} for more detail on the superfield description of minimal supergravity in the present notation.}
 \begin{eqnarray}\label{SGeqmG}
 && G_a =0 \; , \qquad
\\ \label{SGeqmR}
 && R=0 \; ,  \qquad \bar{R}=0\; . \qquad
\end{eqnarray}

Eqs. (\ref{SGeqmG}) and (\ref{SGeqmR}) can be obtained by varying the action (\ref{SGact}) with respect to supervielbein 
obeying the supergravity constraints (\ref{4WTa=}), (\ref{4WTal=}) \cite{WZ78}. Such admissible variations are expressed 
through a vector parameter
$\delta {\cal H}^a$ and complex scalar parameter $\delta {\cal U}= (\delta \bar{{\cal U}})^*$ which enter the variation 
of the supervielbein and spin connection under the symbol of the chiral projector  
$({\cal D}^\alpha{\cal D}_\alpha -\bar{R})$ (see \cite{BdAIL03,IB+CM:2011} for more detail). They  correspond to the 
variations of the so-called {\it prepotentials}, unconstrained superfields which appear in the  general solution of 
the supergravity constraints. The minimal supergravity constraints are solved in terms of the axial vector superfield 
${\cal H}^\mu$ \cite{OS78} and chiral compensator $\Phi$ \cite{Siegel:1978nn}. This latter obeys 
$\bar{{\cal D}}_{\dot{\alpha}}\Phi=0$ and, hence, can be expressed as 
$\Phi =(\bar{{\cal D}}_{\dot{\alpha}}\bar{{\cal D}}{}^{\dot{\alpha}}-R){\cal U}$ with a complex unconstrained superfield
${\cal U}$.

Thus the set of minimal supergravity prepotentials includes ${\cal H}^\mu$, ${\cal U}$ and  $\bar{{\cal U}}=({{\cal U}})^*$ 
which are in one to one correspondence with the set of three independent variations 
$\delta {\cal H}^a$, $\delta {\cal U}$ and  $\delta \bar{{\cal U}}=(\delta  {{\cal U}})^*$ of the Wess--Zumino approach 
to supergravity \cite{WZ77,WZ78} producing the three superfield equations (\ref{SGeqmG}) and  (\ref{SGeqmR}). In short, 
as it had been known already from \cite{WZ78},
 \begin{eqnarray}\label{vSGsf} \delta S_{SG} =  \int
d^8Z E\;  \left[{1\over 6} G_a \; \delta H^a -  2 R\; \delta \bar{{\cal U}} -2 \bar{R}\; \delta {\cal U}\right] \; . \quad
\end{eqnarray}

\subsection{Supermembrane action in minimal supergravity background}\label{sugra2-3}

As it is well known, the supermembrane action \cite{BST87,4Dsuperm}({\it cf. } chapter \ref{chapterscalar}) is given by the sum of the Dirac--Nambu--Goto and 
the Wess--Zumino term,
  \begin{eqnarray}\label{Sp=2:=}
  S_{p=2}= {1\over 2}\int d^3 \xi \sqrt{g} - \int\limits_{W^3} \hat{C}_3 = -{1\over 6} \int\limits_{W^3} *\hat{E}_a\wedge \hat{E}{}^a - \int\limits_{W^3} \hat{C}_3\; . \qquad
\end{eqnarray}
The former is given by the volume of $W^3$ defined as  integral of the  determinant of the induced metric, 
$g=det(g_{mn})$,
\begin{eqnarray}\label{g=EE}
g_{mn}= \hat{E}_m^{a}\eta_{ab} \hat{E}_n^{b}\; , \qquad
\hat{E}_m^{a}:= \partial_m \hat{Z}^M(\xi) E_M^a(\hat{Z})\; .  \qquad
\end{eqnarray}
Here $\xi^m= (\tau,\sigma^1, \sigma^2)$ are local coordinates on $W^3$ and $\hat{Z}^M(\xi)$ are coordinates functions 
which determine the embedding of $W^3$ as a surface in target superspace $\Sigma^{(4|4)}$,
\begin{eqnarray}\label{W3inS44}
W^3\; \subset  \Sigma^{(4|4)}\; : \qquad Z^M= \hat{Z}{}^{{ {M}}}(\xi)= (\hat{x}{}^{\mu}(\xi)\, ,
\hat{\theta}^{\breve{\alpha}}(\xi))\; . \;
\end{eqnarray}
In the second equality  of (\ref{Sp=2:=}) the Dirac--Nambu--Goto term is written  as an integral of the wedge 
product of the pull--back of the $\Sigma^{(4|4)}$ bosonic supervielbein form $E^a$ to $W^3$,
\begin{eqnarray}\label{hEa=dxiE}
\hat{E}{}^a= d\xi^m \hat{E}_m^{a} = d \hat{Z}^M(\xi) E_M^a(\hat{Z})\; ,  \qquad
\end{eqnarray}
and of its Hodge dual two form defined with the use of the induced metric (\ref{g=EE}) and its inverse $g^{mn}$,
\begin{eqnarray}\label{*Ea:=}
*\hat{E}^a:= {1\over 2}d\xi^m\wedge d\xi^n\sqrt{g}\epsilon_{mnk}g^{kl}\hat{E}_l^a \; . \qquad \end{eqnarray}

The second, Wess--Zumino term of the supermembrane action (\ref{Sp=2:=}) describes the  supermembrane coupling 
to a 3--form gauge potential $C_3$ defined on $\Sigma^{(4|4)}$,
\begin{eqnarray} \label{C3:=}
 C_3= {1\over 3!} dZ^M\wedge dZ^N\wedge dZ^K C_{KNM}(Z)= {1\over 3!} E^C\wedge E^B\wedge E^A C_{ABC}(Z)\;  .  \qquad
\end{eqnarray}
Thus, to write a supermembrane action, one has to construct the 3--form gauge potential $C_3$ in the target superspace 
$\Sigma^{(4|4)}$ and take its pull--back to the supermembrane worldvolume
\begin{eqnarray} \label{C3:=}
 \hat{C}_3 &=& {1\over 3!} d\hat{Z}^M\wedge d\hat{Z}^N\wedge d\hat{Z}^K C_{KNM}(\hat{Z}(\xi))=
 {1\over 3!} \hat{E}^C\wedge \hat{E}^B\wedge \hat{E}^A C_{ABC}(\hat{Z})= \qquad \nonumber \\ &=&  {1\over 3!} d\xi^m\wedge d\xi^n\wedge d\xi^k \hat{C}_{knm} =
   d^3\xi \epsilon^{mnk}  \hat{C}_{knm}\;  .  \qquad
\end{eqnarray}

Actually, to study {\it supermembrane in supergravity background}, it is sufficient to know the field strength of 
the above 3--form potential, $H_4=dC_3$. This should be closed, $dH_4=0$, and supersymmetric invariant 4--form.
In flat superspace such a form exists and represents a nontrivial Chevalley--Eilenberg cohomology  of the ${\cal N}=1$ 
supersymmetry algebra \cite{JdA+PKT89,Jose+PaulPRL:89,AI95}({\it cf. } chapter \ref{chapterscalar}).

The minimal supergravity superspace  allows for existence of two closed 4-forms
\begin{eqnarray} \label{H4L}  & H_{4L} =  - {i\over 4} E^b\wedge E^a \wedge E^\alpha \wedge E^\beta \sigma_{ab\; \alpha\beta} -   {1\over 128} E^{d} \wedge E^c \wedge E^b \wedge E^a \epsilon_{abcd} R \, , \qquad
 dH_{4L}=0 \, ,
\end{eqnarray}
and its complex conjugate $H_{4R}=(H_{4L})^*$ (see \cite{IB+CM:2011}).
Its real part,
\begin{eqnarray} \label{H4=HL+HR}  H_4&:=&dC_3= {1\over 4!} E^{A_4}\wedge ... \wedge E^{A_1}
H_{A_1\ldots A_4}(Z)= H_{4L}+H_{4R}\;  ,   \qquad
\end{eqnarray}
is also closed  and  provides  the  4--form field strength associated to the Wess--Zumino (WZ) term of the supermembrane 
action  in the minimal supergravity background  \cite{Ovrut:1997ur}, $\int_{W^3} C_3$ in (\ref{Sp=2:=}). Indeed, the 
WZ term can also be defined as an integral of the closed 4 form $H_4$,  related to $C_3$ by $H_4=dC_3$, over some four 
dimensional space ${W}^4$ the boundary of which is given by the supermembrane worldvolume $W^3$,
\begin{eqnarray} \label{WZ=intH4}
\int\limits_{W^3}{C}_3= \int\limits_{W^4\;:\; \partial W^4=W^3}{H}_4
\;  .    \qquad
\end{eqnarray}
The condition that the form $H_4$ is closed, $dH_4=0$, guaranties that the integral $\int_{{W}^4}H_4$ is independent on 
the choice of ${W}^4$ and, thus, is related to the supermembrane worldvolume $W^3$.

The fact that  the knowledge of $H_4$ is completely sufficient for studying the properties of closed supermembrane in 
a supergravity background is related to that in this case the only dynamical variables are the supermembrane coordinate 
functions $\hat{Z}^M(\xi)$, that the action is written in term of pull--back of differential forms to $W^3$, and that 
the variation of the differential form with respect to the coordinates can be calculated with the use of the Lie derivative 
formula, in particular $\delta_{\delta Z}C_3= i_{\delta Z} H_4 + di_{\delta Z}  C_3=
{1\over 3!} E^{A_4}\wedge ... \wedge E^{A_2} \; \delta Z^{M} \, E_M^{A_1}\,
H_{A_1\ldots A_4}(Z) + d({1/2} E^{C}\wedge E^{B} \; \delta Z^{M} \, E_M^{A}\,
C_{ABC}(Z))$.  \footnote{
For closed supermembrane $\partial W^3= \emptyset$ so that $\int_{W^3}d\alpha_2 =\int_{\partial W^3}\alpha_2 = 0$  for 
any 2-form $\alpha_2$, including for $\alpha_2=i_{\delta Z} C_3$.}

The supermembrane equations of motion in the minimal supergravity background, which are obtained by varying 
the action (\ref{Sp=2:=}) with respect to the coordinate functions $\delta \hat{Z}^M(\xi)$,
\begin{eqnarray} \label{vSp2:=}
\delta S_{p=2}= \int_{W^3} \left({1\over 2} {\cal M}_{3a} E_M{}^a (\hat{Z})+  i \Psi_{3\alpha} E_M{}^\alpha (\hat{Z}) + i \Psi_{3\dot\alpha} E_M{}^{\dot\alpha} (\hat{Z})  \right)
\, \delta \hat{Z}^M(\xi)
\;  ,    \qquad
\end{eqnarray}
read
\begin{eqnarray}\label{SmEqm=b}
 {\cal M}_{3\, a}&:=& {\cal D}* \hat{E}_a + i \hat{E}{}^b\wedge \hat{E}{}^{\alpha}\wedge \hat{{E}}{}^{{\beta}}{\sigma}_{ab{\beta}{\alpha}} -  i  \hat{E}{}^b\wedge \hat{\bar{E}}{}^{\dot{\alpha}} \wedge \hat{\bar{E}}{}^{\dot{\beta}}\tilde{\sigma}_{ab\dot{\beta}\dot{\alpha}} - \qquad \nonumber \\ &&
-{1\over 8} \hat{E}{}^b\wedge \hat{E}{}^c\wedge \hat{E}{}^d\epsilon_{abcd} (R+\bar{R})
=0
\;   \qquad
\end{eqnarray}
and
\begin{eqnarray}\label{SmEqm=f}
 \bar{\Psi}{}_{3\dot{\alpha}}:= *\hat{E}_a\wedge \left(\hat{E}{}^\alpha {\sigma}{}^a_{{\alpha}\dot{\alpha}}-
 (\tilde{\bar{\gamma}} {\sigma}{}^a)_{\dot{\alpha}\dot{\beta}} \hat{\bar{E}}{}^{\dot{\beta}}\right)=0 \; , \qquad
 \\
 \label{SmEqm=bf}
 {\Psi}{}_{3{\alpha}}:= *\hat{E}_a\wedge \left( {\sigma}{}^a_{{\alpha}\dot{\alpha}} \hat{\bar{E}}{}^{\dot{\alpha}}+
  \hat{{E}}{}^{{\beta}} ( {\sigma}{}^a\tilde{\bar{\gamma}})_{{\alpha}{\beta}}\right)=0 \; , \qquad
  \end{eqnarray}
where the matrix $\bar{\gamma}_{\beta\dot{\alpha}}$ is defined by
  \begin{eqnarray}\label{tg:=}
  \bar{\gamma}_{\beta\dot{\alpha}}= \epsilon_{\beta\alpha}\epsilon_{\dot{\alpha}\dot{\beta}}\tilde{\bar{\gamma}}{}^{\dot{\beta}{\alpha}}= {i\over 3!\sqrt{g}} \sigma^a_{\beta\dot{\alpha}}\epsilon_{abcd} \epsilon^{mnk} \hat{E}{}_m^b\hat{E}{}_n^c\hat{E}{}_k^d =- (
  \bar{\gamma}_{{\alpha}\dot{\beta}})^*\;  \qquad
\end{eqnarray}
and obeys
\begin{eqnarray}
  \label{tgtg=I} \bar{\gamma}_{\beta\dot{\alpha}}\tilde{\bar{\gamma}}{}^{\dot{\alpha}{\alpha}}=  \delta_{\beta}{}^{{\alpha}}\; , \qquad\tilde{\bar{\gamma}}{}^{\dot{\alpha}{\alpha}}  \bar{\gamma}_{\alpha\dot{\beta}}=  \delta^{\dot{\alpha}}{}_{\dot{\beta}}\; . \qquad
\end{eqnarray}
Some identities involving the above matrix are
\begin{eqnarray}
  \label{tgs=stg+} \bar{\gamma}\tilde{\sigma}{}^a = -  {\sigma}{}^a \tilde{\bar{\gamma}} +
 {i\over 3!\sqrt{g}} \epsilon_{abcd} \epsilon^{mnk} \hat{E}{}_m^b\hat{E}{}_n^c\hat{E}{}_k^d
\; , \qquad
\\  \label{Etgsa=}
*\hat{E}_a \bar{\gamma}\tilde{\sigma}{}^a\bar{\gamma}= *\hat{E}_a {\sigma}{}^a
\; , \qquad *\hat{E}_a \bar{\gamma}\tilde{\sigma}{}^a=- *\hat{E}_a {\sigma}{}^a \tilde{\bar{\gamma}}
\; , \qquad
\\  \label{EEs=*Esg}
{1\over 2} \;
 \hat{E}{}^b\wedge \hat{E}{}^a\wedge \hat{{E}}{}^{{\beta}}\; {\sigma}_{ab{\beta}{\alpha}} = \; *\hat{E}_a \wedge \hat{{E}}{}^{{\beta}}({\sigma}^a\tilde{\bar{\gamma}})_{{\beta}{\alpha}} \; , \qquad
\\  \label{EEts=*Etsg} {1\over 2}
 \hat{E}{}^b\wedge \hat{E}{}^a\wedge \hat{\bar{E}}{}^{\dot{\beta}}\tilde{\sigma}_{ab\dot{\beta}\dot{\alpha}}=
- *\hat{E}_a \wedge \hat{\bar{E}}{}^{\dot{\beta}} (\tilde{\sigma}{}^a\bar{\gamma})_{\dot{\beta}\dot{\alpha}}
\; . \qquad
\end{eqnarray}
They are useful, in particular, to show that  the fermionic equations of motion obey the Noether identity 
$\bar{\Psi}{}_{3\dot{\alpha}}= {\Psi}{}_{3}^{\alpha}{\bar{\gamma}}_{{\alpha}\dot{\alpha}} $ reflecting the local 
fermionic $\kappa$--symmetry
\begin{eqnarray}
  \label{kappaZ=}
\delta_\kappa \hat{Z}^M = \kappa^\alpha (\xi)(E_\alpha^M(\hat{Z})+ \bar{\gamma}_{\alpha\dot{\alpha}}\epsilon^{\dot{\alpha}\dot{\beta}}E_{\dot{\beta}}^M(\hat{Z})) \; \qquad
\end{eqnarray}
with the local fermionic ``parameter''  $\kappa^\alpha (\xi) =({\bar{\kappa}}{}^{\dot{\alpha}})^*$ obeying
\begin{eqnarray}
  \label{kappa=*}
\kappa^\alpha (\xi)  = -
{\bar{\kappa}}_{\dot{\alpha}}(\xi) \tilde{\bar{\gamma}}{}^{\dot{\alpha}{\alpha}} \; \qquad  \Leftrightarrow \; \qquad
{\bar{\kappa}}_{\dot{\alpha}} (\xi) = - \kappa^\alpha (\xi)  {\bar{\gamma}}_{{\alpha}\dot{\alpha}}\; .
\end{eqnarray}
The relation of the supermembrane $\kappa$--symmetry in curved superspace with the minimal supergravity 
constraints was discussed  in \cite{Ovrut:1997ur}. The flat superspace limit of our equations reproduces 
the equations of the seminal paper \cite{4Dsuperm}.

\subsection{3--form potential in the minimal supergravity  superspace. Special minimal supergravity  }\label{sugra2-4}


Thus, as we have seen in the previous subsection,  to find the equations of motion of supermembrane in supergravity 
background as well as to study its symmetries it is sufficient to know the closed 4-form $H_4=dC_3$ in the background 
superspace.

However, to calculate the supermembrane current(s) describing the supermembrane contribution(s) to the supergravity 
(super)field equations, one needs to vary the Wess--Zumino term $\int \hat{C}_3$ of the supermembrane action with respect 
to the supergravity (super)fields. Thus one arrives at a separate problem of  finding the variation
\begin{eqnarray} \label{vC3:=}
 \delta C_3= {1\over 3!} E^C\wedge E^B\wedge E^A \beta_{ABC}(\delta)\;    \qquad
\end{eqnarray}
such that $d\delta C_3=\delta H_4$ reproduces the variation of $H_4$ from (\ref{H4=HL+HR}), (\ref{H4L}),
written in terms of the basic supergravity variations (we refer to Appendix \ref{appendiceC} for the explicit expression of 
$\delta H_4$).

Studying such a technical problem we have found that it imposes a restriction on the independent 
variations of the supergravity prepotentials, or equivalently, on the independent parameters of the admissible 
supervielbein variations,  thus transforming the generic minimal supergravity into a {\it special minimal supergravity}.
This off--shell supergravity formulation had been described for the first time in \cite{Grisaru:1981xm}, further discussed 
in \cite{Gates:1980az} (see also latter \cite{Kuzenko+05})  and elaborated in  \cite{Ovrut:1997ur} using the elegant 
combination of superfield results and the component 'tensor calculus' approach on the line of \cite{BW}.

In \cite{IB+CM:2011}  we described this {\it special minimal supergravity} in the complete Wess--Zumino superfield formalism. 
Referring to that paper for technical details, we only notice that the existence of the 3-form potential 
imposes a restriction on the prepotential structure of minimal supergravity which in our approach manifests itself 
in that the basic complex variations ${\delta U}$ and   ${\delta \bar{U}}=({\delta U})^*$ are expressed in terms of one 
real variation $\delta V$, essentially
\begin{eqnarray}
 \label{vcU=vV+}
\delta {\cal U} = {i\over 12}\delta {V} \; ,   \qquad \delta \bar{{\cal U}} = - {i\over 12}\delta {V} \;  . \qquad
\end{eqnarray}
As a result, the variation of the special minimal supergravity action  is essentially 
(see Appendix \ref{appendiceC})
\begin{eqnarray}\label{vSGsf=s}  \delta S_{SG} &=&  {1\over 6} \int
d^8Z E\;  \left[ G_a \; \delta H^a + (R-\bar{R}) i\delta {V}  \right]
 \; . \qquad
\end{eqnarray}
Hence the set of superfield equations of special minimal supergravity still includes
 the vector superfield equation (\ref{SGeqmG}),
\begin{eqnarray}\label{SGeqmG=0}
G_a=0\; , \qquad
\end{eqnarray}
but instead of the complex scalar superfield equations (\ref{SGeqmR}), valid in the case of generic minimal supergravity, 
in the case of special minimal supergravity we have only the real scalar equation
\begin{eqnarray}\label{SGeqmR+*=0}
R- \bar{R}=0\; . \qquad
\end{eqnarray}
Clearly, due to chirality of $R$, $\;\bar{{\cal D}}_{\dot\alpha} {R}=0$,  and anti-chirality of 
$\bar{R}$, $\; {\cal D}_\alpha \bar{R}=0$, the above Eq. (\ref{SGeqmR+*=0}) also implies that $d(R+\bar{R})=0$ so that 
on the mass shell the complex superfield $R$ is actually equal to a {\it real}  constant,
\begin{eqnarray}\label{bR=4ic}
R=4c \; , \qquad \bar{R}=4c\; , \qquad c=const=c^*\; .
\end{eqnarray}
Using (\ref{RRicci}), one finds that the superfield equation (\ref{SGeqmR+*=0}) results in Einstein equation with 
cosmological constant
\begin{eqnarray}\label{RRici=c}
R_{bc}{}^{ac}= 3c^2 \delta_b{}^a\; .
\end{eqnarray}
The value of the cosmological constant is proportional to the square of the above arbitrary constant $c$, which has 
appeared as an integration constant, so that the special minimal supergravity is characterized by a 
{\it cosmological constant generated dynamically}.

The above mechanism of the dynamical generation of cosmological constant in special minimal supergravity is the same as 
was observed by  Ogievetski and Sokatchev \cite{OS80} in their theory of axial vector superfield. 
In the language of component spacetime approach to supergravity the dynamical generation of cosmological constant in 
the special minimal supergravity was described in \cite{Ovrut:1997ur} and before it, in purely bosonic perspective in 
\cite{Aurilia:1978qs,Duff1980,Brown:1988kg} and in the context of spontaneously broken $N=8$ supergravity in 
\cite{Aurilia:1980xj}.

\section{Superspace action and superfield equations of motion for the interacting system of dynamical supergravity 
and supermembrane }
\label{SGacSGeq}

The action for interacting system of dynamical supergravity and supermembrane reads
\begin{eqnarray}\label{Sint=SG+Sp2}
 S&=& S_{SG}+ T_2 S_{p=2}= \int d^8 Z E(Z) +
 {T_2\over 2}\int d^3 \xi \sqrt{g} - T_2 \int\limits_{W^3} \hat{C}_3\; ,
\end{eqnarray}
where $S_{p=2} $ is the same as in Eq. (\ref{Sp=2:=}), the supervielbein (\ref{4EAM}) and the 3-form potential (\ref{C3:=}) 
are assumed to be restricted by the minimal supergravity constraints (\ref{4WTa=}), (\ref{4WTal=}), (\ref{H4=HL+HR}), 
(\ref{H4L}). Furthermore, as we have discussed in previous sections (and in more details in Appendix \ref{appendiceC}), 
the existence of the 3-form potential imposes the restrictions (\ref{vcU=vV+}) on the prepotentials of minimal 
supergravity or equivalently on the basic supergravity variations.

As a result, the superfield equations which appear as a result of variation of the interacting action (\ref{Sint=SG+Sp2}) 
read 
\begin{eqnarray}\label{Ga=Ja}
 G_a= T_2J_a\; ,
\end{eqnarray}
and
\begin{eqnarray}\label{R-bR=cX}
R-\bar{R}= -i T_2{\cal X}
\end{eqnarray}
where $J_a$ and ${\cal X}=({\cal X})^*$ are supermembrane scalar superfields. Roughly speaking, they are obtained as a 
result of varying the supermembrane action with respect to the prepotentials of the special minimal supergravity, 
this is to say  as $\delta S_{p=2} /\delta H^a$ and $\delta S_{p=2} /\delta V$, and have the form
\begin{eqnarray}\label{Ja=NG+WZ}
J_a&=&
\int\limits_{W^3} {3\over \hat{E}} \hat{E}{}^b \wedge \hat{E}{}^\alpha \wedge \hat{E}{}^\beta\; {\sigma}_{ab \alpha\beta}  \delta^8 (Z-\hat{Z}) - \qquad \nonumber
\\ && -
\int\limits_{W^3} {3i\over \hat{E}} \left( *\hat{E}_a \wedge \hat{E}{}^\alpha + {i\over 2}   \hat{E}{}^b \wedge \hat{E}{}^c \wedge \hat{\bar{E}}_{\dot{\beta}}\epsilon_{abcd} \tilde{\sigma}^{d\dot{\beta}\alpha}\right) {\cal D}_\alpha \delta^8 (Z-\hat{Z})
+c.c -
\nonumber
\\
&& -  \int\limits_{W^3}  {i\over 8\hat{E}}   \, \hat{E}{}^b \wedge \hat{E}{}^c \wedge \hat{E}{}^d \, \epsilon_{abcd} \left( {\cal D}{\cal D}- {1\over 2}\bar{R} \right) \delta^8 (Z-\hat{Z}) + c.c.  + \qquad \nonumber
\\ && + \int\limits_{W^3}  {1\over 4\hat{E}} *\hat{E}_b \wedge \hat{E}{}^b  \, G_a\,  \delta^8 (Z-\hat{Z})  -  \nonumber  \qquad
\\
&& -\int\limits_{W^3}  {1\over 4\hat{E}}\; *\hat{E}_c \wedge \hat{E}{}^b \tilde{\sigma}^{d\dot{\alpha}\alpha} \left( 3\delta_a^c \delta_b^d-  \delta_a^d \delta_b^c \right)[{\cal D}_\alpha , \bar{\cal D}_{\dot{\alpha}}] \delta^8 (Z-\hat{Z})  \; , \quad
\end{eqnarray}
and
\begin{eqnarray}\label{cX=}
{\cal X}& =&  {6i\over {E}}  \int\limits_{W^3} \hat{E}^a \wedge \hat{E}{}^\alpha  \wedge \hat{\bar{E}}{}^{\dot{\alpha}}\, {\sigma}^a_{\alpha\dot{\alpha}}\; \delta^8 (Z-\hat{Z}) -
\nonumber  \qquad
\\  && - {3\over 2}\int\limits_{W^3} { \hat{E}{}^b \wedge \hat{E}{}^a \wedge \hat{E}{}^\alpha \over \hat{E}}   \; {\sigma}_{ab \alpha}{}^{{\beta}} {\cal D}_{\beta} \delta^8 (Z-\hat{Z})
+c.c + \nonumber  \;
\\ && +  \int\limits_{W^3}  {   \, \hat{E}{}^b \wedge \hat{E}{}^c \wedge \hat{E}{}^d \over 8 \hat{E}}\, \epsilon_{abcd} \tilde{\sigma}^{a\dot{\alpha}\alpha} [{\cal D}_\alpha , \bar{\cal D}_{\dot{\alpha}}] \delta^8 (Z-\hat{Z})+\;
\nonumber  \;
\\ && + i\int\limits_{W^3}  {*\hat{E}_a \wedge \hat{E}{}^a \over 4\hat{E}}
\left( {\cal D}{\cal D}- \bar{R} \right) \delta^8 (Z-\hat{Z}) + c.c. + \;  \qquad \nonumber \;
\\&& +\int\limits_{W^3}  {1 \over 4\hat{E}} \hat{E}{}^b \wedge \hat{E}{}^c \wedge \hat{E}{}^d \epsilon_{abcd} G{}^{a}\; \delta^8 (Z-\hat{Z})\; . \qquad
\end{eqnarray}

Notice that, as a consequence of (\ref{DG=DR}), the supermembrane current superfields obey
\begin{eqnarray}\label{DJ=-iDX}
\bar{{\cal D}}{}^{\dot{\alpha}} J_{\alpha\dot{\alpha}} = i {{\cal D}}_{{\alpha}} {\cal X}
\; , \qquad
{{\cal D}}{}^{{\alpha}} J_{\alpha\dot{\alpha}} = -i \bar{{\cal D}}_{\dot{\alpha}}{\cal X}\; .
\end{eqnarray}
Although at first glance these relations look different from any of listed in \cite{Seiberg+2009,Kuzenko:2010}, they can be 
reduced to the Ferrara--Zumino multiplet \cite{FerraraZumino74} if one takes into account Eq. (\ref{R-bR=cX}). Indeed, 
this states that the real superfield ${\cal X}$ in the {\it r.h.s.} of Eq. (\ref{DJ=-iDX}) is the sum of chiral 
superfield (equal to $iR$) and its complex conjugate, so that only the first (second) one contributes to the 
{\it r.h.s.} of the first (second) equation in (\ref{DJ=-iDX}).

%
%

\section{Spacetime component equations of the $D=4$ ${\cal N}=1$ supergravity--supermembrane interacting system}\label{sugra4}

\subsection{Wess--Zumino gauge plus partial gauge fixing of the  local spacetime supersymmetry (WZ$_{\hat{\theta}=0}$ gauge)}\label{sugra4-1}

The structure of the current superfields (\ref{Ja=NG+WZ}), (\ref{cX=}) is quite complicated. So is the structure of 
their components. To simplify the supercurrent  components which contribute to the equations of physical, spacetime 
component fields, we use the superspace general coordinate invariance to fix the Wess--Zumino (WZ) gauge  on supergravity superfields,
\begin{eqnarray}  \label{WZgauge=f} i_{\underline{\theta}}  E^{
{\alpha}}&:=& \theta^{\breve{\underline{\alpha}}} E_{\breve{\underline{\alpha}}}{}^{\alpha}
= \theta^{ {\alpha}} \; , \qquad i_{\underline{\theta}}  E^{\dot{\alpha}}:=\theta^{\breve{\underline{\alpha}}} E_{\breve{\underline{\alpha}}}{}^{\dot\alpha}
=  \bar{\theta}{}^{ \dot{\alpha}} \; , \qquad \\  \label{thal=} &&  \theta^{ {\alpha}}:= \theta^{\breve{\underline{\beta}}} \delta_{\breve{\underline{\beta}}}^{\,
 {\alpha}}\; , \qquad  \bar{\theta}{}^{ \dot{\alpha}}:= \theta^{\breve{\underline{\beta}}} \delta_{\breve{\underline{\beta}}}^{\,\dot{\alpha}}\; , \qquad \\ \label{WZgauge=b}
i_{\underline{\theta}} E^{\underline{a}}& :=& \theta^{\breve{\underline{\alpha}}} E_{\breve{\underline{\alpha}}}{}^{\underline{a}}=0\; , \quad \\
 \label{WZgauge=w}i_\theta w^{\underline{a}\underline{b}}&:=&  \theta^{\breve{\underline{\beta}}} w_{\breve{\underline{\beta}}}^{\underline{a}\underline{b}}= 0\;  \qquad
\end{eqnarray}
(see \cite{BdAIL03} for references and more detail) and the (pull--back to $W^3$ of the) local spacetime supersymmetry 
to set to zero the fermionic Goldstone field of the supermembrane,
\begin{eqnarray}\label{thGauge}
\hat{\theta}{}^{\underline{\alpha}}(\xi) =0\qquad \Leftrightarrow \qquad \hat{\theta}{}^{{\alpha}}(\xi) =0\; , \qquad \hat{\bar{\theta}}{}^{\dot{\alpha}}(\xi) =0
\; .  \qquad
\end{eqnarray}
A detailed discussion on this  ''WZ$_{\hat{\theta}=0}$'' gauge can be found in
 \cite{BdAI,BdAIL03,B+I03,IB+JdA:05}. We notice only few of its properties.

Firstly,  in the WZ gauge (\ref{WZgauge=f}), (\ref{WZgauge=b}) the leading component of supervielbein matrix has a 
triangular form,
  \begin{eqnarray}\label{WZ0gg2}
 E_N{}^A\vert_{\theta =0} = \left(\begin{matrix} e_\nu^a(x) &
 \psi_\nu^{\underline{\alpha}}(x)\cr
 0 & \delta_{\breve{\beta}}{}^{\underline{\alpha}} \end{matrix}\right) \qquad \Rightarrow \qquad  E_A{}^N\vert_{\theta =0} = \left(\begin{matrix} e_a^\nu(x) &
- \psi_a^{\breve{\beta}}(x)\cr
 0 & \delta_{\underline{\alpha}}{}^{\breve{\beta}}\end{matrix}\right) \; ,
 \end{eqnarray}
which implies, in particular, the following relation between the leading component of $T_{ab}{}^{\alpha}$ and 
the true gravitino field strength ${\cal D}_{[\mu}\psi_{\nu]}^{\alpha}(x)$
\begin{eqnarray}\label{TbbfWZ}
T_{ab}{}^{\alpha}\vert_{\theta =0} & = 2 e_a^\mu e_b^\nu
{\cal D}_{[\mu}\psi_{\nu]}^{\alpha}(x) -  {i\over 4} (\psi_{[a}\sigma_{b]})_{\dot{\beta}}
G^{\alpha\dot{\beta}}\vert_{\theta =0}
 - {i\over 4} (\bar{\psi}_{[a}\tilde{\sigma}_{b]})^{\alpha}
R\vert_{\theta =0} \; .
\end{eqnarray}

Secondly, we would like to comment on symmetries leaving Eqs. (\ref{WZgauge=f})--(\ref{thGauge}) invariant. 
The WZ gauge (\ref{WZgauge=f}), (\ref{WZgauge=b}), (\ref{WZgauge=w}) is preserved by spacetime diffeomorphisms, 
local Lorentz symmetry and supersymmetry. Fixing further the gauge (\ref{thGauge}) we break 1/2 of the local 
supersymmetry {\it on the worldvolume of the supermembrane}. The only restriction on the parameter of the local 
spacetime supersymmetry $\epsilon^{\alpha}(x)$ is the condition that its pull--back to 
$W^3$, $\hat{\epsilon}{}^{{\alpha}}:= {\epsilon}{}^{{\alpha}}(\hat{x}(\xi))$, and its complex 
conjugate $\hat{\bar{\epsilon}}{}^{\dot{\alpha}}:= \bar{\epsilon}^{\dot{\alpha}}(\hat{x}(\xi))$ 
are related by
\begin{eqnarray}\label{1/2susy}
\hat{\epsilon}{}^{{\alpha}}= \hat{\bar{\epsilon}}_{\dot{\alpha}}\tilde{\bar{\gamma}}{}^{{\dot{\alpha}\alpha}}\; ,\qquad
\end{eqnarray}
where $\tilde{\bar{\gamma}}{}^{{\dot{\alpha}\alpha}}$ is the supermembrane $\kappa$--symmetry projector 
(\ref{tg:=}) calculated with $\hat{\theta}(\xi)=0$. Eq. (\ref{1/2susy}) is tantamount to saying that the pull--back 
of the local supersymmetry parameter to $W^3$ is expressed through the $\kappa$--symmetry parameter of the supermembrane. 
There are no restrictions on the local supersymmetry parameter outside the supermembrane worldvolume so that the 
equations (\ref{1/2susy}) can be understood as the boundary condition imposed on the supersymmetry parameter on the 
domain wall $W^3$.

\subsection{Current superfields in the WZ$_{\hat{\theta}=0}$ gauge. Current prepotentials and Rarita--Schwinger equation}\label{sugra4-2}

In the gauge (\ref{WZgauge=f})--(\ref{1/2susy}),
\begin{eqnarray}\label{hEb=gauge}
\hat{E}{}^a= \hat{e}{}^a= d\hat{x}{}^\mu e_\mu{}^a(\hat{x})    \; ,  \qquad
\hat{E}{}^{\alpha} = \hat{\psi}{}^{\alpha}= d\hat{x}{}^\mu {\psi}_\mu{}{}^{\alpha}(\hat{x})    \; ,  \qquad
\end{eqnarray}
and
\begin{eqnarray}\label{fDdel}
&& {\cal D}_\alpha \delta^8(Z-\hat{Z})= {1\over 8} \theta_\alpha\, \bar{\theta}\bar{\theta}\, \delta^4(x-\hat{x}) + \propto \underline{{\theta}}^{\wedge 4}
\, , \nonumber\\ 
&& \bar{{\cal D}}_{\dot\alpha} \delta^8(Z-\hat{Z})= - {1\over 8} \bar{\theta}_{\dot\alpha}\, {\theta}{\theta}\, \delta^4(x-\hat{x}) + \underline{{\theta}}^{\wedge 4}
\; ,  \quad \\ \label{fDfDdel}
&& {\cal D}^\alpha {\cal D}_\alpha \delta^8(Z-\hat{Z})= - {1\over 4}  \bar{\theta}\bar{\theta}\, \delta^4(x-\hat{x}) + \propto {\theta} \, \bar{\theta}\bar{\theta}
\, ,  \quad \nonumber \\ &&  \bar{{\cal D}}_{\dot\alpha} \bar{{\cal D}}{}^{\dot\alpha} \delta^8(Z-\hat{Z})= - {1\over 4} {\theta}{\theta}\, \delta^4(x-\hat{x}) + \propto {\theta}{\theta} \, \bar{\theta}
\; , \quad  \\
\label{fDbfDdel}
&& {}[{\cal D}_\alpha ,  \bar{{\cal D}}_{\dot\alpha} ]\delta^8(Z-\hat{Z})= -{1\over 2} \theta_\alpha\, \bar{\theta}_{\dot\alpha}\, \delta^4(x-\hat{x}) +  \propto \underline{{\theta}}^{\wedge 3}
\; ,  \quad
\end{eqnarray}
where $\underline{{\theta}}^{\wedge 4}:= {\theta}{\theta} \, \bar{\theta}\bar{\theta}$
and
$\underline{{\theta}}^{\wedge 3}$ denotes terms proportional to either  
${\theta}{\theta} \, \bar{\theta}$ or ${\theta} \, \bar{\theta} \bar{\theta} $ (or both, which implies 
$\propto {\theta}{\theta} \, \bar{\theta}\bar{\theta}$). Using these relations and introducing the current pre-potential 
fields
\begin{eqnarray}\label{Kab(x)=}
{\cal P}_a{}^b(x)&:=& \int\limits_{W^3}  {1\over \hat{e}} *\hat{e}_a \wedge \hat{e}{}^b  \, \delta^4 (x-\hat{x})  \; ,  \qquad
 \\ \label{Kab(x)=}
{\cal P}_a(x) &:=& \int\limits_{W^3}   {1\over \hat{e}} \epsilon_{abcd} \hat{e}{}^b \wedge \hat{e}{}^c \wedge \hat{e}{}^d \,
 \, \delta^4 (x-\hat{x}) = \qquad \nonumber \\ &=& e_a^\mu (x) \int\limits_{W^3}    \epsilon_{\mu\nu\rho\sigma} d\hat{x}{}^\nu \wedge  d\hat{x}{}^\rho \wedge  d\hat{x}{}^\sigma  \,
 \, \delta^4 (x-\hat{x}) \;  , \qquad
\end{eqnarray}
we find that the vector and scalar current superfields (\ref{Ja=NG+WZ}), (\ref{cX=}) have the form
\begin{eqnarray}\label{Ja=gauge}
J_{\alpha \dot{\alpha}} \vert_{\hat{\theta}=0} =  {\theta_\beta\, \bar{\theta}_{\dot{\beta}} \over   8}
(\;  3 {\cal P}_a{}^b(x) {\sigma}^a_{\alpha \dot{\alpha}}\tilde{\sigma}{}_b^{\beta\dot{\beta}}- 2\delta_{\alpha}{}^{\beta} \delta_{\dot{\alpha}}{}^{\dot{\beta}} {\cal P}_b{}^b(x)) - i {({\theta}{\theta}  - \bar{\theta}\bar{\theta})\over 32}   {\sigma}^a_{\alpha \dot{\alpha}} {\cal P}_a(x)
+  \propto \underline{\theta}^{\wedge 3}
\qquad
\end{eqnarray}
and
\begin{eqnarray}\label{cX=gauge}
{\cal X} \vert_{\hat{\theta}=0} &=& - {\theta{\sigma}^a\bar{\theta} \over 16} {\cal P}_a+ i {({\theta}{\theta}  - \bar{\theta}\bar{\theta})\over 16}   {\cal P}_a{}^a(x)
+  \propto \underline{\theta}^{\wedge 3}
\; . \quad
\end{eqnarray}
Using (\ref{Ja=gauge}) and (\ref{cX=gauge}) one can easily check that Eqs. (\ref{DJ=-iDX}) are satisfied at lowest order 
in $\underline{\theta}$.

One also sees that there is no explicit supermembrane contributions to the Rarita--Schwinger equations of the 
supergravity--supermembrane interacting system which thus reads
\begin{eqnarray}\label{SG+2p=RS=on}
\epsilon^{\mu\nu\rho\sigma}
e_{\nu}^a (x) {\cal D}_\rho \psi_\sigma^{\alpha}(x)\, \sigma_{a\alpha\dot{\alpha}} =0\; .
\end{eqnarray}

However, such a contribution is actually present in (\ref{SG+2p=RS=on}) implicitly, hidden inside the covariant 
derivative. Indeed, as indicated by Einstein equation, the bosonic vielbein and the spin connection do contain some 
contributions from supermembrane.

\subsection{Einstein equation of the supergravity--supermembrane interacting system in the WZ $_{\hat{\theta}=0}$ gauge}\label{sugra4-3}

The Einstein equation with supermembrane current contributions can be obtained as leading term in the decomposition  
of Eq. (\ref{RRicci}), {\it i.e.}
\begin{eqnarray} \label{RRicci=J}
R_{bc}{}^{ac}\vert_{_{\theta=0}}& =& {1\over 32} ({{\cal D}}^{{\beta}} \bar{{\cal D}}^{(\dot{\alpha}|}
J^{{\alpha}|\dot{\beta})} - \bar{{\cal D}}^{\dot{\beta}} {{\cal D}}^{({\beta}}J^{{\alpha})\dot{\alpha}})\vert_{_{\theta=0}}
\sigma^a_{\alpha\dot{\alpha}}\sigma_{b\beta\dot{\beta}} - {3i\over 64}
(\bar{{\cal D}}\bar{{\cal D}}{\cal X} - {{\cal D}}{{\cal D}}{\cal X})\vert_{_{\theta=0}} \, \delta_b^a + \qquad \nonumber \\ && + {3\over 16} (R\bar{R})\vert_{_{\theta=0}} \delta_b^a\; . \qquad
\end{eqnarray}
The first two terms in the {\it r.h.s.} of Eq. (\ref{RRicci=J}) can be easily calculated from Eqs. 
(\ref{Ja=gauge}), (\ref{cX=gauge}), while the last term, in the light of that the scalar superfield equation has 
the form of Eq. (\ref{R-bR=cX}), is expressed in terms of $(R+\bar{R})^2\vert_{_{\theta=0}}$  and requires a separate 
study. As an intermediate resume let us fix that
\begin{eqnarray} \label{RRicci=Jg1}
R_{bc}{}^{ac}\vert_{_{\theta=0\; , \; \hat{\theta}=0}}& =& - {3\over 32} \, T_2\, \left( {\cal P}_b{}^a(x) - {1\over 2} \delta_b^a  {\cal P}_c{}^c(x)\right) + {3\over 64} (R+\bar{R})^2\vert_{_{\theta=0}} \delta_b^a\; . \qquad
\end{eqnarray}
The last term is the square of $(R+\bar{R})\vert_{_{\theta=0}}$ which, as a result of (\ref{R-bR=cX}), obeys the equation
\begin{eqnarray} \label{d(R+bR)=}
\partial_\mu (R+\bar{R})\vert_{_{\theta=0}}= {T_2\over 16} \int\limits_{W^3}    \epsilon_{\mu\nu\rho\sigma} d\hat{x}{}^\nu \wedge  d\hat{x}{}^\rho \wedge  d\hat{x}{}^\sigma  \,
 \, \delta^4 (x-\hat{x}) \;  . \qquad
\qquad
\end{eqnarray}
The solution of this equation can be written in the form
\begin{eqnarray} \label{(R+bR)=c+int}
R(x)+\bar{R}(x)= 8c +{ T_2\over 16} \int\limits^x_{x_0} d\tilde{x}^\mu \int\limits_{W^3}     \epsilon_{\mu\nu\rho\sigma} d\hat{x}{}^\nu \wedge  d\hat{x}{}^\rho \wedge  d\hat{x}{}^\sigma  \,
 \, \delta^4 (\tilde{x}-\hat{x})  \; ,
\end{eqnarray}
where $c$ is an arbitrary constant which corresponds to the value of $(R+\bar{R})$ at the spacetime point 
$x_0^\mu$ providing the lower limit of the integral in the second term, $c=(R(x_0)+\bar{R}(x_0))/8$.

One can easily check that
\begin{eqnarray}\label{Theta=}
\Theta (x,x_0|\hat{x}):=
\int\limits^x_{x_0} d\tilde{x}^\mu \int\limits_{W^3}     \epsilon_{\mu\nu\rho\sigma} d\hat{x}{}^\nu \wedge  d\hat{x}{}^\rho \wedge  d\hat{x}{}^\sigma  \,
 \, \delta^4 (\tilde{x}-\hat{x}) \; ,  \qquad
\end{eqnarray}
entering the second term in the {\it r.h.s.} of (\ref{(R+bR)=c+int}), obeys
\begin{eqnarray}\label{dTheta=1}
\partial_\mu \Theta (x,x_0|\hat{x})= \int\limits_{W^3}     \epsilon_{\mu\nu\rho\sigma} d\hat{x}{}^\nu \wedge  d\hat{x}{}^\rho \wedge  d\hat{x}{}^\sigma  \,\, \delta^4 (x-\hat{x})\; . \qquad
\end{eqnarray} 
Furthermore, using a convenient local frame in the neighborhood of the worldvolume, one can check that 
$\Theta (x,x_0|\hat{x})$ vanishes if the points $x^\mu$ and ${x}^\mu_0$ are on the same side of spacetime with respect 
to the domain wall provided by the supermembrane worldvolume while it is equal to $\pm 1$  if these points belongs to 
the different branches  of the spacetime separated by this domain wall. This is  to say that Eq. (\ref{Theta=}) defines 
a counterpart of the Heaviside  step function associated to the direction orthogonal to the supermembrane worldvolume. 
The last statement about association implies that $\Theta (x,x_0|\hat{x})$ is a functional of the supermembrane 
coordinate function $\hat{x}^\mu(\xi)$.
Furthermore, as in the case of the standard Heaviside  step function $\Theta (y)$, we can use
 \footnote{In the case of standard standard Heaviside  step function this is equivalent to setting the 
indefinite value $\Theta (0)$ equal to $1/2$. Indeed, calculating the derivative $\partial_y (\Theta(y)\Theta (y))= 2 \Theta(y)\delta (y)= 2 \Theta(0)\delta (y)$  
we find that this coincides with $\partial_y \Theta(y)= \delta (y)$ when  $\Theta (0)=1/2$.  
In our case $\Theta (x,x_0|\hat{x})$ is the counterpart of either $+\Theta(y)$ or $-\Theta(y)$ so that 
$(\Theta (x,x_0|\hat{x}))^2= \pm \Theta (x,x_0|\hat{x})$. However, by a suitable choice of the location of 
the point $x_0$ with respect to $W^3$ one can always arrive at the situation with nonnegative $\Theta (x,x_0|\hat{x})$. 
Below for simplicity we assume this choice is made. }
 $(\Theta (x,x_0|\hat{x}))^2= \Theta (x,x_0|\hat{x})$.

Thus our solution of the auxiliary field equations (\ref{(R+bR)=c+int}) can be written as ({\it cf. } \cite{Aurilia:1978qs})
\begin{eqnarray} \label{(R+bR)=c+T2}
R(x)+\bar{R}(x)= 8c + { T_2\over 16} \Theta (x,x_0|\hat{x})\;
\end{eqnarray}
and implies that Eq. (\ref{RRicci=Jg1}) reads
\begin{eqnarray} \label{RRicci=K+Hev}
&&R_{bc}{}^{ac}(x) = - {3T_2\over 32} \left( {\cal P}_b{}^a(x) - {1\over 2} \delta_b^a  {\cal P}_c{}^c(x)\right)  + \nonumber\\
&&+3 \delta_b^a \, \left( c^2 + \left(\left({T_2\over 128} +  c \right)^2 -  c^2 \right) \Theta (x,x_0|\hat{x})\right)\; , \qquad
\end{eqnarray}
where ${\cal P}_b{}^a(x)$ is the singular contribution  defined in (\ref{Kab(x)=}).

\subsection{Cosmological constant generation in the interacting system and its ``renormalization'' due to supermembrane}\label{sugra4-4}

Let us analyze the supermembrane contribution to the Einstein equations.
These can be separated in two classes, one containing singular contributions and the other containing regular 
contributions proportional to $T_2$.

Being a bit more provocative one can say about three classes, counting also the contribution proportional to square of 
the arbitrary integration constant $c$, as far as this comes from the auxiliary field sector of the special minimal 
supergravity, the off--shell formulation which is 'elected' by the supermembrane. As we have already commented in 
sec. \ref{sugra2-4}, the supermembrane can exist in a background of a generic minimal  supergravity, 
however the supermembrane interaction with dynamical supergravity  requires this to be special minimal supergravity. 
This in its turn, even in the absence of any matter (neither of the field theoretical type nor of branes), produces 
Einstein equations with a cosmological constant generated dynamically. Then this cosmological constant proportional 
to the square of the above  arbitrary integration constant $c$ should also be considered as a(n indirect) contribution 
of the supermembrane to the Einstein equation.

To be more concrete, Eq. (\ref{RRicci=K+Hev}) can be written in the form $R_{bc}{}^{ac}(x) =  3 c^2
\delta_b^a + \propto T_2$,
\begin{eqnarray} \label{RRicci=ta2+T2t}
R_{acb}{}^{c}(x)& =&
\eta_{ab} \, 3c^2 + T_2  \left( {\cal T}_{ab}^{sing}(x) + {\cal T}_{ab}^{reg}(x)\right)\; . \qquad
\end{eqnarray}
 When $T_2$ is set to zero, it contains a nonvanishing cosmological constant contribution with $\Lambda =- {3c^2}$. 
This (AdS-type) cosmological constant is generated dynamically as far as it is proportional to the (minus) square of 
the arbitrary integration constant $c$ which is inevitable in the special minimal supergravity equations due to its 
auxiliary field structure (see \cite{Ovrut:1997ur} and \cite{IB+CM:2011} for references and more discussion). In its turn, 
special minimal supergravity, and not generic minimal supergravity can be included into the action of the 
supergravity--supermembrane interacting system. In this sense the cosmological constant generated dynamically is 
the first 'relict' contribution from the supermembrane to the Einstein equation of the interacting system.

The second type of the supermembrane contributions to the {\it r.h.s.} of the Einstein equation are singular 
terms $\propto {\cal P}_c{}^d(x)$ (\ref{Kab(x)=}),
\begin{eqnarray} \label{T2sing=}
 {\cal T}_{ab}^{sing}(x)&=& - T_2 {3\over 32} \left({\cal P}_{ba}(x) - {1\over 2} \eta_{ba}  {\cal P}_c{}^c(x)\right)= \qquad \nonumber \\
 &=&- {3T_2\over 32} \int\limits_{W^3}  {1\over \hat{e}} *\hat{e}_a \wedge \hat{e}{}_b  \, \delta^4 (x-\hat{x}) + {3T_2\over 64} \eta_{ba}  \int\limits_{W^3}  {1\over \hat{e}} *\hat{e}_c \wedge \hat{e}{}^c  \, \delta^4 (x-\hat{x})
 \;  \qquad
\end{eqnarray}
which are expected when (super)gravity interacts with supermembrane.

In the  third type we collect the regular  supermembrane contributions which are proportional to the supermembrane tension,
\begin{eqnarray} \label{T2reg=}
 {\cal T}_{ab}^{reg}(x)&=& \eta_{ab} {\cal T}^{reg}(x) \; , \qquad  {\cal T}^{reg}(x) =
+ {3T_2\over 64}
\left({T_2\over 256} + {c} \right)   \Theta (x,x_0|\hat{x})\; . \qquad \;  \qquad
\end{eqnarray}

To appreciate the role of this contribution it is instructive to consider the Einstein equation in two pieces of 
the spacetime separated by the supermembrane worldvolume. Let us denote the half-space where $\Theta (x,x_0|\hat{x})=1$ 
by $M^4_+$ and the half-space where $\Theta (x,x_0|\hat{x})=0$ by $M^4_-$. Then the singular terms  (\ref{T2sing=}) do 
not contribute and the Einstein equation reads
\begin{eqnarray} \label{RRicci+=}
M^4_+\; : \qquad R_{acb}{}^{c}(x)& =&   3
\eta_{ab} \, \left({T_2\over 128} +  c \right)^2\; . \qquad
\\ \label{RRicci-=}
M^4_-\; : \qquad R_{acb}{}^{c}(x)& =&   3
\eta_{ab} \, c^2 \; . \qquad
\end{eqnarray}

An evident observation is that, in the general case,
the  cosmological constants in  different branches of spacetime separated
by the worldvolume $W^3$ are different.

One also notices that the cosmological constants in $M^4_+$ and $M^4_-$ coincide if
$c=-{T_2\over 256}$. However, as far as $c$ is an arbitrary integration constant, fixing its value is equivalent 
to imposing a kind of boundary conditions and we do not see any special reason to chose such boundary conditions in 
such a way that $c=-{T_2\over 256}$. Rather we should allow a generic value of  $c$ and thus accept that the cosmological 
constant takes different values in the branches of spacetime separated by the supermembrane worldvolume.

Notice that the solution of the Einstein equation describing membranes separating two $AdS_5$ spaces with different 
values of cosmological constants were studied in \cite{Gogberashvili:1998iu}, as a Brane world alternative to the dark 
energy, and  \cite{JMMS+:2001} in relation with the hypothesis on possible change of signature in the Brane World models. 
See also \cite{shellUni} for the related studies. In the bosonic perspective the appearance of different cosmological 
constants on the different sides of a domain wall interacting with gravity and a 3--form gauge field was known from 
\cite{Aurilia:1978qs}, where it was used as a basis for a bag model for hadrons, and from  \cite{Brown:1988kg} where this 
effect was proposed as a mechanism for damping the cosmological constant. Our present study indicates that the result on 
the different values of cosmological constant on the different sides of the supermembrane domain wall is an imminent 
consequence of the dynamics of the supersymmetric interacting system of the supermembrane and dynamical $D=4$  ${\cal N}=1$ 
supergravity.

\section{On supersymmetric solutions of the interacting system equations}
\label{solutions}
When searching for purely bosonic supersymmetric solutions, setting
$\psi_\mu^\alpha=0$, one studies the Killing spinor equations, which appears as the conditions of supersymmetry preservation, 
$\delta_\epsilon \psi_\mu^\alpha=0$. When starting from superfield formulation of supergravity, 
$\delta_\epsilon \psi_\mu^\alpha$ can be calculated with the use of superspace Lie derivative, this is to say 
$\delta_\epsilon \psi_\mu^\alpha= D_\mu\epsilon^{\alpha} + (E_\mu^C \epsilon^{\underline{\beta}}T_{\underline{\beta} C}{}^\alpha )\vert_{\theta =0} $.
Hence, in a generic off-shell $D=4$ ${\cal N}=1$  minimal supergravity background the Killing equations are
\begin{eqnarray}\label{Killing:=De+}
 {D}\epsilon^{\alpha} + {i\over 8} e^c (\epsilon
\sigma_c\tilde{\sigma}_d)_{\beta} {}^{\alpha}\; G^d\vert_{\theta =0}   +{i\over 8} e^c
\; (\bar{\epsilon}\tilde{\sigma}_{c}){}^{\alpha} \;R\vert_{\theta =0}=0  \; \qquad
\end{eqnarray}
and the complex conjugate equation. Using the superfield equations of motion (\ref{Ga=Ja}), (\ref{R-bR=cX}), 
the explicit form of the current superfields in the WZ$_{\hat{\theta}=0}$ gauge, Eqs. (\ref{Ja=gauge}), (\ref{cX=gauge}), 
and Eq. (\ref{(R+bR)=c+T2}), we find that the Killing equation (\ref{Killing:=De+}) reads
\begin{eqnarray}\label{De=ta+}
{D}\epsilon^{\alpha} + {i\over 2} e^a
\; (\bar{\epsilon}\tilde{\sigma}_{a}){}^{\alpha}  \left( c + { T_2\over 128} \Theta (x,x_0|\hat{x})\right)=0\; .\qquad
\end{eqnarray}
We can split this on two Killing equations valid in two different branches of spacetime separated by the 
supermembrane worldvolume,
\begin{eqnarray}\label{M-=Killing}
M_-^4\, &:& \qquad {D}\epsilon^{\alpha} + {i\over 2} e^a
\; (\bar{\epsilon}\tilde{\sigma}_{a}){}^{\alpha} \, c =0\; , \qquad \\ \label{M+=Killing}
M_+^4\, &:& \qquad {D}\epsilon^{\alpha} + {i\over 2} e^a
\; (\bar{\epsilon}\tilde{\sigma}_{a}){}^{\alpha}  \left( c + { T_2\over 128} \right)=0\; .\qquad
\end{eqnarray}
The supersymmetry parameter should also obey the boundary conditions (\ref{1/2susy}) on the worldvolume $W^3$, which is 
the common boundary of $M_+^4$ and $M_-^4$,
\begin{eqnarray}\label{W3=1/2susy}
W^3= \pm \partial M_\pm^4\; : \qquad  \hat{\epsilon}{}^{{\alpha}}= \hat{\bar{\epsilon}}_{\dot{\alpha}}\tilde{\bar{\gamma}}{}^{{\dot{\alpha}\alpha}}\; ,\qquad
\hat{\epsilon}{}^{{\alpha}}:= {\epsilon}{}^{{\alpha}}(\hat{x}(\xi)) \; , \quad \hat{\bar{\epsilon}}{}_{\dot{\alpha}}:= \bar{\epsilon}_{\dot{\alpha}}(\hat{x}(\xi)) \; .  \qquad
\end{eqnarray}

The detailed study of these system of Killing spinor equations and the search for the supersymmetric solutions  of the 
interacting system equations on their basis is an interesting subject for future.
An intriguing question is whether the supersymmetric solutions of the equations of the interacting system exist in 
the generic case of arbitrary $c$ corresponding to different values of cosmological constants on different sides of 
the supermembrane worldvolume, or supersymmetry selects some particular values of the constant $c$.
Presently we can state that if  obstructions existed, they would occur due to the singular terms with support 
on the worldvolume $W^3$, while the mere fact of different values of cosmological constant on the branches of 
spacetime situated on the different sides of $W^3$ does not prohibit supersymmetry. Indeed, let us study the 
integrability conditions for the Killing spinor equations in $M_\pm^4$. Applying the exterior covariant derivatives 
to  Eqs. (\ref{M-=Killing}) and (\ref{M+=Killing}) and using the Ricci identities 
$DD\epsilon^{\alpha}=- {1\over 4} R^{ab} \epsilon^{\beta} {\sigma}_{ab\,\beta}{}^{\alpha}$ and the equations 
complex conjugate to (\ref{M-=Killing}) and (\ref{M+=Killing}), we find
\begin{eqnarray}\label{M-=DKilling}
M_-^4\, &:& \qquad  R^{ab} \epsilon^{\beta} {\sigma}_{ab\,\beta}{}^{\alpha} =  {1\over 4}|c|^2 e^d\wedge e^c \,  \epsilon^{\beta}{\sigma}_{cd\,\beta}{}^{\alpha} \; ,\qquad \\ \label{M+=DKilling}
M_+^4\, &:& \qquad  R^{ab} \epsilon^{\beta} {\sigma}_{ab\,\beta}{}^{\alpha} =  {1\over 4}\left| c + { T_2\over 128} \right|^2 e^d\wedge e^c \,  \epsilon^{\beta}{\sigma}_{cd\,\beta}{}^{\alpha}
\; . \qquad
\end{eqnarray}
If we search for a purely bosonic  solution preserving all the supersymmetry in $M_-^4$ and $M_+^4$,  
Eqs. (\ref{M-=DKilling}) and (\ref{M+=DKilling}) should be obeyed for an arbitrary $\epsilon^\alpha$. 
This implies
\begin{eqnarray}\label{RM-=susy}
M_-^4\, &:& \qquad R_{cd}{}^{ab}=    {1\over 2}|c|^2 {\delta}_{[c}{}^a{\delta}_{d]}{}^b \; ,\qquad \\ \label{RM+=susy}
M_+^4\, &:& \qquad R_{cd}{}^{ab}={1\over 2} \left| c + { T_2\over 128} \right|^2 {\delta}_{[c}{}^a{\delta}_{d]}{}^b
\; , \qquad
\end{eqnarray}
{\it i.e.} that $M_\pm^4$ are AdS spaces with apparently different cosmological constants. One can easily check that 
(\ref{RM-=susy}) and (\ref{RM+=susy}) solve our equations of motion  (\ref{RRicci-=}) and (\ref{RRicci+=}) and 
thus describe the completely supersymmetric solution of the system of the supergravity equations of the interacting 
system (at least) when these are considered modulo singular terms with the support on $W^3$.

Let us stress that such a system of equations does contain the supermembrane contributions: not only an indirect, 
which comes from an arbitrary cosmological constant generated dynamically due to the structure of the supergravity 
auxiliary fields imposed by the supergravity interaction with supermembrane (see \cite{Ovrut:1997ur} and also 
\cite{Aurilia:1978qs,OS80,Aurilia:1980xj,Duff1980}), but also {\it direct}, which is a shift of cosmological 
constant on one of the sides of the brane worldvolume on the value which is proportional to the supermembrane tension 
(see \cite{Aurilia:1978qs,Brown:1988kg}).
Furthermore, although preserving all 4 supersymmetries in  $M_-^4$ and $M_+^4$, when considered as a solution of 
the equations of interacting system, Eqs. (\ref{RM-=susy}) and (\ref{RM+=susy}) describe the 1/2 BPS state, {\it i.e.} 
the state preserving 1/2 of the supersymmetry. Indeed, when considering the interacting system we have to restrict the 
local supersymmetry parameter by the boundary conditions (\ref{W3=1/2susy}) on $W^3$ and these clearly break 1/2 of 
the supersymmetry on $W^3$.

\chapter[Conformal higher spin theory]
{Conformal higher spin theory in extended tensorial superspace}\label{chapterspin}
\thispagestyle{chapter} 
 \initial{I}n this chapter we present the superfield equations in tensorial ${\cal N}$--extended superspaces to describe 
the ${\cal N}=2,4,8$ supersymmetric generalizations of free conformal higher spin theories.
We obtain them by quantizing a superparticle model in ${\cal N}$--extended tensorial superspace.
We show that no nontrivial generalizations of Maxwell and Einstein equations to tensorial space appear because 
${\cal N}$--extended higher spin supermultiplets just contain additional scalar and spinor fields which obey the standard higher 
spin equations in their tensorial space version.
We find also that these additional fields appear in the basic superfield under derivatives so that the theory is invariant under Peccei--Quinn--like 
symmetries.

\section{Superparticle in $\mathcal{N}$-extended tensorial superspace}
\label{superPart} 

Let us begin by presenting a simple dynamical model in extended tensorial superspace the quantization of which produces the supersymmetric 
higher spin field equations.

\subsection{An action for the $\Sigma^{({n(n+1)\over 2}|{\cal N}\,n)}$ superparticle}

The action for a superparticle in extended tensorial superspace $\Sigma^{({n(n+1)\over 2}|{\cal N}\,n)}$ has the form
\begin{eqnarray}
\label{SNpreon} S =  \int d \tau \, {\cal L}  =  \int d \tau \,
[\dot{\hat{X}}{}^{\alpha\beta}(\tau) - i \dot{\hat{\theta}}{}^{\alpha I}(\tau)
\hat{\theta}{}^{\beta I}(\tau)]\lambda_{\alpha}(\tau)\lambda_{\beta}(\tau)\; ,  \qquad
\begin{cases}\alpha , \beta = 1,...,n \; , \cr I=1,2,..., {\cal N}\; ,\end{cases}
\end{eqnarray}
where the $\lambda_\alpha(\tau)$ are auxiliary {\it commuting} spinor variables,
$\hat{X}^{\alpha\beta}(\tau)=\hat{X}^{\beta\alpha}(\tau)$ and $\hat{\theta}^{\alpha I}(\tau)$
are the bosonic and fermionic coordinate functions which define the superparticle worldline
\begin{eqnarray}
&&W^1\in \Sigma^{({n(n+1)\over 2}|{\cal N}\,n)}:
{Z}^{\cal M}=\hat{Z}^{\cal M}(\tau)=(\hat{X}{}^{\alpha\beta}(\tau)\,,\,\hat{\theta}{}^{\alpha I}(\tau)), 
\end{eqnarray}
and the dot denotes derivative with respect to proper time $\tau$.

It is convenient to write the action (\ref{SNpreon}) in the form
\begin{eqnarray}
\label{SNpr=} S =\int_{W^1} {\hat{\Pi}}{}^{\alpha\beta}
\lambda_{\alpha}(\tau)\lambda_{\beta}(\tau)\; ,  \qquad
\end{eqnarray}
where
\begin{eqnarray}
\label{hPi=}
{\hat{\Pi}}{}^{\alpha\beta}(\tau) = d\tau {\hat{\Pi}}{}_\tau^{\alpha\beta}(\tau)
= d\tau( {\dot{\hat{X}}{}^{\alpha\beta} - i \dot{\hat{\theta}}{}^{I (\alpha} }
\hat{\theta}{}^{\beta ) I} )\; , \quad  \begin{cases}\alpha , \beta = 1,...,n \; ,
\cr I=1,2,..., {\cal N}\; ,\end{cases}
\end{eqnarray}
is the pull-back to $W^1$ of the vielbein $\Pi{}^{\alpha\beta}$
of the flat ${\cal N}$-extended tensorial superspace
$\Sigma^{({n(n+1)\over 2}|{\cal N}n)}\,$ 
\begin{eqnarray}
\label{Pi=} {{\Pi}}{}^{\alpha\beta}= d{{X}}{}^{\alpha\beta} -id{{\theta}}{}^{I (\alpha }
{\theta}{}^{\beta )I}\; .  \quad
\end{eqnarray}

The superparticle action is manifestly invariant under rigid supersymmetry of the ${\cal N}$-extended tensorial 
superspace $\Sigma^{({n(n+1)\over 2}|{\cal N}n)}\,$,
\begin{eqnarray}
\label{susy=} \delta_\epsilon
{{X}}{}^{\alpha\beta} = i{{\theta}}{}^{I (\alpha }{\epsilon}{}^{\beta )I}\quad , \quad
\delta_\epsilon {{\theta}}{}^{I \alpha } = {\epsilon}{}^{\beta I}\quad ,
\end{eqnarray}
which acts on the worldline fields as
\begin{eqnarray}
\label{susy=h}
 \delta_\epsilon {\hat{X}}{}^{\alpha\beta}
 =i{\hat{\theta}}{}^{I (\alpha }{\epsilon}{}^{\beta )I}\; , \quad
\delta_\epsilon {\hat{\theta}}{}^{I \alpha }= {\epsilon}{}^{\beta I}\; ; \quad
\delta_\epsilon {\lambda}_\alpha =0 \; .
\end{eqnarray}
The action (\ref{SNpreon}) is also manifestly invariant  under the $GL(n, \mathbb{R})$
transformations of the $\alpha, \beta=1,...,n$ indices, which reduce to the
$n$-dimensional representation of $Spin(1,D-1)$ when these indices are thought
of as Lorentz-spinorial ones\footnote{In \cite{V01s,V01c} the counterparts of
${\lambda}_\alpha$ were called `$s$-vectors' to avoid their immediate
identification as $GL(n,\mathbb{R})$ vectors or $SO(1,D-1)$ spinors. }.

\subsection{Symplectic supertwistor form of the action}

Actually, the  $\Sigma^{({n(n+1)\over 2}|{\cal N}n)}\,$ superparticle
action (\ref{SNpreon}) is invariant under the larger $OSp({\cal N}|2n)$ supergroup.
To make this manifest as well as to determine easily the number
of physical degrees of freedom it is convenient to use Leibniz's rule\footnote{It is
sufficient to use $\lambda_{\alpha}\lambda_{\beta}d{{X}}{}^{\alpha\beta}=
\lambda_{\alpha}d(\lambda_{\beta}{{X}}{}^{\alpha\beta}) - \lambda_{\alpha}
{{X}}{}^{\alpha\beta}d\lambda_{\beta}$; no integration by parts is needed.} to rewrite the
action (\ref{SNpreon}) in the form
\begin{eqnarray}
\label{SNpr=STw} S = \int_{W^1}   (\lambda_{\alpha}d\mu^\alpha - \mu^\alpha d\lambda_{\alpha}
- id\chi^I\, \chi^I ) = \int_{W^1} d\Upsilon^{\Sigma}\Xi_{\Sigma\Omega}\Upsilon^{\Omega}\; .
\qquad
\end{eqnarray}
This action is written in terms of the bosonic spinor $\lambda_\alpha(\tau)$,
which is present in (\ref{SNpreon}),  a second bosonic spinor $\mu^\alpha$ and $\mathcal{N}$ real
fermionic variables $\chi^I$; these form the $\mathcal{N}$-extended
orthosymplectic supertwistor (see \cite{BL98,BLS99} for ${\cal N}=1$)
\begin{eqnarray}
\label{YS=}
\Upsilon^{\Sigma}= \left(\begin{matrix} \mu^\alpha & \lambda_{\alpha}&  \chi^I \end{matrix}\right)\; , 
\qquad \alpha =1,\ldots, n \; , \qquad I=1,\ldots, {\cal N} \; .
\end{eqnarray}
This generalizes the Penrose twistors \cite{Penrose} (or conformal $SU(2,2)$ spinors)
and the Ferber-Schirafuji supertwistors \cite{Ferber,Shirafuji} (carrying the basic
representation of $D$=4 $SU(\mathcal{N}|2,2)$). The $\Upsilon^{\Sigma}$'s carry the defining
representation of the $OSp({\cal N}|2n)$ supergroup,
the transformations of which preserve the $(2n+\mathcal{N})\times (2n+\mathcal{N})$
orthosymplectic `metric' $\Xi_{\Sigma\Omega}$,
\begin{eqnarray}
\label{YS=} \Xi_{\Sigma\Omega}= \left(\begin{matrix} 0 &   \delta_\alpha{}^\beta & 0 \cr -
\delta^{\alpha}{}_\beta & 0 & 0 \cr  0 & 0 & -i\delta^{IJ} \end{matrix}\right)\; , \quad \alpha
=1,\ldots, n \; , \quad I=1,\ldots, {\cal N} \; .
\end{eqnarray}
In fact, $OSp(1|2n)$ may be considered as a supersymmetric generalization of the superconformal group for $D={n\over 2}+2$
(see \cite{BLSP2000} and \cite{BLS99,vH+vP:82,BdAPV:30-32} and refs. therein).

The relations between the supertwistor components and the variables of the action
(\ref{SNpreon}) that make the transition between both actions are
\begin{eqnarray}
\label{mu=} \mu^\alpha = {\hat{X}}{}^{\alpha\beta} \lambda_{\beta} - {i\over 2}
{\hat{\theta}}{}^{\alpha I} \chi^I  \; , \qquad \chi^I=  {\hat{\theta}}{}^{\alpha I}\,
\lambda_\alpha \; , \qquad
\end{eqnarray}
which generalize the Penrose and the Ferber-Shirafuji incidence relations
\cite{Penrose,Ferber,Shirafuji} (see \cite{BL98} and
\cite{BLS99} for ${\cal N}=1$). Since the action (\ref{SNpr=STw})
does not possess any gauge symmetries, the components of the orthosymplectic
supertwistors are the true, physical degrees of freedom ($2n$ bosonic and ${\cal N}$
fermionic) of our generalized superparticle model.

\subsection{Gauge symmetries of the original action}

By construction, the actions (\ref{SNpr=STw}) and (\ref{SNpreon})  describe the same
dynamical system. Thus, since the action (\ref{SNpreon}) depends on ${n(n+1)\over 2}+n$
bosonic variables and $\mathcal{N}n$ fermionic ones, it should possess $n(n-1)/2$
bosonic gauge symmetries and ${\cal N}\, (n-1)$ fermionic ones to reduce the
number of degrees of freedom to those of the supertwistors $\Upsilon^{\Sigma}$
appearing in the action (\ref{SNpr=STw}). The simplest way to describe these gauge symmetries,
called  fermionic $\kappa$-symmetry and bosonic $b$-symmetry,  is to define
restrictions on the basic variations of the bosonic and fermionic coordinate
functions (see \cite{BLS99} for the ${\cal N}=1$ superparticle case and \cite{BdAPV:30-32} for the case of superstring in tensorial 
superspace.)
\begin{eqnarray}
\label{k-sym=a} \delta_\kappa  {\hat{X}}{}^{\alpha\beta}= i \delta_\kappa
{\hat{\theta}}{}^{I(\alpha } {\hat{\theta}}{}^{\beta) I}\; , \qquad \fbox{$\delta_\kappa
{\hat{\theta}}{}^{\alpha I}\lambda_\alpha =0 $}\; , \qquad \delta_\kappa \lambda_\alpha =0 \;
, \qquad \\ \label{k-sym=b} \fbox{$ \delta_b  {\hat{X}}{}^{\alpha\beta} \lambda_\alpha =0
$}\; , \qquad \delta_b {\hat{\theta}}{}^{I\alpha }\lambda_\alpha =0\; , \qquad \delta_b
\lambda_\alpha =0 \; . \qquad
\end{eqnarray}

\subsection{Constraints and their conversion to first class}

In the hamiltonian formalism, the $\delta_\kappa$ and $\delta_b$ gauge symmetries in
Eqs.~(\ref{k-sym=a}), (\ref{k-sym=b}) are generated by first class constraints which may be
extracted from the following bosonic and fermionic primary constraints of the model
(\ref{SNpreon})
\begin{eqnarray}
 \label{Df=pi-}
 {d}_{\alpha I} := \pi_{\alpha I} + i P_{\alpha\beta}  \theta^{\beta I} \approx 0 \; ,
 \qquad \;  \\
 \label{P-ll=} {} {\cal P}_{\alpha\beta} := P_{\alpha\beta} - \lambda_\alpha\lambda_\beta
\approx 0\; , \qquad P^{\alpha (\lambda )} \approx 0\; , \qquad
\end{eqnarray}
where
\begin{eqnarray}
\label{P:=} P_{\alpha\beta}:= {1\over 2}{\delta {\cal L} \over \delta
\dot{\hat{X}}{}^{\alpha\beta}} \; , \qquad
P_\alpha^{(\lambda )}:= {\delta {\cal L} \over \delta \lambda{}^{\alpha}}\; , \qquad
\pi_{\alpha I}:= {\delta {\cal L} \over \delta \dot{\hat{\theta}}{}^{\alpha I}} \; ,  \qquad
\end{eqnarray}
are the canonical momenta conjugated to the coordinate functions and to the auxiliary bosonic
spinor ($s$-vector).  Using the canonical Poisson brackets
\begin{eqnarray}
 \label{Poisson=b}
{} &&  [\hat{X}{}^{\gamma\delta} ,  P_{\alpha\beta}]_{PB} = - [P_{\alpha\beta} ,
\hat{X}{}^{\gamma\delta}]_{PB} = \delta_{\alpha }{}^{(\gamma}\delta_{\beta}{}^{\delta )} \, ,\\
&& {}[ \lambda{}_{\beta} ,  P^{\alpha (\lambda )}]_{PB}  =
  - [P^{\alpha (\lambda )} , \lambda{}_{\beta}]_{PB} =
\delta_{\beta}{}^{\alpha }
  \, ,  \qquad  \\
\label{Poisson=f}
&&  \{ \pi_{\alpha I} \, , \, \hat{\theta}{}^{\beta J}\}_{PB}=
 \{ \hat{\theta}{}^{\beta J}\, , \, \pi_{\alpha I} \}_{PB}= - \delta_{\alpha }{}^{\beta} \delta_{I}{}^{J}\; ,
\end{eqnarray}
it follows that the nonvanishing Poisson brackets of the above constraints are
\begin{eqnarray}
\label{[C,C]=} {}\{ {d}_{\alpha I} , {d}_{\beta J} \}_{PB}= - 2i P_{\alpha\beta}  \delta_{IJ}
\; , \qquad [{\cal P}_{\alpha\beta}, P^{\gamma (\lambda)}]_{PB} =  -2\lambda_{(\alpha
}\delta_{\beta )} {}^\gamma
  \; . \qquad
\end{eqnarray}
These clearly indicate that the primary constraints above are a mixture of first and second
class constraints. Rather than separating them, we use below the so-called `conversion
procedure' \cite{Faddeev1984,Faddeev1986,Batalin1987,Batalin1989,Egorian1988,Egorian1989,Egorian1993,BLS99} by which a pair of degrees of freedom is added
to each pair of second class constraints to modify Eqs.~(\ref{[C,C]=}) in such a way that
they form a closed algebra. In this way, these modified constraints become first class ones
generating gauge symmetries in the enlarged phase space. In it, all the constraints of the
model are first class and account as well for the original second class constraints. These
can be recovered by gauge fixing the additional gauge symmetries/first class constraints of
the system in the enlarged phase space. For the ${\cal N}=1$ version of (2.1)
this was done in \cite{BLS99}.

As the bosonic sector of all the superparticle models is the same irrespective of ${\cal N}$,
we may use the results of \cite{BLS99} for ${\cal N}=1$ and state
that the conversion in the bosonic sector is effectively reduced to ignoring the
constraints $P^{\alpha (\lambda )}\approx 0$ in the analysis. An easy way
to see that this is indeed consistent is to observe that, as far as
$\lambda_\alpha\not= (0,...,0)$ (the usual configuration space restriction for twistor-like
variables), the second brackets in (\ref{[C,C]=}) show that $ P^{\alpha (\lambda )}= 0$ is a
good gauge fixing condition for $n$ of $n(n+1)/2$ gauge symmetries generated by the
constraints ${\cal P}_{\alpha\beta}$.

To perform the conversion in the fermionic sector, we introduce the ${\cal N}$ fermionic
variables $\chi^I$ and postulate for them the Clifford-type Poisson brackets
\begin{eqnarray}
 \label{chi,chi=}
{}\{ {\chi}^{I} , {\chi}^{J} \}_{PB}= - 2i   \delta^{IJ}  \; . \qquad
\end{eqnarray}
These ${\chi}^{I}$ are then used to modify the fermionic constraints to ${\mathbb D}_{\alpha I}
=d_{\alpha I} + i\chi^I\lambda_\alpha$. Thus, after conversion, the superparticle model
(\ref{SNpreon}) is described by the following set of first class constraints
\begin{eqnarray}
 \label{convC=}
{\mathbb D}_{\alpha I}  := \pi_{\alpha I} + i P_{\alpha\beta}  \theta^{\beta I} + i\chi^I
\lambda_\alpha \approx 0\; , \qquad {\cal P}_{\alpha\beta} := P_{\alpha\beta} -
\lambda_\alpha\lambda_\beta \approx 0\; , \qquad
\end{eqnarray}
which obey the superalgebra of the rigid supersymmetry of $\mathcal{N}$-extended tensorial superspace
$\Sigma^{({n(n+1)\over 2}|{\cal N}n)}$, 
\begin{eqnarray}
 \label{[C,C]=1}
{}\{ {\mathbb D}_{\alpha I} , {\mathbb D}_{\beta J} \}_{PB}= -2i {\cal P}_{\alpha\beta}  \delta_{IJ}
\; , \qquad [ {\cal P}_{\alpha\beta}, {\mathbb D}_{\gamma I} ]_{PB}= 0 \; , \qquad [ {\cal
P}_{\alpha\beta}, {\cal P}_{\gamma\delta}]_{PB} =  0
  \; . \qquad
\end{eqnarray}

\medskip

\section{Quantization of the superparticle in $\Sigma^{({n(n+1)\over 2}|\mathcal{N}n)}$ with even
$\mathcal{N}$ and conformal higher spin equations} \label{q-part}

Quantizing the model in its orthosymplectic-twistorial formulation (\ref{SNpr=STw}) is
straightforward. The canonical hamiltonian is equal to zero and thus the Schr\"odinger
equation simply states that
the wavefunction is independent of $\tau$. Following a procedure similar to that in
\cite{BLS99} one can show that, in the $n=4$ tensorial space corresponding to  $D=4$, the
wavefunction of the bosonic limit of the superparticle model (\ref{SNpr=STw}) describes the
solution of the free higher spin equations. This means that it can be written in terms of an
infinite tower of left and right chiral fields $\phi_{A_1\ldots A_{2s}}(p_\mu)$ and
$\phi_{\dot{A}_1\ldots \dot{A}_{2s}}(p_\mu)$ for all half-integer values of $s$ with $p_\mu
p^\mu=0$.

Let $\mathcal{N}>1$ and even (as it will be henceforth). Quantization {\it \`a la} Dirac of a
dynamical system with first class constraints requires imposing them as equations to be
satisfied by its wavefunction. In the case of our superparticle model (\ref{SNpreon}) this
wavefunction can be chosen to depend on the coordinates of $\mathcal{N}$-extended tensorial
superspace ($X^{\alpha\beta}$, $\theta^{\alpha I}$), on the bosonic spinor
($s$-vector) variable $\lambda_\alpha$ and on {\it a half} of the fermionic variables
$\chi^I$ as far as they are, by (\ref{chi,chi=}), their own momenta. The separation of a half
of the real $\chi^I$ coordinates can be achieved by introducing complex\footnote{A separation
in pairs of conjugate constraints is used in the Gupta-Bleuler
method of quantizing systems with second class constraints as the massive
superparticle \cite{KappaAL,dAz-Luk:82+1,Fryd:84}.} variables $\eta^q$,
$(\eta^q)^*=\bar{\eta}_q$, $q=1,\ldots , {\cal N}/2$, so that $\chi^I=(\chi^q , \chi^{{\cal
N}/2 +q})= ((\eta^q+\bar{\eta}_q) , i(\bar{\eta}_q-{\eta}^q))$,
\begin{eqnarray}
\label{etai=} \eta^q= {{\chi}^q -  i\chi^{{\cal N}/2 +q}\over 2}\; , \qquad \bar{\eta}_q =
{\chi^q + i\chi^{{\cal N}/2 +q}\over 2}\; , \qquad \{ \bar{\eta}_q,
\eta^p\}_{PB}=- i\delta_q{}^p \; . \qquad
\end{eqnarray}
Then, the wavefunction superfield in the coordinates representation depends only on
$\eta^q$,
\begin{eqnarray}
 \label{cW=}
 {\cal W}= {\cal W}(X^{\alpha\beta},\theta^\alpha ; \lambda_\alpha ; \eta^q)\; ,
\end{eqnarray}
the various momenta are given by the differential operators
\begin{eqnarray}
 \label{opP=}
 P_{\alpha\beta}=- i \partial_{\alpha\beta} \; ,\qquad
 \pi_{\alpha I}=-i {\partial\over \partial {\theta}^{\alpha I}} \;,
 \qquad \bar{\eta}_q =  {\partial \over \partial \eta^q} \; ,
\end{eqnarray}
and the Poisson brackets become quantum commutators or anticommutators ($[\;,\;\}_{PB}\mapsto
\frac{1}{i\hbar}[\;,\;\}$; we take $\hbar= 1$). The quantum constraints operators, to be
denoted by the same symbol (although having in mind ${\mathbb D}_{\alpha I}^{classical}\mapsto -i {\mathbb
D}_{\alpha I}^{quantum}$, ${\cal P}_{\alpha\beta}^{classical}\mapsto -i {\cal
P}_{\alpha\beta}^{quantum}$) are then
\begin{eqnarray}
\label{bbDI=} {\mathbb D}_{\alpha I}  := {\partial\over \partial {\theta}^{\alpha I}} + i
\partial_{\alpha\beta}  \theta^{\beta I} - \chi^I \lambda_\alpha \; , \qquad \\ \label{cP=qu}
{\cal P}_{\alpha\beta} := \partial_{\alpha\beta} -i \lambda_\alpha\lambda_\beta\; , \qquad
\end{eqnarray}
and have to be imposed on the wavefunction (\ref{cW=}).

For even $\mathcal{N}$, it is convenient to introduce complex
Grassmann coordinates and complex Grassmann derivatives,
\begin{eqnarray}
\label{qTh=}
\Theta^{\alpha q} = {1\over 2}(\theta^{\alpha q}-i \theta^{\alpha (q+{\cal
N}/2)})= (\bar{\Theta}^{\alpha}_q)^*  \quad \Leftrightarrow \quad \partial_{\alpha q}:=
{\partial\over \partial {\Theta}^{\alpha q}}={\partial\over \partial {\theta}^{\alpha q}} + i
{\partial\over \partial {{\theta}}^{\alpha (q+{\cal N}/2)}}\; ,
\end{eqnarray}
$q=1,\dots,\mathcal{N}/2$, and conjugate pairs of fermionic constraints
\begin{eqnarray}
 \label{nablaq=}
 {\nabla}_{\alpha q} & := & {\mathbb D}_{\alpha q}+ i {\mathbb D}_{\alpha (q+{\cal N}/2)}  =
 \partial_{\alpha q} + 2i \partial_{\alpha\beta}  \bar{\Theta}{}^{\beta}_{q} -
 2  \lambda_\alpha \, {\partial \over \partial \eta^q}= {\cal D}_{\alpha q} -
 2  \lambda_\alpha \, {\partial \over \partial \eta^q}  \; , \qquad \\
\label{bnablaq=} \bar{\nabla}_{\alpha}{}^{q} & := & {\mathbb D}_{\alpha q}- i {\mathbb D}_{\alpha
(q+{\cal N}/2)}  = \bar{\partial}_{\alpha}{}^{q} + 2i \partial_{\alpha\beta}
{\Theta}{}^{\beta q} - 2  \lambda_\alpha \, \eta^q= {\bar{\cal D}}{}_{\alpha}{}^{ q} -2
\lambda_\alpha  \eta^q  \; . \qquad
\end{eqnarray}
Since $\{ {\cal D}_{\alpha q} ,\bar{{\cal D}}_{\beta}^{p} \}= 4i {\partial}_{\alpha\beta}
\delta_{q}^{p}\,$, the above ${\nabla}_{\alpha q}$, $\bar{\nabla}_{\alpha}{}^{q}$ and the
bosonic constraint (\ref{cP=qu}) determine the superalgebra given by the only nonzero
bracket
\begin{eqnarray}
 \label{q[C,C]=1}
\{ {\nabla}_{\alpha q} ,\bar{{\nabla}}_{\beta}^{p} \}= 4i {\cal P}_{\alpha\beta}
\delta_{q}^{p} \quad.
\end{eqnarray}
This shows that it is sufficient to impose on the superwavefunction (\ref{cW=}) the fermionic
constraints,
\begin{eqnarray}
 \label{CfW=0}
{\nabla}_{\alpha q}  {\cal W}:=  {\cal D}_{\alpha q}  {\cal W} - 2  \lambda_\alpha \,
{\partial \over \partial \eta^q}  {\cal W}=0 \, , \qquad \\
 \label{Cf*W=0}
\bar{{\nabla}}_{\alpha}^{p} {\cal W}:=\bar{\cal D}{}_{\alpha}^{ q}{\cal W} - 2  \lambda_\alpha
\, \eta^q {\cal W}=0 \; , \qquad
\end{eqnarray}
since the mass-shell-like bosonic constraint,
\begin{eqnarray}
 \label{CbW=0}
{\cal P}_{\alpha\beta}   {\cal W}:= (\partial_{\alpha\beta} - i \lambda_\alpha\lambda_\beta
)   {\cal W}=0
  \; , \qquad
\end{eqnarray}
will follow as a consistency condition for (\ref{CfW=0}), (\ref{Cf*W=0}).

Decomposing the superwavefunction in a finite power series in the complex Grassmann variable
$\eta^q$,
\begin{eqnarray}
 \label{cW=W0+etaqWq+}
  {\cal W} (X,\Theta^q,\bar{\Theta}_q,\lambda, \eta^q) =
 W^{(0)}(X,\Theta^q,\bar{\Theta}_q,\lambda) +
 \sum\limits_{k=1}^{{\cal N}/2} {1\over k!} \eta^{q_k}\ldots \eta^{q_1} \,
 W^{(k)}_{q_1\ldots q_k} (X,\Theta^q,\bar{\Theta}_q,\lambda)
  \; , \qquad
\end{eqnarray}
we find that Eqs.~(\ref{CfW=0}), (\ref{Cf*W=0}) imply
\begin{eqnarray}
 \label{DW=lW}
 {\cal D}_{\alpha q} W^{(0)}= 2\lambda_\alpha W_q^{(1)}\; , \quad ... \; , \quad {\cal D}_{\alpha q}
 W_{q_1\ldots q_k}^{(k)}=
 2\lambda_\alpha  W_{qq_1\ldots q_k}^{(k+1)}\;  , \quad \ldots \; , \qquad \nonumber \\
 {\cal D}_{\alpha q} W_{q_1\ldots q_{{\cal N}/2}}^{({\cal N}/2)}=  0\; ,
\end{eqnarray}
and
\begin{eqnarray}
 \label{*DW=lW}
 \bar{{\cal D}}{}_{\alpha}^{q} W^{(0)}= 0\; , \quad
 \bar{{\cal D}}{}_{\alpha}^{q}  W_{q_1\ldots q_k}^{(k)}=
 2k \lambda_\alpha  W_{[q_1\ldots q_{k-1}}^{(k-1)}\delta_{q_k]}{}^q\; , \quad \ldots \; ,
 \quad \nonumber \\
 \bar{{\cal D}}{}_{\alpha}^{q} W_{q_1\ldots q_{{\cal N}/2}}^{({\cal N}/2)} =
 {\cal N}  \lambda_\alpha  W_{[q_1\ldots q_{{\cal N}/2-1}}^{({\cal N}/2-1)}\delta_{q_{{\cal
 N}/2}]}^q
  \; .
\end{eqnarray}
Eqs.~(\ref{DW=lW}) show that all the superfields $W_{q_1\ldots q_k}^{(k)}$ can be constructed
from fermionic derivatives of the superfield $W^{(0)} (X,\Theta^q,\bar{\Theta}_q,\lambda)$
which is chiral as a consequence of the first equation in  (\ref{*DW=lW}),
\begin{eqnarray}
\label{DDDW=lllW}
 {\cal D}_{\alpha_{k} q_{k}}\ldots  {\cal D}_{\alpha_1 q_1} W^{(0)}=
 2^k \lambda_{\alpha_1} \ldots \lambda_{\alpha_k} W_{q_1\ldots q_k}^{(k)}
  \; , \qquad  \bar{{\cal D}}{}_{\alpha}^{q} W^{(0)}= 0  \; . \qquad
\end{eqnarray}
Then, the  wavefunction ${\cal W}$ is completely characterized by the chiral superfield
$W^{(0)} (X,\Theta^q,\bar{\Theta}_q,\lambda)$. As a consequence of (\ref{CbW=0}), $W^{(0)}$
obeys
\begin{eqnarray}
 \label{CbW(0)=0}
{\cal P}_{\alpha\beta}  W^{(0)}:= (\partial_{\alpha\beta} - i \lambda_\alpha\lambda_\beta  )
W^{(0)} =0 \; ,
\end{eqnarray}
which is solved by a planewave in tensorial space,
\begin{eqnarray}
 \label{W0=eXll}
 W^{(0)} (X,\Theta^q,\bar{\Theta}_q,\lambda) =
 \tilde{W} (\lambda, \Theta^q,\bar{\Theta}_q)\exp \{ i \lambda_\alpha\lambda_\beta
 X^{\alpha\beta}\}\; .
\end{eqnarray}
The chirality of $W^{(0)}$ and the first equation in  (\ref{DW=lW}), which now implies
$\partial_{\alpha q} W^{(0)}\propto \lambda_\alpha$, show that the general solution for the
superparticle wavefunction is determined by the following chiral plane wave superfield
\begin{eqnarray}
 \label{W0=WeXll}
 W^{(0)} (X,\Theta^q,\bar{\Theta}_q,\lambda) = {w}(\lambda \, , \, \Theta^q\lambda)
 \exp \{ i \lambda_\alpha\lambda_\beta (X+2i\Theta^{p}\bar{\Theta}_p)^{\alpha\beta}\}\; ,
 \qquad
\end{eqnarray}
where $\Theta^q\lambda=\Theta^{\alpha\, q}\lambda_\alpha$ and
\begin{eqnarray}
 \label{w=}
 {w}(\lambda \, , \, \Theta^q\lambda )  = w^{(0)}(\lambda ) +
 \sum\limits_{k=1}^{{\cal N}/2} {1\over k!} (\lambda\Theta^{q_k})\ldots (\lambda\Theta^{q_1})
 \,
 w^{(k)}_{q_1\ldots q_k} (\lambda)\; .  \qquad
\end{eqnarray}

We refer to \cite{BLS99} for a discussion on how the arbitrary function
$w^{(0)}(\lambda_\alpha )$ with $\alpha=1,2,3,4$ encodes all the solutions
of the massless higher spin equations in $D=4$ and to \cite{BLS99,BBdAST05}
for the $D=6, 10$ cases. The key point is that $\lambda_\alpha$
carries the degrees of freedom of a light-like momentum
($\lambda {\gamma}_a\lambda$ is light-like in $D$=4,6,10 which corresponds
to $n$=4,8,16) plus those of spin. The $d.o.f.$ of $\lambda_\alpha$ and
those of the lightlike momenta are encoded, both up to a scale factor,
in the coordinates of the compact manifolds $S^{n-1}=S^{2D-5}$
and $S^{\frac{n}{2}}=S^{D-2}$, respectively. The spheres\footnote{The `celestial
spheres' $S^{D-2}$ are the bases $S^{2,4,8}$ of the Hopf fibrations
$S^{2D-5}\rightarrow S^{D-2}\; (S^{n-1}\rightarrow S^{n \over 2})\,$ of $S^{\,3,7,15}$,
($n,D$)=(4,4), (8,6), (16,10). The fibres $S^{D-3}=S^{1,3,7}$ of these bundles
correspond to the complex, quaternion and octonion numbers of unit modulus. The
remaining $n$=2, $D$=3 case corresponds to the first of the four Hopf fibrations,
$S^1\rightarrow \mathbb{R}P^1$; its fibre is determined by the reals of unit
modulus, $Z_2$, and there are no extra coordinates.} $S^{\frac{n}{2}-1}$
are related to helicity in the $n=4,\,D=4$ case and to its multidimensional
generalizations for $D=6,10$ \cite{BLS99,BBdAST05}.

To obtain a superfield on tensorial superspace describing massless conformal higher spin
theories with extended supersymmetry, the wavefunction  (\ref{W0=WeXll}) has to
be integrated over $\,{\mathbb R}^n- \{0\} \sim {\mathbb S}^{n-1}\times$ ${\mathbb R}_+ \,$,
parametrized by $\lambda_\alpha$, with an appropriate
measure that we denote by $d^{n}\lambda$,
\begin{eqnarray}
\label{Phi=intWeXll}
 \Phi (X,\Theta^q,\bar{\Theta}_q) =
  \int d^{n}\lambda W^{(0)} (X,\Theta^q,\bar{\Theta}_q,\lambda) =
   \int d^{n}\lambda {w}(\lambda \, , \, \Theta^q\lambda)
   e^{ i \lambda_\alpha\lambda_\beta (X+2i\Theta^{p}\bar{\Theta}_p)^{\alpha\beta}}\; .\nonumber\\
\end{eqnarray}
One can easily check that the superfield $\Phi$ is chiral
\begin{eqnarray}
 \label{*DPhi=0}
  {\bar {\cal D}}_\alpha^q\Phi (X,\Theta^{q'},\bar{\Theta}_{p'}) = 0\;   \qquad
\end{eqnarray}
and satisfies the equation
\begin{eqnarray}
 \label{asDDPhi=0}
  {\cal D}_{q[\beta}{\cal D}_{\gamma ]p}\Phi (X,\Theta^{q'},\bar{\Theta}_{p'}) = 0\;  .
  \qquad
\end{eqnarray}
These are the superfield equations for the wavefunction of the superparticle in ${\cal
N}$-extended tensorial superspace for even ${\cal N}$.

\section{From the superfield to the component form of the higher spin equations
in tensorial space} \label{sec4}

\subsection{${\cal N}=2$}
\label{Nigual2}

When ${\cal N}=2$, Eqs.~(\ref{*DPhi=0}) and (\ref{asDDPhi=0}) can be written as 
\begin{eqnarray}
\label{hsSEq=2N1} \bar{{\cal D}}_\alpha \Phi =0\; ,\nonumber\\
 {\cal D}_{[\alpha}{\cal D}_{\beta]} \Phi  = 0 \;  
\end{eqnarray}
and reproduce equations from page XIV.
It is easy to check that Eqs.~(\ref{hsSEq=2N1}) imply the vanishing of all the components of the `chiral' superfield $\Phi (X,\Theta
,\bar{\Theta})= \Phi (X+2i\Theta \cdot \bar{\Theta} ,\Theta )$, except the first two,
\begin{eqnarray}
 \label{Phi=N2}
  \Phi (X,\Theta ,\bar{\Theta})= \phi(X_L )+ i\Theta^\alpha  \psi_\alpha (X_L)\quad ,  \quad
  X_L^{\alpha\beta}=X^{\alpha\beta}+2i\Theta^{(\alpha}\bar{\Theta}{}^{\beta )}
  =X_L^{\beta\alpha}\quad ,
\end{eqnarray}
where $X_L^{\alpha\beta}$ is the analogue of the
bosonic coordinates for the chiral basis of standard $D=4$ superspace.
The above components are the {\it complex} bosonic scalar and the complex
fermionic spinor fields
obeying the free higher spin equations in tensorial space form \cite{V01s},
\begin{eqnarray}
\label{HSpinEq=b+f}
 \partial_{\alpha [\gamma }\partial_{\delta ]\beta } \phi(X)=0\; ,
\qquad   \partial_{\alpha [\beta }\psi_{\gamma ]} (X)=0\;  . \qquad
\end{eqnarray}
Let us recall that the ${\cal N}=1$ supermultiplet contains a real bosonic scalar and
a real fermionic spinor field that obey the same equations (\ref{HSpinEq=b+f}). Hence,
the ${\cal N}=2$ supermultiplet of the conformal fields in tensorial
superspace is given by the complexification of the  ${\cal N}=1$ supermultiplet.

 Clearly, the above results are $n$-independent and thus, besides $n=4$, they
are also valid for the $n=8$ and $n=16$ cases corresponding to the $D=6$ and $D=10$
multiplets of massless conformal higher spin fields.

\subsection{${\cal N}=4$}
\label{secN4}

In contrast with the ${\cal N}=2$ case, spin-tensorial
components are present when ${\cal N}> 2$. For ${\cal N}=4$, the general
solution of the  superfield equations (\ref{*DPhi=0}) and
(\ref{asDDPhi=0}) is given by
\begin{eqnarray}
 \label{Phi=N4}
 \Phi (X,\Theta^q ,\bar{\Theta}_q)=
 \phi(X_L )+ i\Theta^{\alpha q}  \psi_{\alpha q} (X_L) +
 \epsilon_{pq}\Theta^{\alpha q}\Theta^{\beta p}  {\cal F}_{\alpha \beta} (X_L) \; ,  \qquad
 \\
\label{XL= N4} X_L^{\alpha\beta}=X^{\alpha\beta}+2i\Theta^{q(\alpha }\bar{\Theta}{}^{\beta
)}_q \; , \; q=1,2 \; ,  \qquad
\end{eqnarray}
where, again, the complex scalar and spinor fields obey the standard higher spin equations in
their tensorial superspace form,
\begin{eqnarray}
\label{HSpin4N=b+f}
  \partial_{\alpha [\gamma }\partial_{\delta ]\beta } \phi(X)=0\; ,
  \qquad   \partial_{\alpha [\beta }\psi_{\gamma ]q} (X)=0\;  , \qquad
\end{eqnarray}
while the complex symmetric bi-spinor (or `$s$-tensor' \cite{V01s}) ${\cal F}_{\alpha
\beta}={\cal F}_{ \beta\alpha}$ satisfies the tensorial counterpart of the $D=4$ Maxwell
equations (when these are written in spinorial notation \cite{Penrose}, see also below),
\begin{eqnarray}
\label{HSpin4N=cF}
\partial_{\alpha [\gamma }{\cal F}_{\delta ]\beta } (X)=0\; , \qquad
{\cal F}_{\alpha \beta}={\cal F}_{ \beta\alpha}\; . \qquad
\end{eqnarray}

However, one can easily show that the general solution of  Eq.~(\ref{HSpin4N=cF})
is expressed through a new complex scalar superfield $\tilde{\phi}(X)$
satisfying the bosonic tensorial space equation in
(\ref{HSpin4N=b+f}),
\begin{eqnarray}
\label{HSpincF=}
  {\cal F}_{\alpha \beta}= \partial_{\alpha \beta } \tilde{\phi}(X)\; , \qquad \\
\label{HSpintp=}
  \partial_{\alpha [\gamma }\partial_{\delta ]\beta } \tilde{\phi} (X)=0\; . \qquad
\end{eqnarray}

\subsubsection{$n=4\,,\, D=4$}

To prove this when $n=4$ ($\alpha,\beta =1,2,3,4$), we begin by decomposing
the complex symmetric $GL(4)$ tensor ${\cal F}_{\alpha \beta}={\cal F}_{ \beta\alpha}$
in $2\times 2$ blocks, thus keeping only the $GL(2,\mathbb{C})$ manifest symmetry\footnote{Notice that the $SL(2,\mathbb{C})$ indices 
$A,B=1,2$, $\dot{A},\dot{B}=1,2$ are denoted by $\alpha$, $\beta$ and $\dot{\alpha}$, $\dot{\beta}$ in chapters \ref{chapterscalar} 
and \ref{chaptersugra+membrane}.},
\begin{eqnarray}
\label{cF=F-V-V-F}
  n=4: \quad {\cal F}_{\alpha \beta}= \left(\begin{matrix} F_{AB} & V_{A\dot{B}} \cr
  V_{B\dot{A}} & F_{\dot{A}\dot{B}} \end{matrix}\right) \; , \qquad A,B=1,2\; , \quad \dot{A},
  \dot{B}=1,2 \; . \qquad
\end{eqnarray}
Let us first notice that the block components of Eq.~(\ref{HSpin4N=cF}) which contain the
antisymmetric tensors (encoded in the symmetric spin-tensors $F_{AB}$ and $
F_{\dot{A}\dot{B}}$) only,
\begin{eqnarray}
\label{dF=0=d*F}
\partial_{\dot{A}[B}F_{C]D}=0\; ,  \qquad
\partial_{{A}[\dot{B}}F_{\dot{C}]\dot{D}} =0\; ,  \qquad
\end{eqnarray}
are equivalent to the Maxwell equations for the complex selfdual field $F_{ab}={i\over
2}\epsilon_{abcd}F^{cd}\propto \sigma_{ab}{}^{CD}F_{CD}$ {\it i.e.}, they imply
$\partial^aF_{ab}=0$ and $\partial_{[a}F_{bc]}=0$.

Consider now the components of Eq.~(\ref{HSpin4N=cF}) which contain the complex vector
$V_{A\dot{B}}=\sigma^a_{A\dot{B}} V_a$ only, namely
\begin{eqnarray}
\label{dvV=}
\partial_{A [\dot{B}}V_{\dot{C}]D}=0
 \; , \qquad \partial_{\dot{B}[A}V_{D]\dot{C}}=0
 \;  \qquad
\end{eqnarray}
and
\begin{eqnarray}
\label{dtV=}
\partial_{A[B}V_{C]\dot{D}}=0
 \; , \qquad \partial_{\dot{A}[\dot{B}}V_{\dot{C}]D}=0
 \;  . \qquad
\end{eqnarray}
 Eqs.~(\ref{dvV=}) imply $\partial_{[a} V_{b]}=0$ and $\partial^aV_a=0$.
The first is solved by $V_a=\partial_{a}\tilde{\phi}$ and the second implies that the scalar
field $\tilde{\phi}$ obeys the Klein-Gordon equation $\partial^a\partial_a\tilde{\phi}=0$. In
the spin-tensor notation these read
\begin{eqnarray}
\label{V=dtp} V_{A\dot{B}}= \partial_{A\dot{B}}\tilde{\phi}\; , \qquad
\partial_{A[\dot{B}} \partial_{\dot{C}]D}\tilde{\phi}=0\; . \qquad
\end{eqnarray}

Next, the components of Eq.~(\ref{HSpin4N=cF})  which contain both vector and antisymmetric
tensor components, $\partial_{{A}\dot{B}}F_{CD}- \partial_{{A}C}V_{D\dot{B}}=0$ and
$\partial_{{A}\dot{B}}F_{\dot{C}\dot{D}}- \partial_{\dot{C}\dot{D}}V_{A\dot{B}}=0$, can be
written in the form
\begin{eqnarray}
\label{dF-ddP=}
\partial_{{A}\dot{B}}(F_{CD} - \partial_{CD}\tilde{\phi})=0\; , \qquad
\partial_{{A}\dot{B}}(F_{\dot{C}\dot{D}} - \partial_{\dot{C}\dot{D}}\tilde{\phi})=0\; ,
\qquad
\end{eqnarray}
the only covariant solution of which is given by
\begin{eqnarray}
\label{F=dP} F_{CD} = \partial_{CD}\tilde{\phi}\; , \qquad F_{\dot{C}\dot{D}} =
\partial_{\dot{C}\dot{D}}\tilde{\phi}\; . \qquad
\end{eqnarray}
Keeping in mind the Maxwell equations (\ref{dF=0=d*F}), one finds that the scalar field
$\tilde{\phi}(X)$ satisfies, besides the Klein-Gordon equation in (\ref{V=dtp}), also the
remaining components of Eq.  (\ref{HSpintp=}),
\begin{eqnarray}
\label{ddtp=0}
\partial_{A[B}\partial_{C]D}\tilde{\phi}=0\; , \qquad
\partial_{\dot{A}[\dot{B}} \partial_{\dot{C}]\dot{D}}\tilde{\phi}=0\; . \qquad
\end{eqnarray}

\subsubsection{Proof for arbitrary $n$}

We now prove that Eqs.~(\ref{HSpincF=}), (\ref{HSpintp=}) provide the general
solution of the Maxwell-like equation in tensorial space, Eq.~(\ref{HSpin4N=cF}),
for any $n$. The Fourier transform of Eq. (\ref{HSpin4N=cF}) is
\begin{eqnarray}
\label{HSpin4N=cFp}
p_{\alpha [\gamma }{\cal F}_{\delta ]\beta}(p)=0\; . \qquad
\end{eqnarray}
The solution of this equation is nontrivial {\rm iff} the matrix of the
generalized momentum has rank one, this is to say when
$p_{\alpha\beta}=\lambda_\alpha\lambda_\beta$  for arbitrary $\lambda_\alpha\not= (0,...,0)$
or, equivalently, when this matrix obeys $p_{\alpha [\gamma }p_{\delta ]\beta}=0$.
The general solution is characterized by
${\cal F}_{\alpha\beta}(\lambda) =\lambda_{\alpha}\lambda_{\beta} \phi(\lambda )$
and can be equivalently written in the form
${\cal F}_{\alpha\beta}(p) =p_{\alpha\beta} \tilde{\phi}(p)$ if
$ p_{\alpha [\gamma }p_{\delta ]\beta} \tilde{\phi}(p)=0$ \footnote{ More formally, the solution
of this equation is given by a distribution with support on
the subspace of tensorial momentum space defined by the condition
$p_{\alpha [\gamma }p_{\delta ]\beta}=0$, so that ${\tilde \phi}(p)
\propto \delta (p_{\alpha [\gamma }p_{\delta ]\beta } )$. }
of  Eqs.~(\ref{HSpin4N=cF}).
As far as set of equations
\begin{eqnarray}
\label{HSpin4NcF=ppP}
{\cal F}_{\alpha\beta}(p) =p_{\alpha\beta} \tilde{\phi}(p) \; , \qquad
p_{\alpha [\gamma }p_{\delta ]\beta} \tilde{\phi}(p)=0\;  \qquad
\end{eqnarray}
provide the Fourier transforms of Eqs.~(\ref{HSpincF=}), (\ref{HSpintp=}),
these describe the general solution.

\subsubsection{Peccei-Quinn-like symmetry}

Thus, the ${\cal N}=4$ higher spin supermultiplet actually contains two complex scalar fields
and two Dirac spinor fields in tensorial space, $\phi(X), \psi^1_\alpha (X),\psi^2_\alpha
(X), \tilde{\phi}(X)$, which satisfy the free bosonic and fermionic higher spin equations,
Eqs.~(\ref{HSpin4N=b+f}), (\ref{HSpintp=}). They appear in the on-shell scalar superfield
decomposition as
\begin{eqnarray}
\label{Phi(X)4N=} \Phi (X,\Theta^q ,\bar{\Theta}_q)= \phi(X_L )+ i\Theta^{\alpha q}
\psi_{\alpha q} (X_L) + \epsilon_{pq}\Theta^{\alpha q}\Theta^{\beta p} \partial_{\alpha
\beta} \tilde{\phi} (X_L)\; , \qquad q,p=1,2\; . \qquad
\end{eqnarray}
However, as the second complex scalar field $ \tilde{\phi}$ enters the original
superfield with a derivative, its zero mode is not fixed. In other words, this scalar is
axion-like: it possesses the Peccei-Quinn-like symmetry
\begin{eqnarray}
\label{Pe-Qu=sym4N}
 \tilde{\phi} (X)\mapsto  \tilde{\phi} (X) + const\; .
\end{eqnarray}

\subsection{${\cal N}=8$}
\label{N8part}

For higher ${\cal N}>4$ the general solution of the set of superfield equations
(\ref{*DPhi=0}) and (\ref{asDDPhi=0}) is given by
\begin{eqnarray}
\label{Phi=N>4} \Phi (X,\Theta^q ,\bar{\Theta}_q)= \phi(X_L )+ i\Theta^{\alpha q}
\psi_{\alpha q} (X_L) + \sum\limits_{k=2}^{{\cal N}/2} {1\over k!} \Theta^{\alpha_k
q_k}\ldots \Theta^{\alpha_1 q_1} {\cal F}_{\alpha_1 \ldots \alpha_k \; q_1 \ldots  q_k}
(X_L)\;  , \qquad \\ \label{F=Fsa} {\cal F}_{\alpha_1 \ldots \alpha_k \; q_1 \ldots  q_k}
(X_L)= {\cal F}_{(\alpha_1 \ldots \alpha_k ) \; [q_1 \ldots  q_k]} (X_L)\; , \;
X_L^{\alpha\beta}=X^{\alpha\beta}+2i\Theta^{q(\alpha }\bar{\Theta}{}^{\beta )}_q\; ,\;
q=1,...,{{\cal N}/2}\,, \quad
\end{eqnarray}
where $\phi(X_L)$ and $\psi_{\alpha q}(X_L)$ obey the standard higher spin equations
(\ref{HSpinEq=b+f}) while the higher components satisfy
\begin{eqnarray}
\label{HSpinN>4=cF}
\partial_{\alpha [\gamma }{\cal F}_{\delta ]\beta_2\ldots \beta_k \; q_1 \ldots  q_k}
(X_L)=0\; , \qquad
{\cal F}_{\alpha_1\ldots  \alpha_q}={\cal F}_{(\alpha_1\ldots  \alpha_q)}\; . \qquad
\end{eqnarray}

   For instance, for ${\cal N}=8$ the superfield solution of the higher spin equations
(\ref{*DPhi=0}) and (\ref{asDDPhi=0})  reads
\begin{eqnarray}
\label{Phi=N8} \Phi (X,\Theta^q ,\bar{\Theta}_q)= \phi(X_L )+ i\Theta^{\alpha q}
\psi_{\alpha q} (X_L) + {1\over 2} \Theta^{\alpha_2 q_2} \Theta^{\alpha_1 q_1} {\cal
F}_{\alpha_1 \alpha_2 \; q_1 q_2}(X_L ) + \quad \nonumber \\ + {i\over 3!}\Theta^{\alpha_3
q_3}\Theta^{\alpha_2 q_2}\Theta^{\alpha_1
q_1}\epsilon_{q_1q_2q_3q}\psi_{\alpha_1\alpha_2\alpha_3}^q (X_L) + \quad \nonumber \\ +
{1\over 4!}\epsilon_{q_1q_2q_3q_4}  \Theta^{\alpha_4 q_4}\ldots \Theta^{\alpha_1 q_1}{\cal
F}_{\alpha_1 \ldots \alpha_4 }(X_L)\;  .
\end{eqnarray}
Its two lowest components obey Eqs.~(\ref{HSpinEq=b+f}), while its higher
order nonvanishing field components satisfy Eqs.~(\ref{HSpinN>4=cF}),
\begin{eqnarray}
\label{HSpinN8=cF}
\partial_{\alpha [\gamma }{\cal F}_{\delta ]\beta\; q_1q_2} (X)=0\; , \qquad \\
\label{HSpinN8=psi}
\partial_{\alpha [\gamma }{\psi}_{\delta ]\beta_2\beta_3}{}^q (X)=0  \, , \qquad \\
\label{HSpinN8=cF4}
\partial_{\alpha [\gamma }{\cal F}_{\delta ]\beta_2\beta_3\beta_4 } (X)=0\; . \qquad
\end{eqnarray}
It is tempting to identify (\ref{HSpinN8=psi}) with the tensorial space generalization of the
Rarita-Scwinger equations and Eq.~(\ref{HSpinN8=cF4}) with that of the linearized conformal
gravity equation imposed on Weyl tensor. However, similarly to the ${\cal N}=4$ case in
Sec.~\ref{secN4}, it is possible to show that the general solutions of Eqs.~(\ref{HSpinN8=cF}),
(\ref{HSpinN8=psi}) and (\ref{HSpinN8=cF4}) are expressed in terms of a sextuplet of
scalar fields $\phi_{q_1q_2}(X)=\phi_{[q_1q_2]}(X)$, a quadruplet of spinorial fields
$\tilde{\psi}_{\alpha_3}^q$ and  a singlet of scalar field $\tilde{\phi}(X)$ obeying
the standard tensorial space fermionic and bosonic higher spin equations
(\ref{HSpinEq=b+f}),
\begin{eqnarray}
\label{HSpin8N=sc} {\cal F}_{\alpha\beta\; q_1q_2} (X)=\partial_{\alpha \beta }
\phi_{q_1q_2}(X) \; , \qquad  \\ \label{HSpin8N=sp} \psi_{\alpha_1\alpha_2\alpha_3}^q (X)=
\partial_{\alpha_1\alpha_2} \tilde{\psi}_{\alpha_3}^q (X)  \;  , \qquad \\
\label{HSpin8N=sc4} {\cal F}_{\alpha_1 \ldots \alpha_4 } (X)=
\partial_{\alpha_1 \alpha_2 }\partial_{\alpha_3 \alpha_4 }{\tilde \phi}(X)\; . \qquad
\end{eqnarray}

Summarizing, the ${\cal N}=8$ supermultiplet of free higher spin fields is described by a set
of two scalar fields, a sextuplet of scalar fields, a spinor field and a quadruplet of
spinorial fields, all in tensorial superspace, which obey the usual type free higher spin
equations
\begin{eqnarray}
\label{HSpinEq=8N}
\partial_{\alpha [\gamma }\partial_{\delta ]\beta } \phi(X)=0\; , \qquad
\partial_{\alpha [\beta }\psi_{\gamma ]} (X)=0\;  ,  \qquad \nonumber
\\
\partial_{\alpha [\gamma }\partial_{\delta ]\beta } \phi_{qp}(X)=0\; , \qquad    \nonumber
\\
\partial_{\alpha [\beta }\tilde{\psi}_{\gamma ]}{}^q (X)=0\;  ,
\qquad \partial_{\alpha [\gamma }\partial_{\delta ]\beta } \tilde{\phi}(X)=0\; .  \qquad
\end{eqnarray}
Thus, we conclude that, in tensorial superspace, at least all the free field dynamics is
carried by the scalar and spinor fields.  However, the `higher' scalar and spinor
fields appear in the basic superfield under the action of one or two derivatives and,
hence, the model is invariant under the following generalized bosonic and
fermionic Peccei-Quinn-like symmetries
\begin{eqnarray}
\label{Pe-Qu=8Nsym}
 \phi_{qp}(X)&\mapsto & \phi_{qp}(X)+ a_{qp}\; , \qquad \nonumber \\
 \tilde{\psi}_{\alpha}{}^q (X)& \mapsto & \tilde{\psi}_{\alpha}{}^q (X)+ \beta_{\alpha}{}^q
 \; , \qquad \nonumber \\
 \tilde{\phi}(X) & \mapsto & \tilde{\phi}(X) + a + X^{\alpha\beta} a_{\alpha\beta} \;  ,
\end{eqnarray}
with constant bosonic parameters $a_{qp}=-a_{pq}$, $a$, $a_{\alpha\beta}$ and constant
fermionic parameter $\beta_{\alpha}{}^q $. Note that the non-constant shift in
$\tilde{\phi}(X)$ is allowed by the presence of two derivatives in Eq.~(\ref{HSpin8N=sc4}).

\chapter{Covariant action and equations of motion for the 11D system of multiple M$0$-branes}\label{chapter11D}
\thispagestyle{chapter} 
\initial{I}n this chapter we present and study the covariant supersymmetric and $\kappa$--symmetric action for a system of 
$N$ nearly coincident M0--branes (mM0 system) in flat eleven dimensional superspace and the equations of motion obtained from this action.
As far as the mM0 action is written with the use of moving frame and spinor moving frame variables, 
we begin by describing their use in a simpler model of single M$0$--brane. 
We obtain the complete set of mM0 equations of motion and study the symmetries of the mM0 action the most relevant of which 
is the reminiscence of the $K_9$ symmetry of the single M0--brane. This symmetry allows us to write the bosonic equations of motion for 
the mM0 center of energy variables in their final form. The center of energy motion is characterized by a nonnegative constant mass $M$ which is contructed from the matrix fields which 
describe the relative motion of the mM0 constituents.
We show that all supersymmetric solutions of the mM0 equations preserve 16 of 32 supersymmetries {\it i.e.} describe ${1\over 2}$ BPS states, 
and are characterized by vanishing center of energy mass $M^2=0$.
We also present two examples of non supersymmetric solutions with $M^2\not= 0$.

\newpage

\section{Single M0--brane in spinor moving frame formulation}
\label{M0sec}
\setcounter{equation}{0}

\subsection{Twistor--like spinor moving frame action and its irreducible
$\kappa$--symmetry}
\label{IIA}

The spinor moving frame action of M$0$--brane reads (see \cite{IB07:M0} and also
\cite{B90,IB+AN:95,BL98',IB+JdA+DS:2006})
\begin{eqnarray}
\label{SM0=} S_{M0} &=& \int_{W^1} \rho^{\#}\, \hat{E}^{=} = \int_{W^1} \rho^{\#}\,
u_a^{=}  \, {E}^a(\hat{Z}) \qquad
\\ \label{SM0==}
& =& {1/16}\int_{W^1}\rho^{\#}\, (v_{q}^{\; -}{\Gamma}_a v_{q}^{\; -}) \, \hat{E}^{a}
\; .
  \end{eqnarray}
In the first line of this equation, (\ref{SM0=}), $\rho^{\#}(\tau)$  is a Lagrange
multiplier,
\begin{eqnarray}
\label{hEa=} \hat{E}^{a}:= {E}^a(\hat{Z})=d\hat{Z}{}^M(\tau)  {E}_M^{a}(\hat{Z})=:
d\tau \hat{E}_\tau^a(Z)\;
  \end{eqnarray}
is the pull--back of the bosonic supervielbein of the 11D target superspace
($a=0,1,...,10$), $E^a= E^a(Z)= dZ^ME_M^a(Z)$,  to the worldline $W^1$ parametrized by
proper time $\tau$. In the case of flat target superspace the supervielbein
can be chosen in the form \footnote{The action (\ref{SM0=}), (\ref{SM0==}) makes sense
when  supervielbein  $E^a= dZ^ME_M^a(Z)$ obeys the 11D superspace supergravity
constraints \cite{BrinkHowe80}. In this chapter we will restrict ourselves by the
case of flat target superspace, described by Eqs. (\ref{Ea=Pi}).},\footnote{We use the (real)  matrices   $\Gamma^a_{\alpha\beta}=\Gamma^a_{\beta\alpha}=
\Gamma^a_{\alpha}{}^\gamma C_{\gamma\beta}$ and
$\tilde{\Gamma}_a^{\alpha\beta}=\tilde{\Gamma}_a^{\beta\alpha}=
C^{\alpha\gamma}\Gamma^a_{\gamma}{}^{\beta}$  constructed as a product of  11D Dirac
matrices $\Gamma^a_{\beta}{}^\gamma$ (obeying $\Gamma^a\Gamma^b +
\Gamma^b\Gamma^a=2\eta^{ab} I_{32\times 32}$) with, respectively, the 11D charge
conjugation matrix  $C_{\gamma\beta}=- C_{\beta\gamma}$ and its inverse
$C^{\alpha\beta}=- C^{\beta\alpha}$. Both $\Gamma^a_{\beta}{}^\gamma$  and
$C_{\beta\gamma}$ are pure imaginary in our mostly minus notation $\eta^{ab}=diag
(1,-1, ..., -1)$.}
\begin{eqnarray}
\label{Ea=Pi} E^a = \Pi^a = dx^a - i d\theta \Gamma^a\theta\; , \quad
E^\alpha=d\theta^\alpha.
  \end{eqnarray}
Finally, $\hat{E}^{=}= \hat{E}^{a}u_a^{=}$ and $u_a^{=}=u_a^{=}(\tau)$ is a
light--like 11D vector,  $u^{=a}u_a^{=}=0$.

One can write the action (\ref{SM0=}) in a probably more conventional from, extracting
$d\tau$ measure from the pull--back of the supervielbein 1--form (see (\ref{hEa=}))
\begin{eqnarray}
\label{SM0=dt} S_{M0} &=& \int_{W^1} d\tau \rho^{\#}\, \hat{E}_\tau^{=}= \qquad
\nonumber \\ &=&\int_{W^1} d\tau \rho^{\#}\, \partial_\tau \hat{Z}{}^M(\tau)
{E}_M^{a}(\hat{Z}(\tau)) u_a^=(\tau) \; .
  \end{eqnarray}
We however, prefer to hide $d\tau$ inside of differential form, define the
Lagrangian 1-form by ${\cal L}_1=d\tau {\cal L}_\tau$,  and write our actions as
integral of this 1--form over the worldline, $\int_{W^1} {\cal L}_1$, rather than as an integral
over $d\tau$ of a density,  $\int d\tau {\cal L}_\tau$.

If we were stoping at this stage, one can easily observe that the action (\ref{SM0=})
can be obtained from the first order form of 11D version of the Brink--Schwarz action,
\begin{eqnarray}
\label{S'M0=} S_{BS} &=& \int_{W^1} \left( p_a \hat{E}^{a} -{e\over 2} p_ap^a d\tau
\right) \; ,
  \end{eqnarray}
by solving the constraints  $p_ap^a=0$ (equations of motion for Lagrange multiplier
$e(\tau)$) and substituting them back to the action.  Furthermore, one might wonder why
the solution $p_a=\rho^{\#}u_a^=$ is written with a multiplier $\rho^{\#}(\tau)$ instead
of just stating that it has the form of $S= \int_{W^1}  p_a \hat{E}^{a}$ with $p_a$ constrained by
$p_ap^a=0$. We will answer that question a bit later, just announcing now that
$\rho^{\#}$ is a kind of St\"{u}ckelberg variable allowing to introduce an $SO(1,1)$
gauge symmetry; although looking artificial at this stage, this symmetry allows to
clarify the group theoretical meaning of $u_a^=$ and also of the set of $16$
constrained spinors appearing in the second representation of $S_{M0}$, Eq.
(\ref{SM0==}).

The light--like vector $u_a^=$ can be considered as a composite of (any of) the $16$
spinors $v^{-\alpha}_q$  provided these are constrained by
\begin{eqnarray} \label{Iu--=vGv}
 v_q^{-\alpha} (\Gamma^a)_{\alpha\beta} v_p^{-\beta}= \delta_{qp}
u^{=}_{ a} \;  \qquad && (\ref{Iu--=vGv}a)  \nonumber \\  2v_q^{-\alpha} v_q^{-\beta}= u^{=}_{ a}
\tilde{\Gamma}^{a\alpha\beta} \; . \qquad && (\ref{Iu--=vGv}b)
\end{eqnarray}
Notice that the trace of (\ref{Iu--=vGv}a) as well as the $\Gamma$--trace of (\ref{Iu--=vGv}b) give $16 u^{=}_{ a} = v_q^{-\alpha} (\Gamma^a)_{\alpha\beta} v_q^{-\beta}$ which 
can be read off (\ref{SM0==}) and (\ref{SM0=}). The set of spinors $v_q^{-\alpha}$ constrained by (\ref{Iu--=vGv}) are 
called {\it spinor moving frame variables} (hence the name `spinor moving frame' for the formulation of superparticle 
mechanics based on the action (\ref{SM0=}), (\ref{SM0==})).
Before discussing their  origin and nature (in sec. \ref{smovfr}), in the next sec. \ref{sec=irrKap} we would like 
to try to convince the reader in the usefulness of these 'square roots' of the light--like vector $u^{=}_{ a}$.

\subsection{Irreducible $\kappa$--symmetry of the spinor moving frame action}
\label{sec=irrKap}
The action (\ref{SM0=}), (\ref{SM0==}) is invariant under the following local fermionic $\kappa$--symmetry transformations
\begin{eqnarray}
\label{kap=irr} && \delta_\kappa \hat{x}^a =  - i  \hat{\theta}
\Gamma^a\delta_\kappa\hat{\theta}\; , \qquad \delta_\kappa \hat{\theta}^\alpha =
\epsilon^{+q} (\tau)  v_q^{-\alpha} \; , \quad \nonumber \\ && \delta_\kappa
\rho^{\#}=0 \; , \quad \nonumber \\   && \delta_\kappa u_a^{=}=0 \quad \Leftarrow \quad
\delta_\kappa v_q^{-\alpha}=0
 \; .  \quad
\end{eqnarray}
These symmetry is {\it irreducible} in the sense of that each of 16 fermionic parameters\footnote{The
$\kappa$--symmetry was discovered in \cite{Siegel83,KappaAL} and was shown to coincide with the local worldline supersymmetry 
in \cite{stv}. Our notation
$\epsilon^{+q} (\tau) $ for the (irreducible)  $\kappa$--symmetry parameter is an implicit reference on this later result 
which will be useful in the discussion below.}
 $\;\epsilon^{+q} (\tau) $ acts efficiently on the variables of the theory and can be used to remove some component of fermionic 
field
$\hat{\theta}^\alpha(\tau)$ thus reducing the number of the degrees of freedom in it to $16$ (while $\alpha=1,...,32$).

In contrast, the $\kappa$--symmetry of the original Brink--Schwarz superparticle action
(\ref{S'M0=})  \cite{Siegel83}
\begin{eqnarray}
\label{kappa} \delta_\kappa \hat{x}^a =  - i  \hat{\theta}
\Gamma^a\delta_\kappa\hat{\theta}\; , \qquad \delta_\kappa \hat{\theta}^\alpha=p_a
\tilde{\Gamma}^{a\alpha\beta}\kappa_\beta (\tau)\; , \quad \nonumber \\  \delta_\kappa
e=-4i \kappa_\beta d\hat{\theta}^\beta \,
 \; ,  \qquad
  \end{eqnarray}
is infinitely reducible. It is parametrized by 32 component fermionic spinor function $\kappa_\beta (\tau)$ which however is not 
acting efficiently on the variable of the theory.\footnote{Roughly speaking,  due to the constraint $p_ap^a=0$,
$\kappa_\alpha$ and $\kappa_\alpha + p_a
\tilde{\Gamma}^{a}_{\alpha\beta}\kappa^{(1)\beta} (\tau)$ produce the same $\kappa$ variation of the Brink--Schwarz superparticle 
variables. One says that the above transformation has a null-vector $\kappa^{(1)\beta} (\tau)$ and, hence, the symmetry is 
{\it reducible}. But this is not the end of story.
One easily observes that $\kappa^{(1)\beta} (\tau)$ and $\kappa^{(1)\beta} (\tau)+p_a
\tilde{\Gamma}^{a\alpha\beta}\kappa^{(2)}_\beta (\tau)$, with an arbitrary $\kappa^{(2)}_\beta (\tau)$, makes the same change of 
the parameter $\kappa_\alpha$. This implies that there is a null--vector  for null--vector and that the $\kappa$--symmetry 
possesses at least the  second rank of reducibility. Furthermore, one sees that this process of finding higher null--vectors 
can be continued up to infinity (next stages are completely equivalent to the first two ones) so that one speaks about infinite 
reducibility of the $\kappa$--symmetry of the Brink--Schwarz superparticle. The number of the fermionic degrees of freedom which 
can be removed by $\kappa$--symmetry is then calculated as an infinite sum
$32-32+32-32+...= 32\cdot (1-1+1-1+...)= 32\cdot \lim\limits_{q\rightarrow 1} (1-q+q^2-...)= 32\cdot \lim\limits_{q\rightarrow 1}{1\over 1+q}= 16$. }

The irreducible $\kappa$--symmetry of the spinor moving frame formulation
(\ref{kap=irr}) can be obtained from the infinitely reducible (\ref{kappa}) by
substituting for $p_a$ the solution $p_a=\rho^{\#}u_a^=$ of the  constraint $p_ap^a=0$;
furthermore, using (\ref{Iu--=vGv}), we find
\begin{eqnarray}
\label{ep=kappa}
\epsilon^{+q}= 2
\rho^{\#}v_q^{-\alpha}\kappa_\alpha \; . \qquad
  \end{eqnarray}
Let us stress that this relation, as well as the transformation rules of the irreducible $\kappa$--symmetry (\ref{kap=irr}), 
necessarily  involves the constrained spinors $v_q^{-\alpha}$. Thus the covariant irreducible  form of the $\kappa$--symmetry 
is a characteristic property of the spinor moving frame and similar ('twistor--like') formulations of the superparticle 
mechanics.\footnote{ Notice that in D=3,4 and 6 dimensions the counterpart of $v_q^{-\alpha}$ can be chosen to be unconstrained 
spinors; see references in {\it e.g.} \cite{B90,BL98',IB07:M0}.}

The importance of the $\kappa$--symmetry is related to the fact that it  reflects a
part of target space  supersymmetry which is preserved by ground state of the brane
under consideration \cite{Bergshoeff:1997kr,BdAI} thus insuring that it is a  BPS
state. Its irreducible form, reached in the frame of spinor moving frame formulation, is useful not only for clarifying its 
nature as worldline supersymmetry  (\cite{stv}), but also for finding the corresponding induced supergravity multiplet which 
is necessary for constructing the mM$0$ action. To address this issue we need to comment on some properties of moving frame 
and spinor moving frame variables.

\subsection{Moving frame and spinor moving frame}
\label{smovfr}
To clarify  the origin and nature of the set of spinors $v^{-\alpha}_q$ which provide the square root of the light--like 
vector $u_a^{=}$ in the sense of Eqs. (\ref{Iu--=vGv}), and which have been used to present the $\kappa$--symmetry in the 
irreducible form (\ref{kap=irr}), it is useful to complete the null--vector $u_a^{=}$ till the  {\it moving frame} matrix,
 \begin{eqnarray}\label{Uin}
& U_b^{(a)}= \left({u_b^{=}+ u_b^{\#}\over 2}, u_b^{i}, { u_b^{\#}-u_b^{=}\over 2}
\right)\; \in \; SO(1,10)\;  \quad
\end{eqnarray}
($i=1,...,9$). The statement that this matrix is an  element of the $SO(1,10)$, having been made in (\ref{Uin}),  is tantamount 
to saying that \begin{eqnarray}\label{UTetaU} U^T\eta U=I\; , \quad \eta^{ab}=diag (+1,-1,...,-1)\; , \quad
\end{eqnarray} which in its turn implies that the moving frame vectors obey the following set of constraints \cite{Sok}
\begin{eqnarray}\label{u--u--=0}
u_{ {a}}^{=} u^{ {a}\; =}=0\; , \quad    u_{ {a}}^{=} u^{ {a}\,i}=0\; , \qquad u_{
{a}}^{\; = } u^{ {a} \#}= 2\; , \qquad
 \\  \label{u++u++=0} u_{ {a}}^{\# } u^{ {a} \#
}=0 \; , \qquad
 u_{{a}}^{\;\#} u^{ {a} i}=0\; , \qquad  \\  \label{uiuj=-} u_{ {a}}^{ i}
 u^{{a}j}=-\delta^{ij}.  \quad
\end{eqnarray}
The 11D {\it spinor moving frame variables} (appropriate for our case) can be defined as $16\times 32$ blocks  of the 
$Spin(1,10)$ valued
matrix  \begin{eqnarray}\label{harmVin} V_{(\beta)}^{\;\;\; \alpha}=
\left(\begin{matrix}  v^{+\alpha}_q
 \cr  v^{-\alpha}_q \end{matrix} \right) \in Spin(1,10)\;
 \;  \qquad
\end{eqnarray}
double covering the moving frame matrix (\ref{Uin}). This statement implies that the similarity transformations with 
the matrix $V$ leaves the 11D charge conjugation matrix invariant and, when applied to the 11D Dirac matrices, produce 
the same effect as 11D Lorentz rotation with   matrix $U$,
\begin{eqnarray}\label{VGVT=UG} \label{VCV=C}
VCV^T=C \; ,  \qquad \\
V\Gamma_b V^T =  U_b^{(a)} {\Gamma}_{(a)}\; , \qquad \\ \label{VTGV=UG} V^T
\tilde{\Gamma}^{(a)}  V = \tilde{\Gamma}^{b} U_b^{(a)}  \; . \qquad
\end{eqnarray} The two seemingly mysterious constraints (\ref{Iu--=vGv}) appear as a 16$\times$16 block of the second of 
these relations, (\ref{VGVT=UG}),
and as a component   $V^T \tilde{\Gamma}^{=}  V = \tilde{\Gamma}^{b} u_b^{=}$ of the third one, (\ref{VTGV=UG})  
(with an appropriate representation of the 11D Gamma matrices (see Appendix \ref{appendiceE})). The other blocks/components of these constraints involve 
the second set of constrained spinors,
\begin{eqnarray}\label{M0:v+v+=u++}
 v_{q}^+ {\Gamma}_{ {a}} v_{p}^+ = \; u_{ {a}}^{\# } \delta_{qp}\; , \qquad
 v_{q}^- {\Gamma}_{ {a}} v_{p}^+ = - u_{ {a}}^{i} \gamma^i_{qp}\; , \qquad
\end{eqnarray}
\begin{eqnarray}\label{M0:u++G=v+v+}
 2 v_{q}^{+ {\alpha}}v_{q}^{+}{}^{ {\beta}}= \tilde{\Gamma}^{ {a} {\alpha} {\beta}} u_{
 {a}}^{\# }\; , \quad
 2 v_{q}^{-( {\alpha}}v_{q}^{+}{}^{ {\beta})}=-  \tilde{\Gamma}^{ {a} {\alpha} {\beta}}
 u_{ {a}}^{i}\; . \quad
\end{eqnarray}
Here $\gamma^i_{qp}$ are the 9d Dirac matrices; they are real, symmetric, $\gamma^i_{qp}=\gamma^i_{pq}$,  and obey the
Clifford algebra \begin{eqnarray}\label{gigj+=} \gamma^i\gamma^j + \gamma^j \gamma^i=
2\delta^{ij} I_{16\times 16}\; , \qquad
\end{eqnarray}
as well as the following identities
\begin{eqnarray}\label{gi=id1}
&& \gamma^{i}_{q(p_1}\gamma^{i}_{p_2p_3) }= \delta_{q(p_1}\delta_{p_2p_3) }\; , \qquad
\\ \label{gi=id2} && \gamma^{ij}_{q(q^\prime }\gamma^{i}_{p^\prime)p }+
\gamma^{ij}_{p(q^\prime }\gamma^{i}_{p^\prime)q } = \gamma^{j}_{q^\prime
p^\prime}\delta_{qp}-\delta_{q^\prime p^\prime}\gamma^{j}_{qp} \; .
\end{eqnarray}

Thus $ v_{q}^{- {\alpha}}$ and $ v_{q}^{+{\alpha}}$ can be identified as  square roots
of the light--like vectors $u_{ {a}}^{=}$ and $u_{ {a}}^{\# }$, respectively, while to
construct $u_{ {a}}^{i}$ one needs both these sets of constrained spinors.

The first constraint,  Eq. (\ref{VCV=C}), implies that the inverse spinor moving frame matrix
\begin{eqnarray}\label{Vharm=M0}
 V^{( {\beta})}_{ {\alpha}}= \left(
v_{ {\alpha}q}{}^+\, ,v_{ {\alpha}q}{}^- \right)\; \in \; Spin(1,10) \; ,  \qquad  \\ \nonumber  V_{( {\beta})}{}^{ {\gamma}}
V_{ {\gamma}}^{ ({\alpha})}=\delta_{( {\beta})}{}^{ ({\alpha})}=\left(\begin{matrix}
\delta_{qp} & 0           \cr
          0 & \delta_{qp} \end{matrix}\right) \qquad  \\ \nonumber \Leftrightarrow  \quad \begin{cases} v_{q}^{- {\alpha}}v_{ {\alpha}p}{}^+=\delta_{qp}= v_{q}^{+
{\alpha}}v_{ {\alpha}p}{}^-\, , \cr  v_{q}^{- {\alpha}}v_{ {\alpha}p}{}^-= 0\; =
v_{q}^{+ {\alpha}}v_{ {\alpha}p}{}^+\, , \end{cases}\;
\end{eqnarray}
can be constructed from  $ v_{q}^{\mp {\alpha}}$,
\begin{eqnarray}
\label{V-1=CV}  v_{\alpha}{}^{-}_q =  i C_{\alpha\beta}v_{q}^{- \beta }\, ,
\qquad v_{\alpha}{}^{+}_q = - i C_{\alpha\beta}v_{q}^{+ \beta }\, .
 \end{eqnarray}

\subsection{Cartan forms, differentiation and variation of the (spinor) moving frame
variables }
 \label{DuDv}

To vary the action and to clarify the structure of the equations of motion one needs to vary  and to
differentiate the moving frame and spinor moving frame variables.  As these
are constrained, at the first glance this problem might look complicated, but,
actually, this is not the case. The clear group theoretical structure beyond the moving
frame and spinor moving frame variables makes their differential calculus and
variational problem extremely simple.

Referring again for the details to  \cite{IB07:M0,mM0:PLB}, let us just state that the
derivatives of the moving frame and spinor moving frame variables can be expressed in
terms of the ${so(1,10)}$--valued Cartan forms $\Omega^{(a)(b)}= U^{(a)c}dU_c^{(b)}$
the set of which can be split onto the covariant Cartan forms
\begin{eqnarray}
\label{Om++i=} \Omega^{=i}= u^{=a}du_a^{i}\; , \qquad \Omega^{\# i}=  u^{\#
a}du_a^{i}\; , \qquad
  \end{eqnarray}
providing the basis for the coset ${SO(1,10)\over SO(1,1)\times SO(9)}$, and the forms
\begin{eqnarray}
\label{Om0:=} \Omega^{(0)}= {1\over 4} u^{=a}du_a^{\#}\; , \qquad \\ \label{Omij:=}
\Omega^{ij}=  u^{ia}du_a^{j}\; , \qquad
  \end{eqnarray}
which have the properties of the $SO(1,1)$ and $SO(9)$ connection respectively. These
can be used to define the $SO(1,1)\times SO(9)$ covariant derivative $D$. The covariant
derivative of the moving frame vectors is expressed in terms of the  covariant Cartan
forms (\ref{Om++i=})
\begin{eqnarray}\label{M0:Du--=Om}
Du_{ {b}}{}^{=} &:= & du_{ {b}}{}^{=} +2 \Omega^{(0)} u_{ {b}}{}^{=}= u_{ {b}}{}^i
\Omega^{= i}\; , \qquad \\ \label{M0:Du++=Om} Du_{ {b}}{}^{\#}&:=& du_{ {b}}{}^{\#} -2
\Omega^{(0)} u_{ {b}}{}^{\#}=  u_{ {b}}{}^i \Omega^{\# i}\; , \qquad \\
\label{M0:Dui=Om}  Du_{ {b}}{}^i &:=& du_{ {b}}{}^{i} - \Omega^{ij} u_{ {b}}{}^{j} =
{1\over 2} \, u_{ {b}}{}^{\# } \Omega^{=i}+ {1\over 2} \, u_{ {b}}{}^{=} \Omega^{\#
i}\; . \qquad
\end{eqnarray}
The same is true for the spinor moving frame variables,
\begin{eqnarray}
\label{Dv-q}  Dv_q^{-\alpha}&:=& dv_q^{-\alpha} +  \Omega^{(0)} v_q^{-\alpha} - {1\over
4}\Omega^{ij} \gamma^{ij}_{qp} v_p^{-\alpha} = \nonumber \\ &=& - {1\over 2}
\Omega^{=i} v_p^{+\alpha} \gamma_{pq}^{i}\; , \qquad \\ \label{Dv+q}  Dv_q^{+\alpha}
&:=& dv_q^{+\alpha} -  \Omega^{(0)} v_q^{+\alpha} - {1\over 4}\Omega^{ij}
\gamma^{ij}_{qp} v_p^{+\alpha} = \nonumber \\ &=& - {1\over 2} \Omega^{\# i}
v_p^{-\alpha} \gamma_{pq}^{i}\; . \qquad
\end{eqnarray}

The variation of moving frame and spinor moving frame variables can be obtained from
the above expression for derivatives by a formal contraction with variation symbol,
$i_\delta d=\delta $ (this is to say, by taking the Lie derivatives). The independent
variations are then described by $i_\delta$ contraction of the Cartan forms, $i_\delta
\Omega^{(a)(b)}$. Furthermore, $i_\delta \Omega^{(0)}$ and $i_\delta \Omega^{ij}$ are
the parameters of the $SO(1,1)$ and $SO(9)$ transformations, which are manifest gauge
symmetries of the model. Then the essential variation of the moving frame and spinor
moving frame variables, this is to say, variations which produce (better to say, which
may produce) nontrivial equations of motion, are expressed in terms of $i_\delta
\Omega^{=i}$ and $i_\delta \Omega^{\# i}$,
\begin{eqnarray}\label{vu--=iOm}
\delta u_{ {b}}{}^{=} = u_{ {b}}{}^i i_\delta\Omega^{= i}\; , \qquad \label{vu++=iOm}
\delta u_{ {b}}{}^{\#}=  u_{ {b}}{}^i i_\delta\Omega^{\# i}\; , \qquad \\
\label{vui=iOm}  \delta u_{ {b}}{}^i  = {1\over 2} \, u_{ {b}}{}^{\# }  i_\delta
\Omega^{=i}+ {1\over 2} \, u_{ {b}}{}^{=}  i_\delta\Omega^{\# i}\; . \qquad
\\
\label{vv-q}  \delta v_q^{-\alpha}= - {1\over 2} i_\delta \Omega^{=i} v_p^{+\alpha}
\gamma_{pq}^{i}\; , \qquad \\ \label{vv+q}  \delta v_q^{+\alpha} = - {1\over 2}
i_\delta\Omega^{\# i} v_p^{-\alpha} \gamma_{pq}^{i}\; . \qquad
\end{eqnarray}

\subsection{$K_9$ gauge symmetry of the spinor moving frame action of the M$0$--brane}
\label{secK9}
A simple application of the above formulae begins by
observing that the parameter $i_\delta\Omega^{\# i}$ does not enter the variation of neither $u_a^=$ nor $v^{-\alpha}_q$. 
However, the  M$0$--brane (\ref{SM0=}), (\ref{SM0==}) involves only
these (spinor) moving frame variables. Hence the transformation of the spinor moving frame corresponding to $\tau$ dependent 
parameters $k^{\# i}=i_\delta\Omega^{\# i}$ are gauge symmetries
of this M$0$  action. These so--called $K_9$--symmetry transformations
\begin{eqnarray}\label{vK9}
\delta u_{ {b}}{}^{=} = 0  , \quad
\delta u_{ {b}}{}^{\#}=  u_{ {b}}{}^i k^{\# i}, \quad \delta u_{ {b}}{}^i  =  {1\over 2} \, u_{ {b}}{}^{=}  k^{\# i}\; , \quad
\\
\label{vK9v}  \delta v_q^{-\alpha}= 0 , \qquad \delta v_q^{+\alpha} = - {1\over 2}
k^{\# i} v_p^{-\alpha} \gamma_{pq}^{i}\;  \qquad
\end{eqnarray}
should be taken into account when calculating the number of M$0$ degrees of freedom.

Quite interesting remnants of this K9 symmetry survives in the multiple M$0$ case and will be essential to understand the 
structure of mM$0$ equations of motion.

\subsection{Derivatives and variations of the Cartan forms}
\label{secDOm}

One can easily check that the covariant Cartan forms are covariantly constant,
\begin{eqnarray}\label{M0:DOm--=} D\Omega^{= i}=  0\; , \qquad D\Omega^{\# i}  = 0\;  ,  \qquad
\end{eqnarray}
where the covariant derivatives include the induced connection (\ref{Om0:=}), (\ref{Omij:=}) \footnote{ $D\Omega^{= i}:=d\Omega^{= i} +2 \Omega^{= i}\wedge \Omega^{(0)}- \Omega^{= j}\wedge \Omega^{ji}$, see  (\ref{M0:Du--=Om})--(\ref{M0:Dui=Om}).}.   The curvatures of these connections, 
\begin{eqnarray}\label{M0:Gauss}
 && F^{(0)}:= d\Omega^{(0)} =    {1\over 4 } \Omega^{=\, i} \wedge
 \Omega^{\# \, i}\; , \qquad  \\
\label{M0:Ricci} && {G}^{ij}:= d\Omega^{ij}+ \Omega^{ik} \wedge \Omega^{kj} = - \Omega^{=\,[i} \wedge \Omega^{\# \, j]}\;
,  \qquad
\end{eqnarray}
can be calculated, e.g., from the integrability conditions of Eqs. (\ref{M0:Du--=Om})--(\ref{M0:Dui=Om}),
 \begin{eqnarray}
\label{DDu++=} && DDu_{ {a}}^{\# } = \; \; 2 F^{(0)}u_{ {a}}^{\#
}\; , \qquad
 DDu_{ {a}}{}^{i} = u_{ {a}}^{j} {G}^{ji}  \; . \qquad
\end{eqnarray}
As in the case of moving frame variables (see sec. \ref{DuDv}), the variations of the Cartan forms can be obtained from the 
above expressions using the Lie derivative formula. Omitting the transformations of manifest gauge symmetries SO(1,1) and SO(9) 
(parametrized by $i_\delta \Omega^{(0)}$ and $i_\delta \Omega^{ij}$), we present the essential variations:
\begin{eqnarray}
\label{vOm++=} \delta \Omega^{\# i}&=& D i_\delta \Omega^{\# i}\; , \qquad  \delta \Omega^{=i}= D i_\delta \Omega^{=i}\; , \qquad
 \\
\label{vOmij=} \delta \Omega^{ij}\; &=& \Omega^{=[i} i_\delta \Omega^{\# j]} -  \Omega^{\# [i}i_\delta \Omega^{=j]}\; , \qquad
\\
\label{vOm0=}
\delta \Omega^{(0)}&=& \frac {1}{4} \Omega^{=i}i_\delta \Omega^{\# i} -
 \frac {1}{4}  \Omega^{\# i}i_\delta \Omega^{=i}  \; . \qquad
\end{eqnarray}
These equations will be useful  to vary the multiple M$0$--brane action in Sec. \ref{SecEqmM0}. For deriving the equations of motion 
of single M$0$--brane it is sufficient to use Eqs. (\ref{vu--=iOm}), (\ref{vv-q}) and (\ref{M0:Du--=Om})--(\ref{Dv+q}).

\section{Equations of motion of a single M$0$--brane and induced ${\cal N}=16$ supergravity on the worldline $W^1$}
\label{M0eqs}

The moving frame matrix $U_a^{(b)}$ (\ref{Uin}) provides a `bridge' between the 11D Lorentz group and its $SO(9)\otimes SO(1,1)$ 
subgroup in the sense that it carries one index ($_a$) of $SO(1,10)$ and one index ($^{(b)}$) transformed by a matrix from 
$SO(9)\otimes SO(1,1)$ subgroup of $SO(1,10)$. Contracting the pull--back of the bosonic supervielbein form $\hat{E}^b$ we 
arrive at
 \begin{eqnarray}\label{EU=EEE}
\hat{E}^{(a)}=\hat{E}^b U_b^{(a)}= ( \hat{E}^{=},
\hat{E}^{\#}, \hat{E}^i)
\;   \qquad
\end{eqnarray}
which is split covariantly in three types of one forms. These are inert under $SO(1,10)$ but carry the nontrivial SO(9) vector 
index (in the case of  $\hat{E}^i$) or $SO(1,1)$ weights (in the cases of $\hat{E}^{=}$ and $
\hat{E}^{\#}$). The corresponding decomposition of the vector representation of  $SO(1,10)$ with respect to
its $SO(9)\otimes SO(1,1)$ subgroup,
$$ {\bf 11}\mapsto {\bf 1}_{-2}+  {\bf 1}_{+2}+  {\bf 9}_{0}\; , $$
is even better illustrated by the equation $\hat{E}^{(a)}U_{(a)}{}^b=\hat{E}^b $ which, in more detail, reads
\begin{eqnarray}\label{E=E+E+E}
\hat{E}^{a}= {1\over 2} \hat{E}^{=} u^{a\#}+
{1\over 2}\hat{E}^{\#}u^{a=} - \hat{E}^i u^{ai}
\; .  \qquad
\end{eqnarray}
 Thus the  moving frame vectors help to split the pull--back of the supervielbein in a
Lorentz covariant manner. The $SO(9)$ singlet one form with $SO(1,1)$ weight -2,  $\hat{E}^{=}=
\hat{E}^b u_b^{=}$ enters the action (\ref{SM0=}) multiplied by the weight +2 worldline scalar field $\rho^{\#}(\tau)$. This 
clearly has the meaning of the Lagrange multiplier: its variation results in vanishing of $\hat{E}^{=}$,
\begin{eqnarray}\label{E==0}
\hat{E}^{=}:= \hat{E}^a u_{ {a}}^{=}=0\; .  \qquad
\end{eqnarray}
Now, the variation of $\hat{E}^{=}$ contain two different contributions, $\delta \hat{E}^{=}= \delta\hat{E}^a u_a^{=}+ \hat{E}^a \delta u_a^{=}$. The first comes from the variation of the  pull--back of the bosonic supervielbein form which in our case of flat target superspace can be easily calculated with the result
\begin{eqnarray}\label{vhEa=}
\delta\hat{E}^a = - i d\hat{\theta}\Gamma^a\delta\hat{\theta}+ d(\delta \hat{x}^a -i \delta\hat{\theta} \Gamma^a\hat{\theta}) \; .  \qquad
\end{eqnarray}
The second term contains the variation of the light--like vector $u_a^{=}$ which can be written as in Eq. 
(\ref{vu--=iOm}), $\delta u_a^{=}= u_a^{i}\, i_\delta \Omega^{=i}$ with an arbitrary $i_\delta \Omega^{=i}$. The corresponding 
variation of the action (\ref{SM0=}) reads  $\delta_u S_{M0}= \int_{W^1} \rho^{\#}\,
\delta u_a^{=}  \, \hat{E}^a = \int_{W^1} \rho^{\#}\,
u_a^{i}  \, \hat{E}^a i_\delta \Omega^{=i} $ and produce the equation of motion
\begin{eqnarray}\label{Ei=0}
\hat{E}^{i}:=
\hat{E}^a u_{ {a}}^{i}=0\;  .  \qquad
\end{eqnarray}
 Using Eq. (\ref{E=E+E+E}) one can collect Eqs. (\ref{E==0}) and (\ref{Ei=0}) in
\begin{eqnarray}\label{E==0=Ei}
\hat{E}^{a}:= {1\over 2}\hat{E}^{\#} u^{a=} \; .  \qquad
\end{eqnarray}
This equation shows that the M$0$--brane worldline $W^1$ is a light--like line in target
(super)space, as it should be for the massless superparticle.

Furthermore (\ref{E==0=Ei}) suggests to consider $\hat{E}^{\#}$ as einbein on the worldline $W^1$; this composite einbein is 
induced by embedding of $W^1$ into the target superspace. The transformation of  $\hat{E}^{\#}$ under the irreducible  
$\kappa$--symmetry (\ref{kap=irr}) is given by
$\delta_\kappa\hat{E}^{\#}= -2i \hat{E}^{+q}\epsilon^{+q}$. In the light of the identification of $\kappa$--symmetry with local 
worldline supersymmetry \cite{stv}, this equation suggests
to consider the covariant {\bf 16}$_{+}$ projection, $ \hat{E}^{+q}=
\hat{E}^{\alpha}v_\alpha^{+q}$, of the pull--back of the fermionic 1--form $E^\alpha$ as induced `gravitino' companion of the 
induced 1d `graviton'  $\hat{E}^{\#}$. Indeed under the $\kappa$--symmetry  (\ref{kap=irr}) this set of forms  show the typical 
transformations rules of
(1d ${\cal N}=16$) supergravity multiplet,
\begin{eqnarray}\label{v1dSG=}
\delta_\kappa \hat{E}^{+q}= D \epsilon^{+q}(\tau) \; ,  \qquad \delta_\kappa
\hat{E}^{\#}= -2i \hat{E}^{+q}\epsilon^{+q}\; .
\end{eqnarray}
Here $D=d\tau D_\tau$ is the $SO(1,1)\times SO(9)$ covariant derivative which we will
specify below. The connection in this covariant derivative are defined in terms of moving frame variables and, hence, are inert 
under the $\kappa$--symmetry; in this sense the induced 1d ${\cal N}=16$ supergravity multiplet is described essentially by 1 
bosonic and 16 fermionic 1--forms $\hat{E}^{\#}$ and $\hat{E}^{+q}$.
Our  action for the mM$0$ system, which we present in the next section,
will contain the coupling of these induced 1d supergravity to the matter describing the
relative motion of the mM$0$ constituents.

The other, {\bf 16}$_-$ projection $\hat{E}^{-q}=  \hat{E}^{\alpha}v_\alpha^{-q}$ of the pull--back of fermionic supervielbein form 
to $W^1$  vanishes on the mass shell,
\begin{eqnarray}\label{E-q=0}
\hat{E}^{-q}:=  \hat{E}^{\alpha}v_\alpha^{-q}=0  \; .  \qquad
\end{eqnarray}
Indeed, varying the coordinate functions in the action (\ref{SM0=}) we arrive at equation ${\delta S_{M0}\over \delta \hat{Z}^M}=0$  which reads
\begin{eqnarray}
\label{vhZ=>}\partial_\tau(\rho^{\#} u_a^{=}E_M^a(\hat{Z}))=0\; .
\end{eqnarray}
In our case of flat target superspace
$E_M^a(\hat{Z})= \delta_M^a -i \delta_M^\alpha (\Gamma^a\hat{\theta})_\alpha $ and one can easily split  (\ref{vhZ=>}) into the 
bosonic vector and fermionic spinor equations (which we prefer to write with the use of $d=d\tau \partial_\tau$)
\begin{eqnarray}
\label{vhx=>} d( \rho^{\#} u_a^{=})=0\; , \\
\label{vhth=>} \rho^{\#} u_a^{=}( \Gamma^a\partial_\tau\hat{\theta} )_\alpha =0\; .
\end{eqnarray}

Using (\ref{Iu--=vGv}b) and assuming $\rho^{\#}\not= 0$ we find that (\ref{vhth=>}) is equivalent to  Eq. (\ref{E-q=0}). This 
implies that the $d\hat{\theta}^\alpha$ can be expressed through the induced gravitino,
\begin{eqnarray}\label{Ef=E+v-}
\hat{E}^{\alpha}=d\hat{\theta}^\alpha = \hat{E}^{+q} v_q^{-\alpha} \; .  \qquad
\end{eqnarray}

Let us come back to the equation for the bosonic coordinate functions, (\ref{vhx=>}) 
(or equivalently, $\partial_\tau( \rho^{\#} u_a^{=})=0$).
Using  (\ref{M0:Du--=Om}) we can write this in the form $0= D\rho^{\#} \, u_a^{=} +
\rho^{\#}  u_a^i\Omega^{=i}$. Here and below we use the covariant derivatives defined in (\ref{M0:Du--=Om}), 
(\ref{M0:Du++=Om}), (\ref{M0:Dui=Om})). Contracting that equation with  $ u^{a\#}$ gives us
\begin{eqnarray}\label{M0:Drho=0}
 D\rho^{\#}=0 \; ,
\end{eqnarray}
while  the nontrivial part of the bosonic equations of motion of a single M$0$--brane,
which can be read off from the coefficient for $u_a^i$, states that the covariant Cartan
form $\Omega^{= i}$ vanishes,
\begin{eqnarray}
\label{Om--i=0} \Omega^{= i}=0 \; . \qquad
\end{eqnarray}
Coming back to Eq.  (\ref{M0:Du--=Om}), we see that Eq. (\ref{Om--i=0}) can be
expressed by stating that the covariant derivative of the light--like vector $u_a^=$
vanishes,
\begin{eqnarray}
\label{Du--=0} Du_a^{=}=0 \; ,  \qquad
\end{eqnarray}
or, equivalently, by
\begin{eqnarray}
\label{Dv-q=0} Dv_q^{-\alpha}=0 \; .  \qquad
\end{eqnarray}

On the other hand, using \begin{eqnarray}
\label{D++E++i=0}
D=d\tau D_\tau= \hat{E}^\# D_\# \; , \qquad
\end{eqnarray}  we can write Eq. (\ref{Du--=0}) in the
form $D_\# u_a^{=}=0$, and, as far as (\ref{E==0=Ei}) implies $u_a^{=}=2 \hat{E}_\#^a$,
in the following more standard form
\begin{eqnarray}
\label{D++E++i=0} D_\# \hat{E}_\#^a =0 \; ,  \qquad
\end{eqnarray}
or, in more detail,
\begin{eqnarray}
\label{D++D++x=0} D_\# D_\# \hat{x}^a=  i D_\#(D_\# \hat{\theta}\Gamma^a\hat{\theta})\;
.  \qquad
\end{eqnarray}

Two more observations will be useful below. The first is that Eq. (\ref{M0:Drho=0}), 
$0=D\rho^{\#}=d\rho^{\#}-2\rho^{\#}\Omega^{(0)}$, can be solved with respect to the induced $SO(1,1)$ connection,
\begin{eqnarray}
\label{M0:Om0=}
\Omega^{(0)} = {d\rho^{\#}\over 2\rho^{\#}} \; .  \qquad
\end{eqnarray}
Notice that this is in agreement with the statement that one can always gauge away any 1d connection: using the local SO(1,1) 
symmetry to fix the gauge $\rho^{\#}=const$ we arrive at $\Omega^{(0)} =0$.

The second comment concerns the supersymmetric pure bosonic solutions of the above equations of motion.

\subsection{All supersymmetric solutions of the M$0$ equations describe 1/2 BPS
states}
\label{susySOL}

As far as the fermionic coordinate function $\hat{\theta}^\alpha$ is transformed by both spacetime supersymmetry and by the 
worldline supersymmetry ($\kappa$--symmetry), $\delta \hat{\theta}^\alpha=
-{\varepsilon}^\alpha + \epsilon^{+q}(\tau) v^{-\alpha}_q(\tau)$,
the purely bosonic solutions of the M$0$ equations, having
\begin{eqnarray}\label{hth=0}
 \hat{\theta}^\alpha = 0 \; ,
\end{eqnarray}
may preserve  a part of target space supersymmetry. This is characterized by parameter
\begin{eqnarray}\label{vep=epv}
 {\varepsilon}^\alpha = \epsilon^{+q}(\tau) v^{-\alpha}_q(\tau) \; .
\end{eqnarray}
The left hands side of this equation contains a constant fermionic spinor
$d{\varepsilon}^\alpha=0$, so that $d(\epsilon^{+q} v^{-\alpha}_q)=D\epsilon^{+q}
v^{-\alpha}_q+ \epsilon^{+q} Dv^{-\alpha}_q=0$. Furthermore, taking into account that
the equations of motion for the bosonic coordinate function, Eq.  (\ref{D++D++x=0}),
implies  (\ref{Dv-q=0}), one finds that the consistency of (\ref{vep=epv}) is the
covariant constancy of the $\kappa$--symmetry parameter $\epsilon^{+q}(\tau)$,
\begin{eqnarray}\label{Dep+q=0}
 D\epsilon^{+q}=0
\; .
\end{eqnarray}
In 1d system the connection can be gauged away so that this condition can be reduced to the
existence of a constant $SO(9)$ spinor  $\epsilon^q$. For instance gauging away the
$SO(9)$ connection and using Eq. (\ref{M0:Om0=}), we can present (\ref{Dep+q=0}) in
the form  $d(\epsilon^{+q}/\sqrt{\rho^{\#}})=0$ and solve it by $\epsilon^{+q}=
 \sqrt{\rho^{\#}}\, \epsilon^q$ with $d\epsilon^q=0$.

This implies that {\it any purely bosonic solution of the M$0$ equations preserves
exactly $1/2$ of the spacetime supersymmetry}.

\section{Covariant action for multiple M0--brane system}
\label{mM0secAC}

\subsection{Variables describing the mM0 system}

Let us introduce the dynamical variables describing the system of multiple M$0$--branes, which we abbreviate as mM$0$. 
Its dimensional reduction  is expected to produce the system of N nearly coincident D$0$-branes (mD$0$ system) and at very 
low energy this later is described by the action of 1d ${\cal N}=16$ supersymmetric Yang--Mills theory (SYM) with the gauge 
group $U(N)$, which is given by dimensional reduction of the 10D ${\cal N}=1$ U(N) SYM down to d=1. Now,  the set of fields 
of the U(N) SYM  can be split onto the non-Abelian SU(N) SYM and Abelian U(1) SYM multiplets. Roughly speaking, this later 
describes the center of energy motion of the mD$0$ system while the former corresponds to the relative motion of the constituents 
of the mD$0$ system. Then it is natural to assume that the relative motion of the mM$0$ constituents are also described by the 
fields of $SU(N)$ SYM multiplet.

Now let us turn to the center of energy motion. We begin by noticing that the $U(1)$ SYM  fields can be seen in the single 
D$0$ brane action (see \cite{Eric+Paul:1996} and refs therein) after fixing the gauge with respect to $\kappa$--symmetry and 
reparametrization symmetry. Originally the action of a single D$0$ brane is written in terms of 10 bosonic and $32$ fermionic 
coordinate functions, worldline fields  corresponding to the coordinates of type IIA $D=10$ superspace. The above gauge fixing 
reduces the number of fermionic fields to $16$ and the number of bosonic coordinate functions to $9$. These are the same as the 
number of physical fields as in 1d reduction of the 10D SYM theory. This also contains the time component of the gauge field which 
can be gauged away by the U(1) gauge symmetry transformation  and do not carry degrees of freedom. The U(1) SYM multiplet 
describing the center of energy motion of the mD$0$ system can be obtained  by fixing the gauge with respect to  $\kappa$-symmetry 
and reparametrization symmetry on the coordinate functions, the same as in the case of single D$0$ brane.

In the light of the above discussion, it is natural to describe the center of energy motion of the mM$0$ system by the 11 bosonic 
and 32 fermionic  coordinate functions, the same  as used to describe the motion of single M$0$--brane, and to assume that the 
wanted  mM$0$  action possesses $\kappa$--symmetry and reparametrization symmetry, like the single M$0$--brane action does.

To resume, following \cite{mM0:PLB,mM0:PRL,mM0:action} we will describe the center of energy motion of $N$ nearly coincident 
M$0$--branes (mM$0$ system) by the 11 commuting and 32 anti-commuting coordinate functions
\begin{eqnarray}\label{hZ=hx+}
\hat{Z}{}^M(\tau)&=& (\hat{x}{}^\mu (\tau), \hat{\theta}{}^\alpha (\tau))\; , \quad \\ \nonumber && \mu =0,1,...,10; \quad
\alpha = 1,2,..., 32
\end{eqnarray}
(the same as used to describe single M$0$--brane), and the  relative motion of the mM$0$ constituents by the fields of the 
$SU(N)$ SYM supermultiplet. These are the bosonic and fermionic Hermitian traceless $N\times N$ matrices
fields   \begin{eqnarray}\label{NxNXi}
&&{\mathbb X}^i(\tau)\quad  and  \quad   \Psi_q (\tau)
\\ \nonumber && (i=1,...,9\, , \qquad q=1,...,16)
\end{eqnarray}
 depending on a (center of energy)
proper time variable $\tau$. The bosonic ${\mathbb  X}^i(\tau)$ carries the index
$i=1,...,9$ of the vector representation of $SO(9)$, while the fermionic $\Psi_q$
transforms as a spinor under $SO(9)$,  $q=1,...,16$.

\subsection{First order form of the 1d ${\cal N}=16$ SYM Lagrangian as a starting point to build mM$0$ action}

The standard 1d ${\cal N}=16$ SYM Lagrangian (obtained by dimensional reduction of 10D SYM) can be written in the following first 
order form
\begin{eqnarray}
\label{LSYM=1}
d\tau L_{SYM}= tr\left(- {\mathbb P}^i \nabla_\tau {\mathbb X}^i + 4i { \Psi}_q \nabla_\tau
{\Psi}_q  \right) +   d\tau {\cal H}  \;   \qquad
  \end{eqnarray}
where the Hamiltonian
\begin{eqnarray}
\label{HSYM=1}
{\cal H}=   {1\over 2} tr\left( {\mathbb P}^i {\mathbb P}^i \right) + {\cal V} ({\mathbb X}) - 2\,  tr\left({\mathbb X}^i\, \Psi\gamma^i {\Psi}\right) \;   \qquad
  \end{eqnarray}
contains the positively definite scalar potential
\begin{eqnarray}
\label{VSYM=} {\cal V} = - {1\over 64}
tr\left[ {\mathbb X}^i ,{\mathbb X}^j \right]^2 \equiv  +{1\over 64}  tr\left[ {\mathbb X}^i
,{\mathbb X}^j \right] \cdot \left[ {\mathbb X}^i ,{\mathbb X}^j \right]^\dagger \; ,
  \end{eqnarray}
Eqs. (\ref{LSYM=1}) and (\ref{HSYM=1}) involve the  auxiliary `momentum' fields, the nanoplet of traceless $N\times N$ matrices ${\mathbb P}^i$, 
and also the gauge field ${\mathbb A}_\tau (\tau)$ which enters the  covariant derivatives $\nabla=d\tau \nabla_\tau $ of the 
above bosonic and fermionic fields,
\begin{eqnarray}
\label{SYMDX=}
 \nabla {\mathbb  X}^i= d{\mathbb X}^i+ [{\mathbb A},    {\mathbb X}^i]\; , \qquad \nabla {\Psi}_q= d{\Psi}_q+ [{\mathbb A},   {\Psi}_q]\; .
  \end{eqnarray}
The action with the above Lagrangian are invariant under the following
d=1 ${\cal N}=16$  supersymmetry transformations with constant fermionic parameter $\varepsilon^q$
\begin{eqnarray}
\label{SYMsusy-X} \delta_\varepsilon {\mathbb X}^i   = 4i \varepsilon^q (\gamma^i  \Psi)_q \; , \quad
\delta_\varepsilon {\mathbb P}^i   = [\varepsilon^{q} (\gamma^i  \Psi)_q,  {\mathbb X}^j]\; ,\qquad \\
\label{SYMsusy-Psi} \delta_\varepsilon \Psi_q =  {1\over 2} \varepsilon^{p} \gamma^i_{pq}  {\mathbb
P}^i-  {i\over 16} \epsilon^{p} \gamma^{ij}_{pq}  [{\mathbb X}^i, {\mathbb X}^j]\; ,\qquad \\
\label{SYMsusy-A}
 \delta_\varepsilon {\mathbb A}  = - d\tau   \varepsilon^{q}  \Psi_q
 \; .  \qquad
\end{eqnarray}

The mM$0$ action should describe the coupling of the above SYM theory to the center of energy variables (\ref{hZ=hx+}). 
As we have discussed above, such an action should possess  the reparametrization symmetry and a 16 parametric local fermionic 
symmetry, a counterpart of the irreducible $\kappa$--symmetry (\ref{kap=irr}) of the single M$0$ action. It is natural also to 
think on this fermionic gauge symmetry as on the local version of the above rigid d=1 ${\cal N}=16$  supersymmetry of the 
SYM action,  Eqs. (\ref{SYMsusy-X})--(\ref{SYMsusy-A}).

\subsection{Induced supergravity on the center of energy worldline}

The natural way to make a supersymmetry local is to couple it to supergravity multiplet. As a by--product such a coupling 
should guaranty the reparametrization (general coordinate) invariance. Now it is the time to recall about induced supergravity 
multiplet on the worldline of the single M$0$--brane constructed in sec. \ref{M0eqs}.
Similarly, we can associate a moving frame (\ref{Uin}) and spinor moving frame (\ref{harmVin}) to the center of energy motion 
of the mM$0$ system and use these together with center of energy coordinate functions (\ref{hZ=hx+}) to build the composite 
d=1 ${\cal N}=16$ supergravity multiplet including the 1d `graviton' and `gravitino'
\begin{eqnarray}
\label{E++=Eu++}
\hat{E}^{\#}&=& \hat{E}^{a}u_a^{\#}= (d\hat{x}^{a}- id\hat{\theta}\Gamma^a\hat{\theta}) u_a^{\#}\; , \qquad
\\ \label{E+q=Ev+q} \hat{E}^{+q}&=& \hat{E}^\alpha v_\alpha^{+q}=d\hat{\theta}^\alpha v_\alpha^{+q}
\; , \qquad
  \end{eqnarray}
transforming under the local supersymmetry as in (\ref{v1dSG=}),
\begin{eqnarray}\label{vcSG=}
\delta_\epsilon \hat{E}^{+q}= D \epsilon^{+q}(\tau) \; ,  \qquad \delta_\epsilon
\hat{E}^{\#}= -2i \hat{E}^{+q}\epsilon^{+q}\; .
\end{eqnarray}

Notice that the use of such a composite supergravity induced by  embedding of the center of energy worldline into the flat target 
11D superspace implies  that the local supersymmetry parameter carries the weight +1
of the $SO(1,1)$ group transformations defined on the moving frame variables. This implies the necessity to adjust the $SO(1,1)$ 
weight also to the fields describing the relative motion of the mM$0$ constituents. Following \cite{mM0:PLB,mM0:PRL,mM0:action} we 
define the  $SO(1,1)$ weight of the bosonic and fermionic fields to be
 -2 and -3, respectively, so that in a more explicit notation (and using the conventions were the upper $^- $ index indicate the 
same -1 weight as the lower $_+ $ one)
 \footnote{Such a chose of weight of the basic matrix fields is preferable for the
description in the frame of superembedding approach, like developed in \cite{mM0:PLB,mM0:PRL}. Once using the density 
$\rho^{\#}=\rho^{++}$ which enters the spinor moving frame action for single  M$0$, we can easily change the weights of 
the fields  multiplying them  by corresponding power of  $\rho^{\#}$. However we find more convenient  to work with the 
`weighted' fields (\ref{bXi=bXi++}), (\ref{Psi=Psi+++}). }
\begin{eqnarray}
\label{bXi=bXi++} && {\mathbb  X}^i= {\mathbb  X}_{\#}^i:= {\mathbb  X}_{++}^i\, ,\qquad i=1,...,9\, , \qquad \\
\label{Psi=Psi+++} && {{ \Psi}}_q= \Psi_{\# \,+q}:= \Psi_{++ \,+q}= \Psi_{\#}{}_q^-\,  ,\quad q=1,...,16\, .
\qquad
 \end{eqnarray}

As in the case of single M$0$--brane, we expect the $SO(1,1)$ as well as $SO(9)$ transforation to be a gauge symmetry of our 
action. This implies the use of covariant derivative with   $SO(1,1)$ and $SO(9)$
connection. As in the case of single M$0$--brane, we define these connections to be constructed from the moving frame 
variables
\begin{eqnarray}
\label{mM0:Om0=} \Omega^{(0)}= {1\over 4} u^{=a}du_a^{\#}\; , \qquad \Omega^{ij}=
u^{ia}du_a^{j}\;  \qquad
  \end{eqnarray}
(see Eqs.  (\ref{Om0:=}) and (\ref{Omij:=})), which are now associated to the  center of energy motion of the mM$0$
system. The covariant derivatives of the su(N) valued matrix fields (\ref{bXi=bXi++}) are defined by
\begin{eqnarray}
\label{DXi=} D{\mathbb X}^i  &:=& d{\mathbb X}^i  + 2\Omega^{(0)} {\mathbb X}^i  - \Omega^{ij} {\mathbb
X}^j+ [{\mathbb A},    {\mathbb X}^i] \; , \qquad \\ \label{DPsi:=} D\Psi_q  &:=& d\Psi_q
 + 3\Omega^{(0)} \Psi_q   -{1\over 4} \Omega^{ij} \gamma^{ij}_{qp} {\Psi}_p+ [{\mathbb A},
 \Psi_q ] \; . \qquad
  \end{eqnarray}
They also involve the $SU(N)$ connection  ${\mathbb A}= d\tau {\mathbb A}_\tau (\tau)$  on the center of
energy worldline $W^1$. The anti-Hermitian traceless  $N\times N$ matrix gauge field ${\mathbb A}_\tau (\tau)$ is an 
independent variable of our model. Let us stress, however, that, as any 1d gauge field, it can be gauged away and thus 
does not carry any degree of freedom.

The covariant derivative of the supersymmetry parameter in (\ref{vcSG=}) reads
\begin{eqnarray}
\label{Dep+q=}
D\epsilon^{+q}= d\epsilon^{+q}- \Omega^{(0)} \epsilon^{+q}  +{1\over 4} \Omega^{ij} \epsilon^{+p}  \gamma^{ij}_{pq} \; , \qquad
\end{eqnarray}
so that the induced connection (\ref{mM0:Om0=}) are also the members of the composite d=1 ${\cal N}=16$ supergravity multiplet.

\subsection{A way towards mM$0$ action }

Now we are ready to present the action for the system of N nearly coincident  M$0$--branes (mM$0$ system) which was proposed 
in \cite{mM0:action}.
It can be considered as a result of  `gauging' of rigid  d=1 ${\cal N}=16$ supersymmetry (\ref{SYMsusy-X})--(\ref{SYMsusy-A}) of 
the $SU(N)$ SYM action with the Lagrangian (\ref{LSYM=1}) achieved by coupling it to a composite d=1 ${\cal N}=16$ supergravity 
(\ref{E++=Eu++}), (\ref{E+q=Ev+q}), (\ref{mM0:Om0=}) induced by embedding of the center of energy worldline of the mM$0$ system 
into the target 11D superspace.

The natural first step on this way is to make the Lagrangian (\ref{LSYM=1}) covariant by coupling it to a 1d gravity. 
This can be  reached by just replacing $d\tau$ in the right hand side of  (\ref{LSYM=1}) by  the 1-form $\hat{E}^{\#}$ of 
(\ref{E++=Eu++}). Then, to provide also the $SO(1,1)$ and $SO(9)$ gauge symmetries, which play  the role of Lorentz and 
R-symmetries in our induced 1d ${\cal N}=16$ supergravity, we should replace the Yang--Mills covariant derivatives in 
(\ref{SYMDX=}) by the $SO(1,1)\times SO(9)$ covariant derivatives defined in (\ref{DXi=}), (\ref{DPsi:=}), and to multiply 
the Lagrangian 1-form thus obtained by $(\rho^{\#})^3$. The next stage is suggested by the fact that setting N=1 in the action 
for the system of N nearly coincident M$0$--brane one should arrive a single M$0$--brane action.  As the SU(N) SYM Lagrangian, 
and all the matrix fields involved in it, vanish when N=1, this implies the necessity just to add the single M$0$ action to the 
integral of the above described Lagrangian form. Then the coupling to induced gravitino can be restored from the requirement of 
local supersymmetry invariance of the mM$0$ action.

\subsection{mM$0$ action}

In such a way we arrive at the mM$0$ action proposed in
\cite{mM0:action}. It reads
\begin{eqnarray}
\label{SmM0=} && S_{mM0} = \int_{W^1} \rho^{\#}\, \hat{E}^{=} + \qquad \nonumber  \\ &&
+ \int_{W^1} (\rho^{\#})^3\, \left(  tr\left(- {\mathbb P}^i D {\mathbb X}^i + 4i { \Psi}_q D
{\Psi}_q  \right) + \hat{E}^{\#} {\cal H} \right)+ \quad \nonumber \\ &&  +  \int_{W^1}
(\rho^{\#})^3\, \hat{E}^{+q}  tr\left(4i (\gamma^i {\Psi})_q  {\mathbb P}^i + {1\over 2}
(\gamma^{ij} {\Psi})_q  [{\mathbb X}^i, {\mathbb X}^j]  \right) , \; \nonumber \\ && {}
  \end{eqnarray}
where ${\cal H}$ is the relative motion Hamiltonian ({\it cf.} (\ref{HSYM=1}))
\begin{eqnarray}
\label{HmM0=} {\cal H} &:=& {\cal H}_{\#\#\#\#}({\mathbb X}, {\mathbb P}, \Psi ) = \qquad
\nonumber \\ &=& {1\over 2} tr\left( {\mathbb P}^i {\mathbb P}^i \right) + {\cal V} ({\mathbb X}) - 2\,  tr\left({\mathbb X}^i\, \Psi\gamma^i {\Psi}\right)  \qquad
  \end{eqnarray}
including  the scalar  potential ({\it cf.} (\ref{VSYM=}))
\begin{eqnarray}
\label{VmM0=} {\cal V} &:=& {\cal V}_{\#\#\#\#} ({\mathbb X} ) = - {1\over 64}
tr\left[ {\mathbb X}^i ,{\mathbb X}^j \right]^2 \qquad \\ &=& +{1\over 64}  tr\left[ {\mathbb X}^i
,{\mathbb X}^j \right] \cdot \left[ {\mathbb X}^i ,{\mathbb X}^j \right]^\dagger \; ,
  \end{eqnarray}
and  the Yukawa--type coupling  $tr\left({\mathbb X}^i\, \Psi\gamma^i {\Psi}\right)$.

The covariant derivatives $D$ are defined in (\ref{DXi=}), (\ref{DPsi:=}).
Their connection are build from the (spinor) moving frame variables, Eq.  (\ref{mM0:Om0=}), which are related to the center of 
energy motion of the mM$0$ system.  These  are also used to construct the composite graviton and gravitino 1-forms 
$\hat{E}^{\#}$ and $\hat{E}^{+q}$, Eqs.  (\ref{E++=Eu++}), (\ref{E+q=Ev+q}). The 1-form $\hat{E}^{=}$ is the same as in the 
case of single M$0$--brane
\begin{eqnarray}
\label{E--=Eu--} \hat{E}^{=}=\hat{E}^{a}u_a^{=} \, . \qquad
  \end{eqnarray}
For the completeness of this section, let us recall that in these equations  $\hat{E}^{a}$ is the pull--back of the bosonic 
supervielbein to the center of energy worldline $W^1$, Eq. (\ref{hEa=}), (\ref{Ea=Pi}), $\hat{E}^\alpha
=d\hat{\theta}^\alpha(\tau)$, $u_a^{=}$ and $u_a^{\#}$ are light--like moving frame
vectors  (\ref{Uin}), (\ref{u--u--=0}), (\ref{u++u++=0}), and $v_\alpha^{+q}$ is an
element of spinor moving frame  (\ref{harmVin}).

Although the first term in (\ref{SmM0=}) coincides with the  single  M$0$--brane action
(\ref{SM0=}), now the Lagrange multiplier $\rho^{\#}$ and spinor moving frame variables are also present in
the second and third terms. This results in that their equations of motion differ from
(\ref{E==0=Ei}), and, as we discuss in the next section, generically, the center of energy motion of the mM0 system is not 
light-like.

\subsection{Local supersymmetry of the mM$0$ action}

The action (\ref{SmM0=}) is invariant under the transformation of the 16 parametric local worldline supersymmetry 
\begin{eqnarray}
\label{susy-th} \delta_\epsilon \hat{\theta}^\alpha &=& \epsilon^{+q} (\tau)
v_q^{-\alpha} \; , \quad
 \\
\label{susy-x} \delta_\epsilon \hat{x}^a &=& - i \hat{\theta} \Gamma^a\delta_\epsilon
\hat{\theta} +   {1\over 2}   u^{a\#} i_\epsilon \hat{E}^{=}\; , \qquad
\\
\label{susy-rho}   \delta_\epsilon \rho^{\#} &=& 0\; , \qquad \\ \label{susy-v}
 \delta_\epsilon  v_q^{\pm\alpha}&=&0
  \;   \Rightarrow  \quad  \delta_\epsilon
  u_a^{=}= \delta_\epsilon u_a^{\#}= \delta_\epsilon u_a^{i}=0\;
 ,  \qquad
\\
\label{susy-X}  \delta_\epsilon {\mathbb X}^i   &=& 4i \epsilon^{+} \gamma^i  \Psi \; , \quad
\delta_\epsilon {\mathbb P}^i   = [(\epsilon^{+} \gamma^i  \Psi),  {\mathbb X}^j]\; ,\qquad \\
\label{susy-Psi}  \delta_\epsilon \Psi_q &=&  {1\over 2} (\epsilon^{+} \gamma^i)_q  {\mathbb
P}^i-  {i\over 16} (\epsilon^{+} \gamma^{ij})_q  [{\mathbb X}^i, {\mathbb X}^j]\; ,\qquad \\
\label{susy-A}
 && \delta_\epsilon {\mathbb A} = -  \hat{E}^{\#} \epsilon^{+q}  \Psi_q + \hat{E}^{+}\gamma^i
 \epsilon^{+}\;    {\mathbb X}^i
 \; ,  \qquad
\end{eqnarray}
where
\begin{eqnarray}
\label{iE==}
& i_\epsilon \hat{E}^{=}= 6 (\rho^{\#})^2 tr \left(i {\mathbb P}^i\epsilon^{+}
 \gamma^{i}\Psi
 -  {1\over 8} \epsilon^{+} \gamma^{ij}\Psi  [{\mathbb X}^i, {\mathbb X}^j] \right) .   \qquad
 \end{eqnarray}

The local supersymmetry transformations of the  fields describing relative motion of mM$0$ constituents, (\ref{susy-X}), 
(\ref{susy-Psi})  coincide with the SYM supersymmetry (\ref{SYMsusy-X}), (\ref{SYMsusy-Psi}) modulo the fact that now the 
fermionic parameter is an arbitrary function of the center of energy proper time,  $\epsilon^{+q} =\epsilon^{+q}(\tau)$. 
The local supersymmetry transformation of the 1d $SU(N)$ gauge field (\ref{susy-A}) differs from the SYM transformation by 
additional term involving the composite gravitino.

The transformations of the center of energy variables Eqs. (\ref{susy-th})--(\ref{susy-v}) describe a deformation of the 
irreducible  $\kappa$--symmetry (\ref{kap=irr}) of the free massless superparticle. Actually, the
deformation touches the  transformation rule (\ref{susy-x})  for the the bosonic
coordinate function,  $\delta_\epsilon \hat{x}^a$ only. The Lagrange multiplier $\rho^{\#}$ and the (spinor) moving 
frame variables are invariant under the supersymmetry, like they are under the $\kappa$-symmetry of a single superparticle.

\section{${\bf m}$M0  equations of motion}
\label{SecEqmM0}

In this section we present and study the complete set of equations of motion for the
multiple M$0$--brane system which follow from the action (\ref{SmM0=}).

\subsection{Equations of the relative motion}

Varying the action with respect to the momentum matrix field $\mathbb{P}^{i}$ gives us
the equation
\begin{eqnarray}\label{DXi=EPi}
&& D\mathbb{X}^i =\hat{E}^{\#}\mathbb{P}^i+4i\hat{E}^{+q}(\gamma^i\Psi)_q
 \qquad
\end{eqnarray}
which allows to identify $\mathbb{P}^{i}$, modulo fermionic contribution, with the
covariant time derivative of $\mathbb{X}^{i}$,
\begin{eqnarray}\label{Pi=DXi}
 \mathbb{P}^i=
D_\# \mathbb{X}^i - 4i\hat{E}_\#^{+}\gamma^i\Psi \; . \qquad
\end{eqnarray}
Here
\begin{eqnarray}\label{D++=d/E}
D_\# = {1\over \hat{E}_\tau^\#}D_\tau\; , \qquad  \hat{E}_\#^{+q} = {1\over
\hat{E}_\tau^\#}\hat{E}^{+q}_\tau\, , \qquad
\end{eqnarray}
are covariant derivative and the induced 1d gravitino field  corresponding to the induced einbein on the worldvolume, 
$\hat{E}^\#=\hat{E}^au_a^\#=: d\tau \hat{E}^\#_\tau $, in the sense of that
\begin{eqnarray}\label{Pi=DXi}
D= \hat{E}^\# D_\# \; , \qquad \hat{E}^{+q}= \hat{E}^\# \hat{E}_\#^{+q}\; . \qquad
\end{eqnarray}
The variation with respect to the worldline gauge field ${\mathbb A}=d\tau {\mathbb A}_\tau $
gives
\begin{eqnarray}\label{Gauss}
[\mathbb{P}^i,\mathbb{X}^i]= 4i\{ \Psi_q \, , \, \Psi_q \}\, \qquad
\end{eqnarray}
and the variation with respect to ${\mathbb X}^i$ results in
\begin{eqnarray}\label{DPi=}
 D\mathbb{P}^i&=& -
\frac{1}{16}\hat{E}^{\#}[[\mathbb{X}^i,\mathbb{X}^j]\mathbb{X}^j] + 
2\hat{E}^{\#}\, \Psi\gamma^{i}\Psi +\hat{E}^{+q}\gamma^{ij}_{qp}[\Psi_p,
\mathbb{X}^{j}]\, . \qquad
\end{eqnarray}
Using (\ref{DXi=EPi}) we can easily present this equation in the form
\begin{eqnarray}\label{DDXi=}
 D_\#D_\# \mathbb{X}^i&=& -
\frac{1}{16}[[\mathbb{X}^i,\mathbb{X}^j]\mathbb{X}^j] + 2 \Psi\gamma^{i}\Psi + 4i
D_\#(\hat{E}_\#^{+}\gamma^{i}\Psi_p)+\hat{E}_\#^{+}\gamma^{ij}[\Psi, \mathbb{X}^{j}] \,
. \qquad
\end{eqnarray}
Finally, the variation with respect to the traceless matrix fermionic field  $\Psi_q$
produces
\begin{eqnarray}\label{DPsi=}
D\Psi &=& \frac{i}{4}\hat{E}^{\#}[\mathbb{X}^i,(\gamma^{i}\Psi)] + 
\frac{1}{2} \hat{E}^{+}\gamma^{i}\, \mathbb{P}^{i}-\frac{i}{16}
\hat{E}^+\gamma^{ij} \, [\mathbb{X}^i,\mathbb{X}^j]\; .  \qquad
\end{eqnarray}

\subsection{A convenient gauge fixing}

To simplify the above equations,  let us use the fact that 1-dimensional connection can
always be gauged away and fix the gauge where the composed $SO(9)$ connection
(\ref{Omij:=}) and also the $SU(N)$ gauge field vanish
\begin{eqnarray}\label{Omij=0}
 \Omega^{ij}=d\tau \Omega_\tau^{ij}=0\; , \\
 \label{bbA=0}
  {\mathbb A}=d\tau {\mathbb A}_\tau =0 \; .
\end{eqnarray}
This breaks the local $SO(9)$ and $SU(N)$, but the symmetry under the rigid
$SO(9)\otimes SU(N)$ transformations remains.

As far as the $SO(1,1)$ gauge symmetry is concerned, we would not like to fix it but
rather use a part ${1\over 2}u^{a\#}\frac{\delta S_{mM0}}{{\delta} \hat{x}^{a}} =0$ of
the equations of motion for the center of energy coordinate functions $\hat{x}^{a}$
(discussed below in full),
\begin{eqnarray}\label{Drho=0}
&&D\rho^{\#}=0\qquad \; ,
\end{eqnarray}
to find the explicit form of the induced $SO(1,1)$ connection  (\ref{Om0:=}),
$\Omega^{(0)}:= {1\over 4} u^{a=}du_a^\#$. Indeed, as far as
$D\rho^{\#}=d\rho^{\#}-2\rho^{\#}\Omega^{(0)}$, Eq. (\ref{Drho=0}) implies
\begin{eqnarray}\label{Om0=}
 \Omega^{(0)}={d\rho^{\#}\over 2\rho^{\#}}\; .
\end{eqnarray}

In the gauge (\ref{Omij=0}), (\ref{bbA=0}) the set of bosonic gauge symmetries is
reduced to the Abelian $SO(1,1)$, $\tau$--reparametrization and $b$--symmetry (which we  describe below in 
sec. \ref{NoetherId}), and the
covariant derivatives simplify to
\begin{eqnarray}\label{D=dr}D\mathbb{X}^i&=& (\rho^{\#})^{-1}
d(\rho^{\#}\mathbb{X}^i)\, ,   \qquad \nonumber \\  D \mathbb{P}^i&=& (\rho^{\#})^{-2}
d((\rho^{\#})^2\mathbb{P}^i)\; , \qquad \nonumber \\ D\Psi_q &=&(\rho^{\#})^{-3/2}
d((\rho^{\#})^{3/2}\Psi_q)\; . \qquad \end{eqnarray}
As a result, Eqs.  (\ref{DDXi=}) and (\ref{DPsi=}) can be written in the following
(probably more transparent) form:
\begin{eqnarray}\label{dtPsi=}
\partial_\tau \tilde{\Psi} = \frac{i}{4} \,
e\,[\tilde{\mathbb{X}}{}^i,(\gamma^{i}\tilde{\Psi})]
+  \frac{1}{2\sqrt{\rho^{\#}}} \hat{E}_\tau ^{+}\gamma^{i}\, \tilde{\mathbb{P}}{}^{i} -
\frac{i}{16\sqrt{\rho^{\#}}} \hat{E}_\tau^+\gamma^{ij} \,
[\tilde{\mathbb{X}}{}^i,\tilde{\mathbb{X}}{}^j]\;  , \qquad
\\ \label{ddXi=}
 \partial_\tau \left(\frac{1}{e}\partial_\tau \tilde{\mathbb{X}}{}^i\right) = -
\frac{e}{16}\, [[\tilde{\mathbb{X}}{}^i,\tilde{\mathbb{X}}{}^j],
\tilde{\mathbb{X}}{}^j] +2\, e \, \tilde{\Psi}\gamma^{i}\tilde{\Psi}  + \qquad
\nonumber \\
 +  4i \partial_\tau \left( { \hat{E}_\tau^{+}\gamma^{i}\tilde{\Psi}\over
 e\sqrt{\rho^{\#}}} \right) +
{1\over \sqrt{\rho^{\#}}} \hat{E}_\tau^{+}\gamma^{ij}[\tilde{\Psi},
\tilde{\mathbb{X}}{}^{j}]\, . \qquad
\end{eqnarray}
Writing  Eqs.  (\ref{dtPsi=}) and (\ref{ddXi=}) we used the redefined fields
\begin{eqnarray}\label{tX=rX}
\tilde{\mathbb{X}}{}^i&=&  \rho^{\#} {\mathbb{X}}{}^i\; , \qquad
\tilde{\Psi}_q=(\rho^{\#})^{3/2} {\Psi}_q\; , \qquad \\ \label{tP=rP}
\tilde{\mathbb{P}}{}^{i}&=&(\rho^{\#})^2{\mathbb{P}}{}^{i}= {1\over e}
\left(\partial_\tau \tilde{\mathbb{X}}{}^i - {4i\over \sqrt{\rho^{\#}}}
\hat{E}_\tau^{+}\gamma^i\tilde{\Psi}\right) \, , \quad
\end{eqnarray}
which are inert under  the $SO(1,1)$, and
\begin{eqnarray}\label{e=E/rho}
e(\tau)= \hat{E}^\#_\tau /\rho^{\#} \;  \quad
\end{eqnarray}
which has the properties of the einbein of the Brink--Schwarz superparticle action
(\ref{S'M0=}).

\subsection{By pass technical comment on derivation of the equations for the center of energy coordinate
functions}

This is the place to present some comments on the convenient way to derive equations of
motion for the center of energy variables (which was actually used as well when
working with single M$0$ in Sec \ref{M0eqs}).

To find the manifestly covariant and supersymmetric invariant linear combinations of
the equations of motion for the bosonic and fermionic coordinate functions,
$\frac{\delta S_{mM0}}{\delta  \hat{x}^{a}}=0$ and $\frac{\delta S_{mM0}}{\delta
\hat{\theta}^{\alpha}}=0$, we introduce the covariant basis $i_\delta  \hat{E}^{A}$ in
the space of variation such that
\begin{eqnarray}\label{vhZS=}
 \delta_{\hat{Z}^M} S_{mM0} = \int_{W^1} \left(\delta  \hat{x}^{a} \frac{\delta
 S_{mM0}}{\delta  \hat{x}^{a}}+ \delta  \hat{\theta}^{\alpha} \frac{\delta
 S_{mM0}}{\delta  \hat{\theta}^{\alpha}}\right)=\qquad \nonumber \\
 = \int_{W^1} \left(i_\delta  \hat{E}^{a} \frac{\delta S_{mM0}}{i_\delta
\hat{E}^{a}}+ i_\delta  \hat{E}^{\alpha} \frac{\delta S_{mM0}}{i_\delta
\hat{E}^{\alpha}}\right)\; . \qquad
\end{eqnarray}
In  the generic case of curved superspace $i_\delta  \hat{E}^{A}= \delta \hat{Z}^{M}
{E}_M^{A}(\hat{Z})$; in our case of flat  target superspace this implies
\begin{eqnarray}\label{i-dEA=flat}
i_\delta  \hat{E}^{a}= \delta  \hat{x}^{a} - i \delta  \hat{\theta}\Gamma^a
\hat{\theta}\; , \qquad i_\delta  \hat{E}^{\alpha}= \delta  \hat{\theta}^{\alpha} \; .
\qquad
\end{eqnarray}
Furthermore, it is convenient to use the moving frame variables to split covariantly
the set of bosonic  equations $ \frac{\delta S_{mM0}}{i_\delta  \hat{E}^{a}}=0$  into
\begin{eqnarray}\label{vS/ivEb}
 \frac{\delta S_{mM0}}{i_\delta  \hat{E}^{=}}&=&{1\over 2}\, u^{a\#}\frac{\delta
 S_{mM0}}{i_\delta  \hat{E}^{a}}\, , \qquad
 \nonumber \\ \frac{\delta S_{mM0}}{i_\delta
 \hat{E}^{\#}}&=&{1\over 2}\, u^{a=}\frac{\delta S_{mM0}}{i_\delta  \hat{E}^{a}}\, , \qquad
 \nonumber \\
 \frac{\delta S_{mM0}}{i_\delta  \hat{E}^{i}}&=&- u^{a i}\frac{\delta S_{mM0}}{i_\delta
 \hat{E}^{a}}\; , \qquad
\end{eqnarray}
and the set of fermionic equations, $ \frac{\delta S_{mM0}}{i_\delta
\hat{E}^{\alpha}} = \frac{\delta S_{mM0}}{\delta\hat{\theta}^{\alpha}}$, into
\begin{eqnarray}\label{vS/ivEf}
\frac{\delta S_{mM0}}{i_\delta  \hat{E}^{-q}}=v_q^{+\alpha}\frac{\delta
S_{mM0}}{\delta  \hat{\theta}^{\alpha}} \; , \qquad \nonumber \\ \frac{\delta S_{mM0}}{i_\delta
\hat{E}^{+q}}=v_q^{-\alpha}\frac{\delta S_{mM0}}{\delta  \hat{\theta}^{\alpha}} \; .
\qquad
\end{eqnarray}
To resume,
\begin{eqnarray}\label{vhZS=cov}
 \delta_{\hat{Z}^M} S_{mM0} &=& \int\limits_{W^1} \, (\delta\hat{x}^{a}-i \delta
 \hat{\theta}\Gamma^a  \hat{\theta})  \left(u_a^=  \frac{\delta S_{mM0}}{i_\delta
 \hat{E}^{=}} + \right. \qquad \nonumber \\ && \left.
 +u_a^\#   \frac{\delta S_{mM0}}{i_\delta  \hat{E}^{\# }} + u_a^i  \frac{\delta
 S_{mM0}}{i_\delta  \hat{E}^{i}}
 \right)+ \nonumber \qquad\\
&&+ \int\limits_{W^1} \delta  \hat{\theta}^{\alpha} \left(v_\alpha^{-q} \frac{\delta S_{mM0}}{i_\delta  \hat{E}^{-q}}+ v_\alpha^{+q}  \frac{\delta S_{mM0}}{i_\delta  \hat{E}^{+q}}\right) . \nonumber \;\\ {} \qquad
\end{eqnarray}

\subsection{Equations for the center of energy coordinate functions}

As we  have already stated, the bosonic equation $\frac{\delta S_{mM0}}{i_{\delta}
\hat{E}^{=}}:= {1\over 2}u^{a\#}\frac{\delta S_{mM0}}{{\delta} \hat{x}^{a}} =0$ results
in Eq. (\ref{Drho=0}) which  is equivalent to (\ref{Om0=}). This observation is useful
to extract consequences of the next equation, $\frac{\delta S_{mM0}}{i_{\delta}
\hat{E}^{\#}}=0$, which  reads
\begin{eqnarray}\label{Dr3H=0}
&& D((\rho^{\#})^3 {\cal H})=0\; .
\end{eqnarray}
 Using (\ref{Drho=0}) one can write Eq. (\ref{Dr3H=0}) in the form of
 \begin{eqnarray}\label{dr4H=0}
&&d((\rho^{\#})^4 {\cal H})=0\; . \qquad
\end{eqnarray}
or, equivalently, $(\rho^{\#})^4 {\cal H}=const$. Due to the structure of ${\cal H}$, Eq. (\ref{HmM0=}), this constant is 
nonnegative. Furthermore, as it has been shown in \cite{mM0:action} (see also sec. \ref{velocitySec}),
it can be identified (up to numerical multiplier) with the mass parameter $M^2$ characterizing
the  center of energy motion,
\begin{eqnarray}\label{r4H=M2/4}
M^2= 4(\rho^{\#})^4 {\cal H}=const \geq 0 \; .
\end{eqnarray}

The remaining projection of the equation for the bosonic  center of energy
coordinate functions,  $\frac{\delta S_{mM0}}{i_{\delta} \hat{E}^{i}}:= - {1\over 2}u^{ai}\frac{\delta S_{mM0}}{{\delta} \hat{x}^{a}} =0$, 
gives us the relation between covariant ${SO(1,10)\over SO(1,1)\times SO(9)}$ Cartan forms
(\ref{Om++i=}),
\begin{eqnarray}\label{Om--=HOm++}
&&\Omega^{=i}=- (\rho^{\#})^2 {\cal H} \; \Omega^{\#i}= - {M^2\over 4(\rho^{\#})^2}\;
\Omega^{\#i}\;.  \qquad
\end{eqnarray}

The nontrivial part of the fermionic equation of the center of energy motion, $
\frac{\delta S_{mM0}}{i_{\delta} \hat{E}^{-q}}:= v_{q}^{-\alpha}\frac{\delta
S_{mM0}}{i_{\delta} \hat{E}^{\alpha}}=0$, reads
\begin{eqnarray}\label{E-q=Om}
&& \hat{E}^{-q}= -{1\over 2} \, \Omega^{\# i}\, \gamma^i_{qp}\nu_{\# p}^{\; -} \; ,
\end{eqnarray}
where
\begin{eqnarray}\label{nuiq=} \nu_{\# q}^{\; - } :=
(\rho^{\#})^{2}tr \left((\gamma^j \Psi)_q
\mathbb{P}^j-\frac{i}{8}(\gamma^{jk}\Psi)_q[\mathbb{X}^j,\mathbb{X}^k]\right)\, . \quad
\end{eqnarray}

\subsection{Noether identities for gauge symmetries. First look. }
\label{NoetherId}
Actually one can show that Eq. (\ref{Dr3H=0}) is satisfied identically when other
equations are taken into  account. (To be precise, Eqs. (\ref{DXi=EPi}), (\ref{Gauss}),
(\ref{DPi=}), (\ref{DPsi=}), (\ref{Drho=0}) have to be used). This is the Noether
identity for the 'tangent space' copy of the reparametrization symmetry (sometimes it
is called {\it 'b-symmetry'}) with the parameter function $i_\delta \hat{E}^\#$. Similarly, one can find the Noether identity 
reflecting the
existence of the ${\cal N}=16$ 1d gauge supersymmetry (\ref{susy-th})--(\ref{iE==}) with the basic parameter 
$\epsilon^{+q}=i_\delta \hat{E}^{+q}$ .
It states the dependence of the one half of the fermionic equations, namely
$\frac{\delta S_{mM0}}{i_{\delta} \hat{E}^{+q}}:= v_{q}^{-\alpha}\frac{\delta
S_{mM0}}{i_{\delta} \hat{E}^{\alpha}}=0$, which reads
\begin{eqnarray}\label{HE+q=}
D\nu_{\# q}^{\; - } =  {\rho}^2 \hat{E}^{+q}  {\cal H}\;
\end{eqnarray}
or $D\nu_{\# q}^{\; - } =  {\rho^{\#}}^2 \hat{E}^{+q}  {\cal H}_{\#\#\#\#}$ in a more complete notation.

\subsection{Equations which follow from the auxiliary field variations and simplification of the above
 equations}

Variation with respect to the Lagrange multiplier $\rho^{\#}$, $\frac{\delta
S_{mM0}}{\delta \rho^{\#}}=0$, expresses the projection $\hat{E}^{=}:= \hat{E}^au_a^{=}$
of the pull--back $\hat{E}^a$ of the bosonic supervielbein to the center of energy
worldline through the relative motion variables,
\begin{eqnarray}\label{E==mM0}
 \hat{E}^{=}&:=& \hat{E}^au_a^{=}= - 3 (\rho^\#)^2{\cal L}_{\#\#\# } =
\qquad \nonumber \\ & =&  3 (\rho^\#)^2 tr\left( \frac{1}{2}\mathbb{P}^iD\mathbb{X}^i +
\frac{1}{64} \hat{E}^{\#}[\mathbb{X}^i,\mathbb{X}^j]^2 -
\frac{1}{4}(E^+\gamma^{ij}\Psi)[\mathbb{X}^i,\mathbb{X}^j] \right)\; .
\qquad
\end{eqnarray}
The  $\hat{E}^{i}:= \hat{E}^au_a^{i}$  projection of this pull--back is expressed by
equations appearing as a result of variation with respect to the spinor moving frame
variables. According to Eqs. (\ref{vu++=iOm})--(\ref{vv+q}), that should appear as
coefficients for $i_\delta \Omega^{= i}$ and $i_\delta \Omega^{\# i}$ in the variation
of the action. Equation $\frac{\delta
S_{mM0}}{i_{\delta} \Omega^{= i}}=0$ reads
\begin{eqnarray}\label{Ei=mM0}
&&\hat{E}^{i}:= \hat{E}^au_a^{i}= - (\rho^{\#})^{-1}\, \Omega^{\# j} \, \left(J^{ij} +
\delta^{ij} J\right)\; ,  \qquad
\end{eqnarray}
where we have introduced the notation
\begin{eqnarray}\label{Jij=}
J^{ij}:= (\rho^{\#})^{3}\, tr\left( \mathbb{P}^{[i}\mathbb{X}^{j]}- i
\Psi\gamma^{ij}\Psi\right)\; ,\qquad \\ \label{J0=} J:= {(\rho^{\#})^{3}\,\over 2}
tr\left( \mathbb{P}^{i}\mathbb{X}^{i}\right)\; .  \qquad
\end{eqnarray}
The $(\rho^{\#})^{3}$ multipliers are introduced to make $J^{ij}$  and $J$ inert under
the $SO(1,1)$ transformations.

In this notation, equation $\frac{\delta S_{mM0}}{i_{\delta} \Omega^{\# i}}=0$ reads
\begin{eqnarray}\label{HEi=}
(\rho^\#)^3 {\cal H} \hat{E}^{i} =  -  \Omega^{= j} \left(J^{ij} - \delta^{ij} J\right)
- 2i(\rho^\#)  \hat{E}^{-q}(\gamma^{i}\nu_{\#}^{ -})_q\, . \nonumber \\ {}
\end{eqnarray}
Using (\ref{Om--=HOm++}), (\ref{Ei=mM0}), (\ref{r4H=M2/4}) and (\ref{E-q=Om}), one can
rewrite Eq. (\ref{HEi=}) as equation for $\Omega^{\# i}$,
\begin{eqnarray}\label{OmM2J=ff}
 \Omega^{\#  j} \, \left(M^2 J^{ij}
- 2i(\rho^\#)^2 \nu_{\#}^{-}\gamma^{ij}\nu_{\#}^{-} \right)=0 \, . \;\;
\end{eqnarray}
Actually, as we are going to show in the next sec.\ref{NoetherI}, {\it taking into account the remnant of the $K_9$ gauge symmetry 
of single M$0$--brane} (see (\ref{vK9}) and  (\ref{vK9v})) , which is present in the mM$0$ action, {\it one  can  present 
the above equation in the form of}
\begin{eqnarray}\label{Om++i=0}
 \Omega^{\#  i} \,  =0 \, , \qquad
\end{eqnarray}
 or, in terms of component, $\Omega_\tau^{\#  j}=0$.
 Due to (\ref{Om--=HOm++})  Eq. (\ref{Om++i=0}) implies
\begin{eqnarray}\label{mM0:Om--i=0}
 \Omega^{=  i} \,  =0 \,  \qquad
\end{eqnarray}
and (\ref{Ei=mM0}) acquires the same form as in the case of single M$0$--brane,
\begin{eqnarray}\label{hEi=0mM0}
&&\hat{E}^{i}:= \hat{E}^au_a^{i}= 0\; .  \qquad
\end{eqnarray}
Furthermore, the fermionic equation of motion (\ref{E-q=Om}) also becomes homogeneous, of the same form as the equation 
for  single M$0$--brane,
\begin{eqnarray}\label{E-q=0mM0}
&& \hat{E}^{-q}= 0 \; .
\end{eqnarray}
Eqs. (\ref{Om++i=0}) and (\ref{mM0:Om--i=0}) also  imply that all the moving frame and spinor moving frame variables are 
covariantly constant,
\begin{eqnarray}\label{mM0:Du=0}
Du_a^{\#}= 0\; ,  \qquad Du_a^{=}= 0\; ,  \qquad Du_a^{i}= 0\; ,  \qquad \\
\label{mM0:Dv=0}
Dv_q^{+\alpha}= 0\; ,  \qquad Dv_q^{-\alpha}= 0\; .  \qquad
\end{eqnarray}
Notice that in the case of single M$0$--brane such a form of equations for moving frame variables can be reached after gauge 
fixing the $K_9$ gauge symmetry with parameter $i_\delta \Omega^{\# i}$. In the mM$0$ case only a part (remnant) of $K_9$ symmetry 
is present so that a part of variations $i_\delta \Omega^{\# i}$ produce nontrivial equations which, together with the above 
mentioned remnant of $K_9$ symmetry, results in   Eqs. (\ref{mM0:Du=0}), (\ref{mM0:Dv=0}).

\subsection{Noether identity, remnant of the $K_9$ gauge symmetry and the final form of the $\Omega^{\# i}$ equation}
\label{NoetherI}
In this section we present the remnant of $K_9$ gauge symmetry leaving invariant the mM$0$ action  and show that, modulo 
this gauge symmetry, Eq. (\ref{OmM2J=ff}) is equivalent to (\ref{Om++i=0}).

Let us write  Eq. (\ref{OmM2J=ff}) as
\begin{eqnarray}\label{OmbbJ=0}
 \Omega_\tau^{\#  j} \, \gimel^{ij}= 0 \; , \qquad \end{eqnarray}
where \begin{eqnarray}\label{bbJ:=}
 \gimel^{ij}=M^2 J^{ij}
- 2i(\rho^\#)^2 \nu_{\#}^{-}\gamma^{ij}\nu_{\#}^{-} \, . \;\;
\end{eqnarray}
As this 9$\times$9 matrix is antisymmetric, it has rank 8 or lower, $rank (\gimel^{ij}) \leq 8$.
In other words, it has at least one `null vector', this is to say a vector $V^i$ which obey\footnote{This should not be 
confused with light--like vectors which can exist in the space with indefinite metric. In particular, our 11D moving 
frame vectors $u_a^=$ and $u^\#_a$ are light--like. To exclude any confusion, in this chapter we never use the name 
'null-vectors' for the  light--like vectors. }
\begin{eqnarray}\label{VbbJ=0}
\exists \; V^i, i=1,...,9 \; :  \qquad \gimel^{ij}V^j=0\, . \;\;
\end{eqnarray}
Actually, the matrix $\gimel^{ij}$ is constructed from the dynamical variables of our model
in  such a way (according to  Eqs.  (\ref{bbJ:=}) and (\ref{Jij=})) that  the number of its null vectors depends on the 
configuration of the fields describing the relative motion of the mM$0$ constituents.  However, as one `null vector' 
always exists, it is sufficient to consider a configuration with $rank (\gimel^{ij}) = 8$, and $\gimel^{ij}$ having just one 
`null vector',  at some neighborhood $\Delta \tau$  of a proper-time moment $\tau$; the generalization for a more complicated 
configurations/neighborhoods is straightforward.

Then, on one hand, the solution of  Eq. (\ref{OmbbJ=0}) in the neighborhood   $\Delta \tau$ is given by
$\Omega_\tau^{\# i}\propto V^i$, or, equivalently,
 \begin{eqnarray}\label{Om++=fv}
\Omega_\tau^{\# i}= f\, V^i\; , \qquad
\end{eqnarray}
where $f=f(\tau)$ is an arbitrary function of the center of energy proper time $\tau$. [For configurations/neighborhoods 
with several  `null vectors' $V^i_r$, $r=1,..., (9-rank \, \mathbb J)$ the solution will be $\Omega_\tau^{\# i}= f^r\, V_r^i$ 
with arbitrary functions $f^r=f^r(\tau)$].

On the other hand, the existence of null vector, Eq. (\ref{VbbJ=0}), implies that a part of Eqs. (\ref{OmbbJ=0}) is satisfied 
identically
\begin{eqnarray}\label{OmbbJV=0}
 \Omega_\tau^{\#  j} \, \gimel^{ij} V^j \equiv  0 \; , \qquad
\end{eqnarray}
when some other equations are taken into account. This is the {\it Noether identity} reflecting the existence of the gauge 
symmetry with the basic variation\footnote{See sec. \ref{bosonic}  for more details on these Noether identity and gauge symmetry 
in the purely bosonic case. Here let us just recall that Eq. (\ref{OmM2J=ff}) appears as an essential part of the 
coefficient for $i_\delta \Omega^{\# i}$ in the variation of the mM$0$ action.}
\begin{eqnarray}\label{iOm++i=aV}
i_\delta \Omega^{\# i}= \alpha \,  V^i\;  \qquad
\end{eqnarray}
with an arbitrary function $\alpha =\alpha (\tau )$.
This is clearly a remnant of the $K_9$ gauge symmetry (\ref{vK9}) of the action (\ref{SM0=}) for single M$0$--brane.

The generic variation of the Cartan 1--form $\Omega^{\# i}$ can be expressed as in Eq.  (\ref{vOm++=}),
which in our  1d case can also be written as
\begin{eqnarray}\label{vOm=DiOm}
\delta \Omega_\tau ^{\# i} = D_\tau i_\delta \Omega^{\# i} \;  . \qquad
\end{eqnarray}
Applying (\ref{vOm=DiOm}) to the variation of the solution (\ref{Om++=fv}) of Eq. (\ref{OmbbJ=0}) under (\ref{iOm++i=aV}), we find 
that
\begin{eqnarray}\label{vf=da}
\delta f(\tau)= \partial_\tau \alpha (\tau)\;  .   \qquad
\end{eqnarray}
Hence, one can use the local symmetry (\ref{iOm++i=aV}) to set $f=0$ and, thus, to gauge away (to trivialize) the solution 
(\ref{Om++=fv}) of Eq. (\ref{OmbbJ=0}).

This proves that  the gauge fixing version of Eq. (\ref{OmbbJ=0}) is given by Eq. (\ref{Om++i=0}), $\Omega^{\# i}=0$.

In sec. \ref{bosonic} we give more detailed discussion of the above local symmetry and its Noether identities reproducing 
independently the above conclusion for the purely bosonic case.

\section{Ground state solution of the relative motion  equations}
\label{grStSec}

The natural first step in studying the above obtained mM$0$ equations is to address the
sector of
\begin{eqnarray}\label{Psiq=0}
\Psi_q=0 \, .  \;
\end{eqnarray}
As far as the fermionic equations of motion  have the same form (\ref{E-q=0mM0}) as for the single M$0$--brane, 
$\hat{E}^{-}_{q}=0$, the only possible fermionic contribution to the relative motion equations might come from
the induced gravitino $\hat{E}^{+q}=d\hat{\theta}^{\alpha}v_\alpha^{+q}$. However,
 with (\ref{Psiq=0}), the fermionic equation of the relative motion
(\ref{DPsi=}) results in
 \begin{eqnarray}\label{Psi=0=>E+}
 \hat{E}^{+}\gamma^{i}\, \mathbb{P}^{i}-\frac{i}{8} \hat{E}^+\gamma^{ij} \,
 [\mathbb{X}^i,\mathbb{X}^j]=0
\; .
\end{eqnarray}
As it will be clear after our discussion below, for $M^2>0$ this equation has only
trivial solution $\hat{E}^{+q}=0$, while for $M^2=0$ the 1d gravitino $\hat{E}^{+q}$
remains arbitrary.

\subsection{Ground state of the relative motion}

It is easy to see that a particular configuration of the bosonic fields for which Eq. (\ref{Psi=0=>E+}) is satisfied is
\begin{eqnarray}\label{grSt}
\mathbb{P}^{i}=0 \; , \qquad  [\mathbb{X}^i,\mathbb{X}^j]=0 \; .
\end{eqnarray}
Then the fermionic 1-form
$\hat{E}^{+q}$ remains arbitrary (and pure gauge) as it is in the case of single M$0$--brane.

Together with (\ref{Psiq=0}), Eqs. (\ref{grSt}) describe the ground state of the
relative motion. For it the relative motion Hamiltonian (\ref{HmM0=}) and the center of
energy effective mass vanish,
\begin{eqnarray}\label{M2=0mM0}
M^2=0\;
\end{eqnarray}
so that the center of energy motion is light--like. Moreover, when Eqs. (\ref{Psiq=0})
and (\ref{grSt}) hold, all the equations of the center of energy motion  coincide with
the equations for single M$0$--brane.

The ground state of the mM$0$ system is thus described by Eqs. (\ref{Psiq=0}),
(\ref{grSt}) and by a (pure bosonic) ground state solution of the single M$0$
equations. This preserves all 16 worldline supersymmetries, which corresponds (as we
have discussed in sec. \ref{susySOL}) to the preservation of 16 of 32 spacetime supersymmetries.

\subsection{Solutions with $M^2=0$ have relative motion in the ground state sector}
\label{0M2=vacuum}

Curiously enough, being in the ground state of the relative motion is the only
possibility for the mM$0$ system to have the light--like center of energy motion
characterized by zero effective mass
\begin{eqnarray}
\label{M2=0} M^2=0 \; \Leftrightarrow \; {\cal H}= {1\over 2}tr ({\mathbb P}^i{\mathbb P}^i) -
{1\over 64} tr [{\mathbb X}^i, {\mathbb X}^j]^2 =0\; . \qquad
\end{eqnarray}
Indeed, the pure bosonic relative motion Hamiltonian ${\cal H}$ is given by the sum of
two terms both of which are traces of squares of Hermitian operators ($[[{\mathbb X}^i,
{\mathbb X}^j]^\dagger=  [{\mathbb X}^j, {\mathbb X}^i]=- [{\mathbb X}^i, {\mathbb X}^j]$); hence, the sum
vanishes, ${\cal H}=0$, iff both equations in (\ref{grSt}) hold \footnote{We do not
discuss here the possible nilpotent contributions, like the possibility to solve the
equation  $a^2=0$ for a real bosonic $a(\tau)$ by
 $a=\beta_{\alpha_1...\alpha_{17}} \hat{\theta}^{\alpha_1}\ldots
 \hat{\theta}^{\alpha_{17}}$ with  17 center of energy fermions
 $\hat{\theta}^{\alpha}(\tau)$ contracted with  some fermionic
 $\beta_{\alpha_1...\alpha_{17}}= \beta_{[\alpha_1...\alpha_{17}]}$.}, ${\mathbb P}^i= D_\#
 {\mathbb X}^i=0$ and $[{\mathbb X}^i, {\mathbb X}^j]=0$ \footnote{This is true for finite size
 matrices. In the $N\mapsto \infty$ limit (mM$0$ condensate) one can  consider a
 'non--commutative plane' solution with $[{\mathbb X}^i, {\mathbb X}^j]=i\Theta^{ij}$  and
 c-number valued $\Theta^{ij}=-\Theta^{ji}$, see for instance, \cite{GZ+:JHEP11}. In
 the case of finite $N$ this solution cannot be used as far as the right hand side is
 assumed to be proportional to the unity matrix, $I_{N\times N}$ while the trace of the
 commutator vanishes. }.

Thus any nontrivial configuration of the relative motion, with either ${\mathbb P}^i\not=0$
or/and $[{\mathbb X}^i, {\mathbb X}^j]\not=0$,  creates a nonzero effective mass of the center
of energy motion, $M^2\not=0$.

\section{Supersymmetric solutions of  ${\bf m}$M0 equations}
\label{susySolSec}

\subsection{Supersymmetric solutions of the mM$0$ equations have $M^2=0$}

\label{BPS-M2=0}

From Eq. (\ref{susy-Psi}) one concludes that a solution of the mM$0$ equations with
vanishing relative motion fermionic fields, Eq. (\ref{Psiq=0}), can be supersymmetric
if
\begin{eqnarray}
\label{susy-Psi=0}
 (\epsilon^{+} \gamma^i)_q  {\mathbb P}^i-  {i\over 8} (\epsilon^{+} \gamma^{ij})_q  [{\mathbb
 X}^i, {\mathbb X}^j]=0\; . \qquad
\end{eqnarray}
All the 16 worldline supersymmetries (1/2 of the target space supersymmetries) can be
preserved iff this equation is satisfied for arbitrary $\epsilon^{+p}$. This implies
\begin{eqnarray}
\label{susy16-Psi=0} \gamma^i_{qp}  {\mathbb P}^i-  {i\over 8} \gamma^{ij}_{qp}  [{\mathbb
X}^i, {\mathbb X}^j]=0\;  \qquad
\end{eqnarray}
the only solution of which is given by the ground state of the relative motion, Eq.
(\ref{grSt}).

Thus all the bosonic solutions of mM$0$ equations preserving $16$ supersymmetries have
the trivial relative motion sector described by Eq. (\ref{grSt}) which is characterized
by the light--like center of energy motion, $M^2=0$.

This suggests that  $M^2=0$,  is the BPS condition,
{\it i.e.} the necessary condition for the 1/2 supersymmetry preservation. As we are going to show, this is indeed the case, and, 
moreover  \begin{eqnarray}
\label{M2=0BPS} M^2=0  \qquad
\end{eqnarray}
is the BPS equation for preservation of any part of the target space supersymmetry.

Indeed, on one hand, tracing Eq. (\ref{susy-Psi=0}) with $\gamma^j{\mathbb P}^j$ and using
the properties of $tr$ we find $$\epsilon^{+q}  tr({\mathbb P}^i{\mathbb P}^i)=  {i\over 8}
(\epsilon^{+} \gamma^{jk} \gamma^{i})_q  tr({\mathbb P}^i[{\mathbb X}^j, {\mathbb X}^k])\; . $$ On the
other hand, tracing (\ref{susy-Psi=0}) with ${i\over 8}\gamma^{jk}[{\mathbb X}^j, {\mathbb
X}^k]$ and using  the Jacobi identities $[{\mathbb X}^{[i}[{\mathbb X}^j, {\mathbb X}^{k]}]]\equiv
0$ we find $$ {i\over 8} (\epsilon^{+} \gamma^{i}\gamma^{jk})_q  tr({\mathbb P}^i[{\mathbb X}^j,
{\mathbb X}^k])= {1\over 32} (\epsilon^{+q} tr([{\mathbb X}^j, {\mathbb X}^k]^2)\;. $$ Taking the sum
of these two equations and using (\ref{Gauss}) (with fermionic fields set to zero) we find
$\epsilon^{+q}{\cal H}=0$  which, using (\ref{r4H=M2/4}), can be
written as $\epsilon^{+q} \, M^2 =0$,
\begin{eqnarray}
\label{BPS=0} \epsilon^{+q} \, M^2 =0\qquad \Leftarrow \qquad \epsilon^{+q}{\cal H}=0\;
.   \qquad
\end{eqnarray}
For $M^2\not=0 $ this implies $\epsilon^{+q}=0 $, so that the supersymmetry is broken.
Thus all the supersymmetric solutions of mM$0$ equation are characterized by $M^2=0$.

This fact is very important: it means that the existence of our action does not imply
the existence of a  new type  of supersymmetric solutions of the  11D SUGRA
equations\footnote{Although this statement can be done about the solutions preserving
1/2 of the 11D supersymmetry, as it will be clear in a moment, it is universal as far
as a supersymmetric solution of mM$0$ equations can preserve only 1/2 of the tangent
space supersymmetry.}. A BPS solution is in correspondence with the ground state of the
brane or of the multiple brane system; the ground state of mM$0$ system is
characterized by the vanishing effective mass and with the center of energy motion
characteristic for the single M$0$--brane. Thus a supersymmetric solution of 11D SUGRA equations  
corresponding to  single M-wave also describe the mM$0$ (multiple M-wave) ground state.

\subsection{All BPS states of  mM$0$ system are 1/2 BPS}
\label{BPS=1/2}

As we have shown,  a solution of mM$0$ equations can preserve some part of the 16 worldline
supersymmetries (and some part ($\leq 1/2$) of the target space supersymmetry) if and
only if $M^2=0$. Now, in the light of the observation in sec. \ref{0M2=vacuum}, $M^2=0$
implies that the relative motion of the mM$0$ constituents is in its ground state, Eq.
(\ref{grSt}). This has two consequences. Firstly, as the ground state trivially solves the Killing spinor 
equation (\ref{susy-Psi=0}), it preserves all the supersymmetries allowed by the center of energy motion.
Secondly, when the relative motion sector is in its ground state, the center of energy sector of supersymmetric 
solution is described by the same equations as the motion of single M$0$--brane (massless 11D superparticle). Now, as we 
have shown in sec. \ref{susySOL}, the supersymmetric solutions of these M$0$ equations preserve just 1/2 of the target space 
supersymmetry.

{\it This proves that all the supersymmetric solutions of the equations of motion of the mM$0$  system preserve just one 
half of 32 target space supersymmetries. In other words, all the mM$0$ BPS states are 1/2 BPS.}

\section{On  solutions of $\mathbf m$M$0$ equations with $M^2> 0$ }
\label{bosonic}

When $M^2\not=0$, Eq. (\ref{Psi=0=>E+}) has only trivial solutions. (The proof of this fact follows the stages of 
sec. \ref{BPS-M2=0}).  This means that  (\ref{Psiq=0}) results in
\begin{eqnarray}\label{E+q=0}
 \hat{E}^{+q}=0
\; ,
\end{eqnarray}
so that, when $M^2>0$, a configuration with vanishing relative motion fermion is purely bosonic.

\subsection{Purely bosonic equations in the case of $M^2>0$ }

The complete list of nontrivial pure bosonic equations for mM$0$ system with nonvanishing center of energy mass,
$M^2>0$, reads
\begin{eqnarray}\label{Drho0=0}
&& D\rho^{\#}=0\qquad \Leftrightarrow \quad   \Omega^{(0)}={d\rho^{\#}\over
2\rho^{\#}}\; ,
\\ \label{DDXi0=}
&& D_\#D_\# \mathbb{X}^i= - \frac{1}{16}[[\mathbb{X}^i,\mathbb{X}^j]\mathbb{X}^j] \; ,
\\
\label{Gauss0} && [D_\#\mathbb{X}^i,\mathbb{X}^i]= 0\, , \qquad
\qquad
\end{eqnarray}
\begin{eqnarray}\label{E0==mM0}
 \hat{E}^{=}&:=& d\hat{x}^{a}u_a^= =\qquad \nonumber \\
 &=& 3 \hat{E}^{\#}  \left(  (\rho^\#)^2  tr(D_\#\mathbb{X}^i)^2 -
 \frac{M^2}{4(\rho^\#)^2 } \right) ,  \\
\label{Ei0=mM0}
\hat{E}^{i}&:=& d\hat{x}^{a} u_a^{i}= 0 \, \; ,  \qquad
\end{eqnarray}
\begin{eqnarray}\label{OmM2J=0}
&& \Omega^{\#  i} \,=0 \, , \;\; \\
\label{Om--0=M2Om++}
&&\Omega^{=i}= 0\; ,  \qquad
\end{eqnarray}
where $\hat{E}^{\#}= d\hat{x}^{a}u_a^{\#}$ and the center of energy mass $M$ is defined by Eq. (\ref{r4H=M2/4}), 
$M^2= 4(\rho^{\#})^4
{\cal H}$, with the relative motion Hamiltonian
\begin{eqnarray}
\label{HmM00=} {\cal H}  &=& tr\left( {1\over 2}  (D_\# {\mathbb X}^i)^2   - {1\over 64}
\left[ {\mathbb X}^i ,{\mathbb X}^j \right]^2 \right)\; .
  \end{eqnarray}

Notice that (as we have discussed in the general case) the currents
\begin{eqnarray}\label{Jij0=}
J^{ij}&=& (\rho^{\#})^{3}\, trD_\#\mathbb{X}^{[i}\mathbb{X}^{j]}\; ,\qquad  \nonumber \\ J&=&
{(\rho^{\#})^{3}\,\over 2} tr D_\#\mathbb{X}^{i}\mathbb{X}^{i}\; \qquad
\end{eqnarray}
disappear from the final form of equations when one takes into account the presence of the remnants of the $K_9$ symmetry. 
As far as this statement is very important in the analysis of the mM$0$ equations, we are going to give more detail on this 
symmetry and gauge fixing now.

But before let us make an observation that  the current $J^{ij}$ is covariantly constant  on the mass shell (i.e. when the 
above equations of motion are taken into account),
\begin{eqnarray}\label{DJij0=0}
DJ^{ij}=0\; .
\end{eqnarray}
In contrast, in the generic purely bosonic configuration  the scalar current is not a constant, $DJ=dJ\not=0$.

\subsection{Remnant of $K_9$ symmetry in the  bosonic limit of the mM$0$ action and  $\Omega^{\#i}$ equations }

The variation of the bosonic limit of the mM$0$ action (\ref{SmM0=}) can be written in the form
\begin{eqnarray}\label{vSbos=}
\delta S^{^{bosonic}}_{mM0}&=& \int_{W^1} {\cal E}^{= i}_u i_\delta \Omega^{\# i}+ \int_{W^1} {\cal E}^{\# i}_u i_\delta \Omega^{= i}- \qquad \nonumber \\ && - \int_{W^1} {\cal E}_{\hat{x}}^{i} i_\delta \hat{E}^{i}+ \ldots  \; . \qquad \end{eqnarray}
where
 \begin{eqnarray}\label{cE=:=}
 {\cal E}^{= i}_u&=&M^2 \hat{E}^{i}/4 \rho^{\#} + \Omega^{= j}(J^{ij}-\delta^{ij}J)\; , \nonumber \\
  {\cal E}^{\# i}_u&=&\rho^{\#}  \hat{E}^{i}  + \Omega^{\# j}(J^{ij}+\delta^{ij}J)\; , \nonumber \\
{\cal E}_{\hat{x}}^{i}\; &=& \rho^{\#} \Omega^{= i}+ M^2 \Omega^{\# i}/4 \rho^{\#}\; , \qquad \end{eqnarray}
with $J^{ij}$ and $J$ defined in (\ref{Jij0=}) and (\ref{J0=}), and  dots denote the terms involving the other basic 
variations ($\delta \rho^{\#}$, $i_\delta \hat{E}^=$ etc.). Furthermore, one can rearrange the terms in (\ref{vSbos=}) in 
the following way:
\begin{eqnarray}\label{vSbos==}
&& \delta S^{^{bosonic}}_{mM0}= \int_{W^1}  {\cal E}^{\# i}_u \left( i_\delta \Omega^{= i}-{ M^2\over 4(\rho^{\#})^2} i_\delta  \Omega^{\# i}\right)  - \; \nonumber
\\ && \qquad - \int_{W^1} {\cal E}_{\hat{x}}^{i} \left( i_\delta \hat{E}^{i}+ \left(J^{ij}+\delta^{ij}J\right)\, i_\delta \Omega^{\# j} \right) +  \; \nonumber \\ && \qquad   + { M^2\over 2(\rho^{\#})^2}  \int_{W^1} d\tau\,  \Omega_\tau^{\# i} J^{ij} i_\delta \Omega^{\# j}  + \ldots
\, ,  \quad
\end{eqnarray}
In this form it is transparent that the equations of motion corresponding to the $i_\delta \Omega^{\# j}$ variation can be 
written in the form
\begin{eqnarray}\label{Om++iJij=0}
\Omega_\tau^{\# i} J^{ij} =0
\, ,  \quad
\end{eqnarray}
which is the bosonic limit of Eq.  (\ref{OmM2J=ff}). As we have already discussed in the general case, Eq. (\ref{Om++iJij=0}) 
always has a nontrivial solution as far as the antisymmetric $9\times 9$ matrix $J^{ij} = - J^{ji} $ always has at least one 
null vector, a non-zero vector  $V^i$ such that
$V^iJ^{ij}=0$.

Each null--vector generates a nontrivial solution of (\ref{Om++iJij=0}), but also a gauge symmetry of the mM$0$ action. 
Indeed,  as one can easily see from (\ref{vSbos==}), the transformations with $\tau$-dependent parameter $i_\delta \Omega^{\# j}$ 
obeying
\begin{eqnarray}\label{JiOm++==0}
 J^{ij} i_\delta \Omega^{\# j} =0
\, ,  \quad
\end{eqnarray}
completed by
 \begin{eqnarray}\label{iOm--=iOm++}  i_\delta \Omega^{= i}&=& { M^2\over 4(\rho^{\#})^2} i_\delta  \Omega^{\# i} , \qquad \nonumber  \\
i_\delta \hat{E}^{i}\; &=&  - (J^{ij}+J\delta^{ij}) i_\delta \Omega^{\# j}
\; , \qquad
\end{eqnarray}
leave the action invariant, $\delta S^{^{bosonic}}_{mM0}=0$, and, thus define the gauge symmetries of the mM$0$ action.
The transformations of  $\Omega_\tau^{\# i}$ under this gauge symmetry are 
$\delta \Omega_\tau^{\# i}=D_\tau i_\delta \Omega^{\# i}$ (\ref{vOm=DiOm}).  As far as in purely bosonic limit  
$DJ^{ij}=0$ on the mass shell (see Eq. (\ref{DJij0=0})),
\begin{eqnarray}\label{JiOm++=0}
 J^{ij} D_\tau i_\delta \Omega^{\# j} =0\;    \quad
\end{eqnarray}
is also obeyed.  Furthermore, in 1d case all the connection can be gauged away so that the transformation rules of the 
nontrivial solution of Eq. (\ref{Om++iJij=0}) can be summarized as follows
\begin{eqnarray}\label{vOm++i=dtidOm++i}
\delta \Omega_\tau^{\# i}=\partial _\tau i_\delta \Omega^{\# i}\; , \qquad \begin{cases} \Omega_\tau^{\# i} J^{ij} =0 \, \cr  J^{ij} i_\delta \Omega^{\# j} =0 \; ,  \cr
\partial_\tau J^{ij}  =0 \; .
\end{cases}
\end{eqnarray}
This form makes transparent that any  nontrivial solution of Eq. (\ref{Om++iJij=0}) can be gauged away using local 
symmetry (\ref{JiOm++==0}), (\ref{iOm--=iOm++}). Thus, modulo the gauge symmetry,  Eq. (\ref{Om++iJij=0}) is equivalent 
to Eq. (\ref{OmM2J=0}), $\Omega^{\# i}=0$.

\subsection{Center of energy velocity and momentum for $M^2\not=0$}
\label{velocitySec}

Let us notice  one property of the center of energy motion of our M$0$ system which, on the first glance, might looks strange, 
and try to convince the reader that it is rather a natural manifestation of the  influence of relative motion on the center of 
energy dynamics.

Using Eqs.  (\ref{E0==mM0}), (\ref{Ei0=mM0}) we can easily calculate  center of energy velocity  of the bosonic limit of our 
mM$0$ system,
\begin{eqnarray}\label{dhx=mM00}
 && \dot{\hat{x}}^a:= \partial_\tau{\hat{x}}^a=  {1\over 2}\hat{E}_\tau^{=}u^{\# a}+ {1\over 2}\hat{E}_\tau^{\# }u^{= a}-\hat{E}_\tau^{i}u^{i a}= \nonumber \; \\
&&\;= {1\over 2} \hat{E}_\tau^{\# } \left(u^{= a}+ 3u^{\# a}\left(  (\rho^\#)^2  tr(D_\#\mathbb{X}^i)^2 -
 \frac{M^2}{4(\rho^\#)^2 } \right)\right)  .  \nonumber \\ {}
\end{eqnarray}
On the other hand, the canonical momentum conjugate to the center of energy coordinate function $\dot{\hat{x}}^a$ 
is\footnote{${\cal L}^{mM0}_\tau$ is the Lagrangian of the mM$0$ action (\ref{SmM0=}), $S_{mM0}=\int d\tau {\cal L}^{mM0}_\tau$.}
\begin{eqnarray}\label{p=mM0}
p_a= {\partial {\cal L}^{mM0}_\tau\over \partial \dot{x}^a}= \rho^{\#}\left( u_a^{=}+ u^{\#}_{a} \, \frac{M^2}{4(\rho^\# )^2} \right)\; .
\end{eqnarray}
This equation justifies our identification of the constant $M^2$ as a square of the effective mass of the mM$0$ system as it gives
\begin{eqnarray}\label{p2=M2}
p^ap_a=M^2  \; .
\end{eqnarray}
Thus, generically, the center of energy velocity and its momentum are oriented in different directions of 11D spacetime,
\begin{eqnarray}\label{velocity=p-cA}
\dot{\hat{x}}_a &\propto & \left( p_a - {\cal A}_a \right)\; , \qquad \\
\label{cA=}  && {\cal A}_a=  u^{\#}_{ a}\left( \frac{M^2}{\rho^\# } -3  (\rho^\#)^3  tr(D_\#\mathbb{X}^i)^2 \right)
\; . \qquad
\end{eqnarray}

Eq. (\ref{velocity=p-cA}) might look strange if one expects the center of energy motion to be similar to the motion of a free 
particle. However, this relation is characteristic for a charged particle moving in a background Maxwell field (see e.g. 
\cite{Landau}). In our case the counterpart  (\ref{cA=}) of the electromagnetic potential $ {\cal A}_a$ is constructed in terms 
of the relative motion variables. It vanishes when the relative motion is in its ground state.

Thus the seemingly unusual effect of that the mM$0$ center of energy  velocity and momentum are not parallel one to another is 
just one of the manifestations of the mutual influence of the center of energy and the relative motion in mM$0$ system. The 
relative motion variables, when they are not in ground state, generate a counterpart of the 11D background vector potential 
for the center of energy motion.

\subsection{An example of  non-supersymmetric solutions}

Let us fix the gauge (\ref{Omij=0}), (\ref{bbA=0}), ${\Omega^{ij}=0={\mathbb A}}$,  use the $SO(1,1)$ gauge symmetry to 
set  $\rho^{\#}=1$ and the reparametrization symmetry to fix $\hat{E}^{\#}_\tau=1$\footnote{Actually, to be precise, 
there exists an obstruction to fix such a gauge by $\tau$ reparametrization  \cite{e:1}. The best what one can do is to 
fix $\partial_\tau \hat{E}^{\#}_\tau=0$, while the constant value remains indefinite. This is especially important for path 
integral quantization, where the integration over this constant value (mudulus)  should be included in the definition of the 
path integral measure.  As here we do not need in this level of precision, we allow ourselves to simplify the formulas by just 
setting this indefinite constant to unity. },
\begin{eqnarray}\label{gauge=}
\Omega_\tau ^{ij}=0={\mathbb A}_\tau \; , \qquad  \hat{E}^{\#}_\tau=1=\rho^{\#}\, . \quad
\end{eqnarray}
Then \begin{eqnarray}\label{D++=dt}
D_\#=\partial_\tau  \qquad
\end{eqnarray}
and  Eqs. (\ref{DDXi0=}) simplify to
\begin{eqnarray}\label{DDXi00g=}
\ddot{{\mathbb{X}}}{}^i&=& -
\frac{1}{16}\,
[[{\mathbb{X}}{}^i,{\mathbb{X}}{}^j]{\mathbb{X}}{}^j]\, , \quad
\\ \label{Gauss00}
&& [\dot{{\mathbb{X}}}{}^i,{\mathbb{X}}{}^i]= 0\, . \qquad
\end{eqnarray}
These very well known  equations describe the 1d reduction of the 10D $SU(N)$ Yang-Mills gauge theory.

A very simple solution of Eqs.  (\ref{DDXi00g=}) and (\ref{Gauss00}) is provided by
\begin{eqnarray}\label{X=At+B}
&&{\mathbb{X}}^i(\tau)=(A^i \tau +B^i){\mathbb{Y}},\qquad
\end{eqnarray}
where ${\mathbb{Y}}$ is a  constant traceless $N\times N$ matrix, $A^i$ and $B^i$ are constant  $SO(9)$ vectors, and $\tau$ is 
the proper time of the mM$0$ center of energy. The  center of energy effective mass is defined by the trace of ${\mathbb{Y}}^2$ 
and by the length of vector $\vec{A}=\{ A^i\}$,
\begin{eqnarray}
\label{M2=X2} M^2= 4{\cal H}  &=& 2 \vec{A}^2 tr {\mathbb Y}^2 \; , \qquad \vec{A}^2:=A^iA^i \; .
  \end{eqnarray}
Actually, by choosing the initial point of the proper time, $\tau \mapsto \tau - a$,
we can always make the constant SO(9) vectors $A^i$ and $B^i$ orthogonal,
\begin{eqnarray}\label{AB=0}
&& \vec{A}\vec{B}:= A^iB^i=0 .\qquad
\end{eqnarray}
Then the `currents' (\ref{Jij0=})  read
\begin{eqnarray}\label{Jij00=}
J^{ij}=
{A}^{[i}{B}^{j]} tr\mathbb{Y}^2= {{A}^{[i}{B}^{j]} \over 2\vec{A}^2} \; M^2 , \qquad
 J={\tau \over 4}\,  M^2\, .\quad
\end{eqnarray}

Now  the equations for the center of  energy coordinate functions (\ref{E==mM0}), (\ref{Ei0=mM0}) and the gauge fixing 
condition $\hat{E}_\tau^{\#}=1$ imply
\begin{eqnarray}\label{dx==mM0}
 \dot{\hat{x}}{}^{a}u_a^= &=&  {3M^2}/{4} , \qquad
\\ \label{dxi=mM0}
\dot{\hat{x}}{}^{a} u_a^{i}&=& 0  \, \; ,  \qquad
\\ \label{dx++=mM0}
 \dot{\hat{x}}{}^{a}u_a^\#  &=&1\, \; .  \qquad
\end{eqnarray}
With our gauge fixing, Eqs. (\ref{mM0:Du=0}), which follow from (\ref{OmM2J=0}), (\ref{Om--0=M2Om++}), implies that moving 
frame vectors are constant
\begin{eqnarray}\label{mM0:du=0}
\dot{u}_a^{\#}= 0\; ,  \qquad \dot{u}_a^{=}= 0\; ,  \qquad \dot{u}_a^{i}= 0\; . \qquad
\end{eqnarray}
Thus  (\ref{dx==mM0}), (\ref{dxi=mM0}), (\ref{dx++=mM0}) is a simple  system of linear differential equations
\begin{eqnarray}\label{dx===}
 && \dot{\hat{x}}{}^{=}   =  {3M^2}/{4} , \qquad
\\ \label{dxi==0}
&& \dot{\hat{x}}{}^{i}= 0  \, \; ,  \qquad
\\ \label{dx++==}
&& \dot{\hat{x}}{}^{\#}  =1\, \; ,  \qquad
\end{eqnarray}
for the variables
\begin{eqnarray} \label{x=anal}
\hat{x}^= =\hat{x}^a u_a^= \; , \qquad \hat{x}^\# =\hat{x}^a u_a^\# \; , \qquad \hat{x}^i =\hat{x}^a u_a^i \; .  \qquad
\end{eqnarray}
This system can be easily solved for the 'comoving frame' coordinate functions (\ref{x=anal}). The solution in an arbitrary frame
\begin{eqnarray} \label{hx=sol1}
\hat{x}^\mu(\tau) =\hat{x}^\mu(0) + {\tau \over 2}\left(u^{= \mu}+ {3M^2\over 4} u^{\# \mu} \right)
\end{eqnarray}
describe a time-like motion of the center of energy characterized by a nonvanishing effective  mass (\ref{M2=X2}).
The velocity of this motion,
\begin{eqnarray} \label{dhx=sol1}
\dot{x}^\mu = {1\over 2}\left(u^=_\mu+{ 3M^2\over 4} u_\mu^\#\right)
\end{eqnarray}
is not parallel to the canonical momentum (see (\ref{p=mM0}))
\begin{eqnarray} \label{dhx=sol1}
p_\mu =  u^=_\mu+ {M^2\over 4} u_\mu^\# \; .
\end{eqnarray}
As it was discussed in general case in sec. \ref{velocitySec}, this is due to the influence of the relative motion of the mM$0$ 
constituents on the center of energy motion and can be considered as an effect of the induction by the relative motion dynamics 
of a counterpart of the Maxwell background field interacting with the center of energy coordinate functions. In the case under 
consideration this induced Maxwell field is  constant, ${\cal A}_\mu=- u_\mu^\#\, M^2/2 $.

\subsection{Another non-supersymmetric formal solution}

In the case of the system of 2 M$0$--branes, the $2\times 2$ matrix  field $\mathbb X^i$ can be decomposed on Pauli matrices, 
$\mathbb X^i= f^i_J(\tau) \sigma^J$,
 \begin{eqnarray} \label{sigmaI}
\sigma^I\sigma^J = \delta^{IJ} I_{2\times 2} + i\epsilon^{IJK}\sigma^K\; ,\qquad I,J,K=1,2,3. \qquad
\end{eqnarray}
The simplest ansatz which solves the Gauss constraint (\ref{Gauss00}) is $f^i_J(\tau) =\delta^i_Jf (\tau)$  so that 
\begin{eqnarray} \label{sigmaI}
{\mathbb X}^i (\tau) = f (\tau)\delta^i_J \sigma^J , \quad i=1,...,9; \;\;  I,J,K=1,2,3. \quad
\end{eqnarray}
Eq. (\ref{DDXi00g=}) then implies that this function should obey
\begin{eqnarray} \label{ddf=f3}
\ddot{f}+{1\over 2} f^3=0\, .\quad
\end{eqnarray}
The simplest solution of this equation is given by $f(\tau)={2i\over \tau}$ which is complex and thus breaks the condition 
that $\mathbb X^i$ is a Hermitian matrix. Actually one can consider this solution,
\begin{eqnarray} \label{X=2i/tS}
{\mathbb X}^i (\tau) = {2i\over \tau}\, \delta^i_J \sigma^J , \quad J=1,2,3 .\quad
\, \quad
\end{eqnarray}
as an analog of instanton as far as after Wick rotation $\tau\mapsto i\tau$ restores the Hermiticity properties.

Ignoring for a moment  the problem with Hermiticity we can calculate the Hamiltonian and find that it is equal to zero. 
Thus (\ref{X=2i/tS}) is a solution with vanishing center of energy mass, $M^2=0$.

A configuration (\ref{sigmaI}) with nonzero effective center of energy mass can be obtained by observing that 
(\ref{ddf=f3}) has a more general solution given by  the so--called Jackobi elliptic function \cite{Abr}. These functions obey
\begin{eqnarray} \label{df=-f4+C}
\dot{f}^2= -f^4/4+ C
\, \quad
\end{eqnarray}
with an arbitrary constant $C$.  The above discussed particular solution (\ref{X=2i/tS}) of (\ref{ddf=f3}) solves 
(\ref{df=-f4+C}) with $C=0$ which suggests the relation of $C$ with $M^2$.  Indeed, a straightforward calculation shows 
that $C=M^2/12$  so that the instanton--like solution of the mM0 equations of relative motion is given by 2x2 matrices 
(\ref{sigmaI}) with the function $f(\tau)$ obeying
\begin{eqnarray} \label{df=-f4+M2}
\dot{f}^2= {M^2-3f^4\over 12}\, .
\, \quad
\end{eqnarray}

The set of equations for  the center of energy motion includes (\ref{dxi==0}), (\ref{dx++==}) and
\begin{eqnarray}\label{dx==c+f4}
 && \dot{\hat{x}}{}^{=}   =  {3M^2}/{4} - {9(f(\tau))^4}/{2} \; . \qquad
\end{eqnarray}
This equation can be solved numerically, but its detailed study goes beyond the scope of this thesis.

\chapter{Conclusions}
\thispagestyle{chapter} 
In conclusion we list the main results obtained in this thesis. 
\begin{enumerate}
\item The complete set of the superfield equations of motion for the interacting system of four dimensional supermembrane and 
dynamical scalar multiplet have been obtained. Our study has provided the first example of superfield equations for interacting system involving 
matter (not supergravity) superfields and supersymmetric extended object, as well as the first set of superfield equations of motion for 
a dynamical superfield system including supermembrane. Furthermore

\begin{enumerate}
\item The action of supermembrane in a chiral ${\cal N}=1$, $D=4$ superfield background has been presented for the first time.
\item It was shown that the consistency of the interaction with dynamical scalar multiplet requires this to 
be special, namely to be described by chiral superfield of special form: expressed through the real (rather than 
complex) pre-potential superfield. 
\item The equations of motion for spacetime, component fields with supermembrane contributions have been obtained from the 
bulk superfield equations for the simplest case when the bulk part of the action is given by the free kinetic term 
(with K\"{a}hler potential ${\cal K}= \Phi\bar{\Phi}$).  A solution of the dynamical equations for physical fields in the 
leading order on supermembrane tension have been obtained. The effects of inclusion of nontrivial superpotential and relation with 
known supersymmetric domain wall solutions have been discussed.
\end{enumerate}

\item The complete set of superfield equations of motion for the interacting system of dynamical $D=4$ ${\cal N}=1$ 
supergravity and supermembrane has been obtained from the superspace action principle. The supermembrane model is consistent in an 
arbitrary off--shell minimal supergravity background. However the interaction with supemembrane requires the dynamical supergravity to 
be the Grisaru--Siegel--Gates--Ovrut--Waldram special minimal supergravity. The chiral compensator superfield of this are constructed 
from real (rather than complex) prepotential.

\begin{enumerate}
\item We have developed the Wess--Zumino type approach to this special minimal supergravity the characteristic property of which is a 
dynamical generation of the cosmological constant.
\item To see this effect in the interacting system we extract the spacetime component equations from the superfield equations. To 
this end we have fixed the usual Wess--Zumino gauge in the superfield supergravity equations, supplemented by partial 
gauge fixing of the local supersymmetry on the supermembrane worldvolume ($\hat{\theta}{}^{\underline{\alpha}}(\xi)=0$). We have 
shown that the supermembrane current superfields simplify drastically in this ''WZ$_{\hat{\theta}=0}$ gauge''.
\item In the component form of equations obtained in this way it is seen that in the interacting system the supermembrane produces a kind of 
renormalization of the cosmological constant, making its value different in the branches of spacetime separated by the 
supermembrane worldvolume.
\item This allowed us to show that configuration describing a domain wall separating two branches of AdS space with different cosmological 
constants provides a supersymmetric solution of the system of our superfield supergravity equations considered outside the supermembrane 
worldvolume $W^3$.
\end{enumerate}

\item We have obtained the superfield equations in ${\cal N}=2,4$ and $8$
extended tensorial superspaces $\Sigma^{(10|\mathcal{N}4)}$, which describe the
supermultiplets of the $D=4$ massless conformal free higher spin field theory with
$\mathcal{N}$-extended supersymmetry.

\begin{enumerate}
\item The ${\cal N}=2$ supermultiplet of massless conformal higher spin equations is
simply given by the complexification of the ${\cal N}=1$ supermultiplet.
\item For ${\cal N}=4, 8$ no tensorial space generalizations of  the Maxwell, Rarita-Schwinger or linearized conformal gravity 
equations appear. It is shown that ${\cal N}\ge 4$ supermultiplets are built from the scalar and spinor fields in tensorial space which obey the standard higher spin
equations in their tensorial space version. However, some of these appear in the basic superfields under derivatives, so that the ${\cal N}\ge 4$ 
supersymmetric theory is invariant under Peccei--Quinn--like symmetries shifting these fields.
\end{enumerate}

\item We have obtained and studied the equations of motion of multiple M$0$--brane system (mM0) which follow from the 
covariant supersymmetric and $\kappa$--symmetric mM$0$ action proposed in \cite{mM0:action}.

\begin{enumerate}
\item We have found that the mM$0$ action is invariant under an interesting reminiscent of the so--called $K_9$ gauge symmetry 
which is necessary to find the final form of the bosonic equations of motion for the center of energy 
coordinate functions.
\item We have found that, generically, there exists the 'backreaction', the influence of the relative motion on the motion of the center of 
energy, the most important effects of which are that the generic center of energy motion of mM$0$ system is characterized by a 
nonvanishing effective mass $M$ constructed from the matrix field describing the relative motion. Furthermore, when the relative motion is not 
in its ground state, the center of energy velocity and the canonical momentum conjugate to the center of energy coordinate function 
are oriented in different directions of the 11D spacetime. These two effects disappear when $M^2=0$.
\item We have shown that all the mM$0$ BPS states are 1/2 BPS and have the same properties as BPS states of single 
M$0$--brane. In particular, the effective mass of the center of energy motion vanishes for the BPS states. In other words, all the 
supersymmetric purely bosonic solutions of mM$0$ equations preserve just ${1\over 2}$ of 11D supersymmetry, have the relative motion 
sector in its ground state so that all the equations of the center of energy motion acquire the same form as equations for single 
M$0$--brane. 
\end{enumerate}

\end{enumerate}

\appendix
 \chapter{Notation, conventions and some useful formulae in $D=4$}
\thispagestyle{chapter}
\renewcommand{\theequation}{A.\arabic{equation}}
\setcounter{equation}{0}

\initial{I}n chapters \ref{chapterscalar} \ref{chaptersugra+membrane} we use mostly minus Minkowski metric $\eta^{ab}=diag (1,-1,-1,-1)$ and complex Weyl spinor notation.
$D=4$ vector and spinor indices are denoted by symbols from the beginning of Latin and Greek alphabets,
$a,b,c= 0,1,2,3$, $\alpha ,\beta , \gamma =1,2$, $\dot{\alpha}, \dot{\beta}, \dot{\gamma}=1,2$.
In particular, the coordinates of flat $D=4$, ${\cal N}=1$ superspace $\Sigma^{(4|4)}$ are denoted, respectively, by
$x^a$ and  $\theta^\alpha$,  $\bar{\theta}{}^{\dot{\alpha}} =(\theta^\alpha)^*$. 
 The contraction the spinorial indices are raised and lowered by the unit antisymmetric tensors 
$\epsilon^{\alpha\beta}=-\epsilon^{\beta\alpha}=i\sigma^2\equiv \left(\begin{matrix}0 & 1\cr -1 & 0\end{matrix}\right) $, $\epsilon^{\dot{\alpha}\dot{\beta}}=-\epsilon^{\dot{\beta}\dot{\alpha}}$ 
and their inverse
 $\epsilon_{\alpha\beta}=-\epsilon_{\beta\alpha}$, $\epsilon_{\dot{\alpha}\dot{\beta}}=-\epsilon_{\dot{\beta}\dot{\alpha}}$. 
In flat superspace (or within the WZ gauge) this implies $\theta_\alpha= \epsilon_{\alpha\beta}\theta^\beta$, $\theta^\alpha= \epsilon^{\alpha\beta}\theta_\beta$, etc. 
However, to get $\partial^\alpha \theta_\beta =\delta_\beta{}^\alpha$ simultaneously with 
$\partial_\alpha \theta^\beta =\delta_\alpha{}^\beta$ we have to assume that for the derivatives over the fermionic variables
 $\partial^\alpha = - \epsilon^{\alpha\beta}\partial_{\beta}$ so that, when we rise the spinorial index of the covariant fermionic 
derivative (\ref{Dalpha=}), we arrive at  $D^\alpha:=  \epsilon^{\alpha\beta}D_{\beta}=
 - \partial^\alpha - i (\bar{\theta}\tilde{\sigma}{}^a)^\alpha \partial_a$.
In curved superspace we need to introduce world supervector indices $M,N,...$ The coordinates of curved superspace are denoted by $Z^{M}=(x^{\mu},\theta^{\check{\underline{\alpha}}})$ 
with $\check{\underline{\alpha}}=1,2,3,4$; clearly beyond the WZ guage these $\check{\underline{\alpha}}$ are not spinorial indices so we 
do not use their splitting on $\check{\alpha}$ and $\dot{\check{\alpha}}$ (which may however, be useful in the prepotential approach, see \cite{OS78}).
 The star superscript $^*$ denotes complex conjugation of 
bosonic variables and involution on Grassmann algebra (see \cite{Ogievetsky:1975nu} and refs. therein); in practical terms this 
implies that  $(\theta^\alpha\hat{\theta}{}^\beta)^*= \hat{\bar{\theta}}{}^{\dot{\beta}} \bar{\theta}{}^{\dot{\alpha}} =-
\bar{\theta}{}^{\dot{\alpha}}\hat{\bar{\theta}}{}^{\dot{\beta}}$. Then, to keep the plus sign in $(\theta^2)^*= \bar{\theta}{}^2$ 
with $(\theta)^2:=\theta^\alpha\theta_\alpha$, we have to define $\bar{\theta}{}^2:= \bar{\theta}_{\dot{\alpha}} \bar{\theta}{}^{\dot{\alpha}}$. 
The consistency of the Grassmann algebra involution requires that $(\partial_\alpha)^*\equiv  \left({\partial\over \partial {\theta}{}^{{\alpha}}}\right)^*=  -\bar{\partial}_{\dot{\alpha}}\equiv - {\partial\over \partial \bar{\theta}{}^{\dot{\alpha}}}$. 
The covariant spinor derivative defined in (\ref{Dalpha=}) are related by $(D_\alpha)^*=  -\bar{D}_{\dot{\alpha}}$; 
$\bar{D}\bar{D}:= \bar{D}_{\dot{\alpha}}\bar{D}{}^{\dot{\alpha}}= (DD)^*$ where $DD:=D^\alpha D_\alpha$.

The list of properties of relativistic Pauli matrices $\sigma^a_{\beta\dot{\alpha}}= \epsilon_{\beta\alpha} \epsilon_{\dot{\alpha}\dot{\beta}}
 \tilde{\sigma}{}^{a\dot{\beta}\alpha}$ include
\begin{eqnarray}\label{sasb=}
& \sigma^a\tilde{\sigma}{}^b =\eta^{ab} +{i\over 2}\epsilon^{abcd}\sigma_c\tilde{\sigma}_d\; ,\qquad \tilde{\sigma}{}^a{\sigma}^b =\eta^{ab} -{i\over 2}\epsilon^{abcd}\tilde{\sigma}_c{\sigma}_d\; ,\qquad
\\ \label{sab=}
& \sigma^{ab}:={1\over 2}(\sigma^a\tilde{\sigma}{}^b-\sigma^b\tilde{\sigma}{}^a)= {i\over 2}\epsilon^{abcd}\sigma_{cd}\; , \qquad \\ \label{tsab=} & \tilde{\sigma}{}^{ab} :={1\over 2}(
\tilde{\sigma}{}^a{\sigma}^b-\tilde{\sigma}{}^b{\sigma}^a)= -{i\over 2}\epsilon^{abcd}\tilde{\sigma}_{cd}\; ,  \qquad \\
\label{sss=}
& \sigma^{abc}= -i \epsilon^{abcd}{\sigma}_d\; .  \qquad
\end{eqnarray}

\bigskip

The 3-dimensional worldvolume vector indices are denoted by symbols from the middle of Latin alphabet. In particular, the 
local coordinates of the supermembrane worldvolume $W^3$ are denoted by $\xi^m$ with $m=0,1,2$.
The worldvolume Hodge star (denoted by $*$ in the line) operation is defined as  in (\ref{*Ea:=}),
\begin{eqnarray}\label{*Ea:=APP}
*\hat{E}^a:= {1\over 2}d\xi^m\wedge d\xi^n\sqrt{g}\epsilon_{mnk}g^{kl}\hat{E}_l^a \; . \qquad
\end{eqnarray}
In our conventions $d\xi^m\wedge d\xi^n\wedge d\xi^k= - \epsilon^{mnk} d^3\xi \equiv  \epsilon^{knm} d^3\xi$
so that
\begin{eqnarray}\label{*EaEa:=}
*\hat{E}_a\wedge \hat{E}^a =- 3 d^3\xi \sqrt{g} \; , \qquad
*\hat{E}_a\wedge \delta \hat{E}^a =-  d^3\xi \sqrt{g} \hat{E}_{ma} g^{mn}\delta \hat{E}_n^a\;  \qquad
\end{eqnarray}
and
\begin{eqnarray}\label{v*EaEa:=}
\delta(*\hat{E}_a\wedge \hat{E}^a) =3
*\hat{E}_a\wedge \delta \hat{E}^a \; . \qquad
\end{eqnarray}
The superspace generalization of Dirac delta function reads $$\delta^8 (z):= {1\over 16}(\theta)^2(\bar{\theta})^2 \delta^4 (x)$$
and obeys
$$
\int d^8 z\delta^8 (z-\hat{z}) f(z) = f(\hat{z}) \; , \qquad \int d^8 z\delta^8 (z)= \int d^4x \bar{D}\bar{D}DD \delta^8 (z)=1 $$

 \chapter{Supermembrane current superfield which enters the scalar multiplet equations}
\label{appendixB}
\thispagestyle{chapter}
\renewcommand{\theequation}{B.\arabic{equation}}
\setcounter{equation}{0}

\initial{T}he leading component of the Nambu-Goto current superfield (\ref{JNG=}) 
\begin{eqnarray}
&&J^{NG}(Z) = - {1\over 4} \, \int d^3 \xi \, \sqrt{g} \, \sqrt{{\hat{\Phi}\over \hat{\bar{\Phi}}}}\; {D}{D} \delta^8(z-\hat{z})  - {1\over 4} \, \int d^3 \xi \, \sqrt{g}\, \sqrt{{\hat{\bar{\Phi}}\over\hat{ \Phi}}}\; \bar{D}\bar{D}\delta^8(z-\hat{z}) \, , \qquad
\end{eqnarray}
reads
\begin{eqnarray}\label{JNG0=App}
 J^{NG}\vert_{0} &=& + {1\over 16} \,\sqrt{{\phi\over \bar{\phi}}}\; \int d^3 \xi \, \sqrt{g} \, \hat{\bar{\theta}}\hat{\bar{\theta}} \delta^4(x-\hat{x})  + {1\over 16} \sqrt{{\bar{\phi}\over \phi}}\; \int d^3 \xi \, \sqrt{g}\, \hat{{\theta}}\hat{{\theta}}\delta^4(x-\hat{x})   \;  \qquad \end{eqnarray}
The general expression for the fermionic derivative of $ J^{NG}$
\begin{eqnarray} \label{bDJNG=App}
 \bar{D}_{\dot{\alpha}} J^{NG} &=& - {1\over 4} \, \int d^3 \xi \, \sqrt{g}   \,\sqrt{{\hat{\Phi}\over \hat{\bar{\Phi}}}}\; \bar{D}_{\dot{\alpha}}DD \delta^8 (Z-\hat{Z})  \; . \qquad
 \end{eqnarray}
 so that
 \begin{eqnarray} \label{bDJNG0=App}
&\bar{D}_{\dot{\alpha}} J^{NG}\vert_{0} = {1\over 8}\,\sqrt{{\phi\over \bar{\phi}}}\; \int d^3 \xi \, \sqrt{g} \, \hat{\bar{\theta}}_{\dot{\alpha}}\delta^4 (x-\hat{x}) -   {i\over 16} \, \int d^3 \xi \, \sqrt{g}\,\sqrt{{\hat{\phi}\over \hat{\bar{\phi}}}}\; (\sigma^a\hat{{\theta}})_{\dot{\alpha}} (\hat{\bar{\theta}})^2\, \partial_a\delta^4 (x-\hat{x})\,
 \; . \qquad
 \end{eqnarray}

 Furthermore, as
\begin{eqnarray} \label{bDbDJNG=App}
 \bar{D}\bar{D} J^{NG} &=& - {1\over 4}  \int d^3 \xi \, \sqrt{g} \, \,\sqrt{{\hat{\Phi}\over \hat{\bar{\Phi}}}}\; \bar{D}\bar{D} DD \delta^8 (Z-\hat{Z})    \qquad
 \end{eqnarray}
we find 
\begin{eqnarray} \label{bDbDJNG0=App}
&& \bar{D}\bar{D} J^{NG}\vert_0 = \qquad \nonumber \\ &&= \!\!  \int d^3 \xi \, \sqrt{g} \,\sqrt{{\hat{\phi}\over \hat{\bar{\phi}}}}\; \left(-{1\over 4}\delta^4 (x-\hat{x}) -{i\over 4} (\hat{{\theta}}\sigma^a\hat{\bar{\theta}}) \partial_a\delta^4 (x-\hat{x})+{1\over 16} (\hat{{\theta}})^2(\hat{\bar{\theta}})^2 \Box\delta^4 (x-\hat{x}) \right) = \qquad \nonumber \\
&&= -{1\over 4} \,\sqrt{{{\phi}\over {\bar{\phi}}}}\; \int d^3 \xi \, \sqrt{g} \delta^4 (x-\hat{x})
-  \qquad \nonumber \\ &&
-{i\over 4} \,\sqrt{{{\phi}\over {\bar{\phi}}}}\; \int d^3 \xi \, \sqrt{g}  (\hat{{\theta}}\sigma^a\hat{\bar{\theta}}) \left(\partial_a\delta^4 (x-\hat{x})+ \left({\partial_a\phi \over 2\phi }  - {\partial_a\bar{\phi}\over 2 \bar{\phi}}  \right)\delta^4 (x-\hat{x})   \right) +{\cal O}(f^4)\; . \qquad
  \end{eqnarray}
One can also write  Eq. (\ref{bDbDJNG0=App}) in the equivalent but more compact form of
\begin{eqnarray} \label{bDbDJNG0=App-2} \bar{D}\bar{D} J^{NG}\vert_0&=& -{1\over 4}\,\sqrt{{\phi\over \bar{{\phi}}}}\; \int d^3 \xi \, \sqrt{g} \, \delta^4 (x-\hat{x})  + {\cal O}(f^2)_{[\bar{D}\bar{D} J^{NG}]} + {\cal O}(f^4)_{[\bar{D}\bar{D} J^{NG}]}  \; ,  \qquad
 \end{eqnarray}
  where
 \begin{eqnarray} \label{bDbDJNG0=App(f2)}
& {\cal O}(f^2)_{[\bar{D}\bar{D} J^{NG}]} = -{i\over 4} \,\sqrt{{{\phi}\over {\bar{\phi}}}}\; \int d^3 \xi \, \sqrt{g}  (\hat{{\theta}}\sigma^a\hat{\bar{\theta}}) \left(\partial_a\delta^4 (x-\hat{x})+ \left({\partial_a\phi \over 2\phi }  - {\partial_a\bar{\phi}\over 2 \bar{\phi}}  \right)\delta^4 (x-\hat{x})   \right)  \; . \qquad
 \end{eqnarray}

 The leading term of the second, Wess--Zumino (WZ) contribution to the supermembrane current,
\begin{eqnarray}
J^{WZ}(Z) &=& \left(2i \int_{W^3} \hat{E}^c\wedge \hat{E}{}^\alpha \wedge \hat{E}{}^{\dot\alpha} \sigma_{c\alpha\dot\alpha}  + \right. \nonumber \\ && \left. + {1\over 2}  \int_{W^3} \hat{E}^c\wedge \hat{E}^b\wedge \hat{E}{}^\alpha \sigma_{bc\alpha}{}^{\beta}D_\beta  -  {1\over 2}  \int_{W^3} \hat{E}^c\wedge \hat{E}^b \wedge \hat{E}{}^{\dot\alpha} \tilde{\sigma}_{bc}{}^{\dot{\beta}}{}_{\dot\alpha} \bar{D}_{\dot{\beta}} - \right. \nonumber \\ && \left. - {1\over 4!}  \int_{W^3} \hat{E}^c\wedge \hat{E}^b \wedge \hat{E}^a\epsilon_{abcd} \tilde{\sigma}^{d\dot{\alpha}\alpha} [D_\alpha, \bar{D}_{\dot{\alpha}}] \right)  \delta^8(z-\hat{z})\; \qquad
\end{eqnarray}
reads
\begin{eqnarray}\label{JWZ0=App}
  J^{WZ}(x) := J^{WZ}(Z) \vert_0&=&  {1\over 48}  \int_{W^3} \hat{E}^c\wedge \hat{E}^b \wedge \hat{E}^a\epsilon_{abcd}
 \hat{{\theta}}{\sigma}^{d} \hat{\bar{\theta}}\, \delta^4(x-\hat{x})
 - \nonumber \\ &&  - {1\over 16} \int_{W^3} \hat{E}^c\wedge \hat{E}^b\wedge d\hat{{\theta}}{}^\alpha \sigma_{bc\alpha}{}^{\beta}\hat{{\theta}}_\beta \, \hat{\bar{\theta}}\hat{\bar{\theta}} \, \delta^4(x-\hat{x})  -  \nonumber \\ &&  -  {1\over 16} \int_{W^3} \hat{E}^c\wedge \hat{E}^b \wedge d\hat{\bar{\theta}}{}^{\dot\alpha} \tilde{\sigma}_{bc}{}^{\dot{\beta}}{}_{\dot\alpha} \hat{\bar{\theta}}_{\dot{\beta}} \hat{{\theta}}\hat{{\theta}}\,  \delta^4(x-\hat{x}) +  \nonumber \\ && + {i\over 8}\int_{W^3} \hat{E}^c\wedge d\hat{{\theta}}{}^\alpha \wedge d\hat{\bar{\theta}}{}^{\dot\alpha} \sigma_{c\alpha\dot\alpha}  \hat{{\theta}}\hat{{\theta}}\, \hat{\bar{\theta}}\hat{\bar{\theta}} \, \delta^4(x-\hat{x}) = \qquad\nonumber \\ &=&  {1\over 48}  \int_{W^3} \hat{E}^c\wedge \hat{E}^b \wedge \hat{E}^a\epsilon_{abcd}
 \hat{{\theta}}{\sigma}^{d} \hat{\bar{\theta}}\, \delta^4(x-\hat{x}) +{\cal O}(f^4)
 \; . \qquad
\end{eqnarray}

Its closest fermionic partners are
\begin{eqnarray}\label{bDJWZ0=App}
 {D}_{{\alpha}}J^{WZ} \vert_0 &=&  -{1\over 48}  \int_{W^3} \hat{E}^c\wedge \hat{E}^b \wedge \hat{E}^a\epsilon_{abcd}
 ({\sigma}^{d}\hat{\bar{\theta}})_{{\alpha}}\, \delta^4(x-\hat{x})
  +\nonumber \\
  &&   + {1\over 16}  \int_{W^3} \hat{E}^c\wedge \hat{E}^b\wedge \left( 2(d\hat{\bar{\theta}} \tilde{\sigma}_{bc}\hat{\bar{\theta}}) \, \hat{{\theta}}_{{\alpha}} \,+   d\hat{{\theta}}{}^\beta {\sigma}_{bc}{}_{\beta {\alpha}} (\hat{{\theta}})^2\, \right) \delta^4(x-\hat{x}) + \nonumber \\
  && + {i\over 96}  \int_{W^3} \hat{E}^c\wedge \hat{E}^b \wedge \hat{E}^a\epsilon_{abcd}
 ({\sigma}^{de}\hat{{\theta}})_{{\alpha}}  (\hat{\bar{\theta}})^2 \partial_e  \delta^4(x-\hat{x})+ \nonumber \\ &&   + {i\over 32}\int_{W^3} \hat{E}^c\wedge \hat{E}^b\wedge d{\hat{\bar{\theta}}}{}^{\dot{\alpha}}(\sigma^a\tilde{\sigma}_{bc})_{\alpha\dot{\alpha}} (\hat{{\theta}})^2(\hat{\bar{\theta}})^2
 \partial_a \delta^4(x-\hat{x}) -  \nonumber \\ && - {i\over 4}\int_{W^3} \hat{E}^c\wedge d\hat{{\theta}}{}^\beta \wedge d\hat{\bar{\theta}}{}^{\dot\beta}\, \sigma_{c\beta\dot\beta}  \hat{{\theta}}_{\alpha} \,  (\hat{\bar{\theta}})^2 \,  \, \delta^4(x-\hat{x})  =  \qquad
 \nonumber \\ &=&  -{1\over 48}  \int_{W^3} \hat{E}^c\wedge \hat{E}^b \wedge \hat{E}^a\epsilon_{abcd}
 ({\sigma}^{d}\hat{\bar{\theta}})_{{\alpha}}\, \delta^4(x-\hat{x}) +{\cal O}(f^3)
 \; . \qquad
\end{eqnarray}
\begin{eqnarray}\label{bDJWZ0=App}
 \bar{D}_{\dot{\alpha}}J^{WZ} \vert_0 &=&  {1\over 48}  \int_{W^3} \hat{E}^c\wedge \hat{E}^b \wedge \hat{E}^a\epsilon_{abcd}
 (\hat{{\theta}}{\sigma}^{d})_{\dot{\alpha}}\, \delta^4(x-\hat{x})
  -\nonumber \\
  &&   - {1\over 16}  \int_{W^3} \hat{E}^c\wedge \hat{E}^b\wedge \left( 2(d\hat{{\theta}} \sigma_{bc}\hat{{\theta}}) \, \hat{\bar{\theta}}_{\dot{\alpha}} \,+   (d\hat{\bar{\theta}}\tilde{\sigma}_{bc})_{\dot{\alpha}} (\hat{{\theta}})^2\, \right) \delta^4(x-\hat{x}) - \nonumber \\
  && - {i\over 96}  \int_{W^3} \hat{E}^c\wedge \hat{E}^b \wedge \hat{E}^a\epsilon_{abcd}
 (\hat{\bar{\theta}}\tilde{\sigma}^{de})_{\dot{\alpha}}  (\hat{{\theta}})^2 \partial_e  \delta^4(x-\hat{x})- \nonumber \\ &&   - {i\over 32}\int_{W^3} \hat{E}^c\wedge \hat{E}^b\wedge (d\hat{{\theta}}\sigma_{bc}\sigma^a)_{\dot{\alpha}} (\hat{{\theta}})^2(\hat{\bar{\theta}})^2
 \partial_a \delta^4(x-\hat{x}) +  \nonumber \\ && + {i\over 4}\int_{W^3} \hat{E}^c\wedge d\hat{{\theta}}{}^\beta \wedge d\hat{\bar{\theta}}{}^{\dot\beta}\, \sigma_{c\beta\dot\beta}  \hat{\bar{\theta}}_{\dot\alpha} \,  (\hat{{\theta}})^2 \,  \, \delta^4(x-\hat{x})  =  \qquad
 \nonumber \\ &=&  {1\over 48}  \int_{W^3} \hat{E}^c\wedge \hat{E}^b \wedge \hat{E}^a\epsilon_{abcd}
 (\hat{{\theta}}{\sigma}^{d})_{\dot{\alpha}}\, \delta^4(x-\hat{x}) +{\cal O}(f^3)
 \; . \qquad
\end{eqnarray}

Then, as far as
\begin{eqnarray}\label{DDJWZ=}
{D}{D}J^{WZ}(Z) &=& 2i \int_{W^3} \hat{E}^c\wedge \hat{E}{}^\alpha \wedge \hat{E}{}^{\dot\alpha} \sigma_{c\alpha\dot\alpha} {D}{D} \delta^8(Z-\hat{Z})  +  \nonumber \\
&& - {1\over 2}  \int_{W^3} \hat{E}^c\wedge \hat{E}^b\wedge \hat{\bar{E}}{}^{\dot\alpha} \tilde{\sigma}_{bc}{}^{\dot\beta}{}_{\dot\alpha}
\left(\bar{D}_{\dot\beta}{D}{D}+ 4i \sigma^a_{\beta\dot\beta} \partial_a {D}^{{\beta}}\right) \delta^8(Z-\hat{Z}) +  \nonumber \\ &&  + {i\over 6}  \int_{W^3} \hat{E}^c\wedge \hat{E}^b \wedge \hat{E}^a\epsilon_{abcd} \partial^{d}{D}{D} \delta^8(Z-\hat{Z}) \; , \qquad
\\ \label{bDbDJWZ=}
\bar{D}\bar{D}J^{WZ}(Z) &=& 2i \int_{W^3} \hat{E}^c\wedge \hat{E}{}^\alpha \wedge \hat{E}{}^{\dot\alpha} \sigma_{c\alpha\dot\alpha} \bar{D}\bar{D} \delta^8(Z-\hat{Z})  +  \nonumber \\
&& + {1\over 2}  \int_{W^3} \hat{E}^c\wedge \hat{E}^b\wedge \hat{E}{}^\alpha \sigma_{bc\alpha}{}^{\beta}
\left(D_\beta \bar{D}\bar{D}- 4i \sigma^a_{\beta\dot\beta} \partial_a \bar{D}^{\dot{\beta}}\right) \delta^8(Z-\hat{Z}) -  \nonumber \\ &&  - {i\over 6}  \int_{W^3} \hat{E}^c\wedge \hat{E}^b \wedge \hat{E}^a\epsilon_{abcd} \partial^{d}\bar{D}\bar{D} \delta^8(Z-\hat{Z}) \; , \qquad
\end{eqnarray}
one finds that
\begin{eqnarray}\label{DDJWZ0=}
{D}{D}J^{WZ}\vert_0 &=& {1\over 4}\int\limits_{W^3} \hat{E}^c\wedge \hat{E}^b\wedge d\hat{\bar{\theta}} \tilde{\sigma}_{bc}\hat{\bar{\theta}} \delta^4(x-\hat{x}) -
 {i\over 4!}  \int\limits_{W^3} \hat{E}^c\wedge \hat{E}^b \wedge \hat{E}^a\epsilon_{abcd} (\hat{\bar{\theta}})^2\partial^d\delta^4(x-\hat{x}) -
\nonumber \\ && -{i\over 2} \int_{W^3} \hat{E}^c\wedge d\hat{\theta}{}^\alpha \wedge d\hat{\bar{\theta}}{}^{\dot\alpha} \sigma_{c\alpha\dot\alpha} (\hat{\bar\theta}){}^2 \delta^4(x-\hat{x})
  +  \nonumber \\
&& +{i\over 8} \int_{W^3} \hat{E}^c\wedge \hat{E}^b\wedge d\hat{\bar{\theta}}{}^{\dot{\alpha}} (\hat{{\theta}}\sigma^a\tilde{\sigma_{bc}})_{\dot{\alpha}}\,  (\hat{\bar\theta}){}^2 \partial_a\delta^4(x-\hat{x}) =  \nonumber \\  &=& {1\over 4}\int\limits_{W^3} \hat{E}^c\wedge \hat{E}^b\wedge d\hat{\bar{\theta}} \tilde{\sigma}_{bc}\hat{\bar{\theta}} \delta^4(x-\hat{x}) -
 \nonumber \\ && -{i\over 4}  \int\limits_{W^3} \hat{E}^c\wedge \hat{E}^b \wedge \hat{E}^a\epsilon_{abcd} (\hat{\bar{\theta}})^2\partial^d\delta^4(x-\hat{x})
  +  {\cal O}(f^4)\; . \qquad
\end{eqnarray}
\begin{eqnarray}\label{bDbDJWZ0=}
\bar{D}\bar{D}J^{WZ}\vert_0 &=& {1\over 4}\int\limits_{W^3} \hat{E}^c\wedge \hat{E}^b\wedge d\hat{\theta} \sigma_{bc}\hat{\theta} \delta^4(x-\hat{x}) +
 {i\over 4!}  \int\limits_{W^3} \hat{E}^c\wedge \hat{E}^b \wedge \hat{E}^a\epsilon_{abcd} (\hat{\theta})^2\partial^d\delta^4(x-\hat{x}) -
\nonumber \\ && -{i\over 2} \int_{W^3} \hat{E}^c\wedge d\hat{\theta}{}^\alpha \wedge d\hat{\bar{\theta}}{}^{\dot\alpha} \sigma_{c\alpha\dot\alpha} (\hat{\theta}){}^2 \delta^4(x-\hat{x})
  -  \nonumber \\
&& -{i\over 8} \int_{W^3} \hat{E}^c\wedge \hat{E}^b\wedge d\hat{\theta} \sigma_{bc}\sigma^a\hat{\bar{\theta}}\,  (\hat{\theta}){}^2 \partial_a\delta^4(x-\hat{x}) =  \nonumber \\  &=& {1\over 4}\int\limits_{W^3} \hat{E}^c\wedge \hat{E}^b\wedge d\hat{\theta} \sigma_{bc}\hat{\theta} \delta^4(x-\hat{x}) +
\nonumber \\ && +
 {i\over 4}  \int\limits_{W^3} \hat{E}^c\wedge \hat{E}^b \wedge \hat{E}^a\epsilon_{abcd} (\hat{\theta})^2\partial^d\delta^4(x-\hat{x})+ {\cal O}(f^4)\; . \qquad
\end{eqnarray}
To analyze the structure of the auxiliary field equation one needs also to know
\begin{eqnarray}\label{DbDJWZ0=}
[{D}_\alpha\, , \, \bar{D}_{\dot\beta}]J^{WZ}\vert_0 &=& -{1\over 4!}\int\limits_{W^3}
\hat{E}^c\wedge \hat{E}^b\wedge \hat{E}^a \epsilon_{abcd}\sigma^d_{\alpha\dot{\beta}} \delta^4(x-\hat{x}) + {\cal O}(f^2) \; , \qquad
\end{eqnarray}
where
\begin{eqnarray}
\label{DbDJWZ0=f2}
{\cal O}(f^2) &=& {i\over 4!}\int\limits_{W^3}
\hat{E}^c\wedge \hat{E}^b\wedge \hat{E}^a \epsilon_{abcd}\left( (\sigma^{de}\hat{\theta})_{\alpha}\hat{\bar{\theta}}_{\dot{\beta}} + \hat{{\theta}}_{\alpha}(\hat{\bar{\theta}}\tilde{\sigma}{}^{de})_{\dot{\beta}}\right) \partial_e\delta^4(x-\hat{x}) +
\nonumber \qquad \\ &&
 + {1\over 4}  \int_{W^3} \hat{E}^c\wedge \hat{E}^b\wedge \left(( \sigma_{bc}d\hat{\theta})_\alpha \hat{\bar{\theta}}_{\dot\beta}\,  - (d\hat{\bar{\theta}} \tilde{\sigma}_{bc})_{\dot\beta} \hat{{\theta}}_\alpha\,  \right)\delta^4(x-\hat{x}) + {\cal O}(f^4)\; , \qquad \\
\label{DbDJWZ0=f4}
{\cal O}(f^4) &=& {1\over 2\cdot 4!}\int\limits_{W^3}
\hat{E}^c\wedge \hat{E}^b\wedge \hat{E}^a (\hat{\theta})^2(\hat{\bar{\theta}})^2 \epsilon_{abcd}\left( \sigma^d_{\alpha\dot{\beta}}
\Box
\delta^4(x-\hat{x})  - \sigma^e_{\alpha\dot{\beta}} \partial_e \partial^d \delta^4(x-\hat{x})\right) + \nonumber \\
&& +{i\over 8} \int_{W^3} \hat{E}^c\wedge \hat{E}^b\wedge (d\hat{{\theta} }{\sigma}_{bc}{\sigma}^a)_{\dot\beta} \hat{{\theta}}_{\alpha} \,  (\hat{\bar{\theta}}){}^2 \partial_a\delta^4(x-\hat{x})  + \nonumber \\
&& +{i\over 8} \int_{W^3} \hat{E}^c\wedge \hat{E}^b\wedge (d\hat{\bar{\theta} } \tilde{\sigma}_{bc}\tilde{\sigma}^a)_{\alpha} \hat{\bar{\theta}}_{\dot{\beta}} \,  (\hat{\theta}){}^2 \partial_a\delta^4(x-\hat{x}) -   \nonumber \\ &&
 -{i} \int_{W^3} \hat{E}^c\wedge d\hat{\theta}{}^\gamma \wedge d\hat{\bar{\theta}}{}^{\dot\gamma} \sigma_{c\gamma\dot\gamma} \hat{\theta}_{\alpha} \hat{\bar{\theta}}_{\dot{\beta}} \delta^4(x-\hat{x})
 \; . \qquad
\end{eqnarray}

 \chapter{On admissible variations of superfield supergravity}
\label{appendiceC}
\thispagestyle{chapter}
\renewcommand\theequation{C.\arabic{equation}}
\setcounter{equation}{0}

\initial{T}he admissible variations of supervielbein are the variation preserving the superspace constraints. In the case of minimal 
supergravity that read \cite{WZ78,BdAIL03}
\begin{eqnarray}\label{varEa} \delta E^{a} & =& E^a (\Lambda (\delta ) + \bar{\Lambda} (\delta ))
 - {1\over 4} E^b \tilde{\sigma}_b^{ \dot{\alpha} {\alpha} }
[{\cal D}_{{\alpha}}, \bar{{\cal D}}_{\dot{\alpha}}] \delta H^a +
 i E^{\alpha} {\cal D}_{{\alpha}}\delta H^a   - i \bar{E}^{\dot{\alpha}}\bar{{\cal
D}}_{\dot{\alpha}} \delta H^a \; ,\quad \\ \label{varEal}
 \delta E^{\alpha} & = & E^a \Xi_a^{\alpha}(\delta ) +
E^{\alpha} \Lambda (\delta ) + {1\over 8} \bar{E}^{\dot{\alpha}} R \sigma_a{}_{\dot{\alpha}}{}^{\alpha}
\delta H^a \; ,  \end{eqnarray}
where
\begin{eqnarray}
\label{2Lb+*Lb} & 2\Lambda (\delta ) + \bar{\Lambda} (\delta )  = {1\over 4} \tilde{\sigma}_a^{
\dot{\alpha} {\alpha} } {\cal D}_{{\alpha}} \bar{{\cal D}}_{\dot{\alpha}}\delta H^a + {1\over 8} G_a
\delta H^a +   3 ( {\cal D}{\cal D}- \bar{R})\delta {\cal U} .
 \end{eqnarray}
The explicit expression for $\Xi_a^{\alpha}(\delta )$ in (\ref{varEal}), as well as the admissible variations of the spin connection 
superform, can be found in \cite{BdAIL03}.

The variation of the closed 4--form (\ref{H4=HL+HR}), (\ref{H4L}) reads \cite{IB+CM:2011}
\begin{eqnarray} \label{vH4=}  \delta H_4 &=& {1\over 2} E^b\wedge  E^\alpha \wedge  E^\beta\wedge  E^\gamma
\sigma_{ab\; (\alpha\beta} D_{\gamma )}\delta H^a  - {1\over 2} E^b\wedge  E^\alpha \wedge  E^\beta\wedge   \bar{E}{}^{\dot{\gamma}}
\sigma_{ab\; \alpha\beta} \bar{D}_{\dot\gamma}\delta H^a + c.c. - \qquad
 \nonumber \\ && - {i\over 2} E^b\wedge E^a \wedge E^\alpha \wedge E^\beta
\left(\sigma_{ab\; \alpha\beta}\left(2\Lambda(\delta) + \bar{\Lambda}(\delta)\right)+ {1\over 4}\sigma_{c[a|\; \alpha\beta}\tilde{\sigma}_{|b]}{}^{\dot{\gamma}\gamma}[D_\gamma, \bar{D}_{\dot{\gamma}}]\delta H^c
\right) + c.c. + \nonumber \\ && + {i\over 16} E^b\wedge  E^a \wedge  {E}{}^{\alpha} \wedge \bar{E}{}^{\dot\beta} (R\sigma_{ab}\tilde{\sigma}_{c} -\bar{R}\sigma_c\tilde{\sigma}_{ab})_{\alpha\dot\beta}\delta H^c + \propto  \; E^c \wedge E^b\wedge E^a  \; .
\end{eqnarray}
The conditions of that
$\delta H_4$ can be expressed in terms of the variation of the 3--form potential $\delta C_3$,
 \begin{eqnarray} \label{vH4=dvC3}
\delta H_4 = d(\delta C_3)
 \;  \quad
\end{eqnarray}
with $\delta C_3$ decomposed on the basic covariant 3--forms, as in Eq. (\ref{vC3:=}), restrict the set of independent
variations by  \cite{IB+CM:2011}
\begin{eqnarray}
 \label{DDcU=vV+}
({\cal D}{\cal D}- \bar{R})\delta {\cal U} = {1\over 12}( {\cal D}{\cal D}- \bar{R})\left(i\delta {V} + {1\over 2}  \bar{{\cal D}}_{\dot{\alpha}}  \delta \bar{{\kappa}}{}^{\dot{\alpha}} \right)
\; .     \;
\end{eqnarray}
This is equivalent to
\begin{eqnarray}
 \label{vcU=vV+K}
\delta {\cal U} = {i\over 12}\delta {V} + {1\over 24}  \bar{{\cal D}}_{\dot{\alpha}}  \delta \bar{{\kappa}}{}^{\dot{\alpha}}+ {i\over 24} {\cal D}_{{\alpha}} \delta {{\nu}}{}^{{\alpha}}\; ,   \qquad
\end{eqnarray}
where  $\delta {{\nu}}{}^{{\alpha}}$ is an additional independent variation (which does not contribute 
to $({\cal D}{\cal D}-\bar{R})\delta {\cal U}$ and, hence, to the variations of supergravity potentials).

Factoring out the gauge transformations, we can write the variation $\delta C_3$, which produces (\ref{vH4=}) through 
(\ref{vH4=dvC3}),  in the form (\ref{vC3:=}) with \cite{IB+CM:2011}
\begin{eqnarray} \label{bff*f=}
\beta_{\alpha\beta\gamma}(\delta)=0= \beta_{\alpha\beta\dot{\gamma}}(\delta)\; , \quad \beta_{{\alpha}\dot{\beta} a}(\delta)  = i\sigma_{a\alpha\dot{\beta}}\delta V \;
\end{eqnarray}
and
\begin{eqnarray} \label{bffb=}
&&\beta_{\alpha\beta a}(\delta) = - \sigma_{ab\; \alpha\beta} (\delta H^b + \tilde{\sigma}^{b\gamma \dot{\gamma}}D_{\gamma }\delta\bar{\kappa}_{\dot{\gamma}})\; ,    \quad
 \\ \label{bfbb=}
&&\beta_{\alpha ab}(\delta) = {1\over 2}\epsilon_{abcd}\sigma^c_{\alpha\dot\alpha}\bar{D}{}^{\dot\alpha }\delta H^d +{1\over 2} \sigma_{ab\; \alpha}{}^{\beta} D_\beta \delta V  -\nonumber\\
&&-{i\over 4} \tilde{\sigma}_{ab}{}^{\dot{\beta}}{}_{\dot{\gamma}} \bar{D}_{\dot\beta}D_\alpha \bar{\kappa}{}^{\dot{\gamma}}  + {i\over 4} \sigma_{ab\; \alpha}{}^{\beta} \bar{D}_{\dot\beta}D_\beta \bar{\kappa}{}^{\dot{\beta}}   \; ,    \quad \\
\label{bbbb=}
&&\beta_{abc}(\delta)  = {i\over 8} \epsilon_{abcd}\left( \left({\bar{\cal D}}{\bar{\cal D}}-
{1\over 2}R \right)\delta H^d - c.c. \right) + {1\over 4} \epsilon_{abcd}G^d\delta V +\nonumber \quad \\  
&& + {1\over 8} \epsilon_{abcd}\tilde{\sigma}{}^{d\dot{\gamma}\gamma}
[{\cal D}_{\gamma}, {\bar{\cal D}}_{\dot{\gamma}}]\delta V
 - {i\over 16}\epsilon_{abcd}\tilde{\sigma}{}^{d\dot{\gamma}\gamma}\left(\left({\cal D}{\cal D}+{5\over 2}\bar{R}\right) {\bar{\cal D}}_{\dot{\gamma}} {\kappa}_{{\gamma}}- c.c.
  \right)\, .  \quad
\end{eqnarray}

The variation of the {\it special} minimal supergravity action reads
\begin{eqnarray}\label{vSGsf=sK}  \delta S_{SG} &=&  {1\over 6} \int
d^8Z E\;  \left[ G_a \; \delta H^a + (R-\bar{R}) i\delta {V}  \right] - \nonumber \\ &&
 - {1\over 12} \int
d^8Z E\; \left(R {\cal D}_{{\alpha}} \delta {{\kappa}}{}^{{\alpha}}  +
\bar{R} \bar{{\cal D}}_{\dot{\alpha}}  \delta \bar{{\kappa}}{}^{\dot{\alpha}}\right)
 \; . \qquad
\end{eqnarray}
Notice that the variations $\delta {{\kappa}}{}^{{\alpha}}$ and $\delta {\bar{\kappa}}{}^{\dot{\alpha}}$ result in 
equations ${\cal D}_{{\alpha}}R=0$ and $\bar{\cal D}_{\dot{\alpha}}R=0$, which are satisfied identically due to the minimal 
(Eq. (\ref{SGeqmR})) or special minimal supergravity equations of motion (Eq. (\ref{SGeqmR+*=0})). 
In the WZ$_{\hat{\theta}=0}$ gauge (\ref{WZgauge=f})--(\ref{thGauge}) it is also relatively easy to check 
that $\delta {{\kappa}}{}^{{\alpha}}$ does not produce any independent equation for the physical fields of 
the interacting system. This observation has allowed us to simplify the discussion in the main text by neglecting the 
existence $\delta {{\kappa}}{}^{{\alpha}}$ variation.

 \chapter{Supermembrane current superfields entering the superfield SUGRA equations}
\thispagestyle{chapter}
\renewcommand{\theequation}{D.\arabic{equation}}
\setcounter{equation}{0}

\initial{T}he variation of the supermembrane action (\ref{Sp=2:=}) with respect to the vector prepotential of supergravity, $\delta H^a$,  
gives us the vector supercurrent of the form
\begin{eqnarray}\label{D1}
J_a&=&
\int\limits_{W^3} {3\over \hat{E}} \hat{E}{}^b \wedge \hat{E}{}^\alpha \wedge \hat{E}{}^\beta\; {\sigma}_{ab \alpha\beta}  \delta^8 (Z-\hat{Z}) - \qquad \nonumber
\\ && -
\int\limits_{W^3} {3i\over \hat{E}} \left( *\hat{E}_a \wedge \hat{E}{}^\alpha + {i\over 2}   \hat{E}{}^b \wedge \hat{E}{}^c \wedge \hat{\bar{E}}_{\dot{\beta}}\epsilon_{abcd} \tilde{\sigma}^{d\dot{\beta}\alpha}\right) {\cal D}_\alpha \delta^8 (Z-\hat{Z})
+c.c -
\nonumber
\\
&& -  \int\limits_{W^3}  {i\over 8\hat{E}}   \, \hat{E}{}^b \wedge \hat{E}{}^c \wedge \hat{E}{}^d \, \epsilon_{abcd} \left( {\cal D}{\cal D}- {1\over 2}\bar{R} \right) \delta^8 (Z-\hat{Z}) + c.c.  + \qquad \nonumber
\\ && + \int\limits_{W^3}  {1\over 4\hat{E}} *\hat{E}_b \wedge \hat{E}{}^b  \, G_a\,  \delta^8 (Z-\hat{Z})  -  \nonumber  \qquad
\\
&& -\int\limits_{W^3}  {1\over 4\hat{E}}\; *\hat{E}_c \wedge \hat{E}{}^b \tilde{\sigma}^{d\dot{\alpha}\alpha} \left( 3\delta_a^c \delta_b^d-  \delta_a^d \delta_b^c \right)[{\cal D}_\alpha , \bar{\cal D}_{\dot{\alpha}}] \delta^8 (Z-\hat{Z})  \; , \quad
\end{eqnarray}
where ${\hat{E}}= {sdet (E_M{}^A(\hat{Z}))}$ and
\begin{eqnarray}\label{d8:=}
 \delta^8(Z):={1\over 16}\,\delta^4(x) \, \theta\theta \, \bar{\theta}\bar{\theta}\; , \qquad \;
\end{eqnarray}
is the superspace delta function which obeys $\int d^8Z\, \delta^8(Z-Z') f(Z) = f(Z')$ for any superfield $f(Z)$.

The supercurrent (\ref{D1}) enters the {\it r.h.s.} of the vector superfield equation
\begin{eqnarray}
 G_a= T_2J_a\; ,
\end{eqnarray}
which follows from the action (\ref{Sp=2:=}) of the supergravity---supermembrane interacting system.

The scalar superfield equation of the interacting system, which is obtained by varying the interacting action (\ref{Sint=SG+Sp2}) with respect to  the real scalar prepotential of special minimal supergravity, $\delta S/\delta V=0$,  reads
\begin{eqnarray}
R-\bar{R}= -i T_2{\cal X}
\end{eqnarray}
where
\begin{eqnarray}
{\cal X}& =&  {6i\over {E}}  \int\limits_{W^3} \hat{E}^a \wedge \hat{E}{}^\alpha  \wedge \hat{\bar{E}}{}^{\dot{\alpha}}\, {\sigma}^a_{\alpha\dot{\alpha}}\; \delta^8 (Z-\hat{Z}) -
\nonumber  \qquad
\\  && - {3\over 2}\int\limits_{W^3} { \hat{E}{}^b \wedge \hat{E}{}^a \wedge \hat{E}{}^\alpha \over \hat{E}}   \; {\sigma}_{ab \alpha}{}^{{\beta}} {\cal D}_{\beta} \delta^8 (Z-\hat{Z})
+c.c + \nonumber  \;
\\ && +  \int\limits_{W^3}  {   \, \hat{E}{}^b \wedge \hat{E}{}^c \wedge \hat{E}{}^d \over 8 \hat{E}}\, \epsilon_{abcd} \tilde{\sigma}^{a\dot{\alpha}\alpha} [{\cal D}_\alpha , \bar{\cal D}_{\dot{\alpha}}] \delta^8 (Z-\hat{Z})+\;
\nonumber  \;
\\ && + i\int\limits_{W^3}  {*\hat{E}_a \wedge \hat{E}{}^a \over 4\hat{E}}
\left( {\cal D}{\cal D}- \bar{R} \right) \delta^8 (Z-\hat{Z}) + c.c. + \;  \qquad \nonumber \;
\\&& +\int\limits_{W^3}  {1 \over 4\hat{E}} \hat{E}{}^b \wedge \hat{E}{}^c \wedge \hat{E}{}^d \epsilon_{abcd} G{}^{a}\; \delta^8 (Z-\hat{Z})\; . \qquad
\end{eqnarray}

Notice that, as a consequence of (\ref{DG=DR}), the supermembrane current superfields obey
\begin{eqnarray}
\bar{{\cal D}}{}^{\dot{\alpha}} J_{\alpha\dot{\alpha}} = i {{\cal D}}_{{\alpha}} {\cal X}
\; , \qquad
{{\cal D}}{}^{{\alpha}} J_{\alpha\dot{\alpha}} = -i \bar{{\cal D}}_{\dot{\alpha}}{\cal X}\; .
\end{eqnarray}

In the WZ$_{\hat{\theta}=0}$ gauge (\ref{WZgauge=f})--(\ref{thGauge}),
\begin{eqnarray} 
&&i_{\underline{\theta}}  E^{{\alpha}}:= \theta^{\breve{\underline{\alpha}}} E_{\breve{\underline{\alpha}}}{}^{\alpha} = \theta^{ {\alpha}} \; , 
\qquad i_{\underline{\theta}}  E^{\dot{\alpha}}:=\theta^{\breve{\underline{\alpha}}} E_{\breve{\underline{\alpha}}}{}^{\dot\alpha} =  \bar{\theta}{}^{ \dot{\alpha}} \; , \qquad \\  
&&\theta^{ {\alpha}}:= \theta^{\breve{\underline{\beta}}} \delta_{\breve{\underline{\beta}}}^{\, {\alpha}}\; , \qquad  \bar{\theta}{}^{ \dot{\alpha}}:= \theta^{\breve{\underline{\beta}}} \delta_{\breve{\underline{\beta}}}^{\,\dot{\alpha}}\; , \qquad \\ 
&&i_{\underline{\theta}} E^{\underline{a}} :=\theta^{\breve{\underline{\alpha}}} E_{\breve{\underline{\alpha}}}{}^{\underline{a}}=0\; , \quad \\
&&i_\theta w^{\underline{a}\underline{b}}:=  \theta^{\breve{\underline{\beta}}} w_{\breve{\underline{\beta}}}^{\underline{a}\underline{b}}= 0\; 
\hat{\theta}{}^{\underline{\alpha}}(\xi) =0\qquad \Leftrightarrow \qquad \hat{\theta}{}^{{\alpha}}(\xi) =0\; , \qquad \hat{\bar{\theta}}{}^{\dot{\alpha}}(\xi) =0
\; , 
\end{eqnarray}
these current superfields simply drastically,
\begin{eqnarray}
&&J_{\alpha \dot{\alpha}} \vert_{\hat{\theta}=0} =  {\theta_\beta\, \bar{\theta}_{\dot{\beta}} \over   8}
(\;  3 {\cal P}_a{}^b(x) {\sigma}^a_{\alpha \dot{\alpha}}\tilde{\sigma}{}_b^{\beta\dot{\beta}}- 2\delta_{\alpha}{}^{\beta} \delta_{\dot{\alpha}}{}^{\dot{\beta}} {\cal P}_b{}^b(x)) - \nonumber\\
&&-i {({\theta}{\theta}  - \bar{\theta}\bar{\theta})\over 32} {\sigma}^a_{\alpha \dot{\alpha}} {\cal P}_a(x)+\propto \underline{\theta}^{\wedge 3}
\end{eqnarray}
and 
\begin{eqnarray}
{\cal X} \vert_{\hat{\theta}=0} &=& - {\theta{\sigma}^a\bar{\theta} \over 16} {\cal P}_a+ i {({\theta}{\theta}  - \bar{\theta}\bar{\theta})\over 16}   {\cal P}_a{}^a(x)
+  \propto \underline{\theta}^{\wedge 3}
\; , \quad
\end{eqnarray}
where we use the current pre--potential fields defined in (\ref{Kab(x)=})
\begin{eqnarray}
{\cal P}_a{}^b(x)&:=& \int\limits_{W^3}  {1\over \hat{e}} *\hat{e}_a \wedge \hat{e}{}^b  \, \delta^4 (x-\hat{x})  \; ,  \qquad
 \\ 
{\cal P}_a(x) &:=& \int\limits_{W^3}   {1\over \hat{e}} \epsilon_{abcd} \hat{e}{}^b \wedge \hat{e}{}^c \wedge \hat{e}{}^d \,
 \, \delta^4 (x-\hat{x}) = \qquad \nonumber \\ &=& e_a^\mu (x) \int\limits_{W^3}    \epsilon_{\mu\nu\rho\sigma} d\hat{x}{}^\nu \wedge  d\hat{x}{}^\rho \wedge  d\hat{x}{}^\sigma  \,
 \, \delta^4 (x-\hat{x}) \;   \qquad
\end{eqnarray}
and $\underline{{\theta}}^{\wedge 3}$ denotes terms proportional to either 
${\theta}{\theta} \, \bar{\theta}$ or ${\theta} \, \bar{\theta} \bar{\theta} $.

 \chapter{Useful formulae involving 11-dimensional and 9--dimensional Gamma Matrices}
\label{appendiceE}
\thispagestyle{chapter}
\renewcommand{\theequation}{E.\arabic{equation}}
\setcounter{equation}{0}

\initial{W}e use the mostly minus metric convention so that flat spacetime metric reads $\eta_{ab}=diag(1,-1,...,-1)$. 
We choose the following $SO(1,1)\otimes SO(9)$ invariant representation for the 11--dimensional $32\times32$ gamma matrices and charge conjugation 
matrix,
\begin{eqnarray}\label{11DG=1+9+1}
& (\Gamma^{{a}})_{\underline\alpha}{}^{\underline{\beta}}
 \equiv \left({1\over 2}(\Gamma^{\#} + \Gamma^{=}), \Gamma ^{i}, {1\over 2}(\Gamma^{\#} - \Gamma^{=}) \right)\; , \qquad a=0,1,\ldots,9, 10 \; , \qquad i=1,\ldots, 9 \; , \qquad
\nonumber
\\
 \label{11DG=} & (\Gamma ^{\#})_{\alpha}{}^{\beta}
 = \left( \begin{matrix} 0 & 2i\delta_{pq}
 \cr 0 & 0 \end{matrix}  \right)\;  , \qquad  (\Gamma ^{=})_{\alpha}{}^{\beta}
 = \left( \begin{matrix} 0 & 0
 \cr -2i\delta_{pq} & 0 \end{matrix}  \right)\;  , \qquad  (\Gamma ^{i})_{\alpha}{}^{\beta}
 = \left( \begin{matrix} -i \gamma^{i}_{pq} & 0
 \cr 0 & i\gamma^{i}_{pq} \end{matrix}  \right)\;  , \qquad
 \\
\label{11DC=}
&
C_{{\alpha}{\beta}}  =- C_{{\beta}{\alpha}}=
 \left( \begin{matrix} 0 & i\delta_{pq}
 \cr -i\delta_{pq} & 0
\cr\end{matrix} \right)=  (C^{-1}){}^{{\alpha}{\beta}}=:  C^{{\alpha}{\beta}}\; , \qquad
\end{eqnarray}
which are imaginary due to our mostly minus metric convention.
In these representation appear the $16\times16$ 9--dimensional Dirac matrices $\gamma^{i}_{pq}$ which possesses the following 
properties
\begin{eqnarray}
&&\gamma^{(i}\gamma^{j)}=\delta^{ij} {\mathbb I}_{16\times16}, \qquad \gamma^{i}_{pq}= \gamma^{i}_{qp}:=\gamma^{i}_{(pq)}, \nonumber\\
&&\gamma^{i}_{(pq}\gamma^{i}_{r)s}=\delta_{(pq}\delta_{r)s}.
\end{eqnarray}
We do not distinguish upper and lower $SO(9)$ spinor indices because the 9--dimensional charge conjugation matrix is symmetric 
allowing us to chose its representation by Kronecher delta symbol $\delta_{pq}$.
The matrices $\delta_{pq}$,$\gamma^{i}_{pq}$ and $\gamma^{ijk}_{pq}$ provide a complete basis for the set of $16\times16$ symmetric matrices, 
\begin{eqnarray}
 \label{9d=symg}
 \delta_{r(q} \delta_{p)s}={1\over 16} \delta_{pq} \delta_{rs} + {1\over 16} \gamma^i_{pq} \gamma^i_{rs} + {1\over 16\cdot 4!} \gamma^{ijkl}_{pq} \gamma^{ijkl}_{rs}
\; . \qquad
\end{eqnarray}
In our conventions $\gamma^{123456789}_{qp}= \delta_{qp}\,$ and, consequently,
\begin{eqnarray} \label{9d:g5=g4}
\gamma^{i_1\ldots i_7}_{qp}= -{1\over 2} \epsilon^{i_1\ldots i_7jk} \gamma^{jk}_{qp}\; , \qquad \\
 \label{9d:g5=g4}
\gamma^{i_1\ldots i_5}_{qp}= {1\over 4!} \epsilon^{i_1\ldots i_5j_1\ldots j_4} \gamma^{j_1\ldots j_4}_{qp}\; . \qquad
\end{eqnarray}
This, together with (\ref{11DG=1+9+1})  implies that our 11D dirac matrices obey
\begin{eqnarray}
 \label{11Dgammas}
\Gamma^0\Gamma^1\ldots \Gamma^9\Gamma^{(10)}= {1\over 2} \Gamma^{\#} \Gamma^{=}\Gamma^1\ldots \Gamma^9 = - i{\mathbb I}_{32\times 32}\; . \qquad
\end{eqnarray}

 \chapter{Moving frame and spinor moving frame variables}
\label{apendiceF}
\thispagestyle{chapter}
\renewcommand\theequation{F.\arabic{equation}}
\setcounter{equation}{0}

\initial{M}oving frame and spinor moving frame variables are defined as blocks of, respectively, $SO(1,10)$ and $Spin(1,10)$ valued matrices,
\begin{eqnarray}\label{Uin-A}
& U_b^{(a)}= \left({u_b^{=}+ u_b^{\#}\over 2}, u_b^{i}, { u_b^{\#}-u_b^{=}\over 2}
\right)\; \in \; SO(1,10)\;  \quad
\end{eqnarray}
($i=1,...,9$) and    \begin{eqnarray}\label{harmVin-A} V_{(\beta)}^{\;\;\; \alpha}=
\left(\begin{matrix}  v^{+\alpha}_q
 \cr  v^{-\alpha}_q \end{matrix} \right) \in Spin(1,10)\;
 \; . \qquad
\end{eqnarray}
We also use
\begin{eqnarray}\label{Vharm=M0-A}
 V^{( {\beta})}_{ {\alpha}}= \left(
v_{ {\alpha}q}{}^+\, ,v_{ {\alpha}q}{}^- \right)\; \in \; Spin(1,10) \; ,  \qquad
\end{eqnarray}
with \begin{eqnarray}
\label{V-1=CV-A}  v_{\alpha}{}^{-}_q =  i C_{\alpha\beta}v_{q}^{- \beta }\, ,
\qquad v_{\alpha}{}^{+}_q = - i C_{\alpha\beta}v_{q}^{+ \beta }\,
 \end{eqnarray}
 obeying
 \begin{eqnarray}\label{Vharm=M0--A}
 V_{( {\beta})}{}^{ {\gamma}}
V_{ {\gamma}}^{ ({\alpha})}=\delta_{( {\beta})}{}^{ ({\alpha})}=\left(\begin{matrix}
\delta_{qp} & 0           \cr
          0 & \delta_{qp} \end{matrix}\right) \qquad  \Leftrightarrow  \quad \begin{cases} v_{q}^{- {\alpha}}v_{ {\alpha}p}{}^+=\delta_{qp}= v_{q}^{+
{\alpha}}v_{ {\alpha}p}{}^-\, , \cr  v_{q}^{- {\alpha}}v_{ {\alpha}p}{}^-= 0\; =
v_{q}^{+ {\alpha}}v_{ {\alpha}p}{}^+\, . \end{cases}\;
\end{eqnarray}

The algebraic properties of moving frame and spinor moving frame variables are summarized as
\begin{eqnarray}\label{u--u--=0-A}
u_{ {a}}^{=} u^{ {a}\; =}=0\; , \quad    u_{ {a}}^{=} u^{ {a}\,i}=0\; , \qquad u_{
{a}}^{\; = } u^{ {a} \#}= 2\; , \qquad
 \\  \label{u++u++=0-A} u_{ {a}}^{\# } u^{ {a} \#
}=0 \; , \qquad
 u_{{a}}^{\;\#} u^{ {a} i}=0\; , \qquad  \\  \label{uiuj=-A} u_{ {a}}^{ i}
 u^{{a}j}=-\delta^{ij}.  \qquad
\\
\label{M0:v+v+=u++-A}
 v_{q}^- {\Gamma}_{ {a}} v_{p}^- = \; u_{ {a}}^{=} \delta_{qp}\; , \qquad v_{q}^+ {\Gamma}_{ {a}} v_{p}^+ = \; u_{ {a}}^{\# } \delta_{qp}\; , \qquad \nonumber \\
 v_{q}^- {\Gamma}_{ {a}} v_{p}^+ = - u_{ {a}}^{i} \gamma^i_{qp}\; , \qquad
\\ \label{M0:u++G=v+v+-A}
  2 v_{q}^{- {\alpha}}v_{q}^{-}{}^{ {\beta}}= \tilde{\Gamma}^{ {a} {\alpha} {\beta}} u_{
 {a}}^{=}\; , \quad 2 v_{q}^{+ {\alpha}}v_{q}^{+}{}^{ {\beta}}= \tilde{\Gamma}^{ {a} {\alpha} {\beta}} u_{
 {a}}^{\# }\; , \qquad \nonumber \\
 2 v_{q}^{-( {\alpha}}v_{q}^{+}{}^{ {\beta})} =-  \tilde{\Gamma}^{ {a} {\alpha} {\beta}}
 u_{ {a}}^{i}\; . \qquad
\end{eqnarray}

In (\ref{M0:v+v+=u++-A}) and (\ref{M0:u++G=v+v+-A}) we have used  real symmetric $16\times 16$
9d Dirac matrices  $\gamma^i_{qp}=\gamma^i_{pq}$ which obey
Clifford algebra \begin{eqnarray}\label{gigj+=-A} \gamma^i\gamma^j + \gamma^j \gamma^i=
2\delta^{ij} I_{16\times 16}\; , \qquad
\end{eqnarray}
and
\begin{eqnarray}\label{gi=id1-A}
&& \gamma^{i}_{q(p_1}\gamma^{i}_{p_2p_3) }= \delta_{q(p_1}\delta_{p_2p_3) }\; , \qquad
\\ \label{gi=id2-A} && \gamma^{ij}_{q(q^\prime }\gamma^{i}_{p^\prime)p }+
\gamma^{ij}_{p(q^\prime }\gamma^{i}_{p^\prime)q } = \gamma^{j}_{q^\prime
p^\prime}\delta_{qp}-\delta_{q^\prime p^\prime}\gamma^{j}_{qp} \; . \qquad
\end{eqnarray}

Derivatives of the moving frame and spinor moving frame variables are expressed in terms of
covariant ${SO(1,10)\over SO(1,1)\times SO(9)}$ Cartan forms
\begin{eqnarray}
\label{Om++i=-A} \Omega^{=i}= u^{=a}du_a^{i}\; , \qquad \Omega^{\# i}=  u^{\#
a}du_a^{i}\; , \qquad
  \end{eqnarray}
and induced  $SO(1,1)\times SO(9)$ connection
\begin{eqnarray}
\label{Om0:=-A} \Omega^{(0)}= {1\over 4} u^{=a}du_a^{\#}\; , \qquad \\ \label{Omij:=-A}
\Omega^{ij}=  u^{ia}du_a^{j}\; . \qquad
  \end{eqnarray}
It is convenient to use these latter to define covariant derivative. Then
\begin{eqnarray}\label{M0:Du--=Om-A}
Du_{ {b}}{}^{=} &:= & du_{ {b}}{}^{=} +2 \Omega^{(0)} u_{ {b}}{}^{=}= u_{ {b}}{}^i
\Omega^{= i}\; , \qquad \\ \label{M0:Du++=Om-A} Du_{ {b}}{}^{\#}&:=& du_{ {b}}{}^{\#} -2
\Omega^{(0)} u_{ {b}}{}^{\#}=  u_{ {b}}{}^i \Omega^{\# i}\; , \qquad \\
\label{M0:Dui=Om-A}  Du_{ {b}}{}^i &:=& du_{ {b}}{}^{i} - \Omega^{ij} u_{ {b}}{}^{j} =
{1\over 2} \, u_{ {b}}{}^{\# } \Omega^{=i}+ {1\over 2} \, u_{ {b}}{}^{=} \Omega^{\#
i}\; . \qquad \nonumber \\ {}
\end{eqnarray}
\begin{eqnarray}
\label{Dv-q-A}  Dv_q^{-\alpha}&:=& dv_q^{-\alpha} +  \Omega^{(0)} v_q^{-\alpha} - {1\over
4}\Omega^{ij} \gamma^{ij}_{qp} v_p^{-\alpha} = \nonumber \\ &=& - {1\over 2}
\Omega^{=i} v_p^{+\alpha} \gamma_{pq}^{i}\; , \qquad \\ \label{Dv+q-A}  Dv_q^{+\alpha}
&:=& dv_q^{+\alpha} -  \Omega^{(0)} v_q^{+\alpha} - {1\over 4}\Omega^{ij}
\gamma^{ij}_{qp} v_p^{+\alpha} = \nonumber \\ &=& - {1\over 2} \Omega^{\# i}
v_p^{-\alpha} \gamma_{pq}^{i}\; . \qquad
\end{eqnarray}

The Cartan forms obey
\begin{eqnarray}\label{M0:DOm--=-A} && D\Omega^{= i}=  0\; , \qquad D\Omega^{\# i}  = 0\;  ,  \qquad
\\ \label{M0:Gauss-A}
 && F^{(0)}:= d\Omega^{(0)} =    {1\over 4 } \Omega^{=\, i} \wedge
 \Omega^{\# \, i}\; , \qquad  \\
\label{M0:Ricci-A} && {G}^{ij}:= d\Omega^{ij}+ \Omega^{ik} \wedge \Omega^{kj} = - \Omega^{=\,[i} \wedge \Omega^{\# \, j]}\; .   \qquad
\end{eqnarray}
Notice that, {\it e.g.}
 \begin{eqnarray}
\label{DDu++=} && DDu_{ {a}}^{\# } = \; \; 2 F^{(0)}u_{ {a}}^{\#
}\; , \qquad
 DDu_{ {a}}{}^{i} = u_{ {a}}^{j} {G}^{ji}  \; . \qquad
\end{eqnarray}

The essential variations of moving frame and spinor moving frame variables can be written as
\begin{eqnarray}\label{vu--=iOm-A}
\delta u_{ {b}}{}^{=} = u_{ {b}}{}^i i_\delta\Omega^{= i}\; , \qquad \label{vu++=iOm-A}
\delta u_{ {b}}{}^{\#}=  u_{ {b}}{}^i i_\delta\Omega^{\# i}\; , \qquad \\
\label{vui=iOm-A}  \delta u_{ {b}}{}^i  = {1\over 2} \, u_{ {b}}{}^{\# }  i_\delta
\Omega^{=i}+ {1\over 2} \, u_{ {b}}{}^{=}  i_\delta\Omega^{\# i}\; . \qquad
\\
\label{vv-q-A}  \delta v_q^{-\alpha}= - {1\over 2} i_\delta \Omega^{=i} v_p^{+\alpha}
\gamma_{pq}^{i}\; , \qquad \\ \label{vv+q-A}  \delta v_q^{+\alpha} = - {1\over 2}
i_\delta\Omega^{\# i} v_p^{-\alpha} \gamma_{pq}^{i}\; , \qquad
\end{eqnarray}
where  $i_\delta
\Omega^{=i}$ and $i_\delta \Omega^{\# i}$ are independent variations.

The essential variations of the Cartan forms read
\begin{eqnarray}
\label{vOm++=-A} \delta \Omega^{\# i}&=& D i_\delta \Omega^{\# i}\; , \qquad  \delta \Omega^{=i}= D i_\delta \Omega^{=i}\; , \qquad
 \\
\label{vOmij=-A} \delta \Omega^{ij}\; &=& -\Omega^{=[i} i_\delta \Omega^{\# j]} -  \Omega^{\# [i}i_\delta \Omega^{=j]}\; , \qquad
\\
\label{vOm0=-A}
\delta \Omega^{(0)}&=& \frac {1}{4} \Omega^{=i}i_\delta \Omega^{\# i} -
 \frac {1}{4}  \Omega^{\# i}i_\delta \Omega^{=i}  \; . \qquad
\end{eqnarray}

 \chapter{Equations of motion for a single M$0$--brane}
\thispagestyle{chapter}

\renewcommand\theequation{G.\arabic{equation}}
\setcounter{equation}{0}

\initial{I}n this appendix we collect the equations of motion for the single M0--brane obtained  in chapter \ref{chapter11D} from 
the spinor moving frame action (\ref{SM0=}), (\ref{SM0==}).\\ 
They read
\begin{eqnarray}
\hat{E}^{=}&:=& \hat{E}^{a}u_a^{=} = 0,\qquad \\
\hat{E}^{i}&:=& \hat{E}^{a}u_a^{i} =0, \qquad \\
 D\rho^{\#}&=&0 \quad \Leftrightarrow \quad \Omega^{(0)}={d\rho^{\#}\over 2\rho^{\#}}\; ,  \qquad \\
\Omega^{= i}&=&0 \quad \Leftrightarrow \quad Du_a^{=}=0 \quad \Leftrightarrow  \quad
Dv_q^{-\alpha}=0\, , \quad \\ \hat{E}^{-q}&:=& \hat{E}^{\alpha}v_\alpha^{-q}=0\, .
\end{eqnarray}
These equations are formulated in terms of pull--backs of bosonic and fermionic supervielbein forms of flat 11D superspace to the mM$0$ worldline $W^1$
\begin{eqnarray}\label{Ea=Pi-A}
 \hat{E}^a  = d\hat{x}^a - i d\hat{\theta} \Gamma^a\hat{\theta}\; , \qquad a=0,1,...,10\; , \\
E^\alpha=d\hat{\theta}^\alpha
 \;   \qquad \alpha= 1,...,32\; ,
  \end{eqnarray}
which are constructed from the coordinate functions $\hat{x}^a(\tau)$, $\hat{\theta}^\alpha (\tau)$ of the proper time $\tau$, and of the moving frame and spinor moving frame variables $u_b^{=}$, $u_b^{i}$, $v_\alpha^{-q}$. 
The properties of these latter as well as of the Cartan forms $\Omega^{=i}$, $\Omega^{(0)}$ and covariant 
derivatives $D$ are collected in Appendix \ref{apendiceF}.

In (\ref{Ea=Pi-A}) and in the main text we have used the real symmetric $32\times 32$ 11D $\Gamma$--matrices 
$\Gamma^a_{\alpha\beta}= (\gamma^a C)_{\alpha\beta}$ which, together with  
$\tilde{\Gamma}_a^{\alpha\beta}= (C\gamma_a)^{\alpha\beta}$, obey  
$\Gamma^{(a}\tilde{\Gamma}^{b)}=\eta^{ab}{\mathbb{I}}_{32\times 32}$.

 \chapter{ Multiple M0 equations of motion}
\thispagestyle{chapter}
\renewcommand\theequation{H.\arabic{equation}}
\setcounter{equation}{0}

\initial{T}he  mM$0$ system, which is to say an interacting system of N nearly coincident M$0$-branes, is described in terms of 
center of energy variables and the 
traceless $N\times N$ matrices ${\mathbb{X}}^i$ ($i=1,...,9$), $\Psi_q$ ($q=1,...,16$). Our action includes also the 
auxiliary  $N\times N$ matrix fields: momentum ${\mathbb{P}}^i$ and the 1d SU(N) gauge field ${\mathbb{A}}_\tau$
(${\mathbb{A}}=d\tau {\mathbb{A}}_\tau$).

The complete list of equations of motion for the mM0 system splits naturally on the equations for the relative motion variables,
\begin{eqnarray}\label{relEqm}
&& D\mathbb{X}^i =\hat{E}^{\#}\mathbb{P}^i+4i\hat{E}^{+q}(\gamma^i\Psi)_q,\nonumber\\
&&[\mathbb{P}^i,\mathbb{X}^i]= 4i\{ \Psi_q \, , \, \Psi_q \}, \nonumber\\
&&D\mathbb{P}^i= - \frac{1}{16}\hat{E}^{\#}[[\mathbb{X}^i,\mathbb{X}^j]\mathbb{X}^j]
 +2\hat{E}^{\#}\, \Psi\gamma^{i}\Psi
+\hat{E}^{+q}\gamma^{ij}_{qp}[\Psi_p, \mathbb{X}^{j}], \nonumber\\ && D\Psi =
\frac{i}{4} \hat{E}^{\#}[\mathbb{X}^i,(\gamma^{i}\Psi)] +
\frac{1}{2} \hat{E}^{+}\gamma^{i}\, \mathbb{P}^{i}-\frac{i}{16} \hat{E}^+\gamma^{ij} \,
[\mathbb{X}^i,\mathbb{X}^j] \, , \qquad
\end{eqnarray}
and the center of energy equations which can be considered as a deformation of the system of
equations for single M$0$ brane. After fixing the gauge under a reminiscent of the $K_9$ symmetry,
these equation read
\begin{eqnarray}\label{E==mM0-A}
 \hat{E}^{=}&:=& \hat{E}^au_a^{=}=
  3 (\rho^\#)^2 tr\left( \frac{1}{2}\mathbb{P}^iD\mathbb{X}^i +
\frac{1}{64} \hat{E}^{\#}[\mathbb{X}^i,\mathbb{X}^j]^2 -  \frac{1}{4}(E^+\gamma^{ij}\Psi)[\mathbb{X}^i,\mathbb{X}^j] \right)\; ,
\qquad \end{eqnarray}
 \begin{eqnarray}
\label{Ei=0mM0-A}
\hat{E}^{i}&:=& \hat{E}^au_a^{i}= 0 \; ,   \\ \hat{E}^{-q}&:=& \hat{E}^{\alpha}v_\alpha^{-q}=0\, ,\\
\left. \begin{matrix} \Omega^{= i}=0 \cr
  \Omega^{\# i}=0
\end{matrix} \right\}
&& \Leftrightarrow \quad \left\{  \begin{matrix} Du_a^{=}=0,   \quad Du_a^{\#}=0, \cr   \quad Du_a^{i}=0\, , \cr
Dv_q^{-\alpha}=0\, , \quad D v_q^{+\alpha}=0\, . \end{matrix}\right.   \qquad  \\
 D\rho^{\#}&=&0 \quad \Leftrightarrow \quad \Omega^{(0)}={d\rho^{\#}\over 2\rho^{\#}}\; ,  \qquad
\end{eqnarray}

As a consequence of the above equation the effective mass $M$ of the mM$0$ center of energy motion, \begin{eqnarray}\label{M2=-A}
M^2= 4 (\rho^{\#})^4 {\cal H}\; ,  \qquad
\end{eqnarray}
 is a constant
\begin{eqnarray}\label{dM2=0-A}
dM^2=0\; .  \qquad
\end{eqnarray}
Eq. (\ref{M2=-A}) expresses $M^2$ in terms of Lagrange multiplier  $\rho^{\#}$ and the relative motion Hamiltonian (\ref{HSYM=1}) \begin{eqnarray}
\label{HSYM=1-A}
{\cal H}=   {1\over 2} tr\left( {\mathbb{P}}^i {\mathbb{P}}^i \right) - {1\over 64}
tr\left[ {\mathbb{X}}^i ,{\mathbb{X}}^j \right]^2 - 2\,  tr\left({\mathbb{X}}^i\, \Psi\gamma^i {\Psi}\right) .  \quad
  \end{eqnarray}

If we fix the gauge where the composed SO(9) connection and also the SU(N) gauge field
vanish, \begin{eqnarray}
&&\Omega^{ij}=d\tau \Omega_\tau^{ij}=0\; , \qquad {\mathbb A}=d\tau {\mathbb A}_\tau =0,
 \end{eqnarray}
the equations of relative motion and Eq. (\ref{E==mM0-A}) simplify to
\begin{eqnarray}
 &&\partial_\tau
\tilde{\Psi} = \frac{i}{4} \, e\,[\tilde{\mathbb{X}}{}^i,(\gamma^{i}\tilde{\Psi})] +
\frac{1}{2\sqrt{\rho^{\#}}} \hat{E}_\tau ^{+}\gamma^{i}\, \tilde{\mathbb{P}}{}^{i} -
\frac{i}{16\sqrt{\rho^{\#}}} \hat{E}_\tau^+\gamma^{ij} \,
[\tilde{\mathbb{X}}{}^i,\tilde{\mathbb{X}}{}^j]\;  ,\nonumber\\ &&
\partial_\tau
\left(\frac{1}{e}\partial_\tau \tilde{\mathbb{X}}{}^i\right) = - \frac{e}{16}\,
[[\tilde{\mathbb{X}}{}^i,\tilde{\mathbb{X}}{}^j]\tilde{\mathbb{X}}{}^j] +2\, e \,
\tilde{\Psi}\gamma^{i}\tilde{\Psi}  +  4i \partial_\tau \left(
{ \hat{E}_\tau^{+}\gamma^{i}\tilde{\Psi}\over e\sqrt{\rho^{\#}}} \right) + {1\over
\sqrt{\rho^{\#}}} \hat{E}_\tau^{+}\gamma^{ij}[\tilde{\Psi}, \tilde{\mathbb{X}}{}^{j}]\,
, \nonumber\\ &&\partial_{\tau} \tilde{\mathbb{X}}{}^i= e\tilde{\mathbb{P}}{}^i +
{4i\over \sqrt{\rho^{\#}}}\left( \hat{E}_\tau^{+}\gamma^{i}\tilde{\Psi}\right)\, ,
\qquad [\tilde{\mathbb{P}}{}^i, \tilde{\mathbb{X}}{}^i]=4i\{\tilde{\Psi}_q,
\tilde{\Psi}_q\}\, , \nonumber\\
&&\rho^{\#} \hat{E}_\tau^{=}=3 tr \left({1\over 2}\tilde{\mathbb{P}}{}^i\partial_\tau
\tilde{\mathbb{X}}{}^i+ {1\over 64}e
[\tilde{\mathbb{X}}{}^i,\tilde{\mathbb{X}}{}^j]^2-{1\over
4\sqrt{\rho^{\#}}}\left(\hat{E}_\tau^{+}\gamma^{ij}\tilde{\Psi}\right)[\tilde{\mathbb{X}}{}^i,\tilde{\mathbb{X}}{}^j]\right)\,
 .
\end{eqnarray}
These equations are written in terms of  redefined fields,
\begin{eqnarray}\label{tX=rX}
&&\tilde{\mathbb{X}}{}^i=  \rho^{\#} {\mathbb{X}}{}^i\; , \quad
\tilde{\Psi}_q=(\rho^{\#})^{3/2} {\Psi}_q\; , \nonumber\\
&&\tilde{\mathbb{P}}{}^{i}=(\rho^{\#})^2{\mathbb{P}}{}^{i}= {1\over e}
\left(\partial_\tau \tilde{\mathbb{X}}{}^i - {4i\over \sqrt{\rho^{\#}}}
\hat{E}_\tau^{+}\gamma^i\tilde{\Psi}\right) \, ,
\end{eqnarray}
  and
\begin{eqnarray}
e(\tau)= \hat{E}^\#_\tau /\rho^{\#} \;  .
\end{eqnarray}
\addcontentsline{toc}{chapter}{Bibliography}
\thispagestyle{chapter}
\bibliographystyle{hunsrt}
\bibliography{Chapters/carlos}
 

\end{document}